\documentclass[11pt,a4paper]{article}

\usepackage{bm}
\usepackage{amsmath, latexsym}
\usepackage{amssymb}
\usepackage{amsfonts}

\usepackage{graphicx}
\usepackage{caption}
\usepackage{subcaption}

\usepackage[onehalfspacing]{setspace}
\usepackage{natbib}
\usepackage{geometry}                		
\usepackage[bottom]{footmisc}
\usepackage{fancyhdr}
\usepackage{parskip}
\usepackage{graphicx}	
\usepackage{enumitem}
\usepackage{afterpage}
\usepackage[tableposition=t]{caption}

\usepackage{multirow}
\usepackage{float}
\usepackage{setspace}
\usepackage{natbib}
\usepackage{url}
\usepackage{eurosym}
\usepackage{tablefootnote}
\usepackage{longtable}
\AtBeginEnvironment{longtable}{\singlespacing}
\usepackage{array}
\newcolumntype{H}{>{\setbox0=\hbox\bgroup}c<{\egroup}@{}}

\usepackage[T1]{fontenc}
\usepackage{lmodern}
\usepackage{csquotes}

\usepackage{pgfplots} 
\usetikzlibrary{patterns}
\usepackage{lscape}
\usepackage[justification=centering]{caption}
\usepackage{adjustbox}
\usepackage{booktabs}
\usepackage{dcolumn}
\newcolumntype{.}{D{.}{.}{-1}}
{\def\sym#1{\ifmmode^{#1}\else\(^{#1}\)\fi}

\usepackage{multirow}
\usepackage{tikz}
\usetikzlibrary{trees}
\usepackage{hyperref}

\date{}

\begin{document}


\begin{titlepage}
\title{Macroeconomic Effects of Active Labour Market Policies:\\ A Novel Instrumental Variables Approach\thanks{Ulrike Unterhofer: University of Basel  (ulrike.unterhofer@unibas.ch),  Conny Wunsch: University of Basel, CEPR, CESifo, IZA (conny.wunsch@unibas.ch). This research resulted from the project \textit{Betriebliche und makroökonomische Wirkungen der aktiven Arbeitsmarktpolitik in Deutschland}, based on data provided by the Institute for Employment Research: IEB V14.04.00-190927, Nürnberg 2019. We thank David Card, Wolfgang Dauth, Anders Forslund, Peter Kugler, Thomas Kruppe, Rafael Lalive, Enrico Moretti, Felix Rochlitz, Kurt Schmidheiny, Rüdiger Wapler, Katja Wolf and Véra Zabrodina for helpful discussions and comments. All errors are our own.}}
\author{Ulrike Unterhofer and Conny Wunsch}
\maketitle
\date{\vspace{-2ex}}
\begin{abstract}
\noindent 
This study evaluates the macroeconomic effects of active labour market policies (ALMP) in Germany over the period 2005 to 2018.  We propose a novel identification strategy to overcome the simultaneity of ALMP and labour market outcomes at the regional level. It exploits the imperfect overlap of local labour markets and local employment agencies that decide on the local implementation of policies. Specifically, 
we instrument for the use of ALMP in a local labour market with the mix of ALMP implemented outside this market but in local employment agencies that partially overlap with this market. We find no effects of short-term activation measures and further vocational training on aggregate labour market outcomes. In contrast, wage subsidies substantially increase the share of workers in unsubsidised employment while lowering long-term unemployment and welfare dependency. Our results suggest that negative externalities of ALMP partially offset the effects for program participants and that some segments of the labour market benefit more than others.
\vspace{1cm}\\
\noindent\textbf{Keywords:} active labour market policies; macroeconomic evaluation; instrumental variables\\
\vspace{0in}\\
\noindent\textbf{JEL Codes:}  C26, H43, J64, J68\\
\vspace{0.5in}\\
\begin{center}
\end{center}
\bigskip
\end{abstract}
\setcounter{page}{0}
\thispagestyle{empty}
\end{titlepage}
\pagebreak \newpage

\setstretch{1.25}

\newpage
\pagenumbering{arabic} 

\section{Introduction}\label{intro}

Active labour market policies (ALMP) aim at improving employment prospects of jobseekers, at integrating them into the labour market and thereby reducing unemployment. They include a variety of policy instruments such as counselling and employment services, labour market training, and subsidised employment. Many developed countries invested heavily into such programs over the last decades and to this day, public spending on ALMP remains high at 0.2 to 2\% of GDP in EU countries \citep{EC2022}. A large literature analyses individual-level effects of these policies, i.e. whether participation in a program increases the employment probability of participants.  This literature finds small or negative effects in the short run but positive effects in the long run, with larger gains for programs that are human capital intensive (see e.g. \citealp{Card2018}).

Comparing individual gains with the associated investments is not sufficient to judge the overall cost-effectiveness of ALMP. These policies may have the intended effects on program participants but can be accompanied by negative spillover effects on non-participants \citep{Abbring2007}. 
Such spillovers may work through wage-setting and labour-demand incentives. For example, if programs change relative wage costs, firms might create jobs for program participants and replace jobs for other workers, possibly resulting in a net reduction in labour demand (see e.g. \citealp{Calmfors1994}). Recently, a number of contributions provide direct empirical evidence for adverse effects of ALMP on non-participants \citep{Crepon2013,Ferracci2014,Gautier2018,Caria2022}.\footnote{For an earlier contribution, see \citealp{Blundell2004}. Spillover effects have also been documented for passive labour market policies such as unemployment insurance \citep{Lalive2015,Huber2021}.} Furthermore, deadweight losses may occur if program participants would have been hired also in absence of the program. Overall, the existing evidence on the aggregate effects of ALMP remains mixed. An important reason are  methodological challenges inherent to macroeconomic evaluations, in particular the simultaneous determination of policies and labour market outcomes at the aggregate level.

In this paper, we study macroeconomic effects of different types of ALMP in Germany and propose a novel identification strategy to overcome these challenges. It is based on the idea that local labour markets are not perfectly coinciding with administrative areas where policy decisions are being made. In Germany, 156 local employment agencies decide on the regional implementation of these policies with large discretion. At the same time, the country is characterised by a decentralised economic structure with several local labour markets. Together, these features provide an ideal laboratory for our macroeconomic evaluation study. Specifically, we analyse how regional labour market outcomes such as the unemployment and employment rates were affected by the intensity of three types of policies: short-term activation measures, further vocational training and wage subsidies. 

Our analysis proceeds in two steps. First, we construct precisely-delineated local labour markets based on commuting flows between municipalities. Second, we estimate the effect of the policy use on aggregate labour market outcomes in these markets using a dynamic panel data model. To address simultaneity, we instrument the policy use in a labour market with the policy use in municipalities of local employment agencies that partially overlap with this labour market but lie outside its borders. Our estimation is based on rich administrative data at the individual level, which we aggregate to the labour market level and by quarter over the period 2005 to 2018. 

Our findings suggest that the effectiveness of ALMP in reducing unemployment or increasing employment in the aggregate is limited. We find no significant effects for short measures and vocational training programs. 
 Wage subsidies, on the other hand, substantially increase the aggregate employment rate. A 10 percentage point increase in the share of unemployed that receive wage subsidies increases the unsubsidised employment rate by 3 percentage points in the long run. This increase is explained by a lower fraction of workers on welfare and in precarious work. Wage subsidies are thus effective in preventing long-term unemployment and the dependence on state benefits. We also find suggestive evidence for some labour market segments benefiting more from the policies. For example, workers older than 50 experience an increase in unsubsidised employment and a corresponding decrease in unemployment as response to higher intensity of training. This age group also experiences the largest effects of wage subsidies on the employment rate.

We make two main contributions to the literature evaluating ALMP. As our first contribution, we propose a new approach to address the methodological problems in macroeconomic evaluations of ALMP. Existing studies have mostly relied on the variation in the policy use between administrative areas such as countries or smaller entities within a country to identify the aggregate effects of such policies.  This poses two types of challenges for identification. 
First, geographical areas delineated by administrative borders are not necessarily the economic entities of interest. If there exist strong economic ties across administrative borders, the effects of policies introduced in one of the areas might spill over to the other. Some contributions have addressed this issue by controlling for spatially correlated  spillover effects  (e.g. \citealp{Fertig2006b, Hujer2009, Dauth2016}). Second, if policy decisions are made at the same geographical level at which outcomes are measured, this complicates the separate identification of cause and effect. Previous attempts to mitigate the simultaneity problem use transformations of policy variables such as ratios \citep{Calmfors1995}, lagged policy variables (\citealp{Puhani2003, Fertig2006b, Wapler2018a, Wapler2022}), or lagged measures as instrumental variables (\citealp{Nickell1997, Nickell1998, Nickell1999, Johansson2001, Estevao2003, Dahlberg2005, Bassanini2006, Hujer2009, Dauth2016, Escudero2018}). Contributions based on truly exogenous variation are rare and often focus on specific programs. \cite{Crepon2013} use a randomised experiment for measuring spillover effects of a job search program for young university dropouts, \cite{Gautier2018}  rely on a difference-in-differences approach to estimate the effects of a job search assistance program.\footnote{There exist also contributions relying on structural parametric approaches e.g., \citealp{Scarpetta1996, Elmeskov1998, Kraft1998, Blanchard2000, Baker2005, Fertig2006a, Altavilla2013}.}

In our empirical approach we circumvent the issue of regional spillover effects by estimating effects at the local labour market level. These markets form self-contained regions and are less likely to have pronounced economic interdependencies. Relying on local labour markets further allows us to exploit a novel source of exogenous variation in policy use at the regional level.\footnote{Two other papers exploit that the regional variation in policy take-up is not corresponding with the variation in regional economic conditions but answer different research questions. \cite{Heuermann2015} quantify the effects of ALMP on individual employment outcomes for the long-term unemployed in a regression discontinuity design based on the idea that policies change discontinuously along job-center borders in Germany. \cite{Wunsch2015} assess the effects of ALMP on firm outcomes exploiting the incomplete overlap of firms’ hiring regions and employment agencies.} Our instruments rely on the imperfect overlap between local labour markets and administrative regions in charge of policy decisions and are less likely to be endogenously determined compared to lagged policy variables.

As our second contribution we provide a comprehensive evaluation of ALMP. Thereby, we expand the knowledge on the macroeconomic effects of ALMP more generally.  Existing studies often derive their econometric specification from the standard matching function \citep{Pissarides2000} and focus on a single outcome: the outflows of jobseekers into employment. These papers find little or no significant effects of ALMP (\citealp{Boeri1996, Puhani2003, Fertig2006b, Hujer2009, Dauth2016, Wapler2022}). Other papers that consider the effects of ALMP on the unemployment rate and employment rate also  remain largely inconclusive about the effectiveness of these policies (\citealp{Nickell1999, Blanchard2000, Dahlberg2005, Bassanini2006,  Altavilla2013, Estevao2003}). More recent evidence suggests that the effects of ALMP might depend on the implementation of the policies \citep{Escudero2018} and on regional labour market conditions \citep{Altavilla2013,Wapler2022}. Furthermore, the microeconomic evidence on ALMP suggests that program effects vary significantly with the characteristics of participants (e.g. \citealp{Heckman1997, Bitler2006, Bergemann2008, Behaghel2014}). These heterogeneities might also materialise at the macroeconomic level. \cite{Escudero2018} for example finds ALMP to be more effective for the population of low-skilled workers. Our reduced-form approach in combination with a unique administrative dataset allows us to shed light on new aspects of the effectiveness of ALMP. We use quarterly variation over a long time period of  14 years and consider a variety of outcomes that have not been studied in the literature so far. Furthermore, this gives us the opportunity to explore possible heterogeneity in the program effectiveness for different segments of the labour market. 


The remainder of the paper is structured as follows. Section \ref{institutions} introduces the institutional setting in Germany and the policies of interest. Section \ref{theory} provides a theoretical background on possible effects. Section \ref{emp_strat} describes the empirical strategy and the data. Section \ref{results} presents the main results and Section \ref{robustness} provides a number of robustness checks. Section \ref{conclusion} concludes.

\section{Institutional Setting and Policies}\label{institutions}

We evaluate the macroeconomic effects of ALMP in Germany for the years 2005 to 2018. Prior to this period, German labour market institutions were heavily reformed by the Job-AQTIV legislation and the Hartz reforms.\footnote{See e.g. \cite{Wunsch2005} for details about the institutional changes. 
} Our analysis focuses on policies that are defined in the German Social Code (SGB) III and are directed at jobseekers receiving unemployment insurance (UI) benefits. The programs are implemented by local employment agencies (shortly referred to as employment agencies or agencies in the following) and financed by mandatory contributions of employers and employees to national unemployment insurance.\footnote{Since 2005, the provision of ALMP in Germany is organised in a two layer system. In addition to policies for unemployment insurance recipients, there exist also policies for jobseekers receiving the means-tested unemployment benefits II. The latter are tax-financed and implemented by job-centers. While it is in part similar programs that are directed at these types of jobseekers, they will not be the target of our analysis.} We only consider so-called discretionary measures, which cover the main instruments of ALMP. In particular, we cover short-term activation measures, training and wage subsidies.\footnote{In parallel, there exist mandatory measures such as benefits for the participation of disabled people in working life, subsidies to promote the start of self-employment and short-time working allowances. 
} These three types of programs are largest in scope and most likely to generate direct and indirect effects at the aggregate level (\citealp{Abbring2007, Imbens2009}). 

\begin{figure}
                 \centering
                 \caption{Expenditures for ALMP over Time \label{fig:exp}}
                 \vspace{10pt}
                  \begin{subfigure}[b]{0.8\textwidth}
                                 \centering \caption*{(a) Total expenditures}
                                 \includegraphics[clip=true, trim={0cm 0cm 0cm 0cm},scale=0.87]{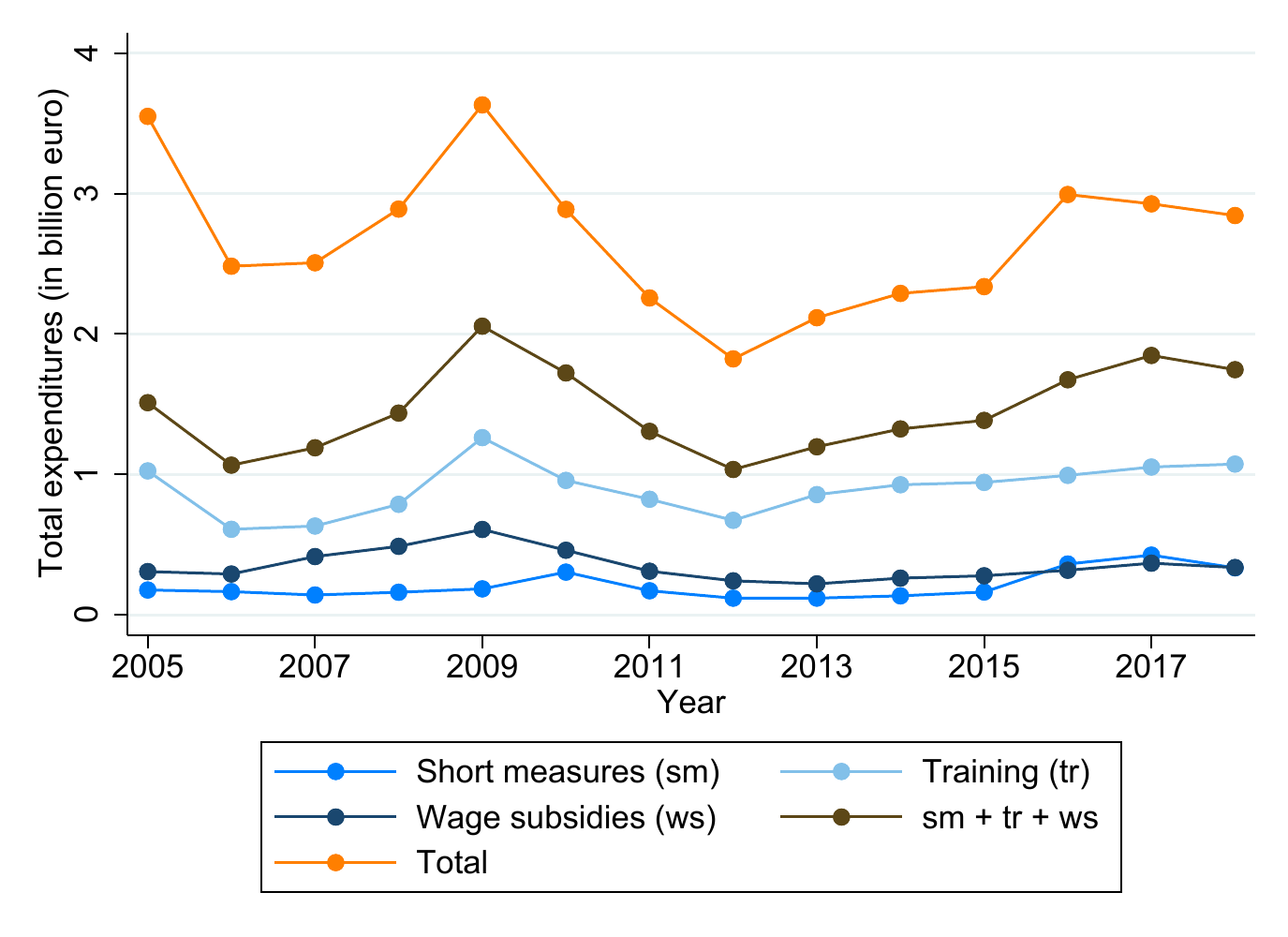}
                 \end{subfigure}
                 \vspace{10pt}  
                 \begin{subfigure}[b]{0.8\textwidth}
                                 \centering \caption*{(b) Expenditure shares}
                                 \includegraphics[clip=true, trim={0cm 0cm 0cm 0cm},scale=0.87]{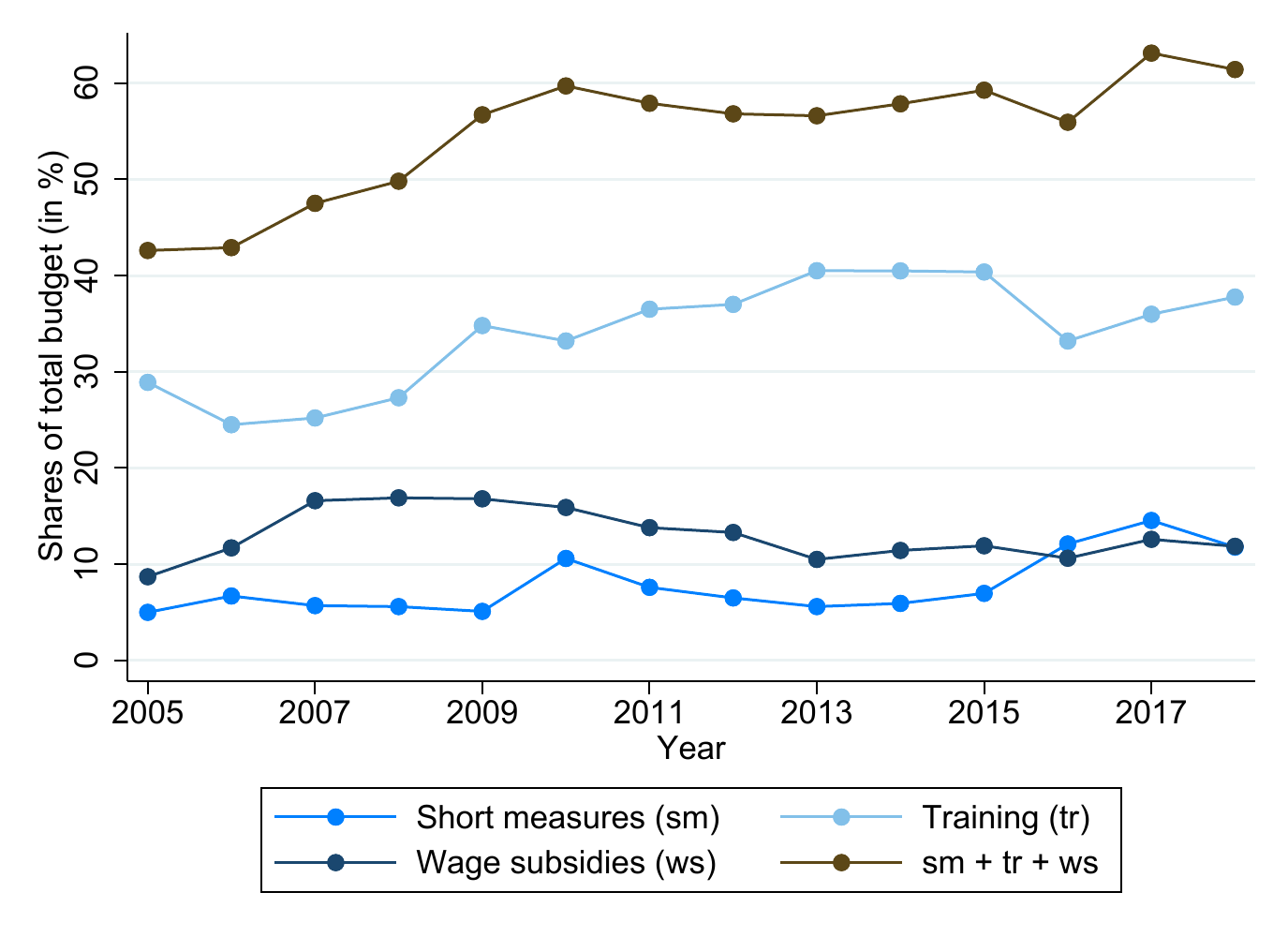}
                 \end{subfigure}
                 \begin{minipage}{\textwidth}
                   \footnotesize \textit{Notes:} Panel (a) plots the German expenditures for ALMP (discretionary measures) in billion euro. Panel (b) plots the corresponding expenditure shares by program type. Short measures (sm), training (tr), wages subsidies (ws). Source: \href{https://statistik.arbeitsagentur.de/DE/Statischer-Content/Statistiken/Themen-im-Fokus/Eingliederungsbilanzen/Archiv-Eingliederungsbilanzen.html?nn=25322}{https://statistik.arbeitsagentur.de}  (accessed in June 2022).
                 \end{minipage}
           \end{figure}

Figure \ref{fig:exp} plots the expenditures for ALMP over time and the allocation of the total budget to different program types. In the time period we consider, short activation measures, further vocational training and wage subsidies jointly comprise 40 to 60\% of the total budget for discretionary measures amounting to 1 to 2 billion Euro yearly. The remaining budget is primarily spent on basic vocational training, start-up subsidies and employment creation schemes.  Expenditures roughly follow trends in the aggregate unemployment rate and are negatively correlated with the business cycle. They decrease until 2006, peak with the great recession in 2009, decrease again until 2012 and are slowly raising slowly since then. 

\textbf{Types of policies.} \quad \textit{Short activation measures} (henceforth short measures) comprise the lowest share of expenditures amongst the programs considered but have the highest inflow of participants (see Figure \ref{fig:exp} and Appendix Figure \ref{fig:inflows}). This is related to their short duration of 2 to 12 weeks\footnote{Short measures at providers are limited to a maximum of three months, in-firm-measures to six weeks.} and high turnover rates. 
These programs 
combine elements of vocational training and job search assistance such as the assessment of skills and employability, aptitude tests for specific jobs, as well as application and skill training. There are two types of measures; class-room training at training providers and on-the-job training at firms.  Employment agencies either directly purchase courses at providers or give out vouchers to jobseekers which they can redeem at certified providers or firms. A reform in 2009 changed the legal foundations of job search measures.\footnote{Before the reform, short measures were known as training measures and as measures for activation and vocational integration after.}  Primarily, it granted training providers more leeway in tailoring the program content to individual needs. Since then, employment agencies have merely set the goal of an integration of jobseekers into the labour market but providers are free in setting the duration and content of specific measures.\footnote{ We run a sensitivity analysis restricting our observation window to the time after 2010 to test whether our results might be driven by these institutional changes and do not find qualitatively different results.}

\textit{Further vocational training programs} (henceforth training)  
represent the largest programs in terms of expenditure shares. They rather steadily increased from about 30\% in 2005 to about 40\% in 2018. The programs 
cover classic vocational training programs with a duration of several months up to two years and  retraining programs (also called degree courses) which generally last 2 to 3 years. While classic vocational training programs provide an extension and adjustment of professional and practical skills, retraining programs offer training for a new vocational degree according to the German system of vocational education.  The allocation of jobseekers to training programs is regulated by a voucher system.\footnote{See \cite{Kruppe2009} or \cite{Doerr2017a} for more details.} Caseworkers review the need for qualification of jobseekers and recommend participation in training by issuing a voucher with a specific educational target. 
\footnote{These vouchers can be redeemed within a period of three months at certified providers within a geographical area that can be reached by public transport in a reasonable amount of time \citep{Doerr2017a}. Due to this restriction, we argue that it is unlikely that a large share of program participants commutes to training providers outside of the labour market.}

As third type of programs we consider \textit{wage subsidies}, in particular the so-called integration subsidies. Expenditures for wage subsidies comprise to about 10 percent of the global budget with a slightly higher share in the years around the great recession. The subsidies are paid to employers at a rate of up to 50\% of the wage if they hire jobseekers with employment impediments, such as long-term unemployed and older workers as well as jobseekers with severe disabilities. 
Firms are required to hire formerly subsidised workers for the same length as regular employees once the subsidy runs out.\footnote{The features of the program remain largely unchanged during the observation period. An exception is the introduction of additional variants of the subsidy, e.g. a subsidy targeting younger workers introduced in 2007 and abolished in 2012. Nevertheless, this variant only covers around 1 percent of the total wage subsidy recipients per year. It should thus not cause any substantial differences in effects.}


\textbf{Allocation of funds to policies.} \quad Our identification strategy relies on the idea that labour markets do not perfectly coincide with employment agencies at which the decision of the local policy mix is made. This decision is shaped by a multi-layer process of budget distribution which we outline below (see also \citealp{Blien2005, Yankova2010, Wunsch2015, Wapler2018a}).

Every year, the Federal Employment Agency (FEA) decides on a global budget for all discretionary measures (see Figure \ref{fig:exp}) which is distributed to 10 regional headquarters and 156 local employment agencies. The rules for allocation are regulated in the Social Code IV (§71b) which states “the regional development of employment, the demand  for labour, the type and extent of unemployment as well as the particular expenditure development in the preceding financial year are to be taken into account.” In practice, the distribution is done in three steps. First, a fixed proportion of the budget is allocated to East Germany. Second, the budgets for East and West Germany are distributed among the regional headquarters in the respective regions. Third, the regional headquarters decide on the funds for each employment agency.\footnote{ Steps two and three are based on an allocation rule that considers the employment growth, the unemployment rate (including program participants), the rate of unemployed with bad labour market prospects such as the long-term unemployed, disabled people and the elderly, as well as the outflows from unemployment into regular employment (See \citealp{Blien1998, Blien2005}).} 


While the FEA and the regional headquarters set some overall guidelines of for the use of different types of policies, local employment agencies have a great deal of autonomy in the ultimate allocation of the budget to specific programs. Program-specific budgets, policy objectives and guidelines are fixed at the agency level in the final quarter of every year for the respective forthcoming year. If the budget is exhausted, the corresponding measure can no longer be offered \citep{Yankova2010}. Unused funds can be transferred to the following year. At the beginning of every year, each agency additionally determines the type of training measures to be carried out in a educational target plan (\textit{Bildungszielplanung}).\footnote{Notice that there exists no uniform procedure for this planning process. See e.g. \cite{Matysik2014}.} The aim is to further align training offers with the demand on the labour market.



\textbf{Reform of the employment agencies in 2012.} \quad From 2012 to 2013 around two-thirds of employment agencies changed their regional area of responsibility. Agencies gave up parts of their territories to other agencies, absorbed parts of territories from other agencies, merged in their entirety or were newly created. The reform took place in three steps in July 2012, October 2012 and January 2013. Overall, around 21 percent of municipalities changed their agency and the total number of agencies was reduced from 178 to 156. Since then, all agency borders correspond with district borders. Similar to the pre-reform period, new agencies are primarily defined as administrative and not functional areas \citep{Hirschenauer2013}.

\section{What Macroeconomic Effects do we Expect?}\label{theory}
A traditional motivation for ALMP has been to improve the matching process in the labour market and to reduce unemployment. From a theoretical point of view, these policies can have ambiguous effects on aggregate labour market outcomes (see e.g. \citealp{Calmfors1994,Calmfors2002} and \citealp{Layard2005} for seminal discussions). They might affect matching efficiency, productivity, job competition as well as job creation and destruction, in part through wage-setting and labour demand incentives. Positive effects on participants may be offset by negative externalities on non-participants such as substitution and displacement effects \citep{Calmfors1994}.  Substitution effects occur when ALMP induce employers to hire from the stock of program participants instead of from the stock of non-participants do due to a change in relative labour costs. 
Displacement effects are closely related and refer to a crowding out of regular labour demand.\footnote{There exists no uniform definition of the terms substitution and displacement effects and they are sometimes used interchangeably in the literature. Some papers understand displacement effects in a narrow sense of  a crowding out of regular employment through competition in the goods market (e.g. \citealp{Layard1991, Calmfors1994}). Others, like us, take on a broader definition that include substitution effects (e.g. \citealp{Bjoerklund1991, Dahlberg2005, Crepon2013}).} 
Furthermore, the effectiveness of the policies might be compromised by deadweight effects. They arise if employers would have hired program participants even in absence of the measure. 

In what follows, we consider each program separately and discuss expected net effects on employment, which is the focus of our analysis. In doing so, we take program-specific unintended effects into account. Deadweight losses and substitution effects are likely to affect all programs under consideration, while displacement are particularly relevant for wage subsidies. Our discussion will be guided by existing empirical evidence on the impact of ALMP at the individual level and the developments in labour demand. For the past decades, pronounced changes in labour demand for different jobs and skills have been documented globally (for Germany, see e.g. \citealp{Hutter2021, Dustmann2009}). In particular, so-called skill-biased technological change shifted demand towards (high) skilled labour \citep{Acemoglu2011} and routine-biased technological change towards non-routine work \citep{Autor2013, Autor2015}. These trends could have important implications for the effectiveness of the programs we study.


\textit{Short measures} primarily aim at reducing frictional unemployment by improving matching efficiency in the market. Specifically, their objective is to remove information asymmetries regarding suitable vacancies, and to improve workers' job search skills and search efficiency. 
If unemployment is characterised by structural mismatches due to an increasing demand for high-skilled and non-routine labour, short measures are unlikely to overcome them. 
The literature evaluating short measures in the 1990s and early 2000s in Germany, has found no or only small effects of these programs on individual employment \citep{Wunsch2008, Wunsch2013}. The programs are more effective after a period of unsuccessful job-search \citep{Biewen2014, Osikominu2013}, if they have a vocational training component \citep{Fitzenberger2013} or are executed in firms \citep{Kopf2013, Gordo2011, Stephan2011}. Lock-in effects are short but the programs are not very successful in increasing employment stability and earnings in the long run \citep{Osikominu2013,Osikominu2021}. 
Considering the at most minor employment effects at the individual level, we might also not expect large effects in the aggregate, especially if only a small part of unemployment is frictional. 

\textit{Training programs} aim at reducing skill mismatch in the labour market by improving the occupational skills of jobseekers and  adapting their qualifications to requirements of vacancies. They are expected to counteract human capital depreciation and to improve the match quality between workers and firms. These programs have been shown to reduce the search intensity of participants and to have considerable lock-in effects in the short run \citep{Fitzenberger2008,Lechner2011,Biewen2014}. How this affects the aggregate unemployment rate in the short run is unclear from a theoretical point of view. On the one hand, it might increase if the employment probability of non-participants remains unaffected. On the other hand, non-participants could benefit from less competition for the same stock of job vacancies, which could increase exits from unemployment in turn. In the long run, training programs have been shown to effectively increase individual employment \citep{Card2018, Lechner2011, Biewen2014}. They are thus likely to generate positive employment effects in the aggregate, unless hiring of participants comes at the expense of non-participants or incumbent workers.

Aggregate employment effects may differ for low-skilled and high-skilled workers and they may also depend on the presence of skill mismatch and labour market tightness. If targeted well, training programs can overcome structural mismatch and remove potential labour shortages. A shift in the workforce towards more skilled workers could then lead to positive employment effects for both the high-skilled and the low-skilled. Low-skilled workers might benefit if the competition in the low-skilled market gets smaller. However, it is also possible that public training crowds out firm investments in training. If firms do not find workers with suitable skills on the market and the gains from filling a vacancy are sufficiently large, they might turn to workers with insufficient skills, which the firms train themselves to obtain the missing skills. The availability of public training makes these investments unnecessary while the corresponding vacancies would still be filled. However, the reduction in hiring costs may additionally incentivise job creation. In the absence of skill mismatch or if training programs are poorly matching labour demand, aggregate employment effects are more ambiguous. The programs might change the composition of the labour force by increasing the proportion of high-skilled workers relative to the proportion of low-skilled. This might lower the bargaining power of the high-skilled and put downward pressure on their wages \citep{Pissarides2000, Mortensen1994}.  Furthermore, it might affect job creation and lead firms to hire trained in the place of untrained jobseekers. 
Such substitution effects might be aggravated if the stock of available vacancies is limited and smaller than the stock of jobseekers.

\textit{Wage subsidies} have the objective of reducing barriers to hiring by acting as a screening device for employers and reducing information asymmetries. They also aim at increasing participants' labour market attachment and productivity through on-the-job training. With that, they are an alternative to formal training to counteract the devaluation of human capital and to reduce the mismatch between skill supply and demand. The literature evaluating wage subsidies in Germany found large positive effects on unsubsidised employment at the individual level  (see \cite{Wolff2013} for a review).\footnote{Notice that investigating the effects of wages subsidies on individual employment is methodologically challenging.  In a quasi-experimental design, the program effect on taking up a job can be hardly disentangled from labour market outcomes conditional on employment.} On aggregate, wage subsidies are expected to generate positive employment effects. In a static labour demand framework, wage subsidies reduce employer’s labour costs and are expected to shift the labour demand curve upwards, thereby increasing employment and wages \citep{Bell1999}. In the search and matching framework they are expected to reduce hiring frictions and thereby increase matching efficiency \citep{Pissarides2000}. Furthermore, they are likely to increase job creation by lowering hiring and wage costs in the short run. If successful, they can achieve long-term integration of jobseekers into unsubsidised employment through higher productivity of workers.

However, unintended effects might attenuate the overall effectiveness of these programs. In particular, displacement effects are an often discussed risk for wage subsidies \citep{Calmfors2002, Dahlberg2005}. Firms might hire subsidised workers but jobs might be crowded out elsewhere in the economy \citep{Calmfors1995}. This might happen at the expense of other jobseekers as regular jobs are replaced with subsidized jobs. Furthermore, wage subsidies could depress the overall wage-level in the types of jobs that are subsidized. Overall, negative externalities depend largely on the exact design of the subsidies. A careful targeting to vulnerable groups of jobseekers with poor employment prospects as well as a mandatory period of employment after the subsidy runs out can attenuate unintended effects. 

In summary, all three programs might generate positive effects on aggregate employment. The net effects will depend on their effectiveness in increasing employment prospects of the participants themselves and their negative externalities on non-participants through displacement and substitution effects. Furthermore, we might not expect the same aggregate effects for all segments of the labour market. On the one hand, there is substantial evidence for heterogenity in the effectiveness of the programs at the individual level \citep{Card2018,McCall2016}. On the other hand, not all workers might be equally affected by potential negative externalities. Low-skilled or medium-skilled workers who do not obtain wage subsidies, for example might suffer more from displacement effects than high-skilled workers.

\section{Empirical Strategy}\label{emp_strat}

In this section we first introduce the general idea of our identification strategy and discuss the identifying assumptions. Second, we describe how we define labour markets and instrumental variables. Third, introduce the main sample restrictions for which the identifying assumptions are most likely fulfilled. Fourth, we present the data and finally our estimation method.

\subsection{Exploiting Incomplete Overlap for Identification}\label{identific_idea}

\subsubsection{General Idea}\label{gen_idea}

The endogeneity problem in macroeconomic evaluation studies of ALMP results from policies and aggregate labour market outcomes being determined simultaneously. As outlined in Section \ref{institutions}, in Germany regional labour market conditions determine policy implementation, which in turn determines regional labour market conditions. We propose a novel empirical strategy to overcome this problem. We estimate aggregate effects at the local labour market level and exploit the incomplete overlap of labour markets and administrative units that decide on the use of ALMP (in our context employment agencies, referred to as agencies in what follows).  In contrast to the agencies whose borders evolved historically \citep{Kropp2012}, labour markets reflect spatial aspects of economic activity. They are functional regions that are spatially bounded by places of work and residency. We define such markets for Germany following \citep{Tolbert1996} based on home-to-work commuting flows between municipalities.

\begin{figure}[h!]
\begin{center}
                 \caption{Illustration of the Identification Strategy} 
                  \label{fig:illustration_id}
               \vspace{-5pt}
               \centering\includegraphics[scale=0.4, clip=true, trim=5cm 0cm 4cm 2cm,page=4]{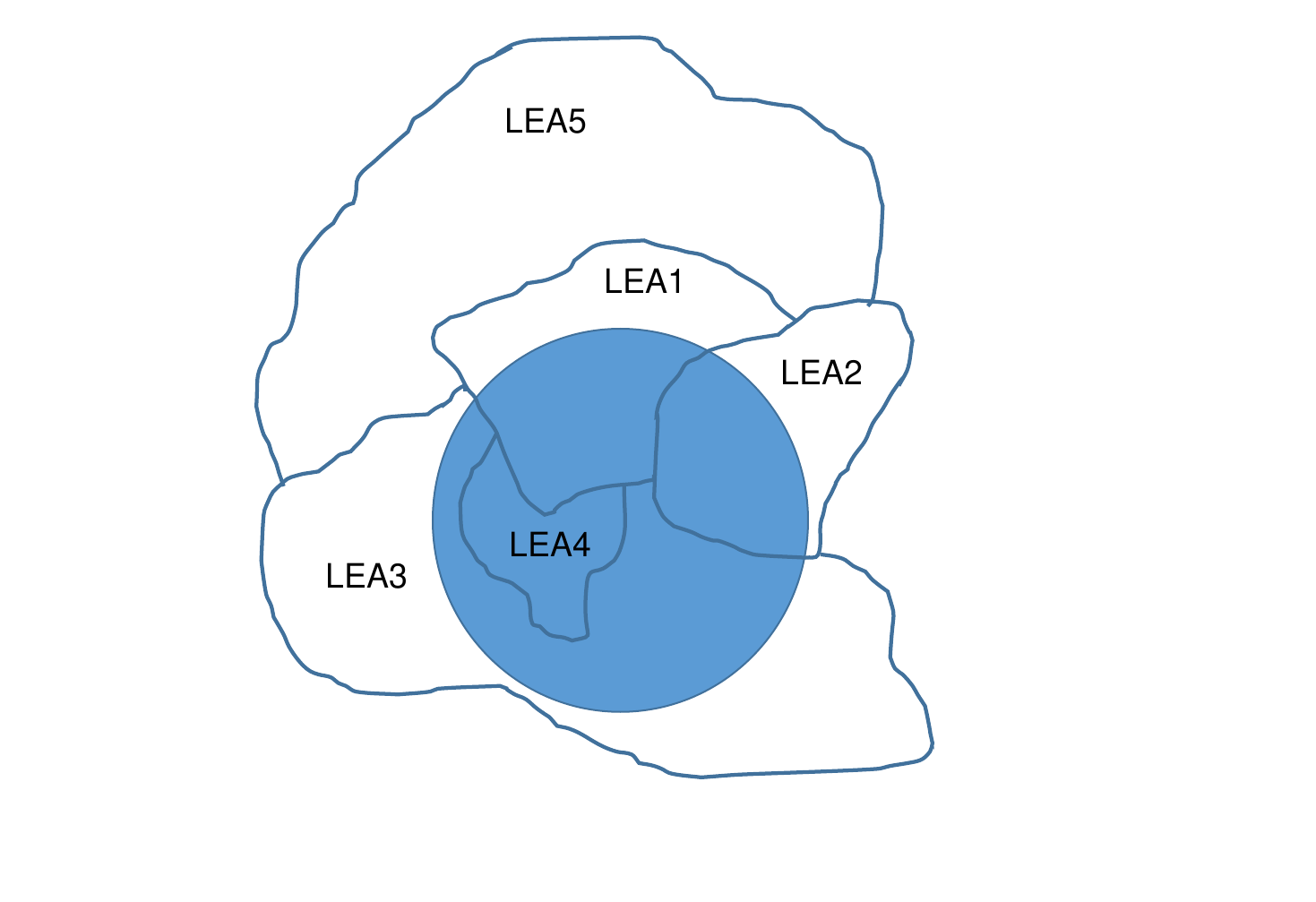}
               \end{center}
    \vspace{0.3cm}           
               \footnotesize 
\end{figure}

We find that labour markets are not perfectly coinciding with agencies and exploit the imperfect overlap of borders as source of exogenous variation. The idea is illustrated in Figure \ref{fig:illustration_id} for an exemplary labour market. The blue circle represents one labour market, which overlaps with four agencies (LEA1, LEA2, LEA3, LEA4). We are interested in measuring the effects of the policies implemented in this labour market, e.g. on the aggregate unemployment rate there. The policies might be in part endogenously determined by the labour market conditions in the blue area. To eliminate endogeneity, we instrument the policy take-up in the labour market with the policy take-up in the municipalities in the white areas outside the blue circle of the three employment agencies that partially overlap with the blue circle (LEA1, LEA2, LEA3). We define this area as instrument area.\footnote{Note that municipalities in the agencies LEA4 and LEA5 would not be considered for the construction of the instrument: LEA4 because it is completely contained in the labour market (no exogenous variation) and LEA5 because it does not overlap with the local labour market (no correlation with policy use in the local market).} We argue that the policies in the white areas are correlated with the policies deployed in the market but exogenous to the labour market outcomes there, once we condition on the characteristics of the market.

\subsubsection{Identifying Assumptions}\label{identific_ass}

For this approach to be valid, we need to make two main identifying assumptions. First, we assume that there are no spillovers across labour markets and across areas considered for instrumentation. To lend support to this assumption, we base our analysis on self-contained labour markets that do not overlap (meaning little commuting flows between the markets).

Second, we assume that the policy use in the municipalities outside the labour market is correlated with policy use in the labour market, but exogenous to the outcome variables of interest, such as the unemployment rate and employment rate, given the characteristics of the market. 
This assumption consists of two parts: First, the instruments need to be relevant, i.e. there needs to be some correlation in the policy take-up in the labour markets and in the instrument areas. Second, the instruments need to be exogenous, i.e. there is no direct effect of the policy use in the instrument areas on the labour market outcomes in a labour market after controlling for the characteristics of the market. Furthermore, the instruments need to be uncorrelated with any other determinants of the labour market outcomes. 

In order to assess the validity of the instruments, we need to understand how the policy mix in each labour market and agency is determined. Recall that each agency has discretionary power over how to allocate their budget to different policies (see Section \ref{institutions}). 
Policies implemented in a labour market are thus a mix of the policy making decisions of all overlapping agencies. The exogeneity of the instruments is coming from two sources. First, we claim that the strategies of the agencies are not only related to labour market conditions but also reflect different policy preferences and management styles \citep{Fertig2006a}. Second, the policies implemented in each labour market depend to a large extent on economic conditions and the composition of the labour force in regional jurisdictions of the overlapping agencies that lie outside the market. These conditions are likely to differ from the ones in the market, since the labour market situation is likely to change at the borders of self-contained labour markets.
The exogeneity assumption is thus more likely to be valid i) if several agencies overlap with a labour market and we have more variation in policy styles, ii)  if the labour force in the labour market represents only a small share of the labour force in the overlapping agencies, i.e. the agencies' policy mix is more likely to be determined by conditions outside of the local labour market, iii) if we are able to characterize the economic situation in the labour market well.
 
Instrument relevance is based on the idea that there is some similarity in the policies implemented in the entire jurisdiction of the agency. If agencies follow a particular strategy independent of the economic conditions and composition of jobseekers in their area of responsibility, e.g. have a strong preference for training programs, this strategy will be reflected in the municipalities inside and outside of the labour market. The policy use in the instrument area and the labour market will then be correlated. This should be true independent of how large the labour force in the market is compared to the labour force in all partially overlapping agencies. Jointly, all overlapping agencies determine the policies deployed in the market. Nevertheless, it is likely that the relevance assumption is more likely to hold if the overlap between single agencies and the labour market is not too small.

\subsection{Definition of Local Labour Markets}\label{llm}

We construct local labour markets based on commuting flows between all German municipalities in 2016 provided by the German Federal Employment Agency (Bundesagentur für Arbeit).\footnote{We check this definition for consistency over time, since there might be reasons to believe that labour markets change over time e.g. because of changes in the regional economic structure 
or differences in commuting behaviour \citep{Dauth2018}. We compare our main definition based on data from 2016 to two alternative definitions based on commuting flows from the years 2009 and 2002. We conclude that labour markets stayed rather stable from 2002 to 2016. See Appendix \ref{app_llm} for more details.}
Commuters are captured by  employment statistics and comprise all employees that were subject to social security contributions on June 30 and whose place of work is different from their place of residence.\footnote{Cross border commuting is only partially registered in the data (only inflows into the country). We disregard this information since we are primarily interested in commuting patterns within German municipalities.} The employment statistics further allow us to derive the resident labour force (RLF), i.e. the sum of all employees that live in a region and local labour force (LLF), i.e. the sum of all employees that work in a region for each municipality.  We complement these data with driving distances between all German municipalities from the Leibniz Institute for Economic Research (RWI). These distances record the driving time by car (in seconds) between the population centers of each municipality for the year 2016.

We follow the clustering approach developed by \cite{Tolbert1987} and \cite{Tolbert1996} who derive commuting zones based on county to county commuting flows in the US.\footnote{These definitions have recently been used contributions by \cite{Autor2015} or \cite{Acemoglu2020}. The approach is also related to some methods used for the definition of labour markets in Germany, e.g. \cite{Fertig2006a}, \cite{Kropp2008, Kropp2011a, Kropp2011b, Kropp2016}.}
Compared to US counties, German municipalities are substantially smaller and vary considerably more in size.\footnote{There were 11084 municipalities in Germany in 2016 with an average RLF of 2800 and a standard deviation of 17871. Municipalities in Rhineland-Palatinate and Schleswig-Holstein are often very small whereas municipalities characterised by a similar population density in North Rhine-Westphalia are considerably larger.} Since this heterogeneity negatively affects the performance of the clustering algorithm,
we first create a more homogeneous database by aggregating the approximately 11000 municipalities into fewer municipality regions. For this, we follow an approach similar to \cite{Kropp2016} and merge sparsely populated municipalities that are close to each other. We calculate a fusion coefficient $F_{ij} $ for every municipality pair $(i,j)$ according to the formula $F_{ij} = distance_{i,j}^2 *(RLF_i+RLF_j)$ which is based on the squared distance between the municipalities and their RLF. 

We then aggregate adjacent municipalities using a hierarchical clustering algorithm with complete linkage. This algorithm first merges the two municipalities with the lowest fusion coefficient and then subsequently adds up similar clusters by considering the largest dissimilarity between the municipalities in two clusters. We generate three different versions of pre-processed data stopping the clustering at 5000, 6500 and 7500 municipality regions. This gives us the chance to check the sensitivity of our results to this first aggregation step. This first step is characterised by a trade-off between homogeneity and size. A higher number of municipality regions results in more homogeneous regions but also in a higher aggregation level as input for the definition of commuting zones.\footnote{ Note that the resulting municipality regions are still comparatively small for all three versions. The average (meadian) RLF amounts to 4167 (1511) for 7500 municipality regions and to 6251 (3077) for 5000 municipality regions. The average number of grouped municipalities amounts to 1.4 and 2.2, respectively. See Appendix Figure \ref{app_fig:munreg}.}

In a second step, we cluster the municipality regions based on the strength of commuting ties between them, starting with 5000. First, we calculate a proportional commuting measure for each pair of municipality regions as proposed by \cite{Tolbert1987}:

$$S_{ij} = \frac{(P_{ij} +P_{ji})}{min(RLF_j,RLF_i)},$$

where $P_{ij}$ denotes the commuting flows from  municipality region $i$ to $j$ and $P_{ji}$ the commuting flows in the other direction. The denominator corresponds to the minimum of the resident labour force in region$i$ or $j$. This  measure captures the commuting relationship between two regions with respect to the smaller of the two regions and has a large value if the commuting ties are strong. We then collect these similarity measures $S_{ij}$ in a symmetric matrix and set the diagonals to 0. Since the clustering algorithm that we apply requires a dissimilarity measure, we derive: $D_{ij} = 1-S_{ij}$.

We use a hierarchical clustering algorithm with average linkage and cluster municipality regions with strong commuting ties bottom up until an average between cluster distance of $c$ is reached. Since there is no clear ex-ante guideline of where to stop the clustering, we run the algorithm several times and stop at different values of $c \in \{0.98, 0.99, 0.991, 0.993, 0.995, 0.997, 0.999 \}$. For every stopping value we obtain a different number of clusters (or labour markets) with higher values resulting in larger and fewer regions.\footnote{Note that in hierarchical clustering, clusters obtained by stopping at a given value are nested within the clusters obtained by stopping at a greater value.} \cite{Tolbert1996} choose an average cluster distance of 0.98 as stopping rule, which we take as minimum value. Given the smaller size of underlying municipality regions, this cut-off leads to a total of 462 labour markets of a relatively small size (average RLF equal to 67700). The maximum stopping value that we choose lies at 0.999 and results in 59 labour markets with an average RLF equal to 530000). We store the number of clusters corresponding to the respective stopping values and repeat the clustering exercise based on 6000 and 7500 municipality regions. In contrast to the first round based on 5000 municipality regions, we now use the stored total number of clusters as stopping rule in order to obtain the same number of labour markets for each underlying number of municipality regions. In the following, we will present the results obtained based on 6000 municipality regions. As described in Appendix \ref{app:munireg}, the characteristics of the resulting labour markets are not very sensitive to this first aggregation step.

\textbf{Characterizing the labour market definitions} \quad Table \ref{tab:char_llm} gives an overview of our labour market definitions based on different stopping values (column 2).\footnote{Note that the stopping values are set based on 5000 municipality regions as described above.} Column (3) reports the number of labour markets that we obtain, column (4) the average RLF in the markets and columns (5)-(9) describe two measures of self-containment. The commuter ratio (CR) in column (5) measures the share of all commuters between labour markets as percent of the total labour force. This value will be small if labour markets are self-contained and there is little commuting across borders. Columns (6)-(9) show the average, the standard deviation, the minimum and maximum of the so-called employment self-containment ratio (ESC) introduced by \cite{vanderLaan2001}.  The ESC measures the share of people living in a region who also work there. 

\begin{table}[htbp]\centering\caption{Labour Market Definitions}\label{tab:char_llm} 
\resizebox{\linewidth}{!}{%
\begin{tabular}{l*{12}{c}}
\hline\hline
    (1)        &         (2)&        (3)&     (4)&         (5)&     (6)&      (7)&     (8)&     (9) \\
         Def. &    Cut-off&        LLM (N)&     RLF (mean)&          CR&     ESC (mean)&      ESC (sd)&     ESC (min)&     ESC(max) \\
\hline
1          &            0.98&         462&    67652.63&       20.22&       65.77&       16.46&       11.11&       98.27  \\
2          &            0.99&         274&   114071.22&       15.91&       74.03&       13.07&       11.11&       98.27 \\
3          &           0.991&         251&   124523.96&       14.89&       75.50&       12.24&       11.11&       98.27  \\
4          &           0.992&         229&   136486.96&       14.08&       76.63&       11.75&       11.11&       98.27 \\
5          &           0.993&         209&   149547.91&       13.57&       78.03&       10.44&       16.56&       98.27 \\
6          &           0.994&         187&   167141.79&       12.78&       79.11&       10.43&       16.56&       98.27 \\
7          &           0.995&         162&   192935.27&       12.06&       80.71&        8.34&       45.88&       98.27 \\
8          &           0.996&         138&   226489.23&       11.10&       82.35&        7.78&       54.43&       98.27 \\
9          &           0.997&         121&   258310.03&       10.35&       83.41&        7.61&       54.43&       98.27 \\
10         &           0.998&          92&   339733.85&        8.21&       86.39&        5.44&       65.34&       98.27 \\
11         &           0.999&          59&   529754.47&        5.83&       89.68&        3.81&       76.56&       98.27 \\
\hline\hline
\end{tabular}}
\end{table}

Both measures of self-containment increase with higher stopping values. Labour market definitions with fewer and larger labour markets are characterised by a higher ESC and a lower commuting ratio. All definitions are characterised by a high level of variation in self-containment. The heterogeneity is less pronounced for higher cut-off values. Based on this evidence, we can conclude that labour market definitions based on higher stopping values are thus more likely to represent closed, functional entities between which regional spillovers are less likely and which satisfy  the identifying assumptions best.\footnote{\citealp{Wicht2020} show that functional labour markets are not necessarily characterised by more intra-regional homogeneity with respect to their economic conditions and that self containment and homogeneity are contradictory goals when defining functional regions. For the purpose of our analysis a high level of self-containment is favourable in order to rule out spill-over effects to neighbouring regions.} The final choice of definition depends on the imperfect overlap of labour markets with agencies, which we discuss in the next section.

Appendix Figure \ref{fig:map_pref} shows a map of the resulting 162 labour markets obtained by the clustering approach described above, based on the cut-off value of 0.995 (our preferred version). Labour markets are shown as coloured areas, the borders of employment agencies (as from 2013) through black solid lines. Labour markets and employment agencies are relatively heterogeneous in size and reflect the distribution of agglomerations in Germany. Comparatively small employment agencies usually represent rather large cities. Often these cities are the center of a large labour market, as for example in the case of Berlin.

\textbf{Comparison to existing definitions.} \quad There exist a number of local labour market definitions based on home-to-work commuting flows for Germany, e.g. 50 labour markets by \cite{Kropp2011a, Kropp2016} or 141 labour markets by \cite{Kosfeld2012}. Two additional definitions for Germany that are frequently used are the 258 labour market regions of the Joint Task of the Federal Government and the federal states dedicated to the "Improvement of Regional Economic Structure" (GRW) and the 96 spatial planning units by the Federal Institute for Research on Building, Urban Affairs and Spatial Development (BBSR).

However, none of the existing definitions are suitable for our methodological approach. The majority is based on aggregations of around 400 districts. Compared to municipalities, districts constitute relatively highly aggregated administrative areas and are thus not able to capture the subtleties in spatial structures that we want to exploit. The resulting labour markets are likely to perfectly overlap with employment agencies.\footnote{See e.g. \cite{Kropp2012} for an overview. Appendix Figure \ref{fig:map_dist} depicts German districts in plain colours and local employment agencies as black solid lines. It becomes evident that district borders often coincide with the borders of local employment agencies and that one local employment agency is comprised of relatively few and predominantly large districts. As a consequence an aggregation of districts to local labour markets will likely result in either coinciding borders of economic and administrative entities or in large deviations.}  Only one labour market definition \citep{Kropp2011a, Kropp2016} is based on commuting flows between municipalities but it contains too few and too large regions for our purposes. We compare our labour market definitions to these established definitions of local labour markets in Appendix \ref{app:otherdef} and find that our approach results in qualitatively comparable labour markets in terms of self-containment.

\subsection{Instrument Areas and Sample Selection}\label{instrument_reg}

We define an instrument area for each labour market. It comprises those municipalities that are outside of a labour market but within the jurisdiction of agencies that partially overlap with the market. The policy use in this area is then jointly considered for the construction of the instruments. 
As outlined in Section \ref{identific_ass}, the instruments are more likely to be valid if the overlap between a labour market and the employment agencies meets certain criteria. For our main analysis we impose the following restrictions. They have to hold for the entire observation window\footnote{The reform in 2012/13 changed the overlap between labour markets and agencies in some cases. If a labour market does not fulfil the overlap criteria for a subset of years, it is excluded from the entire analysis. Given our data, we cannot distinguish between territorial changes between July and October 2012. Thus, we distinguish between three periods for the construction of our instruments: i) January 2005 to July 2012, ii) July 2012 to January 2013, iii) January 2013 to December 2018. Since the majority of territorial changes happened in October 2012, this imprecision affects only a small number of municipalities. Moreover, since we work with quarterly data, we anticipate the new agency borders by only one quarter. }: First, a labour market should overlap with at least two agencies. Second, the resident labour force (RLF) in the market should not constitute more than 50\% of the total RLF in all agencies that partially overlap with the market ($s_{tot} = RLF_{LLM}/\sum_j RLF_{LEA,j} < 0.5 $).\footnote{Notice that this total excludes all agencies that are fully included in the market.} This ensures that the labour market situation in the labour market does not drive the policy mix in overlapping agencies too strongly and that the policy mix deployed in the instrument areas is likely to represent an exogenous source of variation.
For the main analysis, we do not impose a minimum criterion on the overlap. The reasoning behind this is that instrument relevance can be tested and that it should not depend on how large the instrument area is in relation to the size of the market. But we set restrictions on the overlap of single agencies and the labour market in a robustness check.

\begin{figure}
                 \centering
                 \caption{Labour Market Definitions, Overlap Restrictions and Self-containment\label{fig:llm_choice}}
		\vspace{10pt}
                  \begin{subfigure}[b]{0.8\textwidth}
                                 \centering \caption*{(a) Number of labour markets by definition}
                                 \includegraphics[clip=true, trim={0cm 2.2cm 0cm 0cm},scale=0.87]{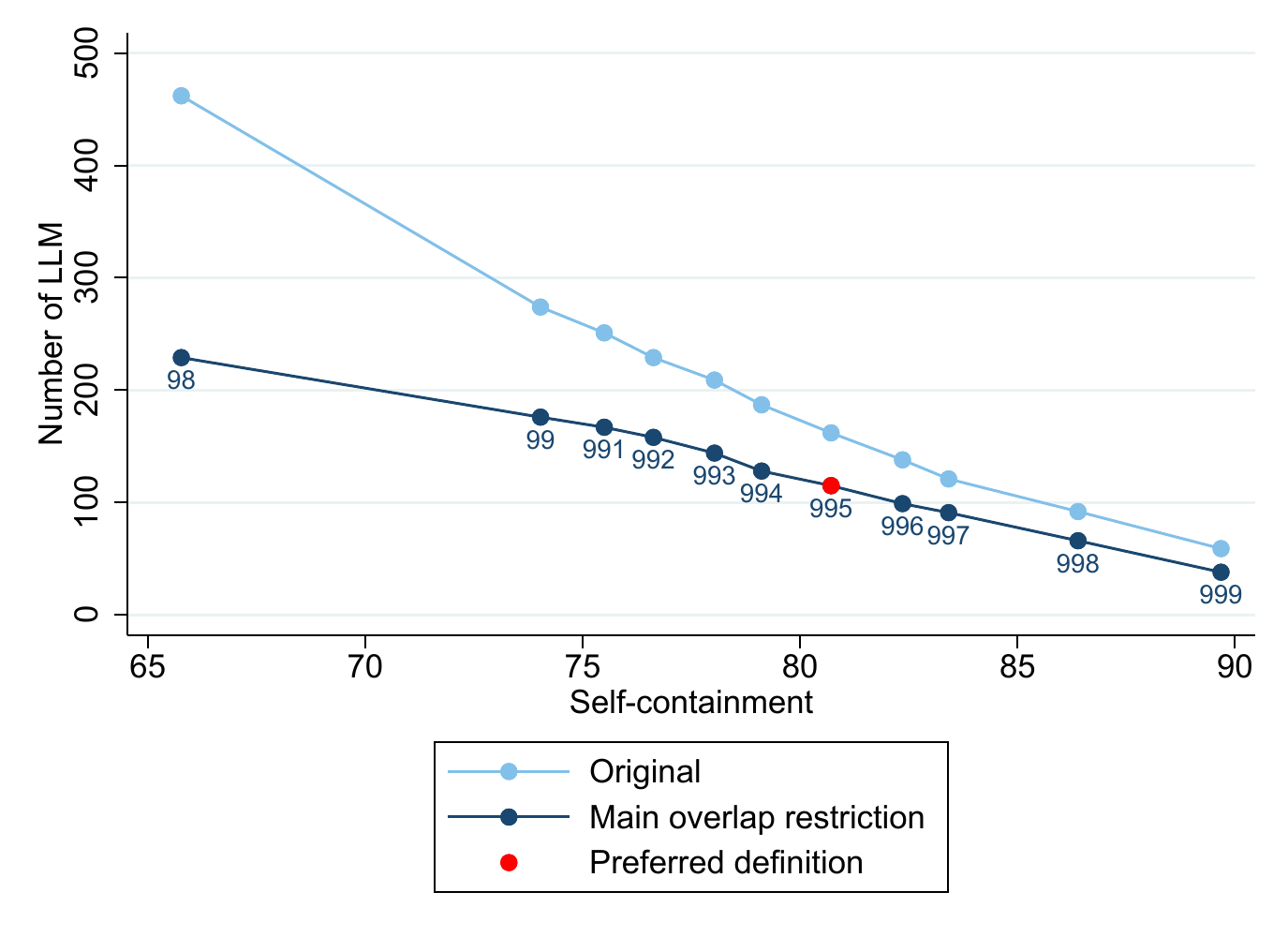}
                 \end{subfigure}
      \vspace{10pt}  
               \begin{subfigure}[b]{0.8\textwidth}
                                 \centering \caption*{(b) Average number of overlapping agencies}
                                 \includegraphics[clip=true, trim={0cm 0cm 0cm 0cm},scale=0.87]{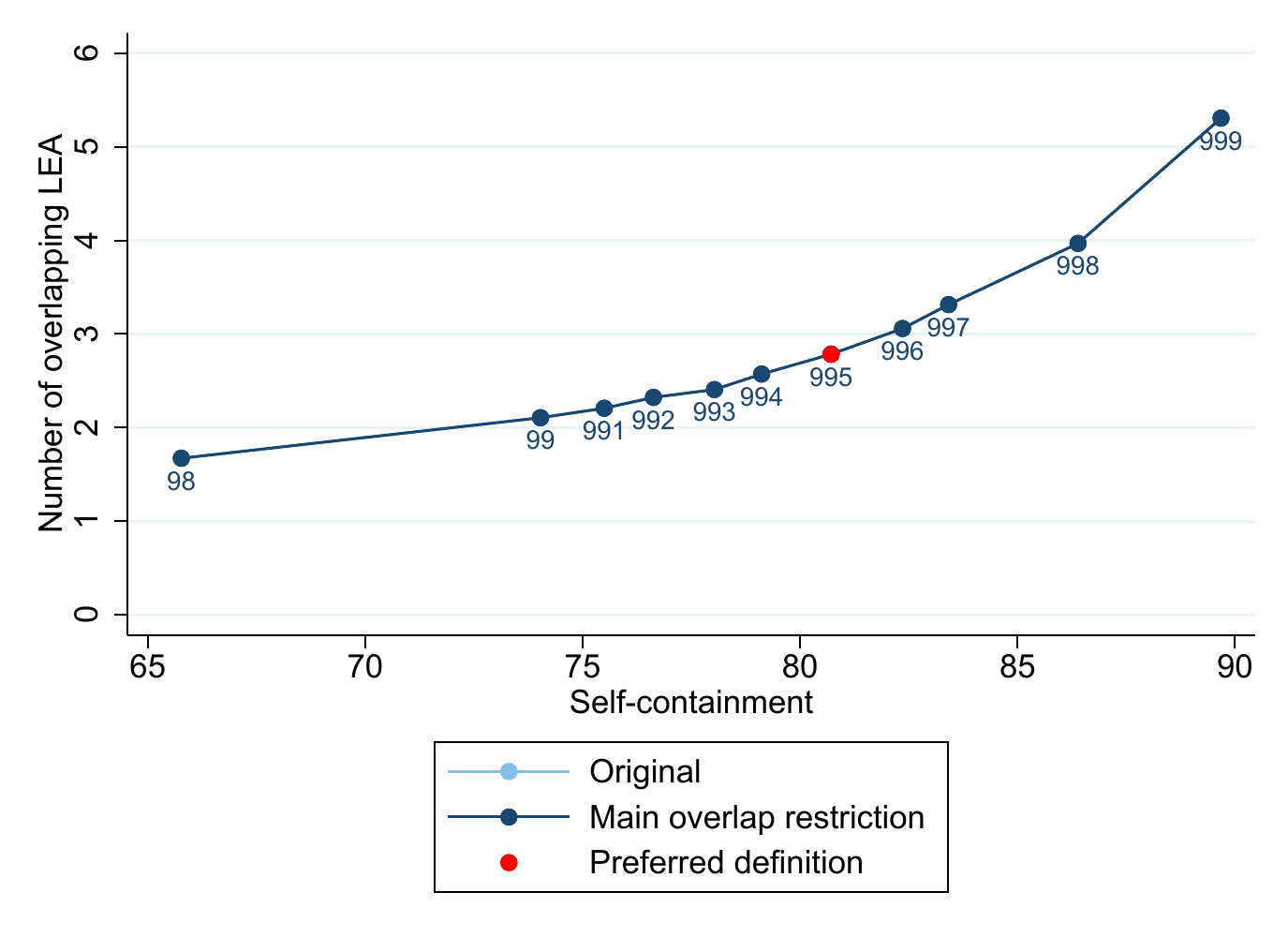}
                 \end{subfigure}

                 \begin{minipage}{\textwidth}
                  \footnotesize \textit{Notes:} Panel (a) plots the number of labour markets, panel (b) the average number of overlapping employment agencies for different levels of self-containment. The original number of labour markets is shown in light blue, the number of markets (and overlapping agencies) after imposing the main overlap restrictions in dark blue. Each dot represents a labour market definition based on a specific stopping value in the clustering step. Our preferred definition is marked in red. 
                 \end{minipage}
           \end{figure}

In order to guide the choice of a labour market definition, we investigate how imposing these restrictions affects both sample size and the number of overlapping agencies across definitions. Figure \ref{fig:llm_choice} (a) plots the number of labour markets as a function of the average level of self-containment. The definitions are based on different stopping values in the clustering step. The light blue line shows the original number of markets, the mid blue line plots the same relationship after conditioning on the overlap criteria for our main specification. Consistent with what we discussed in Section \ref{llm}, higher stopping values result in fewer labour markets with a higher level of self-containment. Restricting overlap reduces the number of labour markets that remain in the sample across all labour market definitions. The reduction is more substantial for the definitions based on larger stopping values. 
Figure \ref{fig:llm_choice} (b) shows that the average number of overlapping agencies is increasing with larger and more self-contained labour markets, i.e.\ higher stopping values. An increase in self-containment from about 75 to 85\% is associated with only a moderate increase in the number of overlapping agencies of about 0.5.

This illustrates that we face a trade-off between the representativeness of the analysis and a clean identification strategy. Our preferred labour market definition should guarantee a sufficient level of representativeness while maximizing the self-containment of the markets as well as the number of overlapping agencies.  The labour market definition that is based on the stopping value of 0.995 achieves this objective. After imposing our main sample restrictions, we obtain an average level of self-containment above 80\%, while maintaining a total number of 115 labour markets. This corresponds to 71
\% of the original number of labour markets and comprises 64\% of the total labour force in Germany (in 2016). The average number of overlapping agencies is larger than two. Appendix Figure \ref{fig:var_space} shows how the labour markets are distributed geographically. Notice that imposing restrictions on the overlap leads to the exclusion of several labour markets around big cities, such as Berlin and Munich. Given that the labour market situation in these markets is likely to be strongly determined by the respective cities, a clean identification of effects in these markets is difficult. We test the sensitivity of our results to the choice of different labour market definitions and overlap criteria in Section \ref{robustness}.


\subsection{Data}\label{data}

For the main analysis we rely on social security records for the years 2000 to 2018 from the Integrated Employment Biographies
(IEB) provided by the Institute of Employment Research (IAB) in Germany. 
Jointly, it covers the universe of program participants in ALMP, a 10\% random sample of unemployed individuals\footnote{We draw from all individuals that are UI or unemployment assistance recipients in the years 2000 to 2004. From 2005 to 2018, we only consider UI recipients. Unemployment assistance was abolished with the Hartz reform in 2005.} and a 2\% random sample of employed individuals. For all of these individuals we observe the labour market histories from 1994 to 2018 with daily accuracy coming from the integrated employment biographies (IEB).

From these individual-level data we construct a panel dataset of labour markets at the quarterly level.\footnote{To construct aggregate measures,  we first sum over all individuals in a specific quarter and labour market. Based on these totals, we calculate quarterly averages  by relating these totals to the total number of employed, unemployed, or resident workers. For characteristics of the employed and unemployed workforce, we base ourselves on spell-level data.}  It contains aggregate endogenous policy and instrumental variables, a set of variables characterizing the workforce in the labour market and aggregate labour market outcomes. We restrict our attention to individuals who are between 20 and 64 years old, since we are interested in macroeconomic effects for the main working population (excluding students and pensioners). The sample of unemployed is used to characterise the unemployed workforce in a labour market and to define aggregate labour market outcomes. The sample of employed is used for the characterization of the employed workforce and aggregate labour market outcomes. The variables measuring the use of ALMP are based on the full population of program participants. We follow the literature (e.g. \citealp{Calmfors2002, Wapler2018a}) and measure the policy variables as so-called accommodation rates. They are defined as the share of participants in a specific program among the jobseekers drawing UI benefits in a labour market.\footnote{Note that since our definition of unemployment is based on UI receipt, we count all program participants receiving UI as unemployed. We count wage subsidy recipients as employed once they enter the program.
} For the instrumental variables, we construct the same shares by dividing the sum of program participants in all instrument areas of a labour market by the total number of jobseekers on UI in the same areas. Our main outcomes of interest are the aggregate unemployment rate (share of UI recipients in the RLF), the unsubsidised employment rate (the share of workers in unsubsidised employment in the RLF), the rate of welfare recipients (the share of unemployed workers receiving welfare benefits in the RLF) and the rate of employed workers on benefits (the share of workers with small jobs that receive UI or welfare benefits).\footnote{Notice that these shares do not exactly add up to one since we do not consider workers in subsidised employment.} Since not all segments of the workforce might be similarly affected by ALMP, we also consider the corresponding rates for different sub-populations such as females, males, individuals below 30, above 50, low, medium and high-skilled. 



\textbf{Censoring.} \quad For data protection reasons, we are obliged to censor quantities based on less than three observations (excluding zero). This entails a significant share of missing values in some of the variables, in particular for small labour markets. We pursue different strategies to deal with the missing values. First, our main sample restrictions naturally lead to the exclusion of small labour markets. Additionally, they ensure that no outcome variables are censored in the final dataset. Second, for all censored control variables characterizing the employed and unemployed workforce, we replace the original variable with 0 and create separate variables flagging censored values. For all variables measuring program participation, we impute censored counts of program participants with the value one. This yields a lower bound on the true count of program participants but the margin of error remains small.\footnote{We check our results for the sensitivity of this imputation using the values 2 or 3 for imputation. The results are qualitatively the same and are available upon request.} 

\begin{figure}[h!]
                \centering
                \caption{Variation in Outcomes over Time \label{fig:variation_time_out}}
              \vspace{10pt}    
                \begin{subfigure}[b]{0.49\textwidth}
                                \centering \caption{Unemployment rate} \label{fig:var_uerate}
                                \includegraphics[clip=true, trim={0cm 0cm 0cm 0cm},scale=0.50]{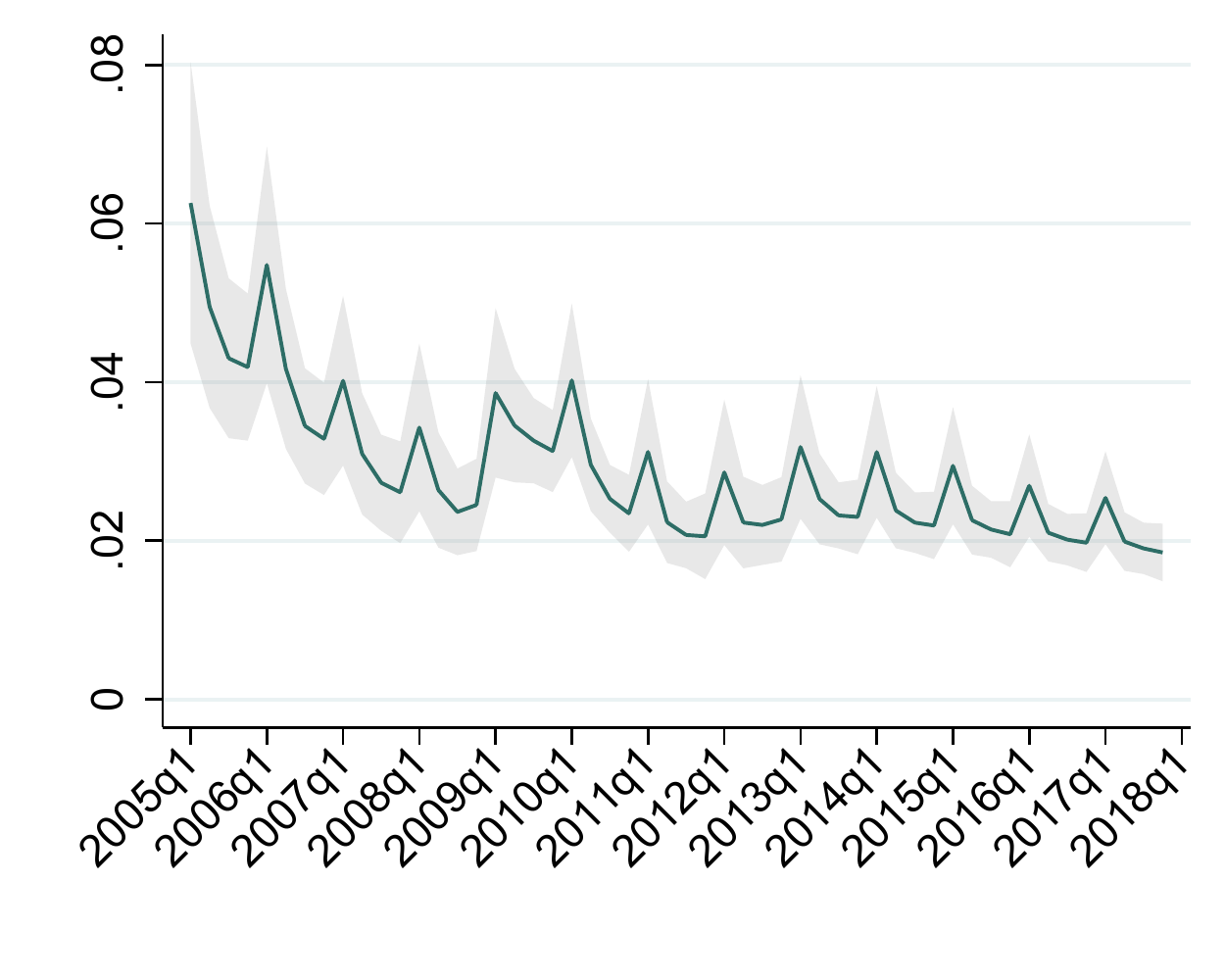}
                \end{subfigure}
                \vspace{10pt}    
                \begin{subfigure}[b]{0.49\textwidth}
                                \centering \caption{Unsubsidised employment rate} \label{fig:var_emprate}
                                \includegraphics[clip=true, trim={0cm 0cm 0cm 0cm},scale=0.50]{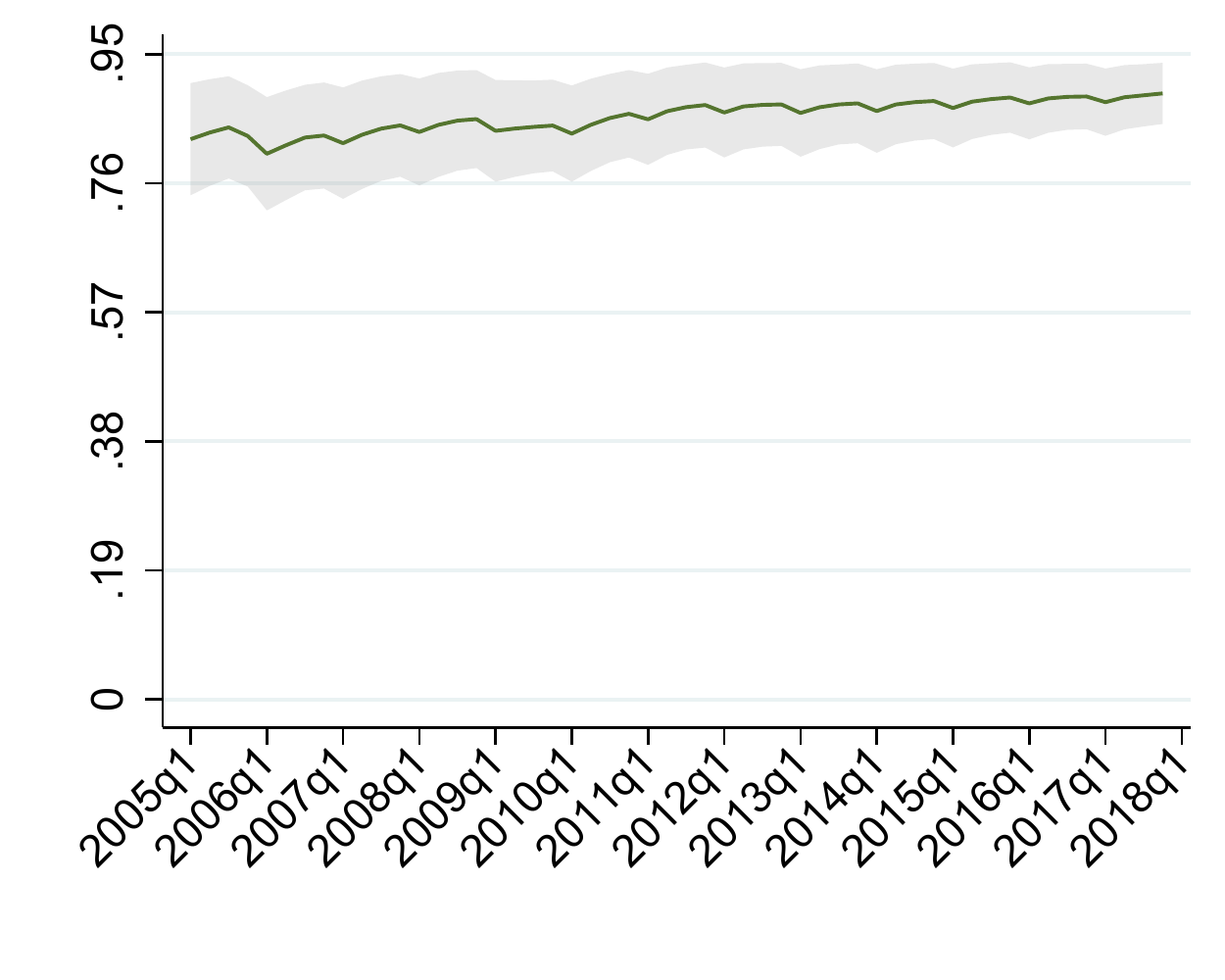}
                \end{subfigure}
                 \vspace{10pt}
                \newline
                \begin{subfigure}[b]{0.49\textwidth}
                                \centering \caption{Rate of welfare recipients} \label{fig:var_wf_rate}
                                \includegraphics[clip=true, trim={0cm 0cm 0cm 0cm},scale=0.50]{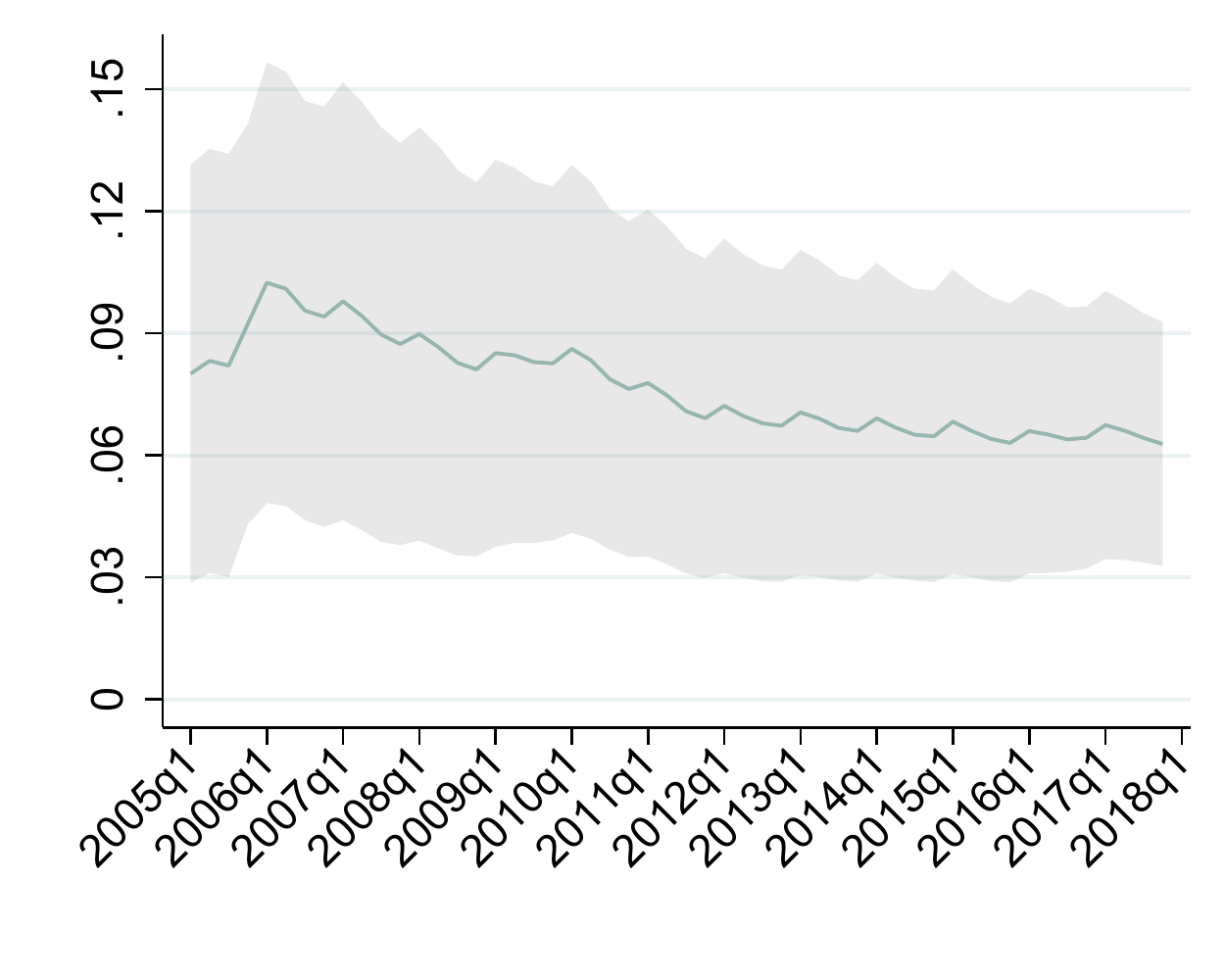}
                \end{subfigure} 
                  \begin{subfigure}[b]{0.49\textwidth}
                                \centering \caption{Rate of employed workers on benefits} \label{fig:prec_work}
                                \includegraphics[clip=true, trim={0cm 0cm 0cm 0cm},scale=0.50]{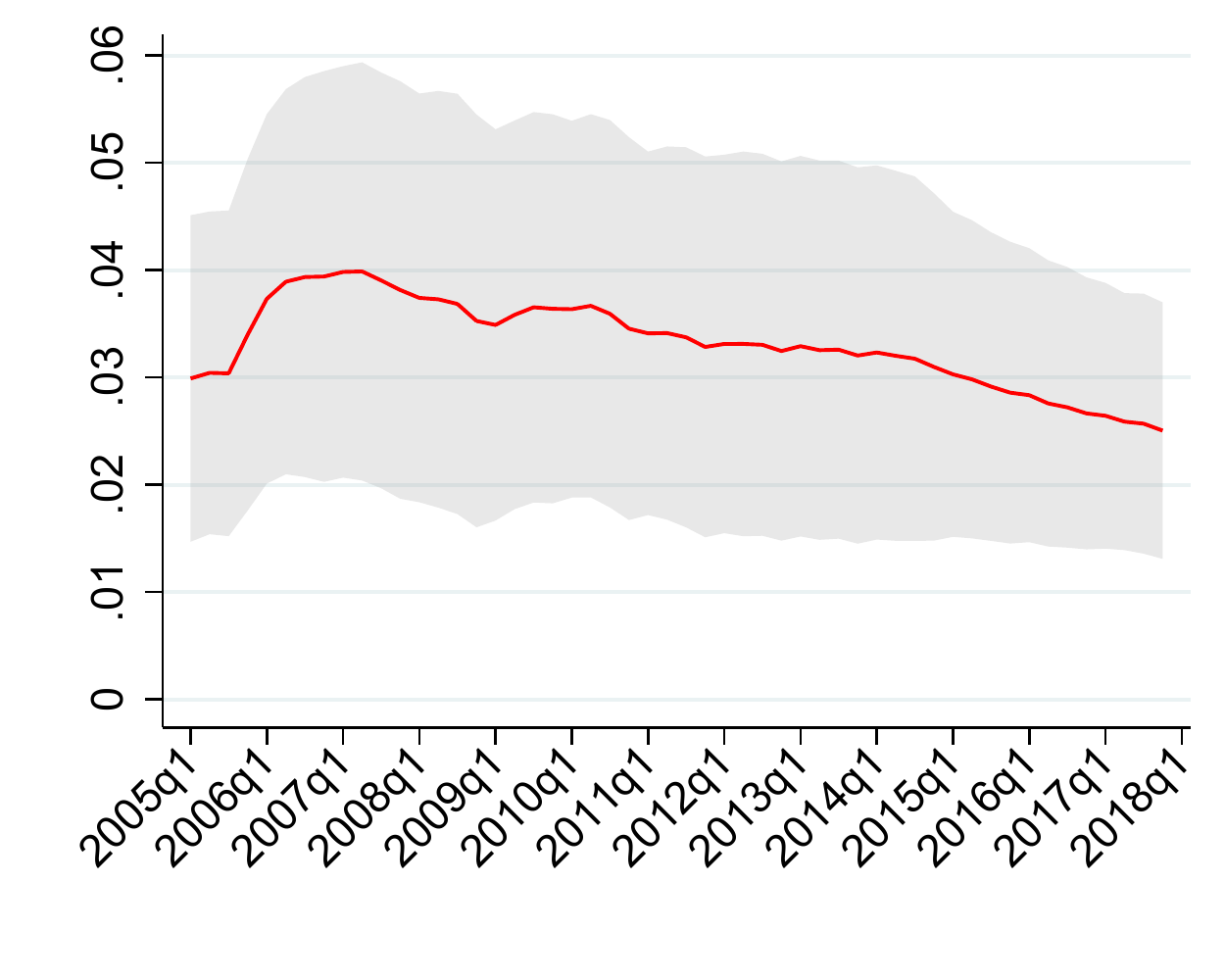}
                \end{subfigure}
                \begin{minipage}{\textwidth}
                                \footnotesize \textit{Notes:} The figures depict the average quarterly variation of the outcomes (a) unemployment rate, (b) unsubsidised employment rate, (c) rate of welfare recipients (d) rate of employed workers on benefits for the period 2005 to 2018. The standard deviation is depicted as grey area.
                \end{minipage}
\end{figure}

\begin{figure}[h!]
                \centering
                \caption{Variation in Programs over Time \label{fig:variation_time_prog}}
                \begin{subfigure}[b]{0.49\textwidth}
                                \centering \caption{Training} \label{fig:var_training}
                                \includegraphics[clip=true, trim={0cm 0cm 0cm 0cm},scale=0.50]{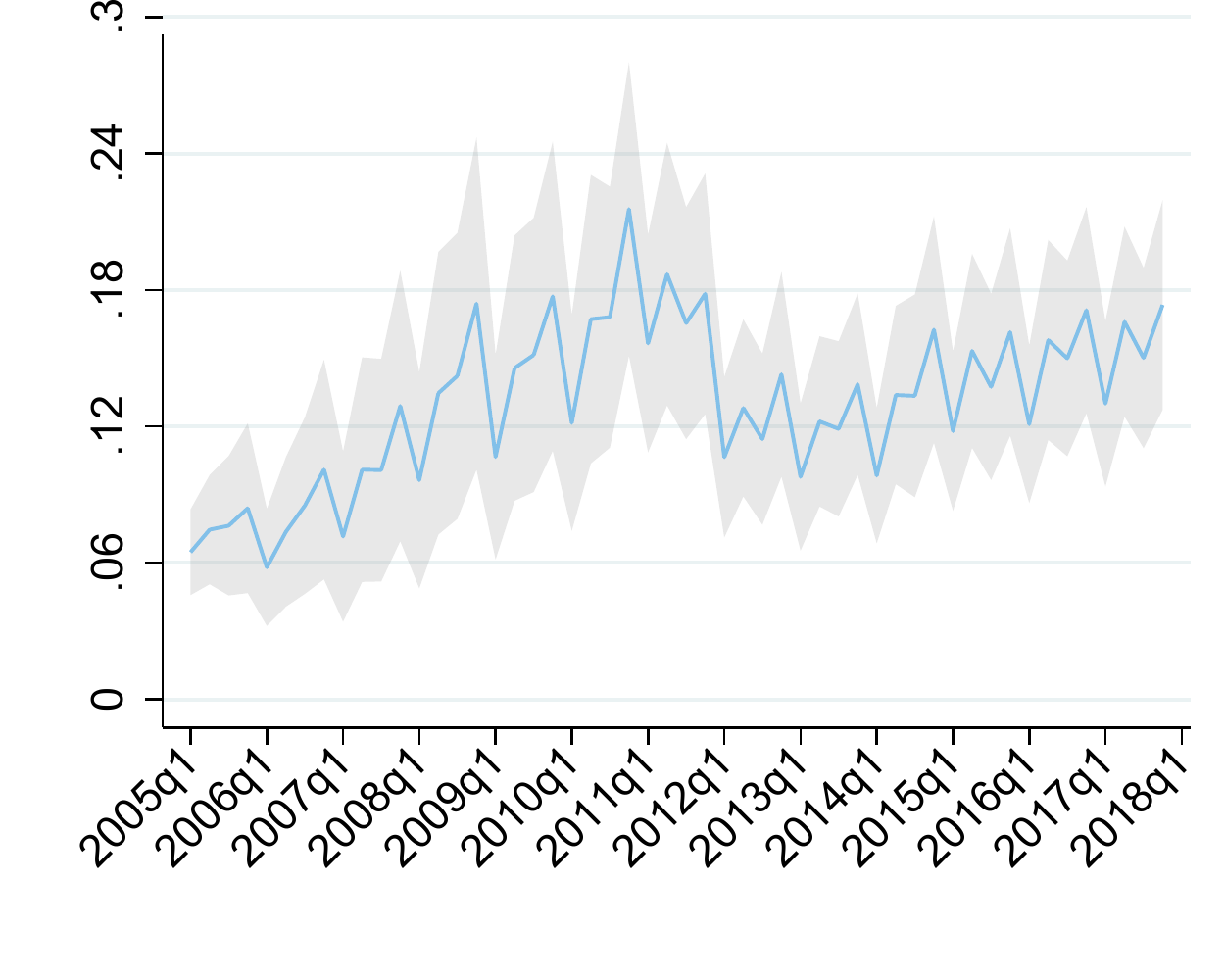}
                \end{subfigure} 
                  \begin{subfigure}[b]{0.49\textwidth}
                                \centering \caption{Short measures} \label{fig:var_sm}
                                \includegraphics[clip=true, trim={0cm 0cm 0cm 0cm},scale=0.50]{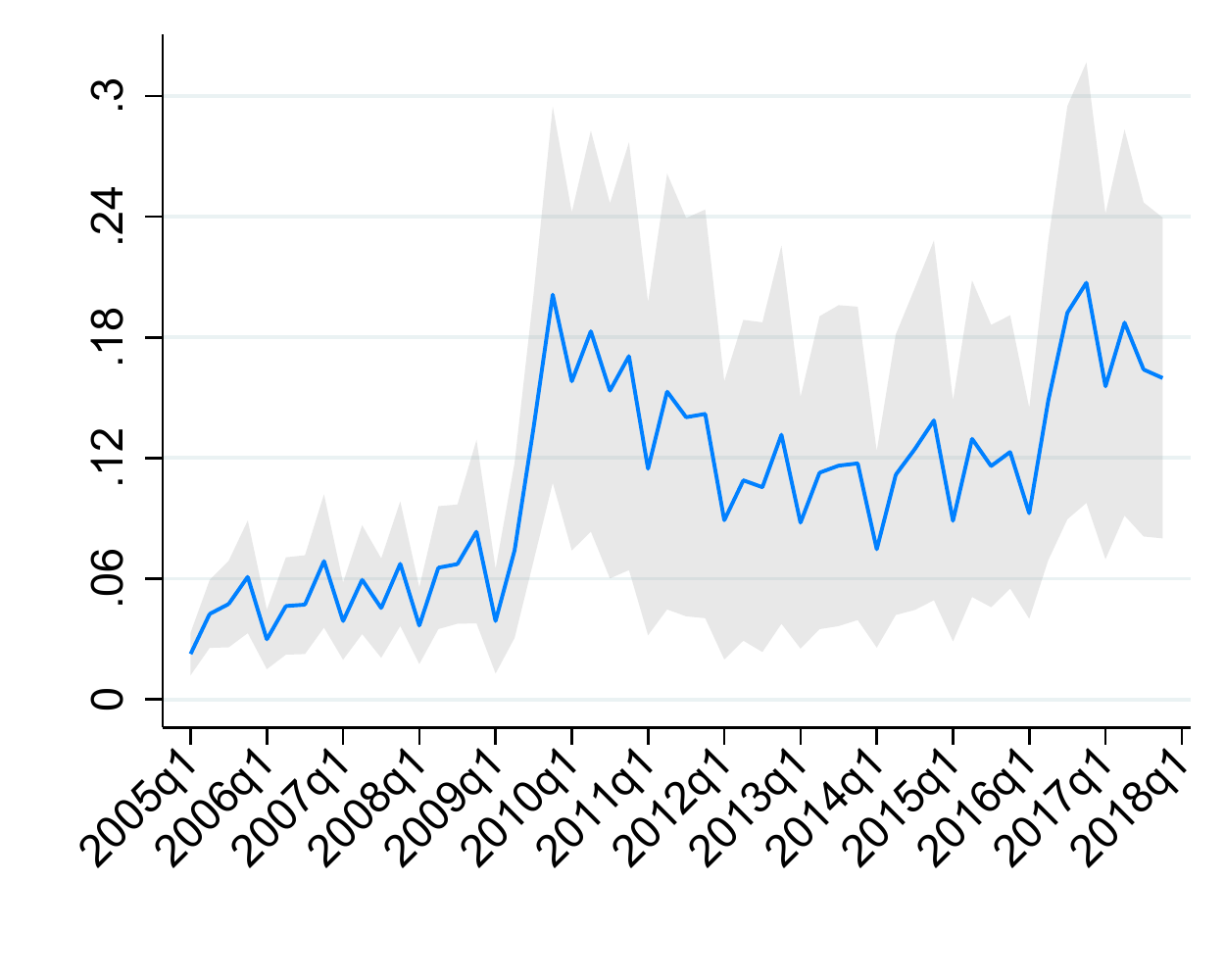}
                \end{subfigure}
                \vspace{10pt}
                \newline
                \begin{subfigure}[b]{0.49\textwidth}
                                \centering \caption{Wage subsidies} \label{fig:var_ws}
                                \includegraphics[clip=true, trim={0cm 0cm 0cm 0cm},scale=0.50]{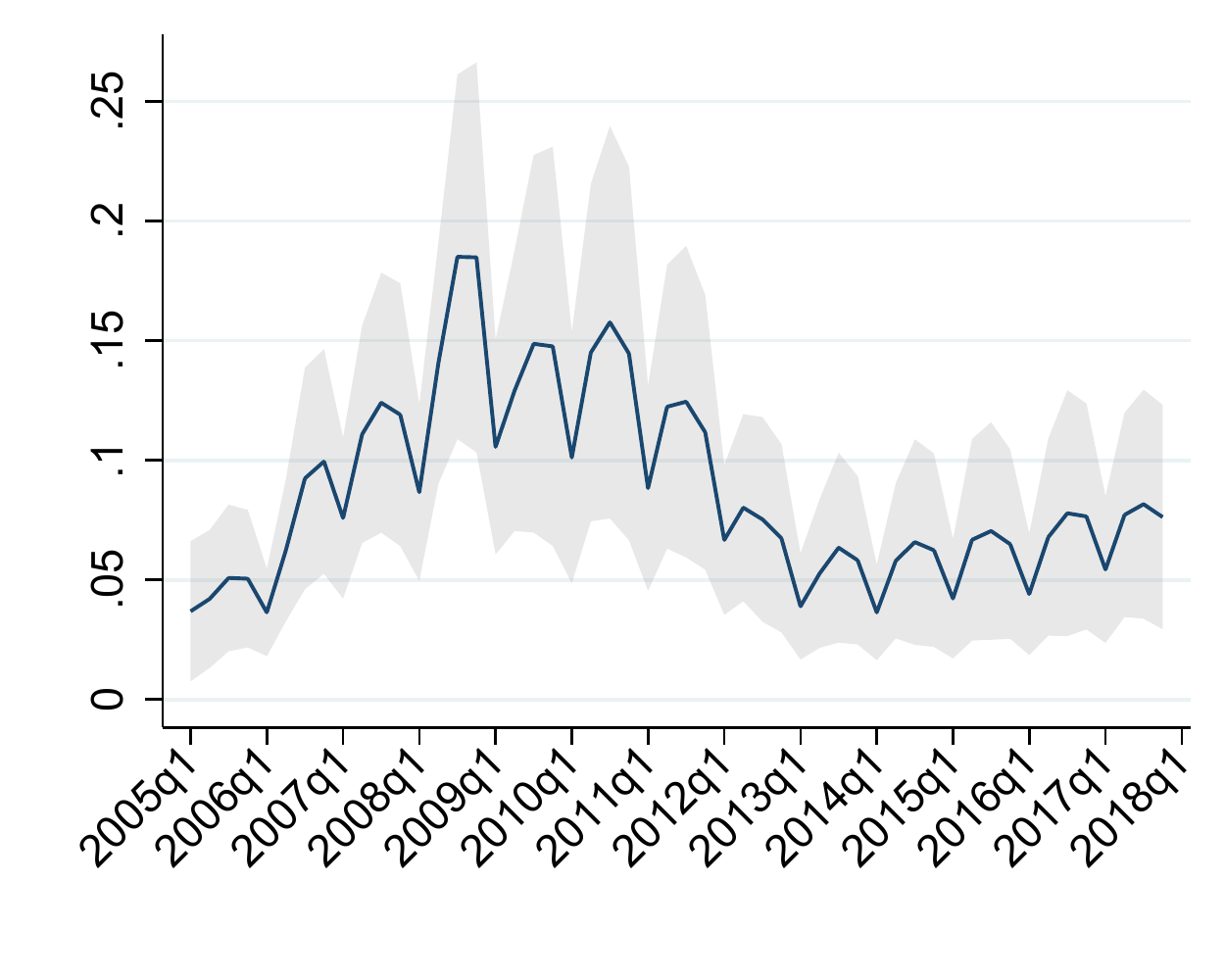}
                \end{subfigure}
                \vspace{10pt}    
                \begin{minipage}{\textwidth}
                                \footnotesize \textit{Notes:} The figures depict the average quarterly variation of the three programs  (a) training, (b) short measures and (c) wage subsidies for the period 2005 to 2018. The standard deviation is depicted as grey area.
                \end{minipage}
\end{figure}

\textbf{Description of the final dataset.} \quad  Based on our preferred definition of labour markets and main sample restriction, we obtain a balanced panel of 115 labour markets which we observe for 52 quarters, from the first quarter of 2005 to the first quarter of 2018. We use the period from 2000 to 2004 merely for the construction of lagged variables in our dynamic panel data model (see Section \ref{estimation}). Appendix tables \ref{tab:sumstats1} and \ref{tab:sumstats2} present summary statistics for all outcome, policy and control variables. Figure \ref{fig:variation_time_out} shows the variation of the main outcome variables over time, Figure \ref{fig:variation_time_prog} the variation of the policy variables. The solid line depicts the sample average in each quarter and the grey area the corresponding standard deviation.

The average unemployment rate (Figure \ref{fig:var_uerate}) is decreasing from a level of 6 to 2\% over the time period that we consider, with a temporary increase in the years around the great recession.\footnote{Notice that this rate is lower compared to the unemployment rate recorded by the Federal Statistical Office which amounts to around 11\% in 2005. This is due to our rate only recording UI recipients and the age restrictions that we impose.} It is characterised by strong seasonal fluctuations which are driven by a high number of inflows into unemployment at the beginning of each year. 
The employment rate (Figure \ref{fig:var_emprate}) is mirroring the trends in unemployment and increases from a level of 82\% to  89\% over time. The rate of welfare recipients (Figure \ref{fig:var_wf_rate}) and employed workers (Figure \ref{fig:prec_work}) on benefits increases from 2005 to 2006 and follows a slight downward trend thereafter.\footnote{This increase relates to institutional changes of the Hartz reforms and an under-reporting of cases in 2005. Our sensitivity analysis based on a shorter time period (starting from 2010) shows that this measurement problem does not drive our results.} The standard deviation is smallest for the unemployment rate and comparatively large for the other outcomes, suggesting that there is substantial variation in the share of employed, welfare recipients and employed workers on benefits across labour markets.

The seasonality in the unemployment rate is reflected in the accommodation rates of program participants. Apart from these fluctuations, the shares follow trends that are similar to the expenditure shares presented in Figure \ref{fig:exp}. The average share of unemployed in training  (Figure \ref{fig:var_training}) increases over time, with a slight drop in 2012. This drop can be explained by a larger number of individuals leaving the stock of participants who started their program during the years of the great recession.\footnote{As a matter of fact, inflows into training are relatively high in the years 2008 to 2010 but they decrease thereafter. See Figure \ref{fig:inflows}.} Further, the average share of unemployed in short measures (Figure \ref{fig:var_sm}) follows an upward trend. Additionally, it jumped from 5 to 20\% with the change in legislation in 2009. Wage subsidies (Figure \ref{fig:var_ws}) were more heavily used around the years of the great recession and remained at a level of around 5\% after 2013.
As Appendix Figure \ref{fig:var_space} shows, there is also substantial variation in the policy use and across labour markets. Wage subsidies, for example, were more heavily deployed in East Germany. This is where the unemployment rate was highest on average during our observation period.

\subsection{Estimation}\label{estimation}

To implement our empirical strategy, we rely on an estimation method that accounts for the dynamic nature of the policy effect on the outcome. The ALMP that we study range from a few weeks up to 3 years. We thus expect the effects to occur with some time lag and to potentially accumulate over time. Moreover, programs might affect intermediate outcomes. We estimate the following autoregressive distributed lag model ARDL(1,q):
\begin{center}
\begin{equation}\label{eq:ardl}
 \Delta y_{it}=\theta \Delta y_{i,t-1} + \sum_{m= 1}^{M}\sum_{j= 0}^{q}\phi_j^m \Delta X_{i,t-j}^m + \gamma W_{i,t-q} + \mu_t  +\Delta \epsilon_{it},
\end{equation}
\end{center}
where $\Delta y_{it} = y_{it}-y_{i,t-4}$ is the yearly change in the outcome $y$ in market $i$ and quarter $t$. This change is regressed on the yearly change in the lagged outcome ($\Delta y_{i,t-1}$) and the  yearly changes in the endogenous accommodation rates for the $M=3$ programs ($\Delta X_{i,t}^m, \ldots,\Delta X_{i,t-q}^m$). In our main specification accommodation rates are included with $q=6$ lags.\footnote{This choice is guided by the fact that many programs have a duration that goes beyond one year. In a robustness check we try different lag orders and find that our results are not sensitive to these model variations (see Section \ref{rob_model}).}

We estimate the model in year differences to deal with integrated time series and serial correlation in the residuals. This is necessary since our outcome and policy variables are characterised by strong yearly fluctuations (see Figures \ref{fig:variation_time_out} and \ref{fig:variation_time_prog} ). At the same time, we directly difference out labour market fixed effects and thereby all aggregate effects that are market specific and time-constant. A vector of exogenous labour market characteristics measured in $t-q$ ($W_{i,t-q}$) additionally takes care of possible changes in the labour market composition over time. It includes the gender, age, nationality, education, training and industry composition of the employed and unemployed workforce\footnote{In the specifications that consider the employment and unemployment rates of different sub-populations, we only include these characteristics as share of the total RLF.}, the share of welfare recipients (as percent of the RLF) and the shares of participants in other ALMP.\footnote{Specifically, we differentiate by programs for the long-term unemployed, programs for the young and all other programs directed at UI recipients and include them as shares of the number of UI recipients.} Finally, quarter-year fixed effects ($\mu_t$) control for aggregate time-trends such as the business cycle. Jointly this rich set of controls makes our instruments conditionally exogenous. For identification, we exploit the residual variation in policy use within labour markets and corresponding instrument areas.

The coefficient $\phi_0^m$ measures the contemporaneous (or short-term) effect of an increase in the share of program participants in program $m$ relative to all jobseekers on aggregate labour market outcomes. The long-run effect is given by $\sum_{j= 0}^{q}\phi_j^m/(1-\theta)$.  We also derive the marginal and cumulative marginal effects of a permanent increase in the shares of unemployed in a specific program type for different quarters.\footnote{Notice that the effects beyond the maximum lag length entirely depend on the auto-regressive parameter $\theta$.} The effects can be interpreted as aggregate effects of an increase in the intensity of a specific ALMP. They measure the net effect of these policies on labour market outcomes by jointly considering the direct effect on participants and possible positive or negative spillover effects on non-participants.

We estimate the model both using OLS and two-stage least-squares (2SLS). For the latter, we instrument the endogenous variables $\Delta X_{i,t}^{m} \ldots,\Delta X_{i,t-q}^{m} $ with the same set of instruments $\Delta Z_{i,t}^{m},\ldots,\Delta Z_{i,t-q}^{m}$. $\Delta Z_{i,t}^{m}$ that measure the accommodation rates in the instrument regions. 
 Our instrumental variable model is just-identified, with 21 endogenous regressors and instruments in our main specification of lag order 6. We do not instrument for the lagged dependent variable but argue that any bias from $Cov(\Delta y_{i,t-1},\Delta\epsilon_{it}) \neq 0$ would be negligible since we have a long time series with $T>50$.\footnote{We show that the lagged dependent variable and the error term are correlated in Appendix \ref{bias}. \cite{Nickel1981} showed that the bias in the least square dummy variable estimator is decreasing with the time dimension.} We assess this potential bias in a robustness check by comparing our results to a distributed lag model that does not include a lagged depended variable. The results are qualitatively comparable across models. 
We estimate standard errors using a cross-sectional bootstrap procedure where we re-sample entire labour markets.\footnote{ \cite{Kapetanios2008} and \cite{Galvao2014} show that this procedure is asymptotically valid for fixed effects estimators if the data do not exhibit cross-sectional but temporal dependence.}

\section{Results}\label{results}


\subsection{First Stage}\label{sec:fs}

Before turning to the results, we test whether our proposed instruments are relevant. In a single endogenous variable model it is common to use the standard first-stage F-statistic to test for weakness of the instruments. Since we instrument for 21 endogenous variables in our model, this statistic is no longer sufficient. Thus, we perform two alternative tests instead, which were developed for the setting with multiple endogenous variables. \cite{Sanderson2016} propose a test that is constructed for individual endogenous regressors by partialling-out linear projections of the remaining endogenous regressors. A similar approach by \cite{Angrist2009} derives conditional F-statistics by replacing other endogenous variables by their reduced form prediction. The corresponding F-statistics and p-values for the first-stage regressions are reported in Table \ref{tab:fs_ue_rate} exemplarily for the model including the lagged dependent unemployment rate.\footnote{We find similar F-statistics based on the models with different outcome variables. See Appendix Tables \ref{tab:fs_emp_rate} to \ref{tab:fs_prec_work_rate}.} We find large and highly significant F-statistics for all endogenous policy variables suggesting that instrument relevance is given. The F-statistics are highest for wage subsidies.\footnote{Notice that the F-statistics increase once we impose stricter criteria on the minimum overlap between the labour market and the employment agencies, as described in Section \ref{rob_llm_ov}.}

\subsection{Effects on Aggregate Labour Market Outcomes}\label{sec:main_results}

\begin{figure}[H]
                \centering
                \caption{Cumulative Marginal Effects on the Unemployment Rate and Unsubsidised Employment Rate \label{fig:cum_effects_iv_alo}}
                \vspace{10pt}
                \begin{subfigure}[b]{0.49\textwidth}
                                \centering \caption*{Training} \subcaption*{Unemployment rate} 
                                \includegraphics[clip=true, trim={0cm 0cm 0cm 0cm},scale=0.50]{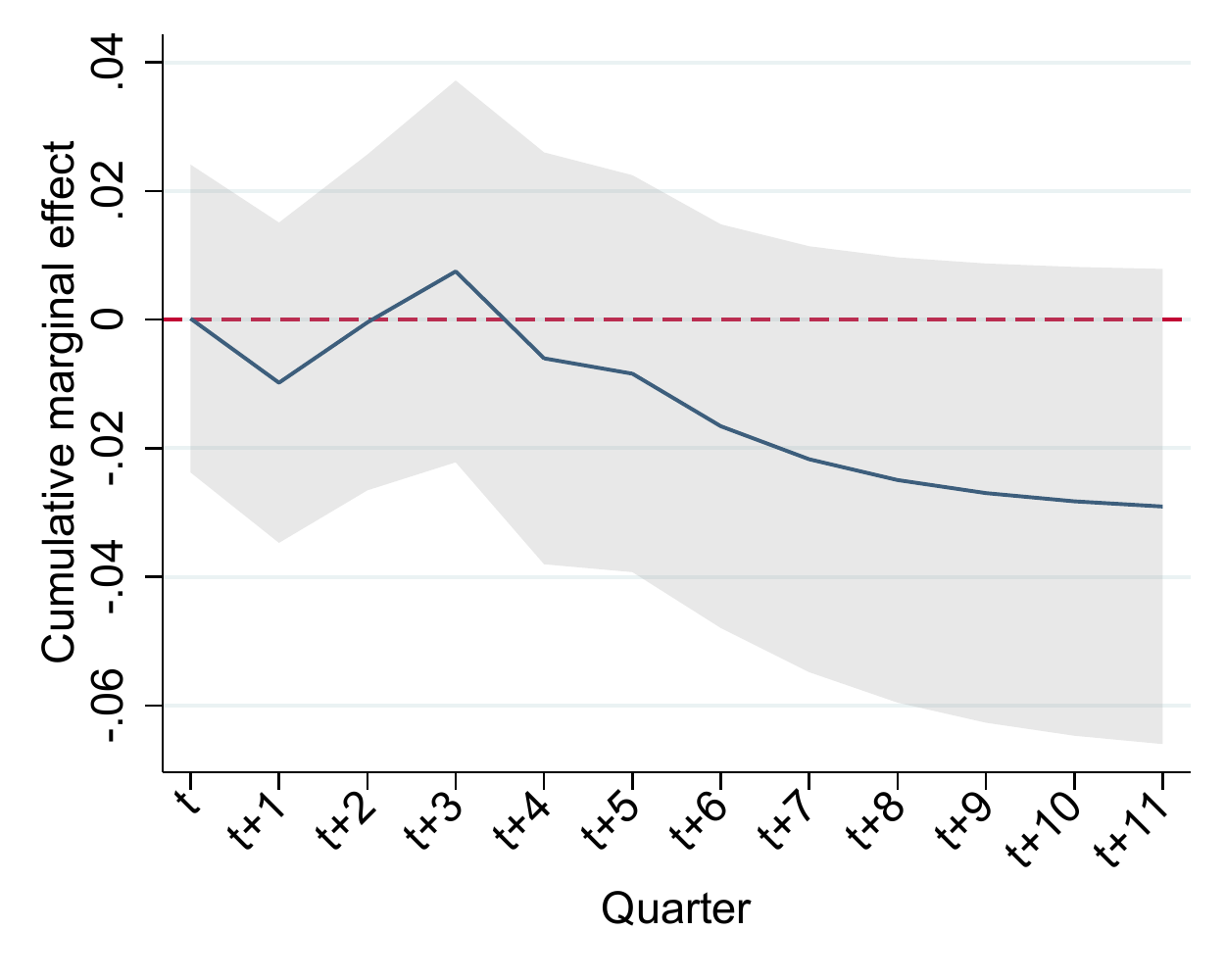}
                \end{subfigure}
                \begin{subfigure}[b]{0.49\textwidth}
                                \centering \caption*{Training} \subcaption*{Unsubsidised employment rate} 
                                \includegraphics[clip=true, trim={0cm 0cm 0cm 0cm},scale=0.50]{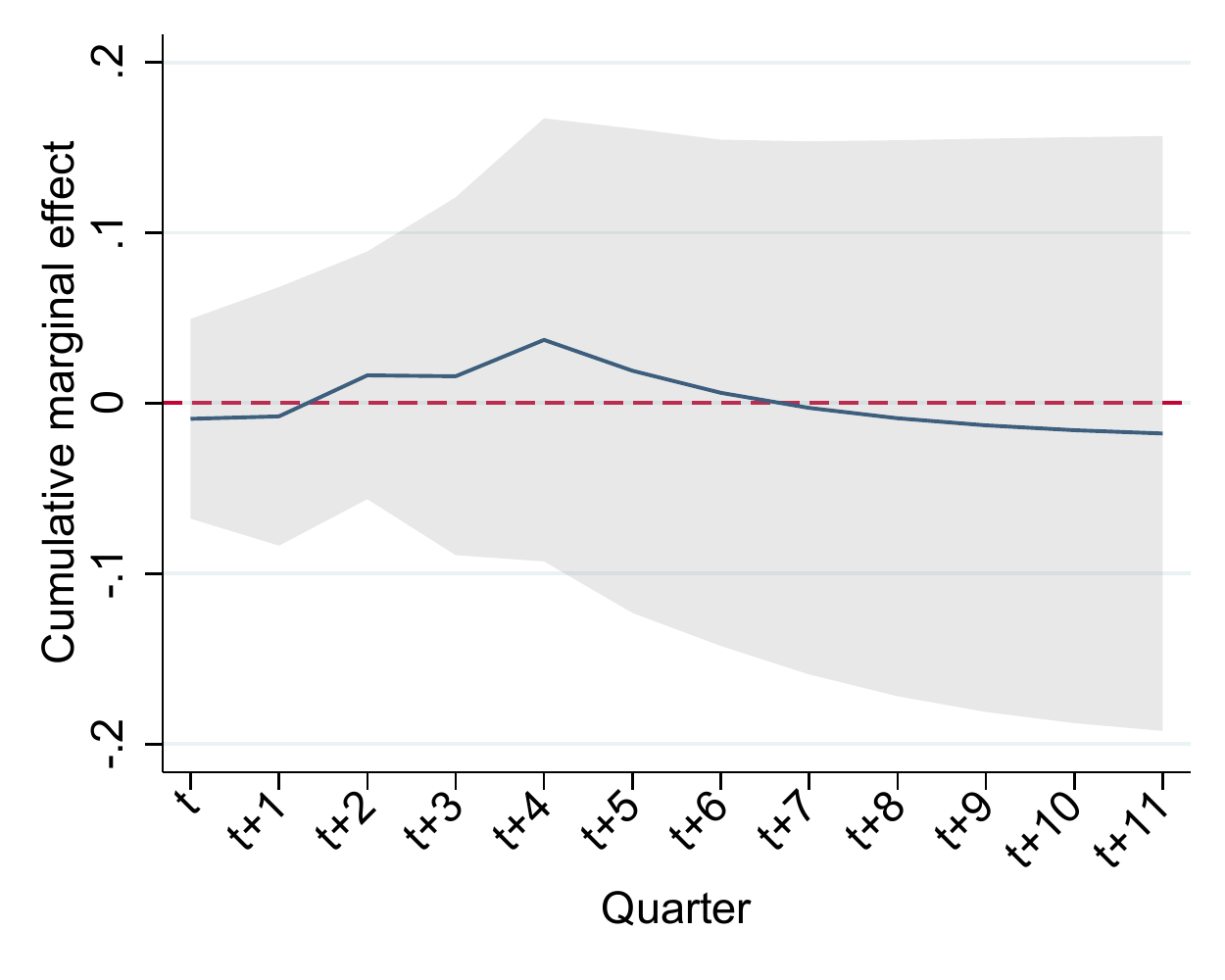}
                \end{subfigure}
                \vspace{10pt}    
                \newline
                \begin{subfigure}[b]{0.49\textwidth}
                                \centering \caption*{Short measures} \subcaption*{Unemployment rate} 
                                \includegraphics[clip=true, trim={0cm 0cm 0cm 0cm},scale=0.50]{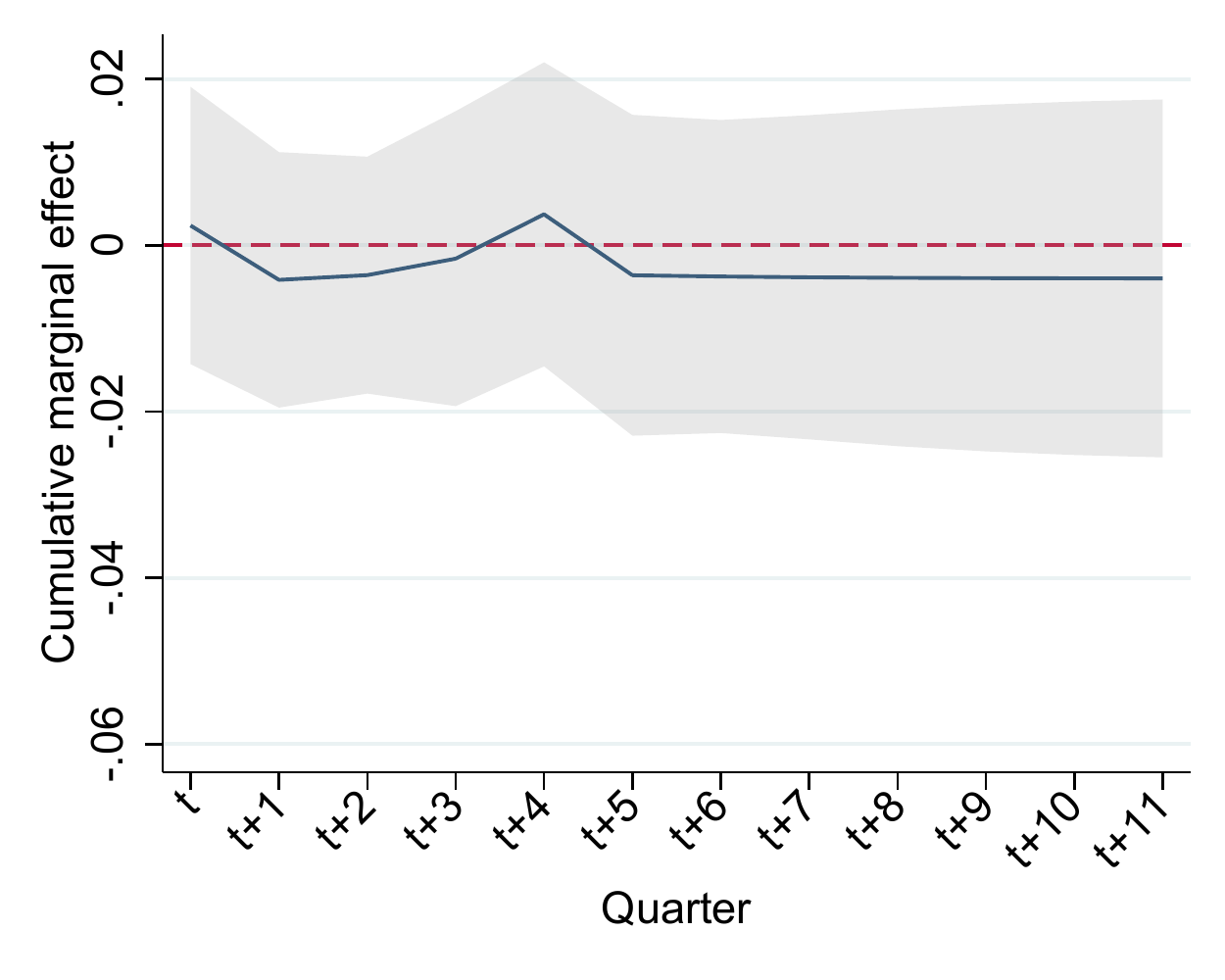}
                \end{subfigure}
                \vspace{10pt}    
                \begin{subfigure}[b]{0.49\textwidth}
                                \centering \caption*{Short measures}  \subcaption*{Unsubsidised employment rate} 
                                \includegraphics[clip=true, trim={0cm 0cm 0cm 0cm},scale=0.50]{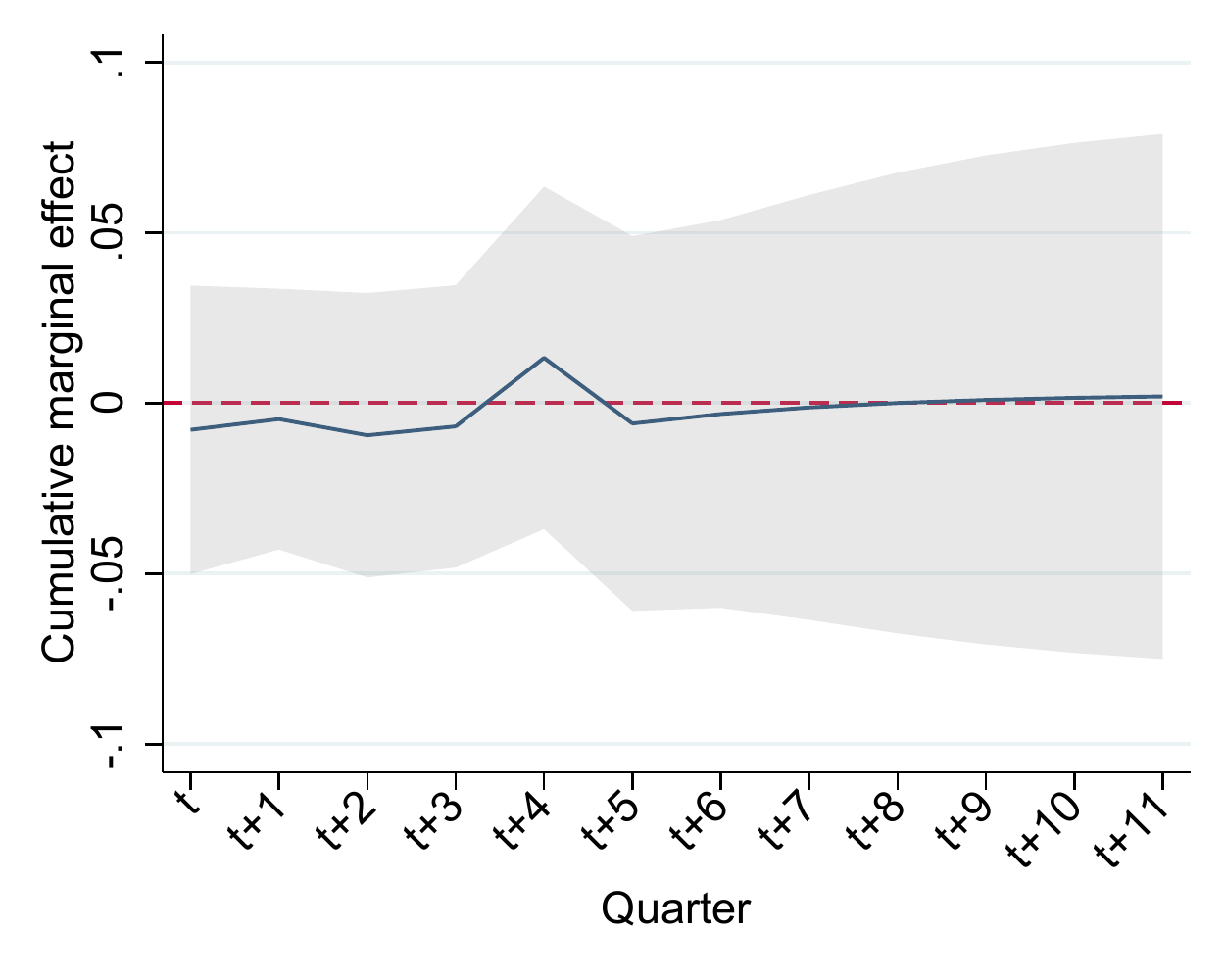}
                \end{subfigure}
                \vspace{10pt}
                \newline
                \begin{subfigure}[b]{0.49\textwidth}
                                \centering \caption*{Wage subsidies} \subcaption*{Unemployment rate} 
                                \includegraphics[clip=true, trim={0cm 0cm 0cm 0cm},scale=0.50]{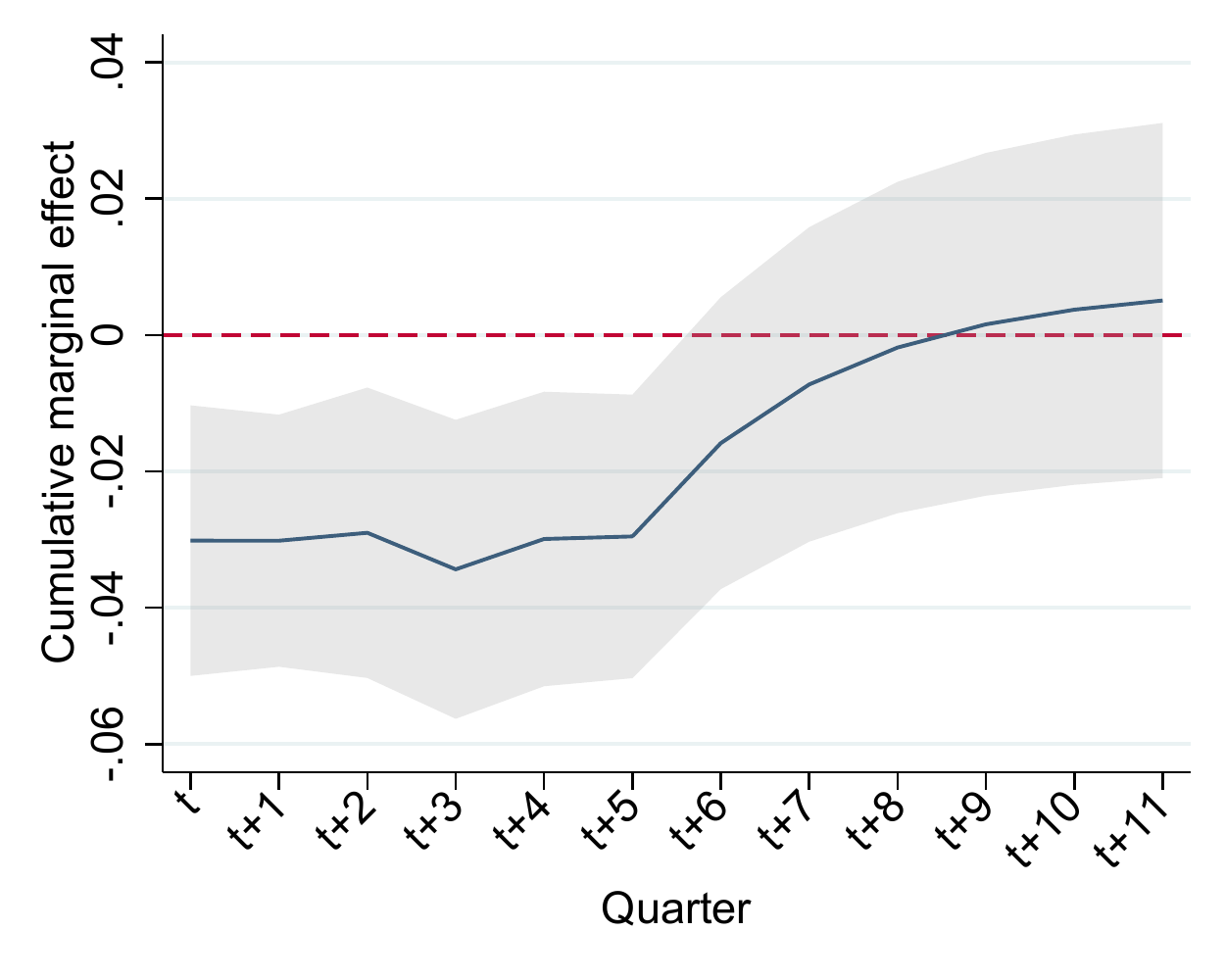}
                \end{subfigure}
                \vspace{10pt}    
                \begin{subfigure}[b]{0.49\textwidth}
                                \centering \caption*{Wage subsidies}  \subcaption*{Unsubsidised employment rate} 
                                \includegraphics[clip=true, trim={0cm 0cm 0cm 0cm},scale=0.50]{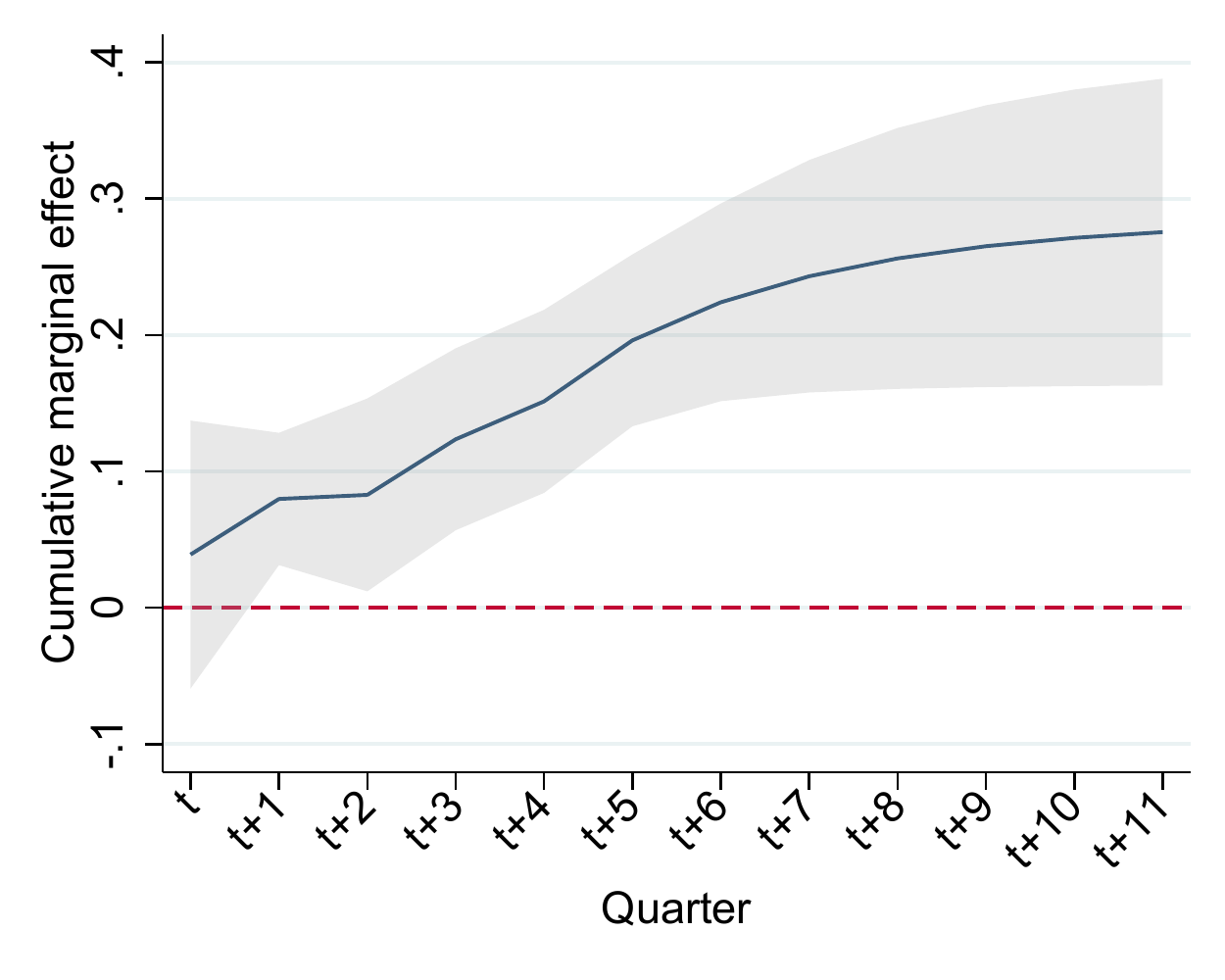}
                \end{subfigure}
                \vspace{10pt}
                \begin{minipage}{\textwidth}
                                \footnotesize \textit{Notes:} These graphs show the marginal and cumulative marginal effects of the three types of ALMP, i.e. training, short measures and wage subsidies on the unemployment rate and unsubsidised employment rate by quarter. 95\% confidence intervals are shown as grey areas. The effects are based on the ARDL model estimated by 2SLS. Program variables are included with 6 lags, main sample restrictions apply.  Standard errors obtained by a cross-sectional bootstrap (499 replications).
                \end{minipage}
\end{figure}

\begin{figure}[H]
                \centering
                \caption{Cumulative Effects on the Rate of Welfare Recipients and Employed Workers on Benefits \label{fig:cum_effects_iv_prec_work}}
                \vspace{10pt}
                \begin{subfigure}[b]{0.49\textwidth}
                                \centering \caption*{Training} \subcaption*{Rate of welfare recipients} 
                                \includegraphics[clip=true, trim={0cm 0cm 0cm 0cm},scale=0.50]{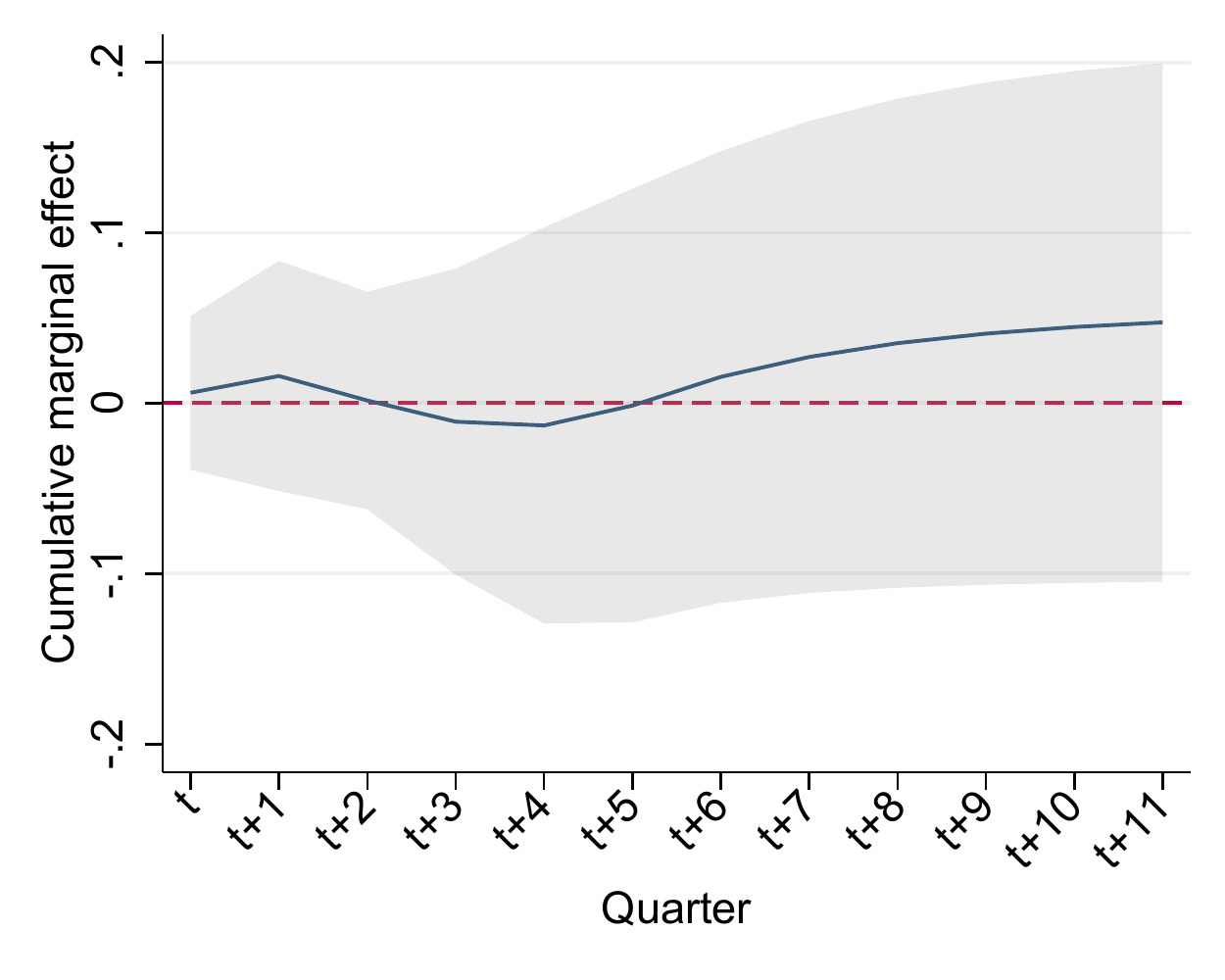}
                \end{subfigure}
                \begin{subfigure}[b]{0.49\textwidth}
                                \centering \caption*{Training} \subcaption*{Rate of employed workers on benefits} 
                                \includegraphics[clip=true, trim={0cm 0cm 0cm 0cm},scale=0.50]{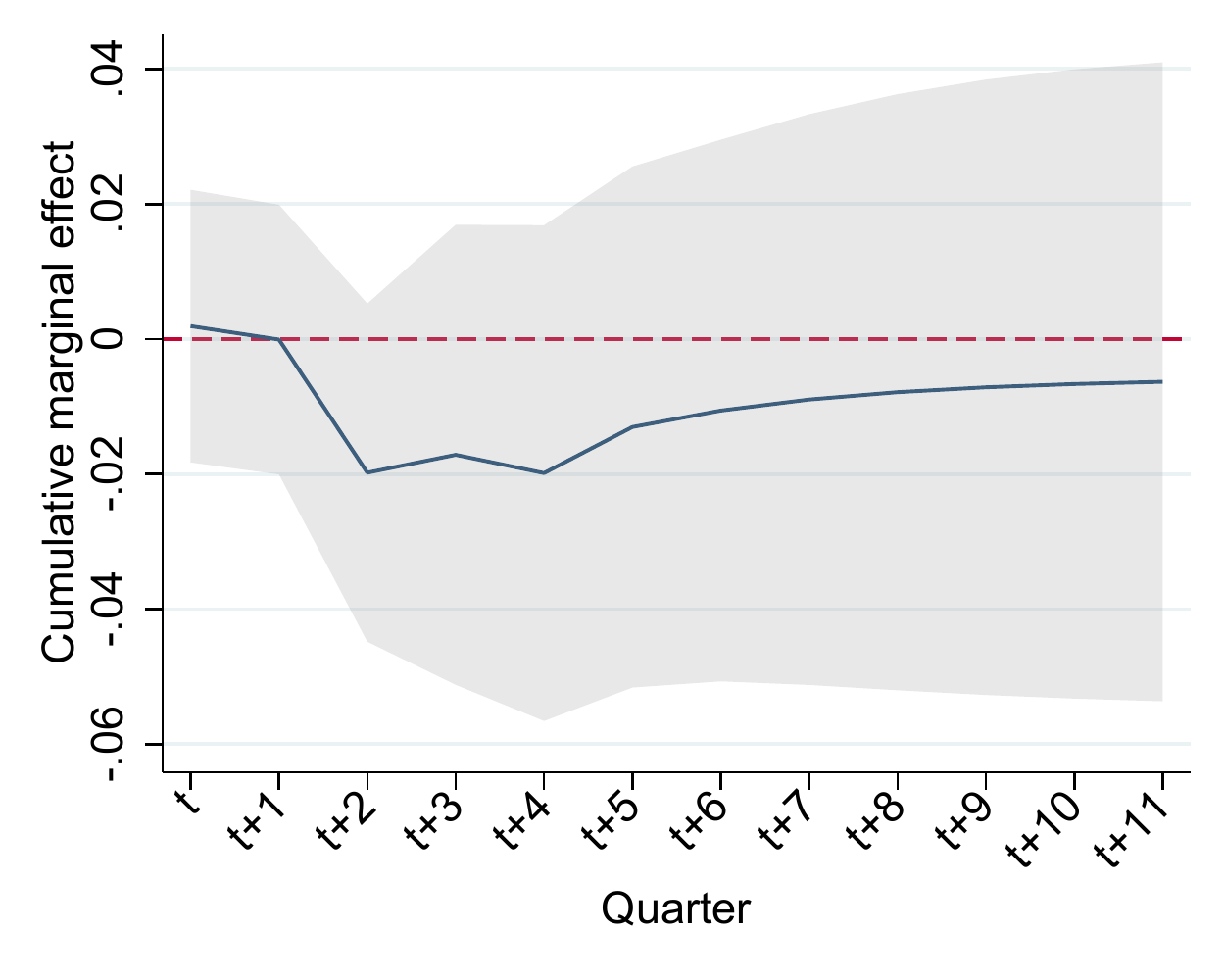}
                \end{subfigure}
                \vspace{10pt}    
                \newline
                \begin{subfigure}[b]{0.49\textwidth}
                                \centering \caption*{Short measures} \subcaption*{Rate of welfare recipients} 
                                \includegraphics[clip=true, trim={0cm 0cm 0cm 0cm},scale=0.50]{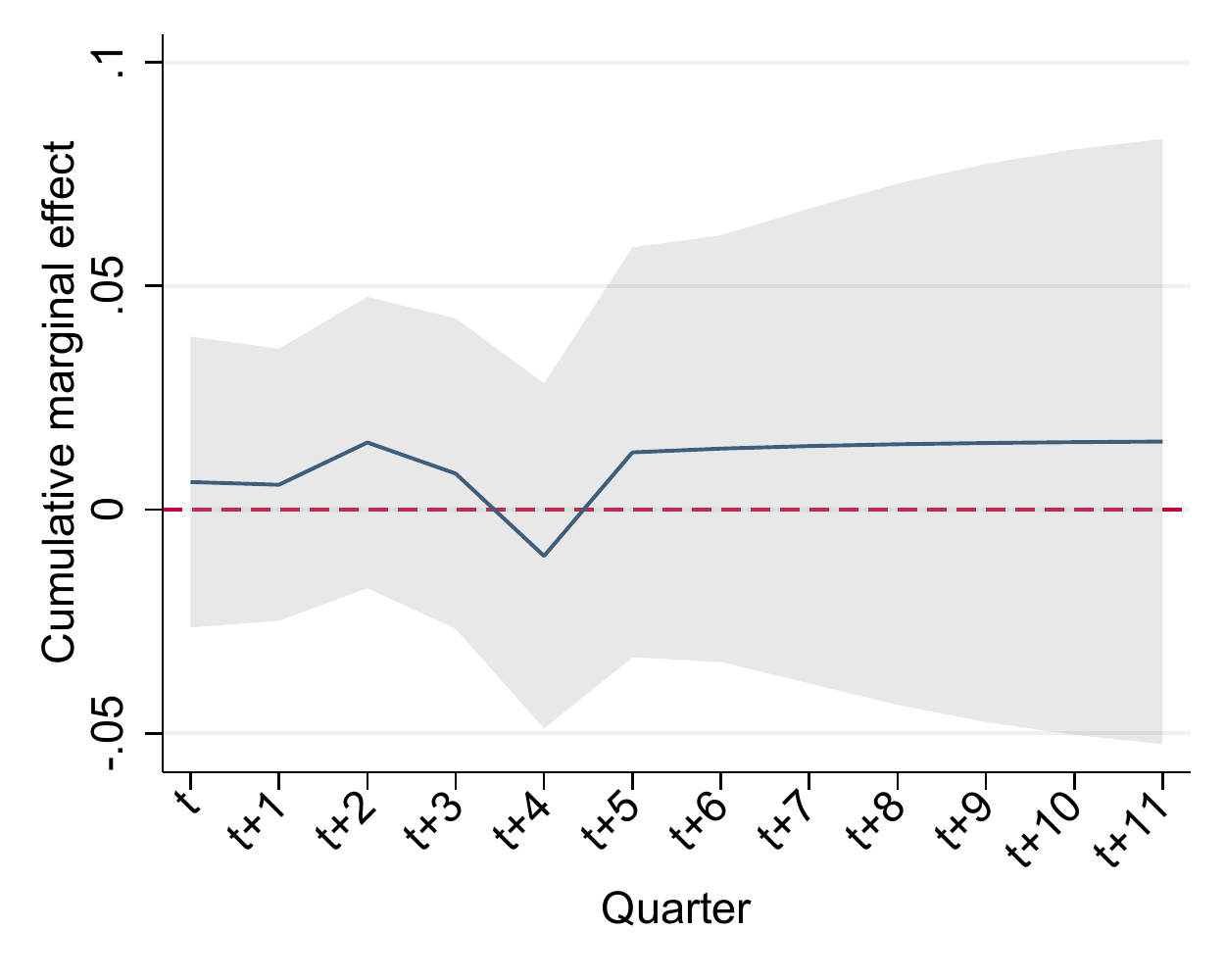}
                \end{subfigure}
                \vspace{10pt}    
                \begin{subfigure}[b]{0.49\textwidth}
                                \centering \caption*{Short measures}  \subcaption*{Rate of employed workers on benefits} 
                                \includegraphics[clip=true, trim={0cm 0cm 0cm 0cm},scale=0.50]{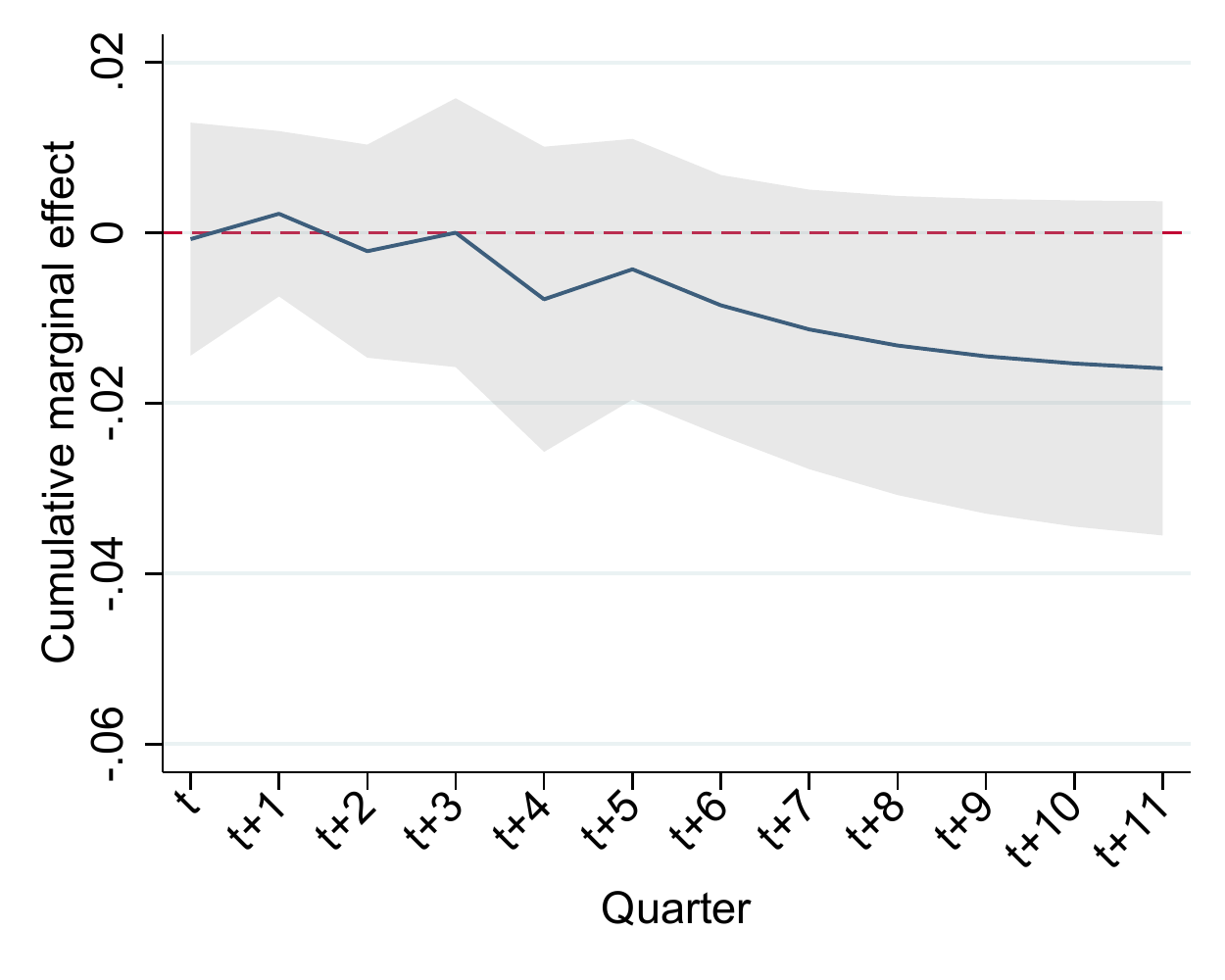}
                \end{subfigure}
                \vspace{10pt}
                \newline
                \begin{subfigure}[b]{0.49\textwidth}
                                \centering \caption*{Wage subsidies} \subcaption*{Rate of welfare recipients} 
                                \includegraphics[clip=true, trim={0cm 0cm 0cm 0cm},scale=0.50]{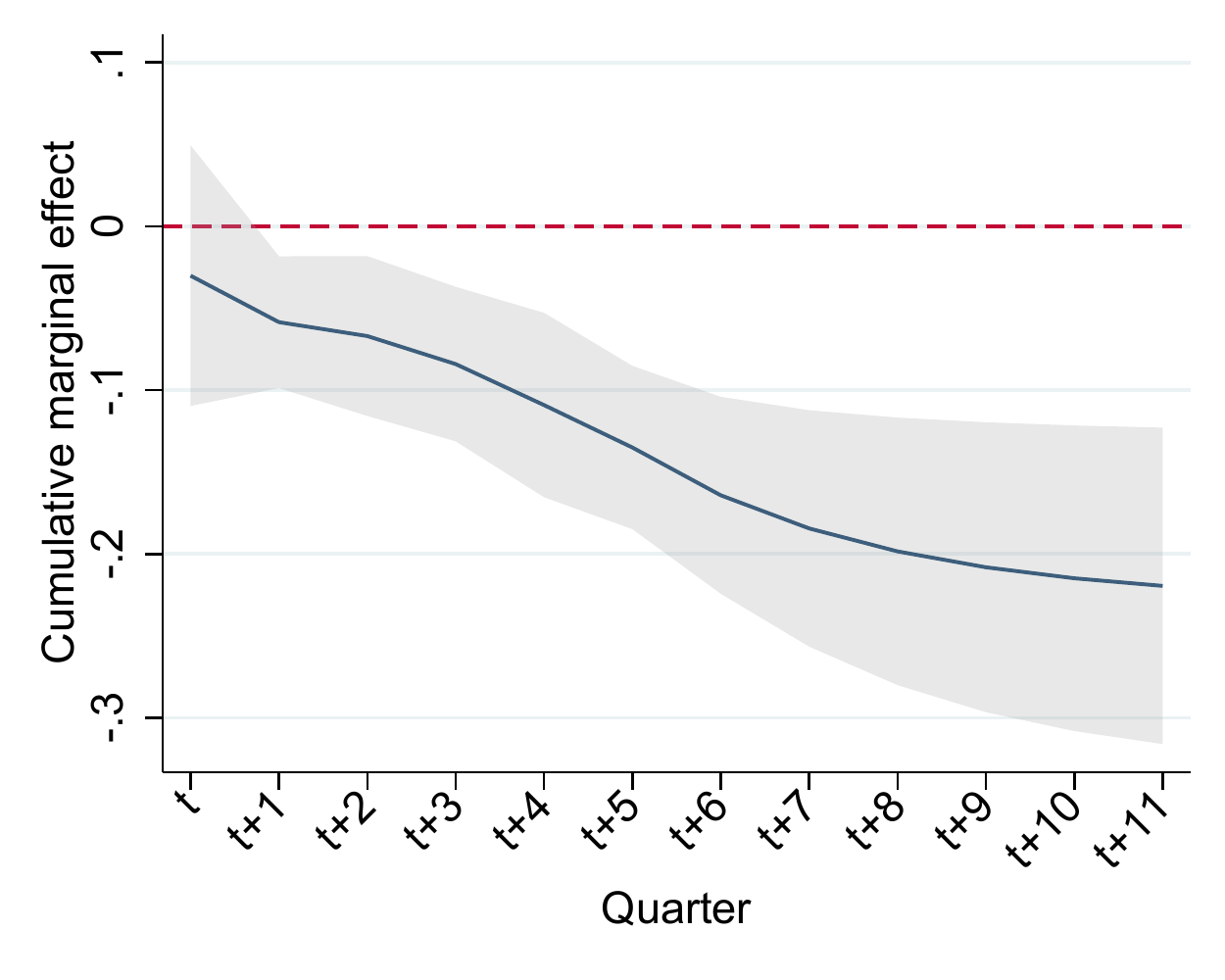}
                \end{subfigure}
                \vspace{10pt}    
                \begin{subfigure}[b]{0.49\textwidth}
                                \centering \caption*{Wage subsidies}  \subcaption*{Rate of employed workers on benefits} 
                                \includegraphics[clip=true, trim={0cm 0cm 0cm 0cm},scale=0.50]{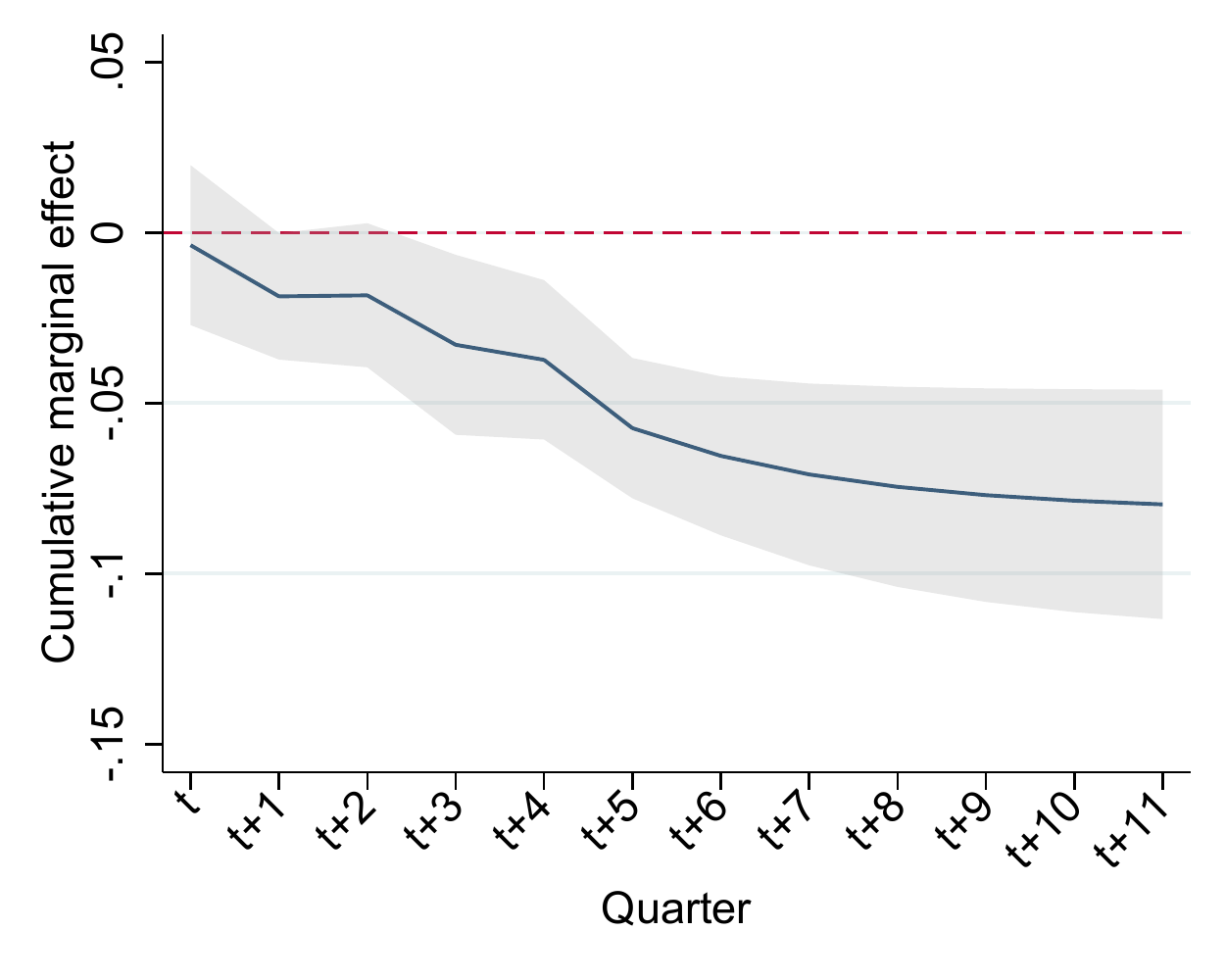}
                \end{subfigure}
                \vspace{10pt}
                \begin{minipage}{\textwidth}
                                \footnotesize \textit{Notes:} These graphs show the marginal and cumulative marginal effects of the three types of ALMP, i.e. training, short measures and wage subsidies on the rate of welfare recipients and rate of employed workers on benefits by quarter. 95\% confidence intervals are shown as grey areas. The effects are based on the ARDL model estimated by 2SLS. Program variables are included with 6 lags, main sample restrictions apply.  Standard errors  are obtained by a cross-sectional bootstrap (499 replications).
                \end{minipage}
\end{figure}

We present our main results for the unemployment rate and the unsubsidised employment rate\footnote{We will use the terms employment rate and unsubsidised employment interchangeably in the discussion of the results since we only measure the effects on unsubsidised employment.}  in Table \ref{tab:main_alo_emp} and for the rate of welfare recipients and the rate of employed workers that receive state benefits (the so-called working poor) in Table \ref{tab:main_wf_prec_work}. Columns (1) to (3) show the effects, standard errors and p-values based on the estimation of the ARDL model using OLS.  Columns (4) to (6) show the same quantities estimated by 2SLS. The tables present the short-term and long-term effects of a ceteris paribus increase in the shares of participants in training, short measures and wage subsidies.  The  first row of each panel shows  the coefficient on the lagged dependent variable.  Figures  \ref{fig:cum_effects_iv_alo} and \ref{fig:cum_effects_iv_prec_work} further inform about the dynamics behind the effects. They show the cumulative marginal effects on the labour market outcomes up to 12 quarters after an increase in the shares of program participants.

\begin{table}[H]
\centering
\caption{First Stage - Unemployment Rate}
\label{tab:fs_ue_rate}
{\small
{
\def\sym#1{\ifmmode^{#1}\else\(^{#1}\)\fi}
\begin{tabular}{l*{4}{c}}
\hline\hline
            &   SW F-stat&    SW p-val&   AP F-stat&    AP p-val\\
\hline
AR training &      108.26&       0.000&      123.08&       0.000\\
AR training (lag 1)&       79.51&       0.000&      142.96&       0.000\\
AR training (lag 2)&       91.79&       0.000&       97.79&       0.000\\
AR training (lag 3)&       85.31&       0.000&       69.23&       0.000\\
AR training (lag 4)&       83.68&       0.000&       97.21&       0.000\\
AR training (lag 5)&       72.93&       0.000&       98.37&       0.000\\
AR training (lag 6)&      107.81&       0.000&      265.05&       0.000\\
AR short measures&       61.26&       0.000&      364.19&       0.000\\
AR short measures (lag 1)&       56.50&       0.000&      146.69&       0.000\\
AR short measures (lag 2)&       83.31&       0.000&      155.23&       0.000\\
AR short measures (lag 3)&       47.92&       0.000&      143.34&       0.000\\
AR short measures (lag 4)&       63.36&       0.000&      146.18&       0.000\\
AR short measures (lag 5)&       59.32&       0.000&      150.48&       0.000\\
AR short measures (lag 6)&       67.61&       0.000&      225.94&       0.000\\
AR wage subsidies&       98.99&       0.000&      398.67&       0.000\\
AR wage subsidies (lag 1)&      102.54&       0.000&      418.28&       0.000\\
AR wage subsidies (lag 2)&       95.11&       0.000&      187.66&       0.000\\
AR wage subsidies (lag 3)&      103.43&       0.000&      245.01&       0.000\\
AR wage subsidies (lag 4)&      107.53&       0.000&      211.31&       0.000\\
AR wage subsidies (lag 5)&       83.56&       0.000&      225.16&       0.000\\
AR wage subsidies (lag 6)&      120.17&       0.000&      490.66&       0.000\\
\hline\hline
\end{tabular}

}
}
\centering
     \begin{minipage}{\textwidth}
          \vspace{6pt}
          \footnotesize \textit{Notes:} The table reports two types of diagnostic test of the first stage for the 21 excluded instruments for the model including the lagged dependent unemployment rate. Each line refers to a separate first-stage regression for the respective accommodation rate (AR). The first two columns show the Sanderson-Windmeijer (2016) F statistics (SW F-stat) and p-values. Columns 3 and 4 the Angrist-Pischke (2009) F-statistics and p-values (AP F-stat).
     \end{minipage}
\end{table}

In both estimation methods, we find the lagged dependent variable to be significant. The estimates range from 0.63 to 0.7 and are lowest for the the unemployment rate. This suggests that there is indeed a temporal adjustment process. 
Based on the OLS estimation we find small negative long-term effects of all programs on the unemployment rate, the rate of welfare recipients and the rate of employed workers on benefits and a simultaneous positive effect on the employment rate. The effects are small in magnitude for short measures and training and larger for wage subsidies. The effects estimated by 2SLS are deviating in sign and magnitude, in particular for training and short measures. OLS seems to overestimate the effects of these policies on the employment rate. As expected, standard errors based on 2SLS are larger compared to the OLS model. 

In the following, we only discuss the results based on the instrumental variables model. We do not find a significant effect of short measures on aggregate labour market outcomes. If anything, the results suggest that short measures slightly reduce the share of employed workers on benefits. For training we find no effects in the short run but a small marginally significant effect on the unemployment rate in the long run.
The cumulative marginal effects in Figures \ref{fig:cum_effects_iv_alo} and \ref{fig:cum_effects_iv_prec_work} suggest that this effect starts to materialise after approximately four quarters. The employment rate does not increase. Instead, our results suggest that training reduces unemployment of UI recipients at the expense of unemployed welfare recipients, although the corresponding estimates in Table \ref{tab:main_wf_prec_work} are too imprecise for statistical significance. For wage subsidies we find a small negative short-term effect on the unemployment rate, which most likely picks up the effect of jobseekers moving out of unemployment and entering subsidised employment when starting this program. As can be seen from Figure 
\ref{fig:cum_effects_iv_alo}, this negative effect only derives from the first quarter since the marginal effect on unemployment is zero and mostly positive thereafter. In the long run, wage subsidies have an economically meaningful and statistically significant positive effect on the employment rate. The unsubsidised employment rate increases by 2.9 percentage points for a 10 percentage point increase in the share of jobseekers (on UI) in wage subsidies. Of this increase, 80\% is explained by a decrease in the rate of welfare recipients and 20\% by a decrease in the rate of employed workers on benefits. Thus, wage subsidies increase employment, prevent long-term unemployment after exhaustion of UI and reduce the number of working poor.

Overall, our results are in line with our expectations in Section \ref{theory}.  
First, we did not expect to find large aggregate effects for short measures, given that documented individual-level effects have been small. This was confirmed by our empirical analysis. 
Second, we find no evidence of an aggregate net effect of training programs on employment. Microeconometric studies document large lock-in effects for program participants in the short-term and positive employment effects in the long-term \citep{McCall2016}. Our results suggest that there are unintended effects on non-participants that are particularly likely if the number of vacancies that can be filled is limited and much lower than the number of jobseekers, which was the case in Germany in our observation period (\citealp{Seele2018}, Figure 1.1.) In the short run, firms might hire non-participants in place of program participants who are not available to the market during the program. In the long run, training participants may seem to find jobs at the expense of non-participants. Our results suggest that unemployed welfare recipients may suffer from these substitution effects.  
Third, the positive effects of wage subsidies on unsubsidised employment are in line with the effects that have been documented on the individual-level (see e.g. \citealp{Stephan2010, Jeanichen2011}). This does not rule out the existence of displacement effects, but we find that positive employment effects dominate on the aggregate. This suggests that wage subsidies succeeded in creating new jobs and in increasing matching efficiency regarding existing jobs. A higher intensity of wage subsides does not lower the rate of jobseekers in UI but reduces the share of long-term unemployed on welfare and the share of workers that are stuck in precarious work. Thus, this program effectively prevents long-term dependence of jobseekers on state benefits. 


\subsection{Effects on Unemployment and Employment Rates for Different Segments of the Labour Market}\label{sec:main_results_subgroup} 

In addition to looking at average outcomes for the entire labour market, we examine whether specific segments are differentially affected by ALMP.  There are several reasons why such an analysis is interesting. First, the literature on individual-level effects of  ALMP has documented substantial effect heterogeneity across participants with different characteristics. Regarding the gender of participants, females were found to benefit more both from training and wage subsidies (e.g. \citealp{Card2018, Biewen2014, Kruppe2018}). The evidence on heterogeneity with regard to age and skill levels is mixed \citep{Card2018, Osikominu2013}. Generally, jobseekers with relatively bad labour market prospects have been shown to benefit more from programs \citep{Wunsch2008, Cockx2020, Knaus2020}.

This heterogeneity might also materialise at the regional level. Second, if negative externalities exist, they might not affect all individuals on the labour market equally. For instance, more vulnerable groups such as low-skilled workers are more likely to suffer from indirect effects. Finally, there is selectivity in the assignment of programs to jobseekers. If certain groups of jobseekers have no or less access to ALMP, they are unlikely to draw direct benefits from the programs but might be affected by indirect spillover effects.

\begin{table}[H]
\centering
\caption{Effects on the Unemployment Rate and Unsubsidised Employment Rate}
\label{tab:main_alo_emp}
{\small
\def\sym#1{\ifmmode^{#1}\else\(^{#1}\)\fi}
\begin{tabular}{l c c c c c c }
\hline\hline
&\multicolumn{3}{c}{OLS} &\multicolumn{3}{c}{2SLS} \\ 
&\multicolumn{1}{c}{(1)} &\multicolumn{1}{c}{(2)} &\multicolumn{1}{c}{(3)}&\multicolumn{1}{c}{(4)} &\multicolumn{1}{c}{(5)} &\multicolumn{1}{c}{(6)} \\ 
\hline
\multicolumn{1}{l}{\textit{Unemployment rate}} \\
            &      effect&          se&       p-val&      effect&          se&       p-val\\
Lagged dependent variable&       0.656&       0.010&       0.000&       0.629&       0.035&       0.000\\
Training(st)&      -0.017&       0.002&       0.000&       0.000&       0.012&       0.989\\
Training(lt)&      -0.010&       0.005&       0.032&      -0.030&       0.019&       0.116\\
Short measures (st)&      -0.007&       0.001&       0.000&       0.002&       0.009&       0.780\\
Short measures (lt)&      -0.006&       0.003&       0.022&      -0.004&       0.011&       0.721\\
Wage subsidies (st)&      -0.034&       0.002&       0.000&      -0.030&       0.010&       0.003\\
Wage subsidies (lt)&      -0.017&       0.006&       0.009&       0.007&       0.014&       0.589\\

\\ \hline
\multicolumn{1}{l}{\textit{Unsubsidised employment rate}} \\
            &      effect&          se&       p-val&      effect&          se&       p-val\\
Lagged dependent variable&       0.701&       0.014&       0.000&       0.685&       0.013&       0.000\\
Training(st)&       0.025&       0.006&       0.000&      -0.009&       0.030&       0.759\\
Training(lt)&       0.050&       0.019&       0.008&      -0.022&       0.092&       0.811\\
Short measures (st)&       0.004&       0.004&       0.240&      -0.008&       0.022&       0.718\\
Short measures (lt)&       0.025&       0.011&       0.032&       0.003&       0.042&       0.945\\
Wage subsidies (st)&       0.022&       0.008&       0.006&       0.039&       0.050&       0.436\\
Wage subsidies (lt)&       0.141&       0.020&       0.000&       0.285&       0.062&       0.000\\
\\
\hline
Quarter-Year FEs&      $\checkmark$ & & &  $\checkmark$ & &       \\
Additional Controls&      $\checkmark$ & & &  $\checkmark$ & &       \\
Observations& 5980 & & &  5980 & &       \\
LLM & 115 & & &  115 & &       \\
\hline\hline
\end{tabular}%
}
\centering
     \begin{minipage}{\textwidth}
          \vspace{6pt}
          \footnotesize \textit{Notes:} This table shows the coefficient on the lagged dependent variable, the short-term (st) and long-term (lt) effects of the three types of ALMP, i.e. training, short measures (SM) and wage subsdies (wagesub) on the unemployment rate (upper panel) and unsubsidised employment rate (lower panel). Columns (1)-(3) show the effects, standard errors (se) and p-values (p-val) for the OLS specification. Columns (4)-(6) show effects, standard errors and p-values for the IV specification. Program variables included with 6 lags, main sample restrictions.  Standard errors obtained by a cross-sectional bootstrap (499 replications).  \end{minipage}
\end{table}

\begin{table}[H]
\centering
\caption{Effects on the Rate of Welfare Recipients and Individuals in Precarious Work}
\label{tab:main_wf_prec_work}
{\small
\def\sym#1{\ifmmode^{#1}\else\(^{#1}\)\fi}
\begin{tabular}{l c c c c c c }
\hline\hline
&\multicolumn{3}{c}{OLS} &\multicolumn{3}{c}{2SLS} \\ 
&\multicolumn{1}{c}{(1)} &\multicolumn{1}{c}{(2)} &\multicolumn{1}{c}{(3)}&\multicolumn{1}{c}{(4)} &\multicolumn{1}{c}{(5)} &\multicolumn{1}{c}{(6)} \\ 
\hline
\multicolumn{1}{l}{\textit{Rate of welfare recipients}} \\
            &      effect&          se&       p-val&      effect&          se&       p-val\\
Lagged dependent variable&       0.701&       0.014&       0.000&       0.693&       0.012&       0.000\\
Training(st)&      -0.007&       0.005&       0.176&       0.006&       0.023&       0.790\\
Training(lt)&      -0.029&       0.017&       0.082&       0.053&       0.080&       0.503\\
Short measures (st)&       0.002&       0.003&       0.448&       0.006&       0.017&       0.710\\
Short measures (lt)&      -0.016&       0.008&       0.061&       0.016&       0.037&       0.676\\
Wage subsidies (st)&      -0.010&       0.006&       0.128&      -0.030&       0.041&       0.457\\
Wage subsidies (lt)&      -0.092&       0.018&       0.000&      -0.230&       0.053&       0.000\\

\\ \hline
\multicolumn{1}{l}{\textit{Rate of employed workers on benefits}} \\
            &      effect&          se&       p-val&      effect&          se&       p-val\\
Lagged dependent variable&       0.681&       0.011&       0.000&       0.669&       0.015&       0.000\\
Training(st)&       0.001&       0.002&       0.726&       0.002&       0.010&       0.852\\
Training(lt)&      -0.012&       0.006&       0.047&      -0.006&       0.025&       0.820\\
Short measures (st)&       0.002&       0.001&       0.067&      -0.001&       0.007&       0.915\\
Short measures (lt)&      -0.002&       0.003&       0.603&      -0.017&       0.011&       0.105\\
Wage subsidies (st)&       0.000&       0.003&       0.948&      -0.004&       0.012&       0.761\\
Wage subsidies (lt)&      -0.047&       0.007&       0.000&      -0.082&       0.018&       0.000\\
\\
\hline
Quarter-Year FEs&      $\checkmark$ & & &  $\checkmark$ & &       \\
Additional Controls&      $\checkmark$ & & &  $\checkmark$ & &       \\
Observations& 5980 & & &  5980 & &       \\
LLM & 115 & & &  115 & &       \\
\hline\hline
\end{tabular}%
}
\centering
     \begin{minipage}{\textwidth}
          \vspace{6pt}
          \footnotesize \textit{Notes:} This table shows the coefficient on the lagged dependent variable, the short-term (st) and long-term (lt) effects of the three types of ALMP, i.e. training, short measures (SM) and wage subsdies (wagesub) on the rate of welfare recipients (upper panel) and rate of employed workers on benefits (lower panel). Columns (1)-(3) show the effects, standard errors (se) and p-values (p-val) for the OLS specification. Columns (4)-(6) show effects, standard errors and p-values for the IV specification. Program variables included with 6 lags, main sample restrictions.  Standard errors obtained by a cross-sectional bootstrap (499 replications).  \end{minipage}
\end{table}

\begin{figure}[H]
                \centering
                \caption{Longterm Effects for Different Subpopulations I\label{fig:lt_effects_subgroups}}
                \vspace{10pt}
                \begin{subfigure}[b]{0.49\textwidth}
                                \centering \caption*{Training} \subcaption*{Unemployment rate} 
                                \includegraphics[clip=true, trim={0cm 0cm 0cm 0cm},scale=0.50]{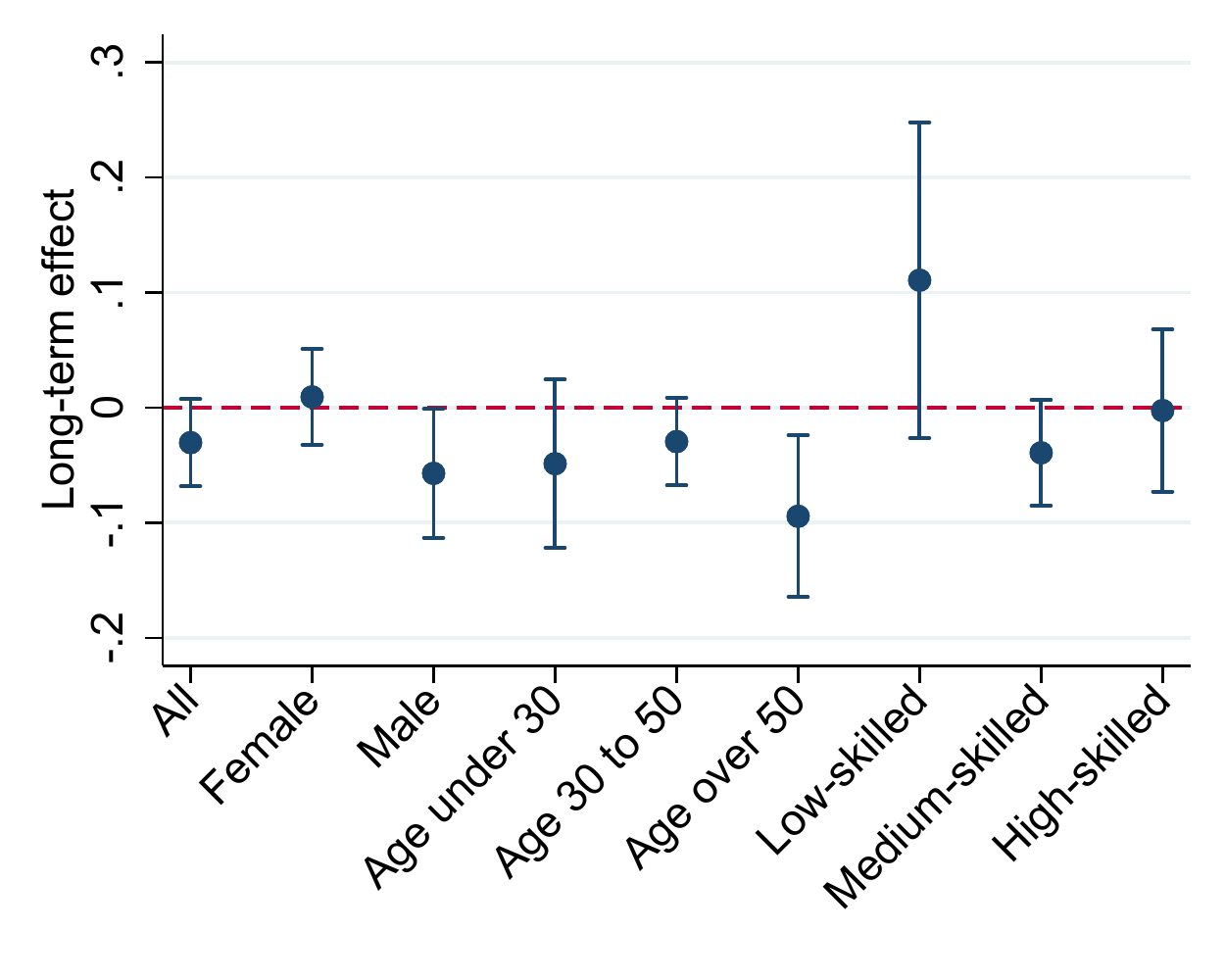}
                \end{subfigure}
                \begin{subfigure}[b]{0.49\textwidth}
                                \centering \caption*{Training} \subcaption*{Unsubsidised employment rate} 
                                \includegraphics[clip=true, trim={0cm 0cm 0cm 0cm},scale=0.50]{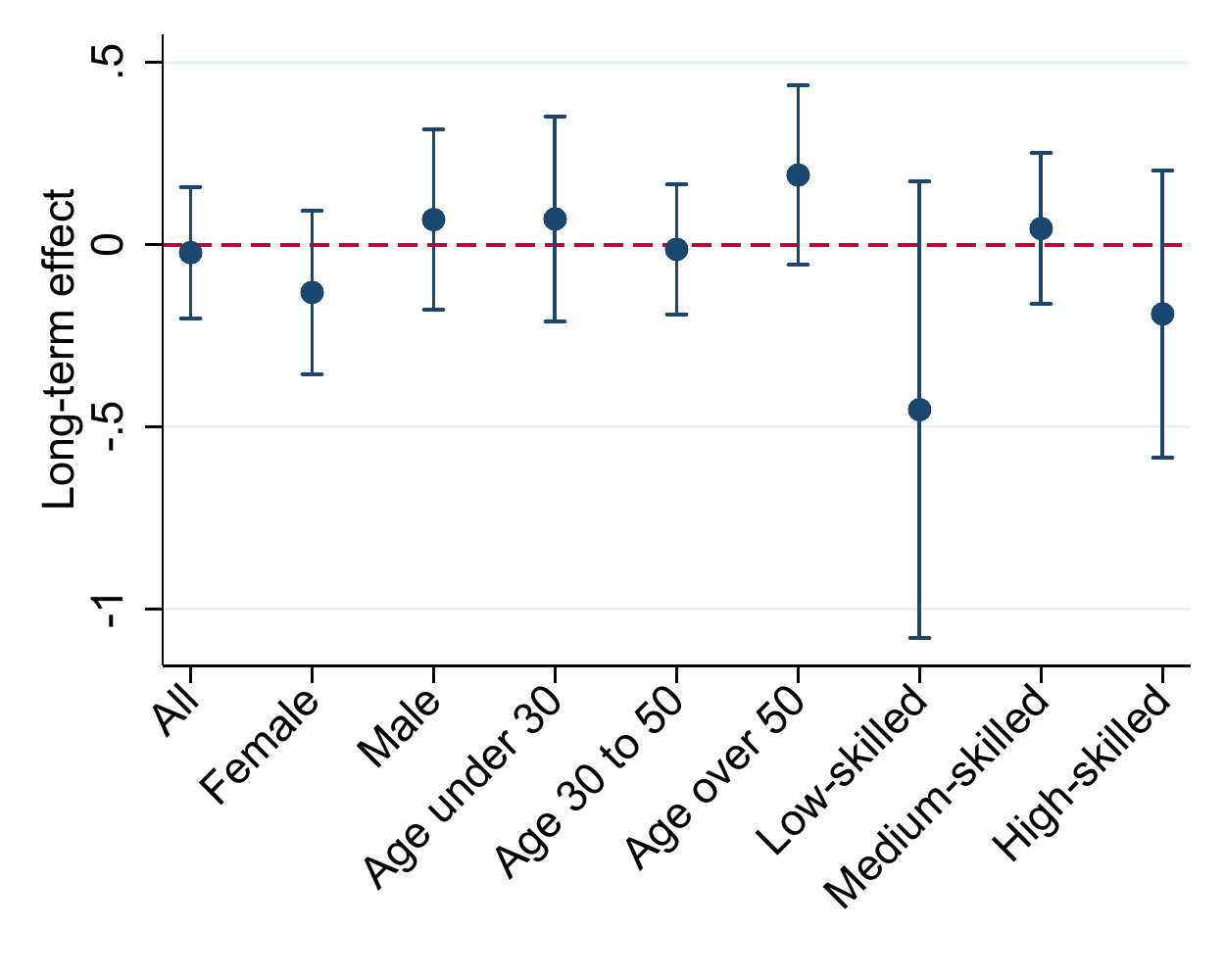}
                \end{subfigure}
                \vspace{10pt}    
                \newline
                \begin{subfigure}[b]{0.49\textwidth}
                                \centering \caption*{Short measures} \subcaption*{Unemployment rate} 
                                \includegraphics[clip=true, trim={0cm 0cm 0cm 0cm},scale=0.50]{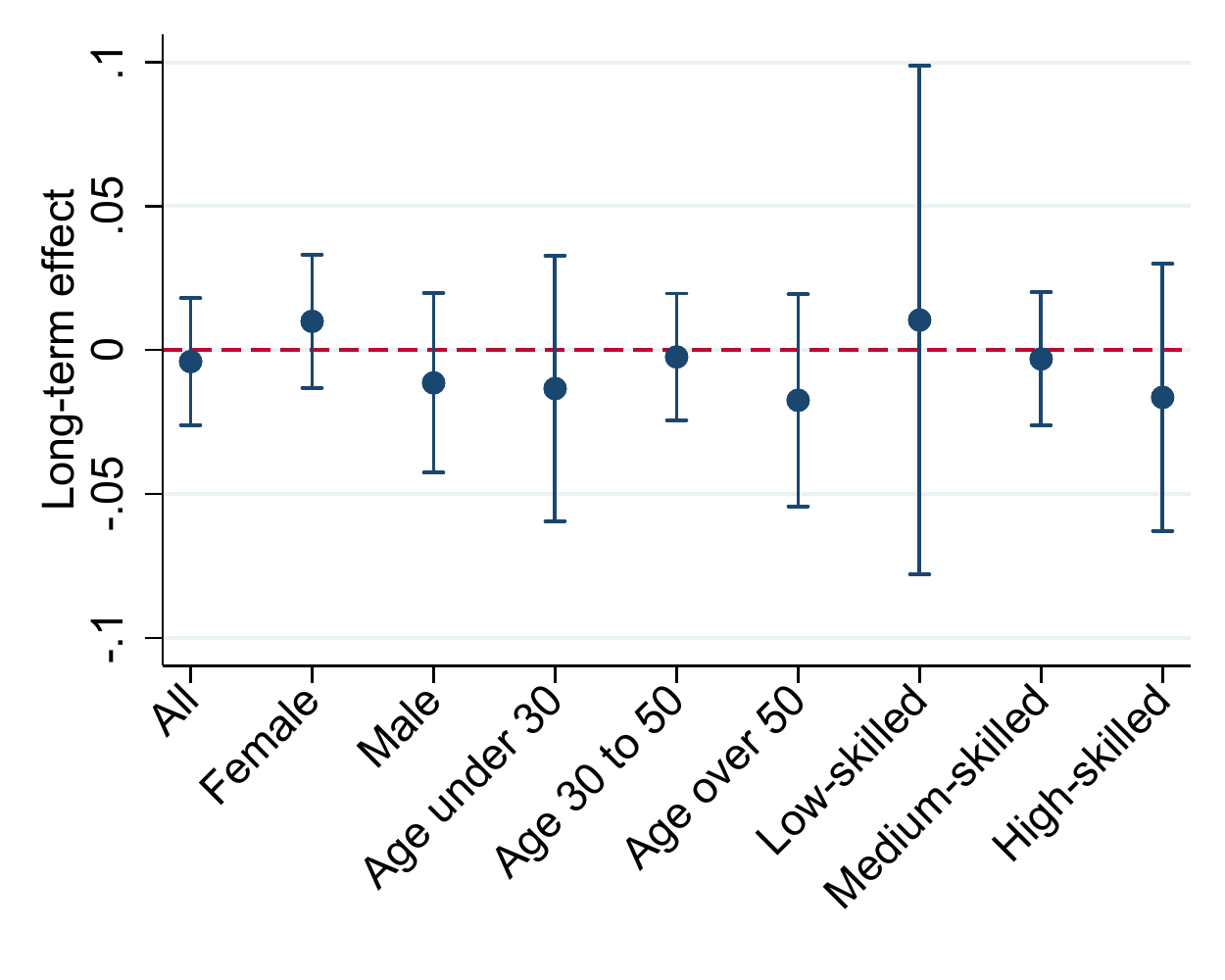}
                \end{subfigure}
                \vspace{10pt}    
                \begin{subfigure}[b]{0.49\textwidth}
                                \centering \caption*{Short measures}  \subcaption*{Unsubsidised employment rate} 
                                \includegraphics[clip=true, trim={0cm 0cm 0cm 0cm},scale=0.50]{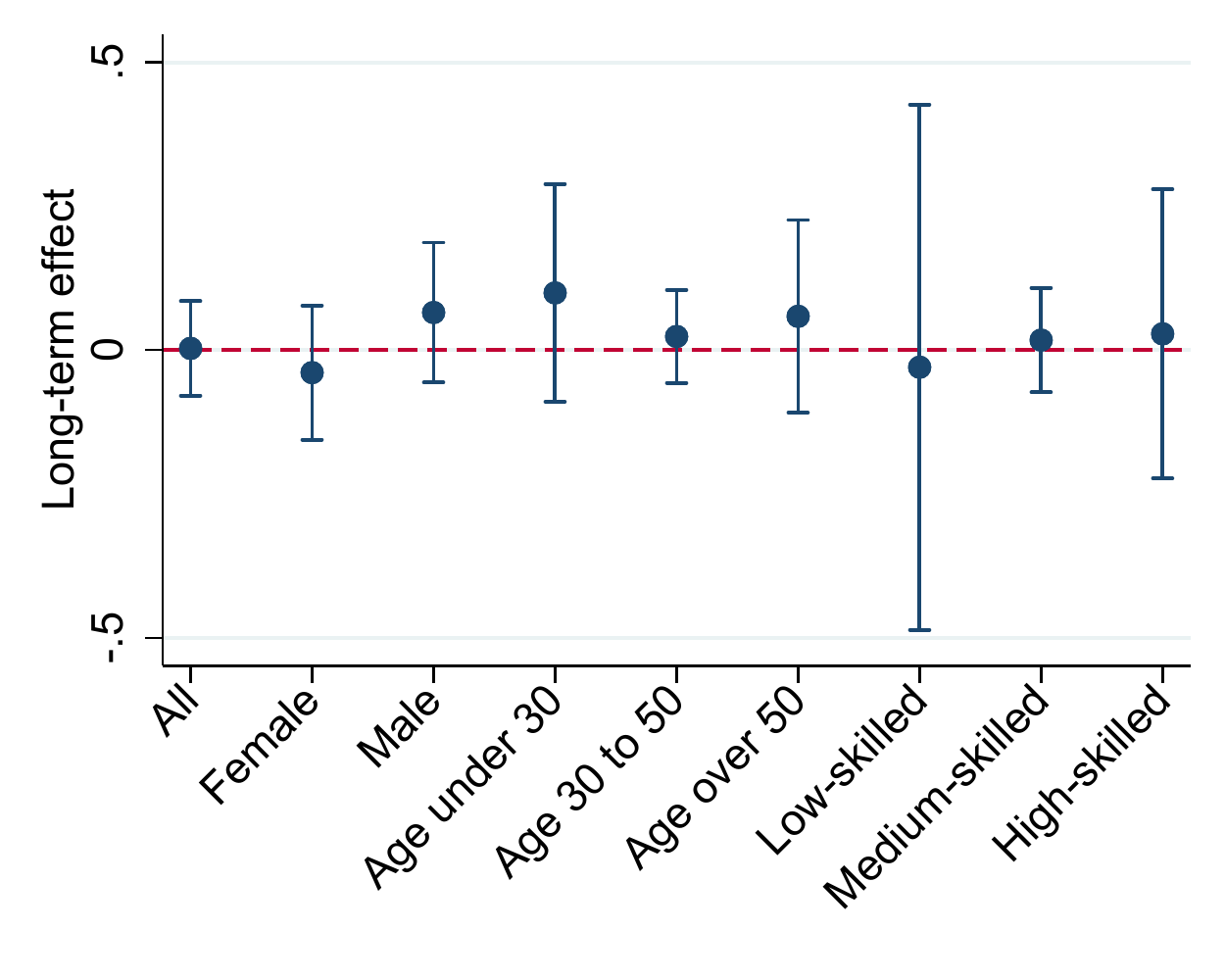}
                \end{subfigure}
                \vspace{10pt}
                \newline
                \begin{subfigure}[b]{0.49\textwidth}
                                \centering \caption*{Wage subsidies} \subcaption*{Unemployment rate} 
                                \includegraphics[clip=true, trim={0cm 0cm 0cm 0cm},scale=0.50]{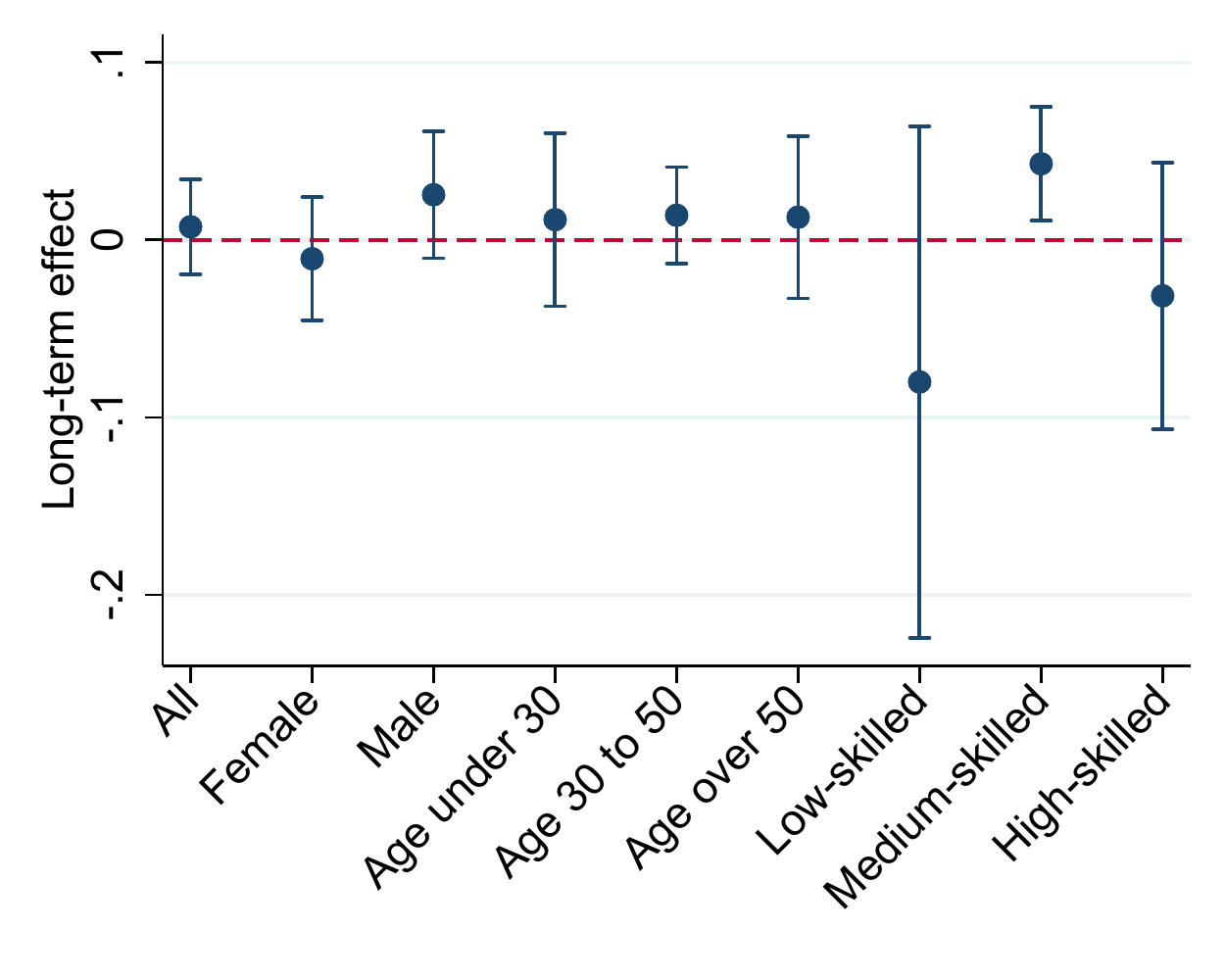}
                \end{subfigure}
                \vspace{10pt}    
                \begin{subfigure}[b]{0.49\textwidth}
                                \centering \caption*{Wage subsidies}  \subcaption*{Unsubsidised employment rate} 
                                \includegraphics[clip=true, trim={0cm 0cm 0cm 0cm},scale=0.50]{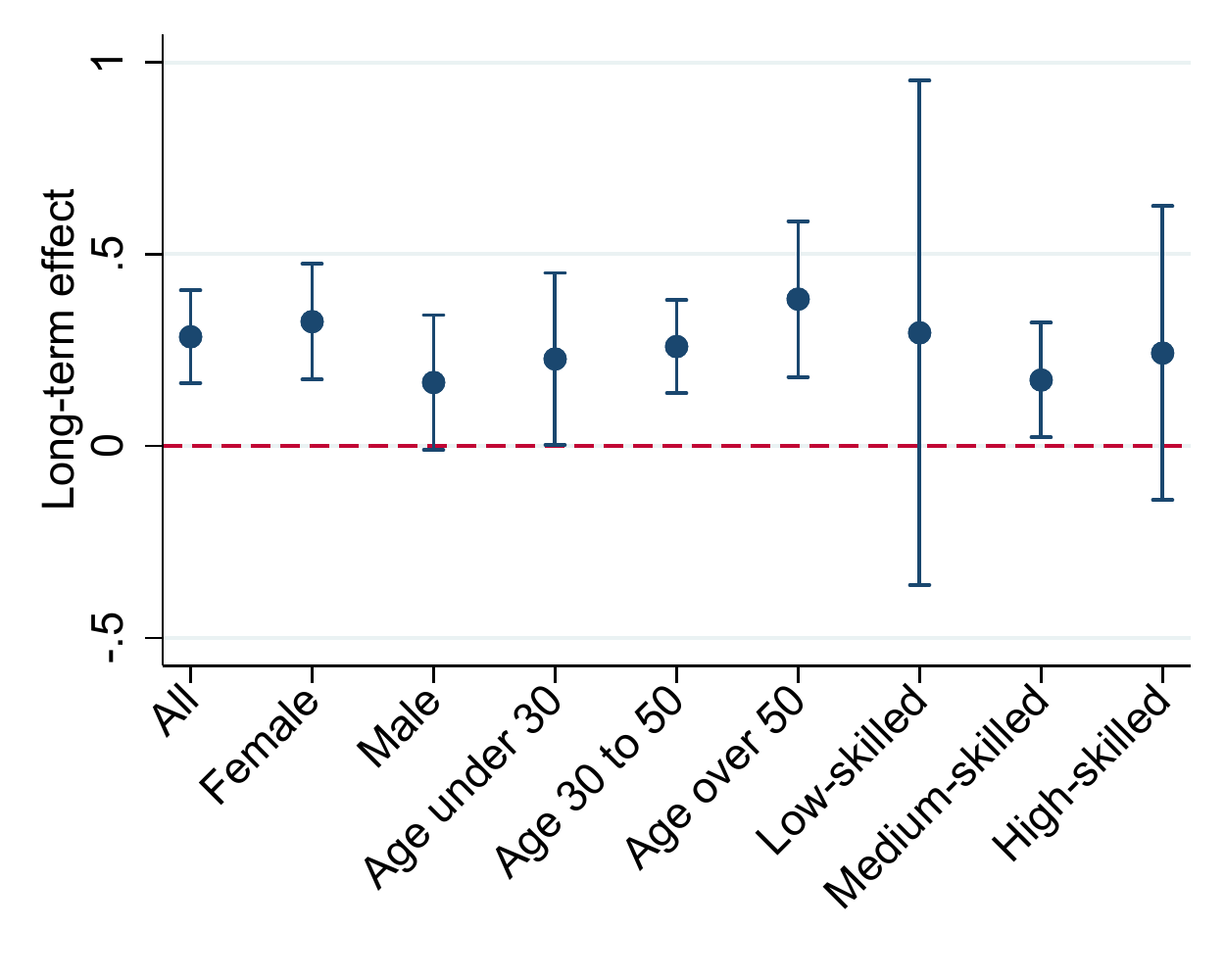}
                \end{subfigure}
                \vspace{10pt}
                \begin{minipage}{\textwidth}
                                \footnotesize \textit{Notes:} This graphs show the long-term effects and the corresponding 96\% CI of the three types of ALMP, i.e. training, short measures and wage subsidies on the unemployment and unsubsidised employment rates.  The effects are based on the ARDL model estimated by 2SLS. Program variables are included with 6 lags, main sample restrictions apply.  Standard errors obtained by a cross-sectional bootstrap (499 replications).
                \end{minipage}
\end{figure}

\begin{figure}[H]
                \centering
                \caption{Longterm Effects for Different Subpopulations II \label{fig:lt_effects_subgroups2}}
                \vspace{10pt}
                \begin{subfigure}[b]{0.49\textwidth}
                                \centering \caption*{Training} \subcaption*{Rate of welfare recipients} 
                                \includegraphics[clip=true, trim={0cm 0cm 0cm 0cm},scale=0.50]{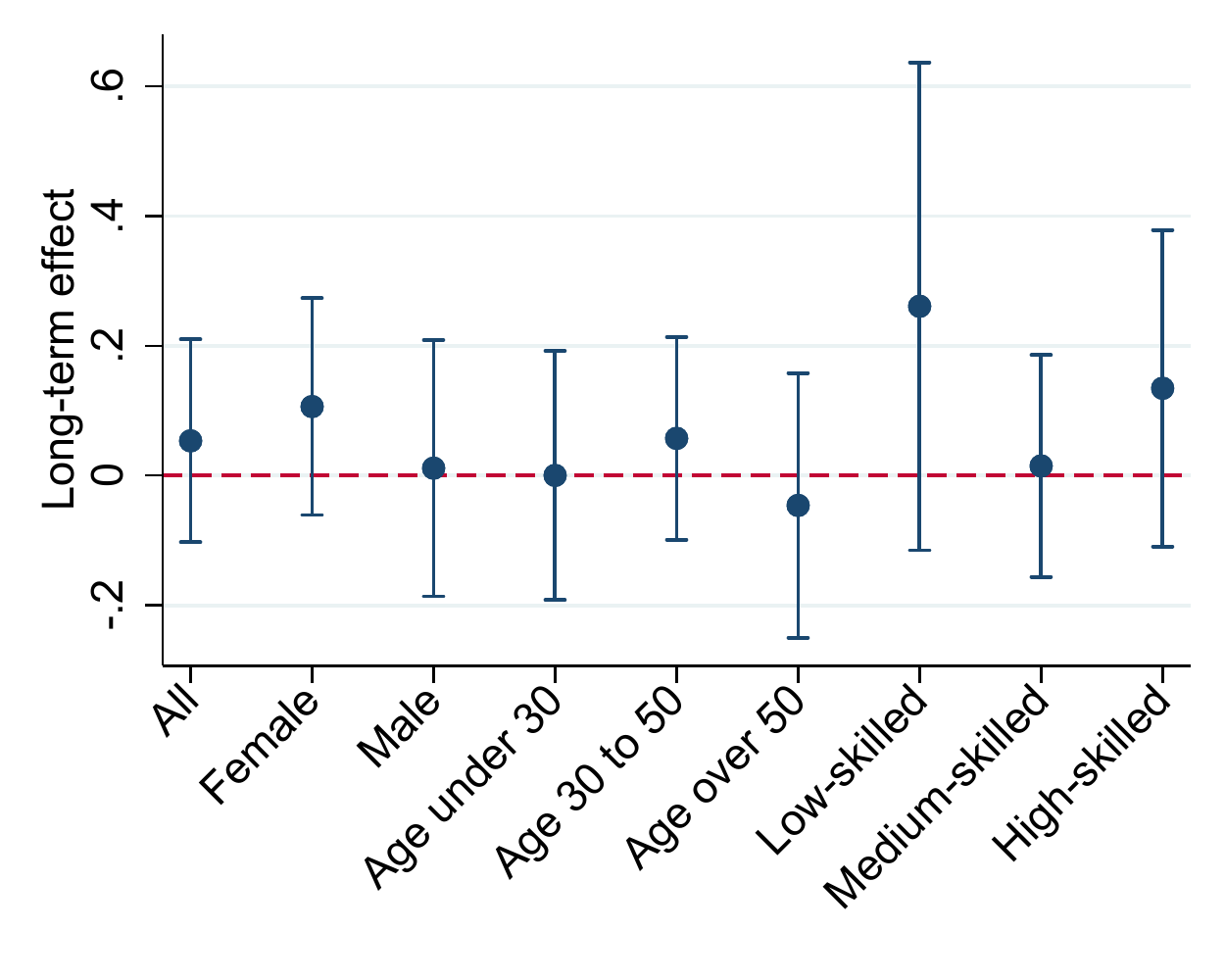}
                \end{subfigure}
                \begin{subfigure}[b]{0.49\textwidth}
                                \centering \caption*{Training} \subcaption*{Rate of employed workers on benefits} 
                                \includegraphics[clip=true, trim={0cm 0cm 0cm 0cm},scale=0.50]{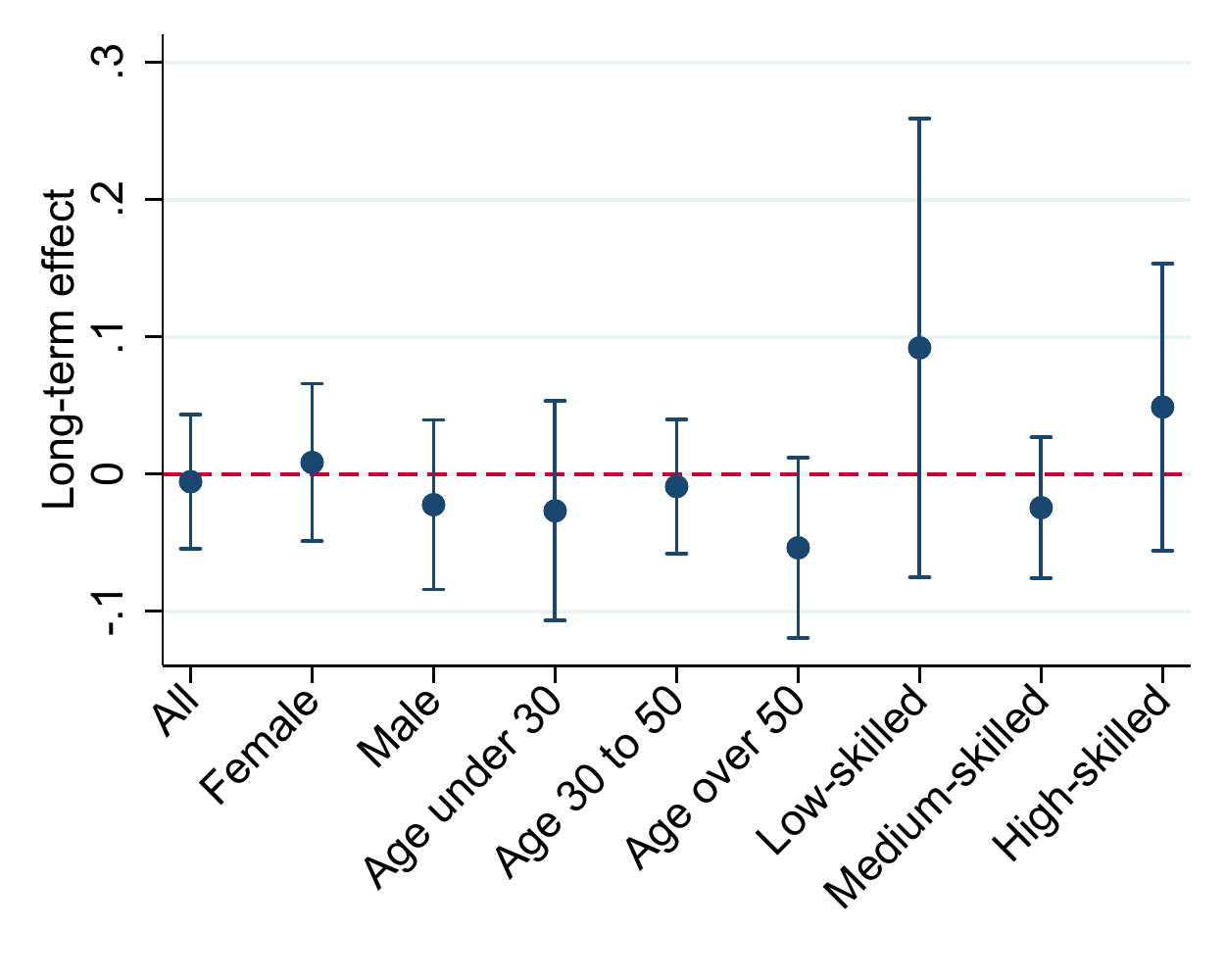}
                \end{subfigure}
                \vspace{10pt}    
                \newline
                \begin{subfigure}[b]{0.49\textwidth}
                                \centering \caption*{Short measures} \subcaption*{Rate of welfare recipients} 
                                \includegraphics[clip=true, trim={0cm 0cm 0cm 0cm},scale=0.50]{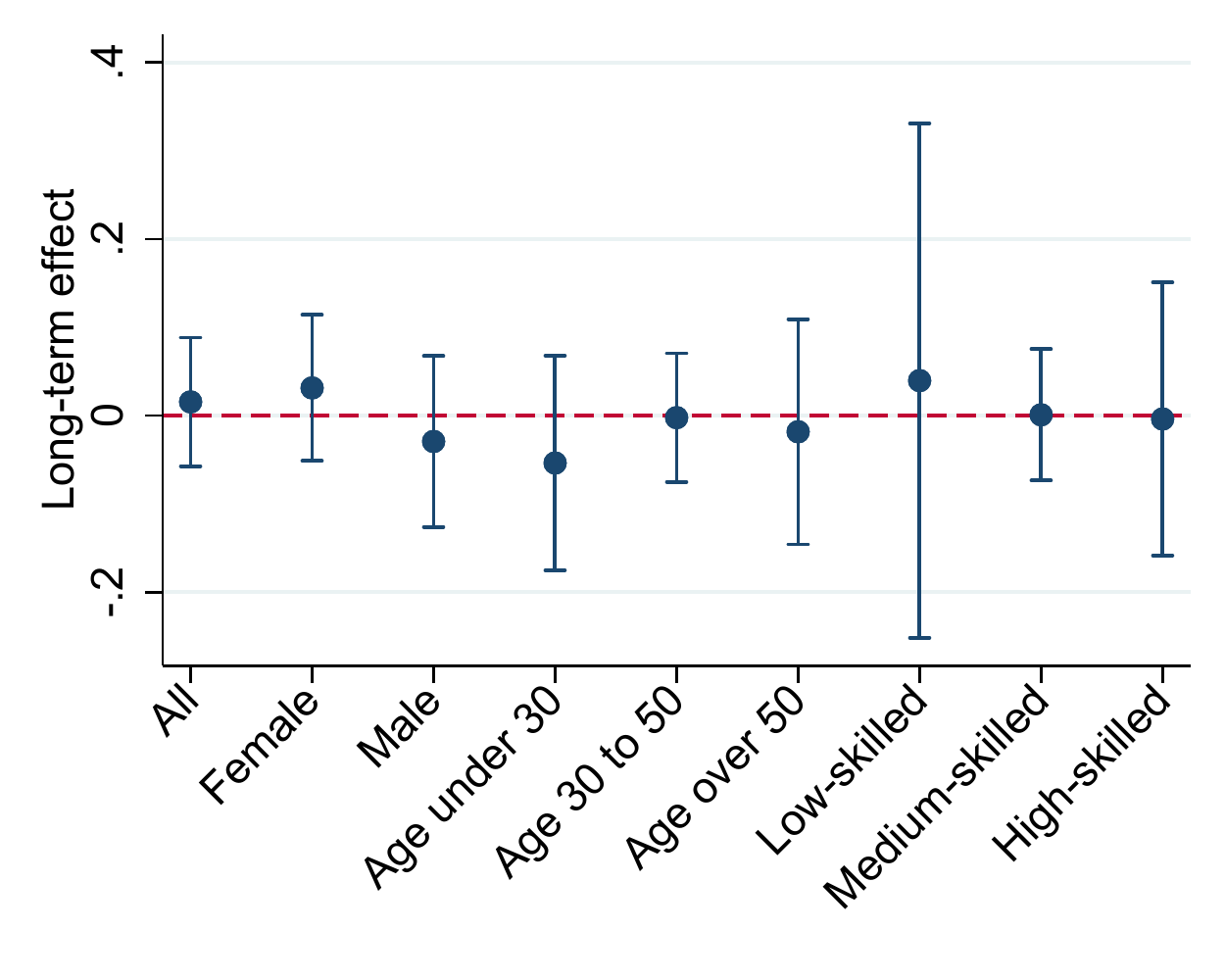}
                \end{subfigure}
                \vspace{10pt}    
                \begin{subfigure}[b]{0.49\textwidth}
                                \centering \caption*{Short measures}  \subcaption*{Rate of employed workers on benefits} 
                                \includegraphics[clip=true, trim={0cm 0cm 0cm 0cm},scale=0.50]{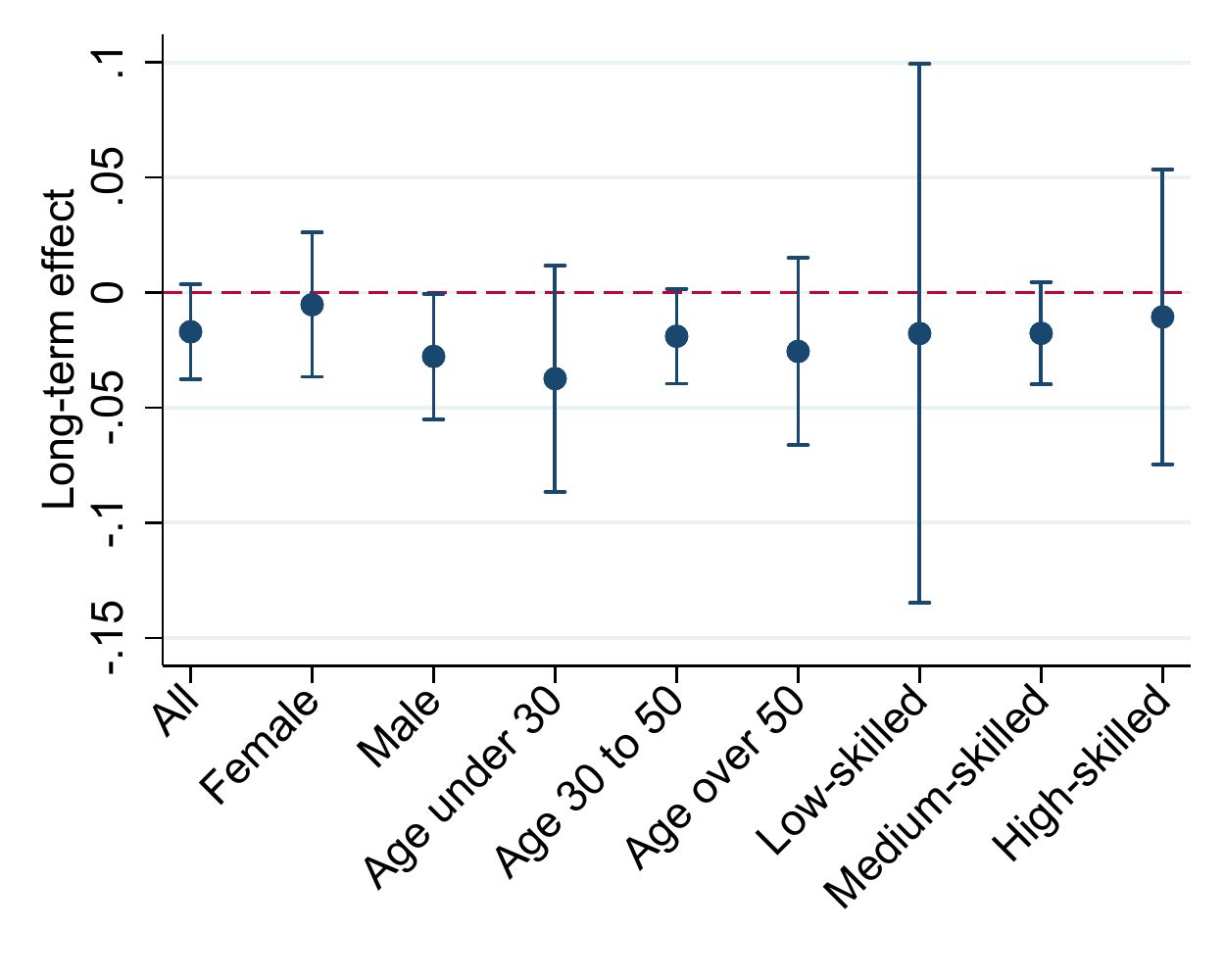}
                \end{subfigure}
                \vspace{10pt}
                \newline
                \begin{subfigure}[b]{0.49\textwidth}
                                \centering \caption*{Wage subsidies} \subcaption*{Rate of welfare recipients} 
                                \includegraphics[clip=true, trim={0cm 0cm 0cm 0cm},scale=0.50]{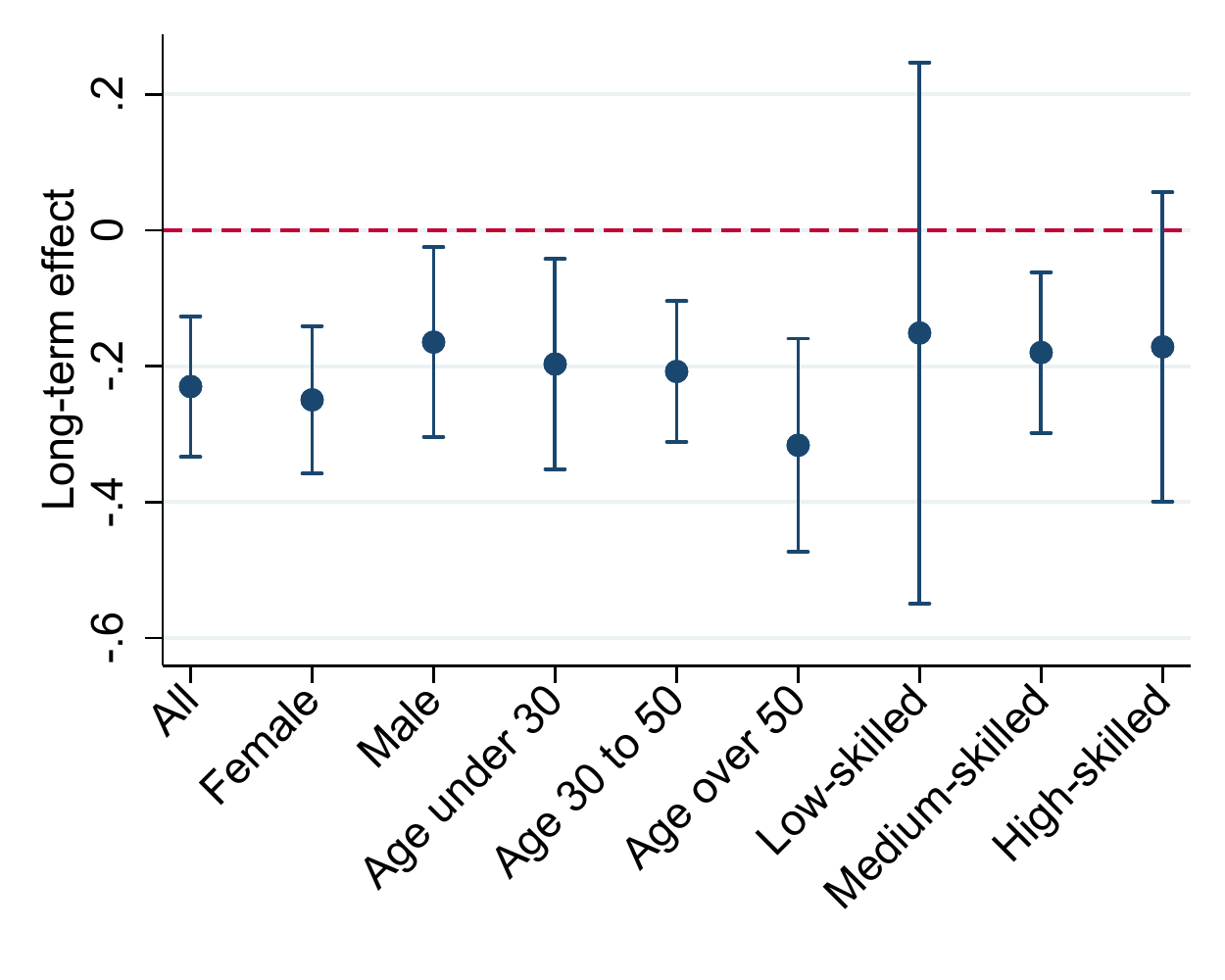}
                \end{subfigure}
                \vspace{10pt}    
                \begin{subfigure}[b]{0.49\textwidth}
                                \centering \caption*{Wage subsidies}  \subcaption*{Rate of employed workers on benefits} 
                                \includegraphics[clip=true, trim={0cm 0cm 0cm 0cm},scale=0.50]{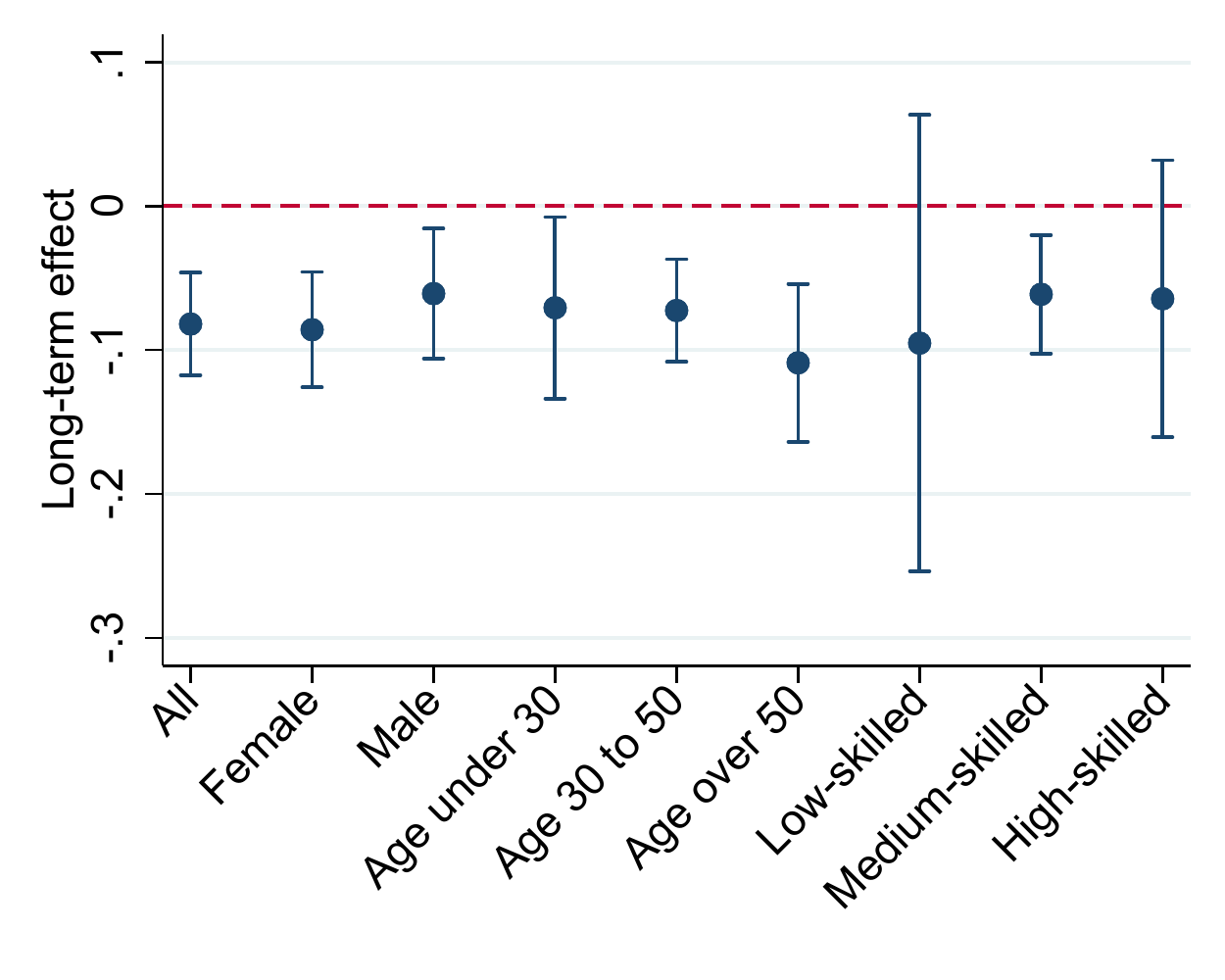}
                \end{subfigure}
                \vspace{10pt}
                \begin{minipage}{\textwidth}
                                \footnotesize \textit{Notes:} This graphs show the long-term effects and the corresponding 96\% CI of the three types of ALMP, i.e. training, short measures and wage subsidies on the rate of welfare recipients and employed workers on benefits.  The effects are based on the ARDL model estimated by 2SLS. Program variables are included with 6 lags, main sample restrictions apply.  Standard errors obtained by a cross-sectional bootstrap (499 replications).
                \end{minipage}
\end{figure}

Figures  \ref{fig:lt_effects_subgroups} and \ref{fig:lt_effects_subgroups2} depict the long-run effects of training, short measures and wage subsidies on the labour market outcomes of different subgroups. Appendix Figures \ref{fig:cum_effects_iv_alo_rate_female} to \ref{fig:cum_effects_iv_alo_rate_o_50}
 show the corresponding cumulative marginal effects on the unsubsidised employment and unemployment rates. Specifically, we consider males, females, workers of different ages (below the age of 30, ages 30 to 50 and above the age of 50) as well as low, medium and high-skilled workers. Similar to our control variables, we define low-skilled as workers without a vocational or academic degree, medium skilled workers as workers with a vocational degree and high-skilled workers as workers with an academic degree.\footnote{Notice that the skill variable is measured with some error, since the share of individuals with missing information about their skill level strongly varies over time.} 

We find some suggestive evidence for heterogeneous aggregate effects across subgroups, even though the confidence intervals largely overlap. When differentiating by gender, the results suggest that males benefit more from training compared to females. While the effect on the unemployment rate is zero for females, it is negative for males. Nevertheless, the effect for males is small and not clearly reflected in a higher employment rate, while for females the estimates lean more towards negative effects on employment. Thus, the overall conclusions about training do not change but the net gains and losses show a tendency of being distributed unequally across gender. Moreover, the effect heterogeneity found at the individual level with more gains for females is not reflected in the aggregate. The same tendency is visible for short measures but the effects are mostly insignificant. If anything, men seem to benefit from short measures by being less dependent on welfare. In contrast, females seem to benefit more from wage subsidies in the long run, which is in line with the effect heterogeneity documented on the individual level.

With respect to age, we find relatively large and positive effects of training for workers above the age of 50 but not for younger workers. These workers experience a long-term decrease in their unemployment rate of around 0.10 percentage points, which is mirrored in a marginally significant increase in the employment rate and a decrease in the rate of employed workers on state benefits. Thus, we do find net benefits of training but only in this segment of the labour market. Also for wage subsidies the effects on the employment rates are higher for older workers. This age gradient is mirrored in lower rates of workers on welfare and a lower share of employed workers on benefits. 

The effects on the labour market outcomes of different skill groups are very noisy. This is partially driven by the low shares of low and high-skilled individuals amongst the unemployed and employed in some labour markets. 
Nevertheless, the results suggest that training might decrease the employment rate for low-skilled workers and increase the unemployment rate in the long run. If anything, medium skilled-workers are the only group to benefit from this type of policy. This provides some suggestive evidence that the substitution effects that are likely to explain the lack of positive net effects of training are more concentrated among the low-skilled segment of the labour market. The effects of short measures are close to zero and insignificant for all skill groups. In contrast to the average effect and other groups, wage subsidies increase the unemployment rate among medium-skilled workers, and the effect on the employment rate is smaller compared to the average effect. This suggests that either the programs are less effective for medium-skilled program participants, or that displacement effects are concentrated in this group. I.e., firms may substitute jobs for medium skilled workers by subsidised workers. 

Overall, the results suggest that the net effects of ALMP are larger for some subpopulations and that desired and undesired effects are distributed unequally across different segments of the labour market. Yet, the estimates are not precise enough to draw strong conclusions.

\section{Robustness}\label{robustness}

\subsection{Sensitivity to the Model Specification}\label{rob_model}

We run two types of robustness checks to test for the sensitivity of the results to the model specification. First, we vary the lag order in the ARDL model and second, we estimate a distributed lag model, leaving out the lagged dependent variable.

\textbf{The lag order.} \quad Additionally to our main specification which includes all policy variables with 6 lags (corresponding to 6 quarters), we estimate two alternative specifications with 4 and 8 lags respectively. The optimal lag length is difficult to know ex-ante and usually a data-based choice. Choosing a lag length that is too short might potentially lead to biased estimates, a lag length that is too long merely decreases precision. Commonly used tests based on the Akaike information criterion (AIC) and the Schwartz/Bayesian information criterion (SBIC) (see e.g. \citealp{Greene2012}) are inconclusive in our setting. We therefore rely on a different approach and compare models based on a different lag order.  The results are depicted in Appendix Figures \ref{fig:lt_effects_rob_lags} and \ref{fig:lt_effects_rob_lags2} for our main outcomes. The figures depict the cumulative marginal effects for different lag orders and the 95\% confidence interval for the main specification. All models yield qualitatively comparable results, with only small deviation in effect sizes. For all outcomes and policy variables, the cumulative effects curves lie within the confidence interval of the main specification. The effects of training slightly deviate from the main specification in the long run. Overall, the results suggest that the lag order does not play a major role in driving the effects.

\textbf{Distributed lag model.} \quad Next, we estimate an alternative model specification to test whether our results are sensitive to the inclusion of a lagged dependent variable in the model. For this, we estimate the same model as in equation (\ref{eq:ardl}) but leave out the yearly change in the lagged outcome. The distributed lag model still allows for a dynamic effect of the policy variables on the outcome but does not assume that the policies affect have an effect through intermediate outcomes. Similar to the ARDL model, the coefficient $\phi_0^m$ measures the short run effects of a policy. The long-run effect is given by the sum of lag weights $\sum_{j=0}^q \phi_j^m$. Both effects estimated by 2SLS are presented in Appendix Table \ref{tab:dl_alo_emp}. We find similar results as for the ARDL model with only small deviations in effect sizes. There is no evidence for an effect of short measures on labour market outcomes, the long-term effect of training on the unemployment rate is smaller in absolute value and insignificant. Also this model shows that wage subsidies significantly increase the unsubsidised employment rate in the long run, with the effect being slightly smaller. The effect is mostly driven by a reduction in the rate of welfare recipients. In sum, the results suggest that the policy effect that goes through intermediate outcomes is small. In turn, it is unlikely that our main effects suffer from a large bias introduced by the correlation of the error term with the lagged dependent variable. If anything, they represent a lower bound of the true effect.

\subsection{Sensitivity to the Labour market Definition and Overlap Criteria}\label{rob_llm_ov}

In a series of robustness checks, we test the sensitivity of our results to the choice of the labour market definition and overlap criteria. 

\textbf{Labour market definitions.} \quad As introduced in Section \ref{llm} different stopping values in the clustering process lead to labour markets of different sizes and self-containment levels and also affect the overall number of markets in the estimation sample. Thus, we check whether our results change for different labour market definitions. Specifically, we compare the effects for labour market definitions depending on the stopping values 0.991, 0.993 and 0.997 with our main definition based on the stopping value 0.995. The results are displayed in Appendix Figure \ref{fig:lt_effects_rob_llm} for the unemployment rate and the unsubsidised employment rate. Appendix Figure \ref{fig:lt_effects_rob_llm2} shows the results for the rate of welfare recipients and the rate of employed workers on benefits. The figures plot the cumulative marginal effects of different programs on the respective labour market outcome as separate lines for each labour market definition. The grey area depicts the 95\% confidence interval for the main definition. We find very similar effect patterns for all labour market definitions and only small deviations in the long-term effects of training and wage subsidies. The effects of these programs are found to be slightly smaller for the definitions based on lower stopping values.

\textbf{Overlap criteria.} \quad In the following, we show that our results are robust to varying the strictness of overlap criteria which strongly supports our identification strategy. We test a number of different sample selection criteria based on the overlap between labour markets and employment agencies.  For all checks, we impose that a labour market needs to overlap with at least two employment agencies and vary the restrictions on the size of the overlap on top. The overlap criteria are defined based on two shares. First, we consider the total overlap $s_{tot}$, which measures the RLF in the labour market as share of the total RLF in all partially overlapping agencies. Second, we consider the overlap between a market and single agencies, $s_{LEA} = RLF_{LLM,LEA}/RLF_{LEA}$. It measures the share of the agency's labour force that overlaps between the market and the agency. The distribution of these two shares and our overlap criteria are shown in Appendix Figure \ref{fig:overlap_crit}. Panel (a) plots the distribution of $s_{tot}$ for all labour markets in our preferred definition that overlap with at least two agencies. Panel (b) plots the distribution of $s_{LEA}$ for these markets and their overlapping agencies. Vertical lines depict how the sample is restricted by different criteria that we explain below.

We test three alternative criteria that target the validity of the exogeneity assumption.  First, we impose a stricter exogeneity criterion and drop all labour markets for which $s_{tot} \geq $ 40\% (exo1). Second, we relax this criterion and keep all labour markets for which $s_{tot} <$ 60\% (exo2). As third restriction (exo3), we assess whether the overlap between the labour market and each agencies is not too large. If this is the case, our identification strategy is more credible, since policy mix in the agency is not fully determined by the economic situation in the market. Thus, we check for each labour market and agency whether the overlapping RLF comprises less than 50\% of the RLF in the respective agency and only keep those labour markets in the sample for which the agencies which violate this restriction jointly represent less than 50\% of the total labour force in the instrument area. 

In order to assess the validity of the relevance assumption, we test four alternative criteria which we impose on top of the main sample restriction that limits $s_{tot}$ at 50\%.  First, we only consider labour markets that do not have fully enclosed agencies or in which enclosed agencies account for 50\% or less of the RLF in the market (rel1). Second, we impose two minimum criteria on  $s_{tot}$. The RLF in the labour market should constitute at least 5\% (rel2) or 10\% (rel3) of the overlapping agencies respectively. Finally, we review the overlap between a labour market and single agencies. Optimally, this overlap is large enough for all agencies that partially overlap with a market. Then the policy mix inside is most likely to correlate with the policy mix outside, e.g. through agency-specific strategies. Since imposing a minimum criterion on the overlap of each agency would be too restrictive, we propose the following two sample restrictions.  First, the labour market must overlap sufficiently with at least one agency, such that $s_{LEA} >$ 20\% (rel4). Second, the labour market must overlap sufficiently with at least two agencies, such that for these agencies $s_{LEA} > $ 10\% (rel5). 

Appendix Figure \ref{fig:N_rob_IV} shows how our estimation sample changes after imposing these alternative sample selection criteria. With most of the criteria, we keep more than 90 labour markets. The reduction in the sample is largest for \textit{rel5}. 
We plot the results of the ARDL model estimated in these different samples in Appendix Figures \ref{fig:lt_effects_rob_IV} and \ref{fig:lt_effects_rob_IV2}. They show the long-term effects of the different training programs separately for each of the main labour market outcomes. Overall, the point estimates are very robust to the variation in overlap criteria. Training and short measures do not have a robust effect on aggregate labour market outcomes, but we find a robust positive employment effect for wage subsidies that is explained by a reduction in the rate of welfare recipients or employed workers on benefits. The results strongly support the validity of our identification strategy and are not sensitive to stricter overlap criteria.

\subsection{Sensitivity to the Reform in 2009}\label{rob_postref}

Finally, we test the sensitivity of our results to a change in the observation period. Motivated by the institutional changes of short measures in 2009, we estimate the effects of ALMP using a shorter time frame, i.e. 2010 to 2018. Appendix Table \ref{tab:post2010_alo_emp} shows the effects of the three policy types on our main set of outcomes estimated by 2SLS. As in our main analysis, we do not find any effect of short term measures on aggregate labour market outcomes. This suggests that absence of effects in our main analysis is not driven by the institutional framework. The long-term effect of training is negative but not statistically significant. Compared to the main analysis, we find a slightly smaller long-term effect of wage subsidies on the employment rate and a negative and significant effect on the unemployment rate. These changes might be partially driven by a different macroeconomic situation in the years 2005 to 2010, including the Great Recession in 2008.

\section{Conclusion}\label{conclusion}

Research on macroeconomic effects of ALMP confronts two important empirical challenges: First,  policies and labour market outcomes are determined simultaneously on the aggregate level. Second, there might be spillover effects across neighbouring regions that are closely linked economically.
This makes it difficult to identify policy effects on the aggregate level. This paper develops a new empirical strategy that allows us to address these challenges by studying effect of the level of self-contained local labour markets and by exploiting their imperfect overlap with the administrative regions that are in charge of policy decisions.

We apply this strategy to estimate the macroeconomic effects of ALMP in Germany over the period 2005 to 2018. We consider three types of policies: short activation measures, vocational training and wage subsidies and estimate their effects on aggregate labour market outcomes at the labour market level. We instrument the policy use in the labour market by the policy use in neighbouring municipalities that are part of the jurisdiction of local employment agencies that partially overlap with the labour market. We find no aggregate effects of a higher program use intensity of short activation measures and further vocational training on the aggregate employment rate. Wage subsidies in contrast, are effective in increasing employment and reducing long-term unemployment and welfare dependency.  A 10 percentage point increase in the share of wage subsidy recipients increases the unsubsidised employment rate by 3 percentage points. Our findings further suggest that not all segments of the labour market are equally affected by the policies. Workers older than 50 years benefit  more from a higher intensity of training programs and wage subsidies compared to younger workers.  We also find differences in the effectiveness of programs across gender and skill groups, even though our estimates lack precision. Our results are robust to a variety of robustness checks such as changes in the model specification, different labour market definitions and overlap criteria.


Our findings bear important policy implications. A major part of the multi-billion euro investment in ALMP in recent decades did not result in a reduction of aggregate unemployment, which is the primary goal of these policies. Even if the policies are effective in increasing employment probabilities of program participants, our results suggest that these individual-level effects are partially offset by negative spillover effects on non-participants. The empirical strategy we apply in this paper does not allow us to directly quantify which workers and jobs are affected negatively.  Nevertheless, our estimates inform about the net gain for all jobseekers in the labour market, which is a primary policy objective. Moreover, our suggestive evidence on heterogeneity in the effectiveness of policies for different labour market segments offers insights on potentially undesired redistribution of limited numbers of jobs across different labour market segments, and it might allow for a better targeting of policies to jobseekers. However, a better understanding of the link between individual-level effects and aggregate effects would be required in order to derive such policy recommendations. 

\newpage

\vspace{12pt}
{\small
\bibliographystyle{econ}
\bibliography{library}
}

\pagebreak

\newpage
\appendix
\section{Appendix}
\setcounter{table}{0}
\setcounter{figure}{0}  
\renewcommand{\thetable}{A.\arabic{table}}
\renewcommand{\thefigure}{A.\arabic{figure}}

\subsection{Local Labour Markets}\label{app_llm}

\subsubsection{Municipalities and Municipality Regions}\label{app:munireg}

In this section we show how the three different aggregations of municipalities into municipality regions and their geographical distribution across Germany. Moreover, we assess whether the choice of the number of municipality regions in the range of 5000 to 7500 affects the self-containment of labour market definitions.

\begin{figure}[htbp]
      \caption{Municipality Regions}\label{app_fig:munreg}
\begin{center}
  \includegraphics[width=0.45\linewidth, clip=true, trim=1cm 0cm 1cm 1cm]{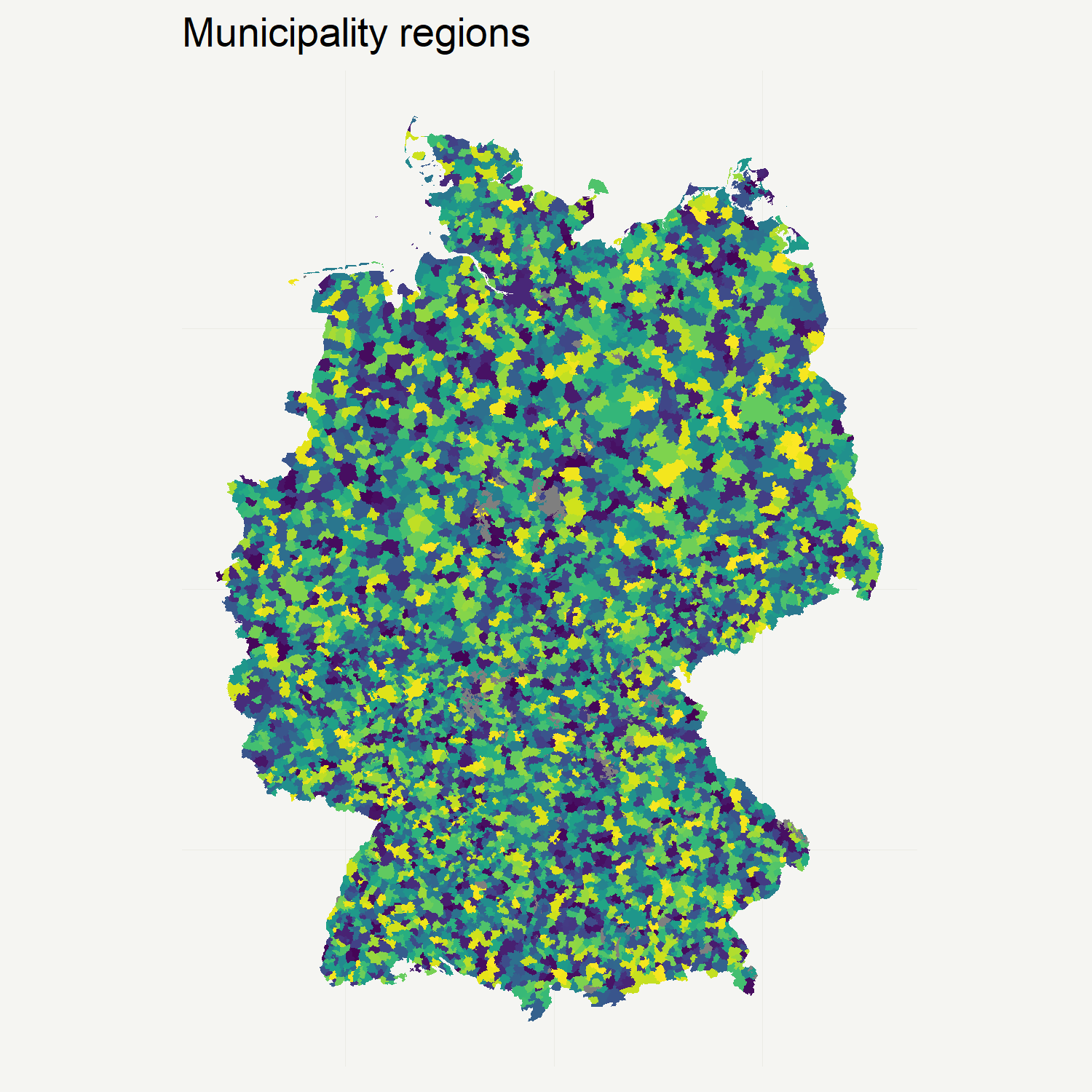}
 \caption*{(a) 5000}
  \includegraphics[width=0.45\linewidth, clip=true, trim=1cm 0cm 1cm 1cm]{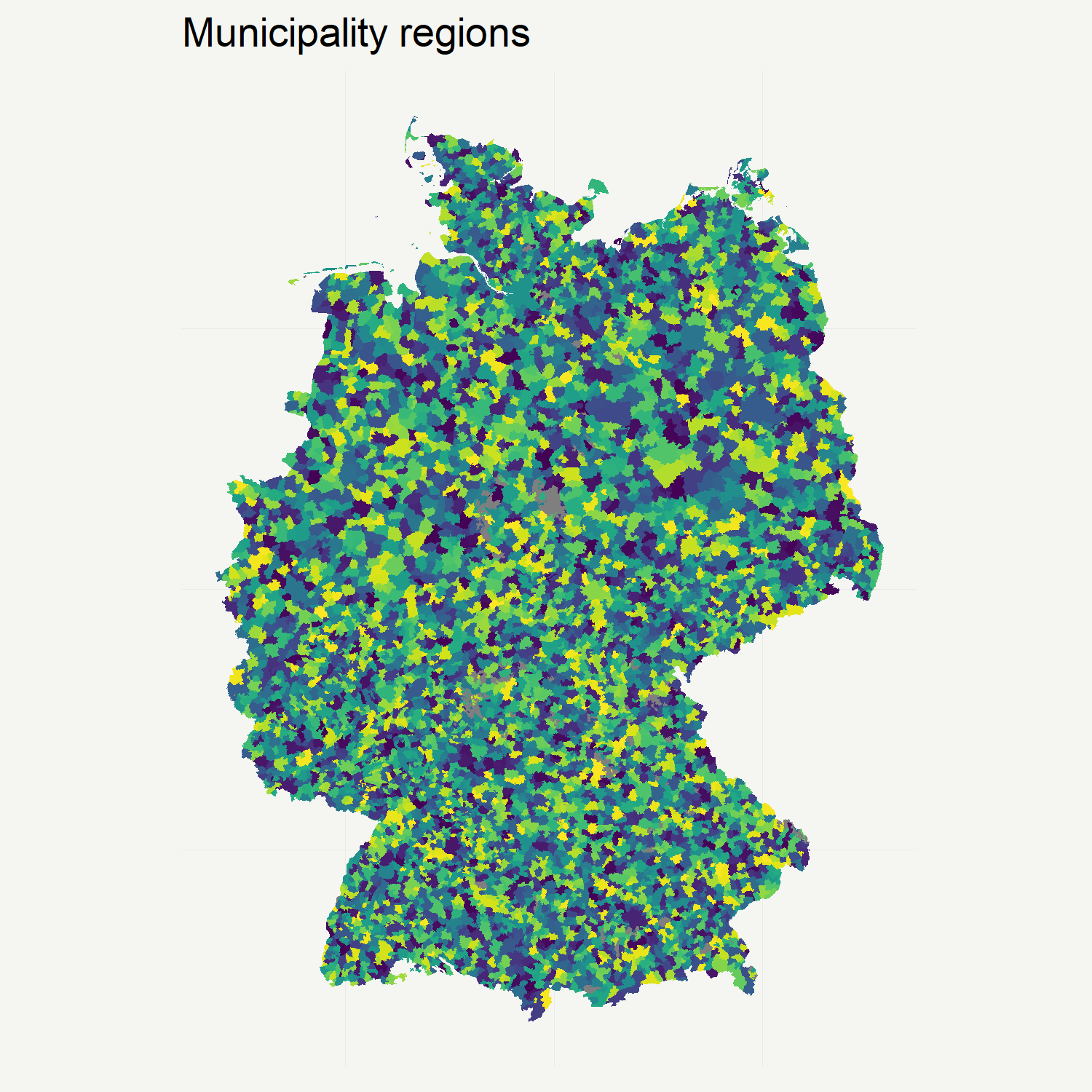} 
  \caption*{(b) 6000}
   \includegraphics[width=0.45\linewidth, clip=true, trim=1cm 0cm 1cm 1cm]{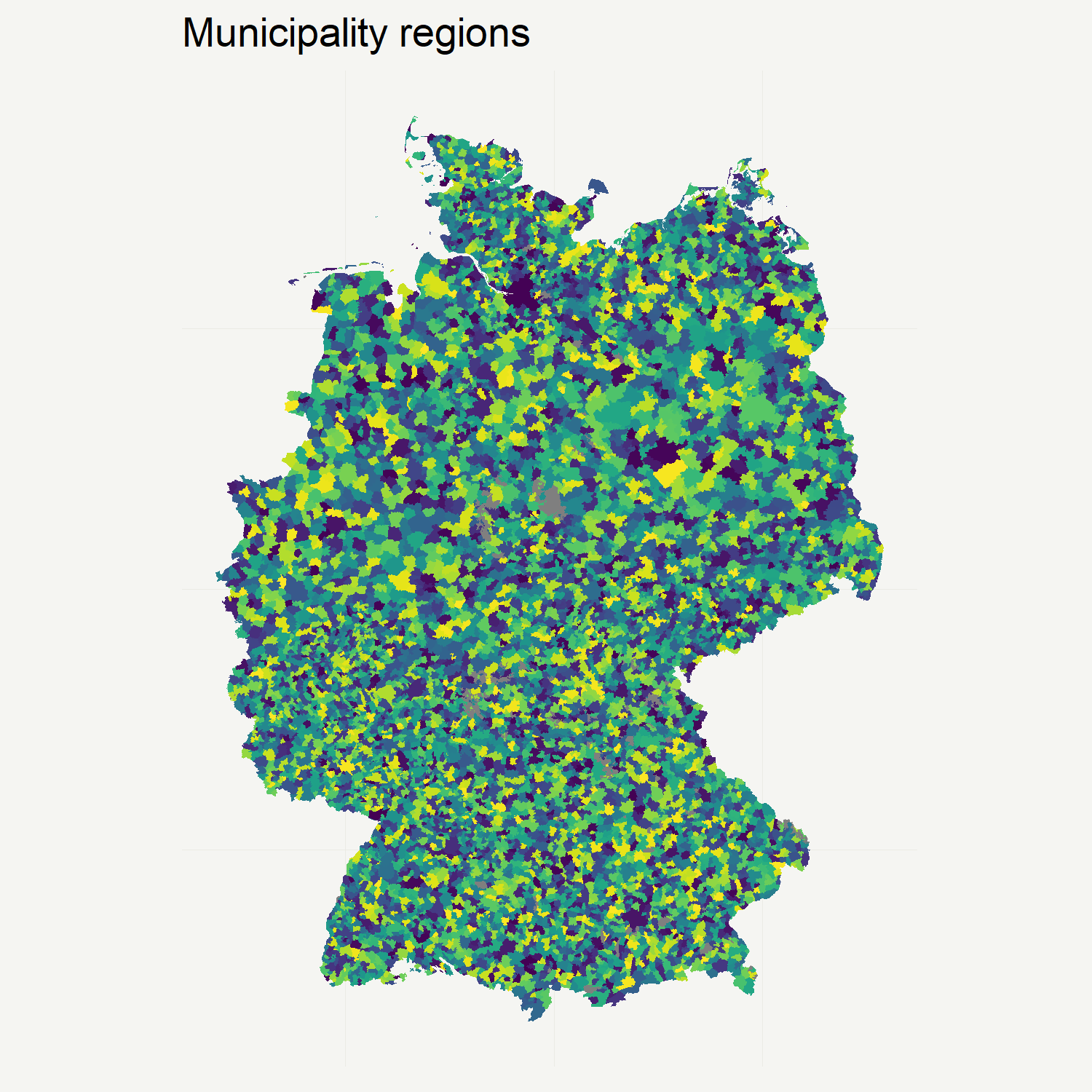} 
   \caption*{(c) 7500}
   \end{center}
     \footnotesize \textit{ Notes:} Municipality regions (5000, 6000, 7500) are depicted as coloured areas. Territorial status of 2016.
\end{figure}

\begin{table}[htbp]\centering\caption{Characterizing Labour Market Definitions (Across Municipality Regions)}\label{app:char_llm} 
\resizebox{\linewidth}{!}{%
\begin{tabular}{l*{12}{c}}
\hline\hline
    (1)        &         (2)&        (3)&     (4)&         (5)&     (6)&      (7)&     (8)&     (9) & (10)\\
         Def. &  Municip. Reg.&   Cut-off&        N LLM&     RLF (mean)&          CR&     ESC (mean)&      ESC (sd)&     ESC (min)&     ESC(max) \\
\hline
1          &        6000&          0.98&         462&    67652.63&       20.22&       65.77&       16.46&       11.11&       98.27\\
2          &        6000&          0.99&         274&   114071.22&       15.91&       74.03&       13.07&       11.11&       98.27\\
3          &        6000&         0.991&         251&   124523.96&       14.89&       75.50&       12.24&       11.11&       98.27\\
4          &        6000&         0.992&         229&   136486.96&       14.08&       76.63&       11.75&       11.11&       98.27\\
5          &        6000&         0.993&         209&   149547.91&       13.57&       78.03&       10.44&       16.56&       98.27\\
6          &        6000&         0.994&         187&   167141.79&       12.78&       79.11&       10.43&       16.56&       98.27\\
7          &        6000&         0.995&         162&   192935.27&       12.06&       80.71&        8.34&       45.88&       98.27\\
8          &        6000&        0.996&         138&   226489.23&       11.10&       82.35&        7.78&       54.43&       98.27\\
9          &        6000&         0.997&         121&   258310.03&       10.35&       83.41&        7.61&       54.43&       98.27\\
10         &        6000&         0.998&          92&   339733.85&        8.21&       86.39&        5.44&       65.34&       98.27\\
11         &        6000&         0.999&          59&   529754.47&        5.83&       89.68&        3.81&       76.56&       98.27\\
\hline
12         &        5000&          0.98&         462&    67652.63&       21.01&       65.32&       16.39&       10.39&       98.27\\
13         &        5000&          0.99&         274&   114071.22&       16.51&       74.27&       11.61&       16.70&       98.27\\
14         &        5000&         0.991&         251&   124523.96&       16.03&       75.40&       11.17&       16.70&       98.27\\
15         &        5000&         0.992&         229&   136486.96&       14.90&       76.46&       10.31&       39.95&       98.27\\
16         &        5000&         0.993&         209&   149547.91&       14.35&       77.47&        9.86&       44.82&       98.27\\
17         &        5000&         0.994&         187&   167141.79&       13.15&       79.01&        9.10&       44.82&       98.27\\
18         &        5000&         0.995&         162&   192935.27&       12.16&       80.08&        9.38&       44.82&       98.27\\
19         &        5000&         0.996&         138&   226489.23&       11.45&       82.11&        7.86&       48.88&       98.27\\
20         &        5000&         0.997&         121&   258310.03&       10.55&       83.48&        7.25&       48.88&       98.27\\
21         &        5000&         0.998&          92&   339733.85&        8.92&       86.07&        5.69&       67.81&       98.27\\
22         &        5000&         0.999&          59&   529754.47&        6.72&       89.07&        4.71&       72.93&       98.27\\
\hline
23         &        7500&          0.98&         462&    67652.63&       18.84&       64.34&       19.20&        5.24&       98.27\\
24         &        7500&          0.99&         274&   114071.22&       14.83&       73.80&       14.08&        8.73&       98.27\\
25         &        7500&         0.991&         251&   124523.96&       14.02&       75.35&       12.77&       11.11&       98.27\\
26         &        7500&         0.992&         229&   136486.96&       13.39&       76.52&       12.31&       11.11&       98.27\\
27         &        7500&         0.993&         209&   149547.91&       12.91&       77.61&       11.66&       11.11&       98.27\\
28         &        7500&         0.994&         187&   167141.79&       12.06&       79.13&       10.44&       11.11&       98.27\\
29         &        7500&         0.995&         162&   192935.27&       11.09&       80.94&        8.90&       41.93&       98.27\\
30         &        7500&        0.996&         138&   226489.23&       10.20&       82.40&        8.47&       53.74&       98.27\\
31         &        7500&         0.997&         121&   258310.03&        9.40&       83.92&        7.65&       56.41&       98.27\\
32         &        7500&         0.998&          92&   339733.85&        7.87&       85.87&        6.89&       60.42&       98.27\\
33         &        7500&         0.999&          59&   529754.47&        5.65&       89.03&        5.44&       64.71&       98.27\\
\hline\hline
\end{tabular}}
\begin{minipage}{\textwidth}
          \vspace{6pt}
          \footnotesize \textit{Notes:} The table compares different statistics for labour market definitions based on a different number of municipality regions (Column 2) and clustering cut-offs (Column 3). It displays the number of local labour markets (N LLM) in column 4, the average resident labour force (RLF (mean)) in column 5, the commuting ratio (CR) in column 6 and the mean, standard deviation and minimum and maximum  employment self-containment ratio (ESC) in columns 7 to 10.
     \end{minipage}
\end{table}

Table \ref{app:char_llm} compares self-containment across labour market definitions for a different number of underlying municipality regions. It shows that the commuting ratio tends to be slightly lower for a higher number of underlying number of municipality regions. Nevertheless, the ESC ratio is not following a clear trend with regard to first level of aggregation. Overall, the choice of municipality regions in the first level of aggregation does not have a strong impact on the closedness of the labour markets.

\subsubsection{Districts and Local Employment Agencies}\label{app:districts}

\begin{figure}[H]
\begin{center}
                 \caption{Districts and Local Employment Agency Borders} 
                  \label{fig:map_dist}
               \vspace{-5pt}
               \centering\includegraphics[scale=1, clip=true, trim=1cm 0cm 1cm 1cm]{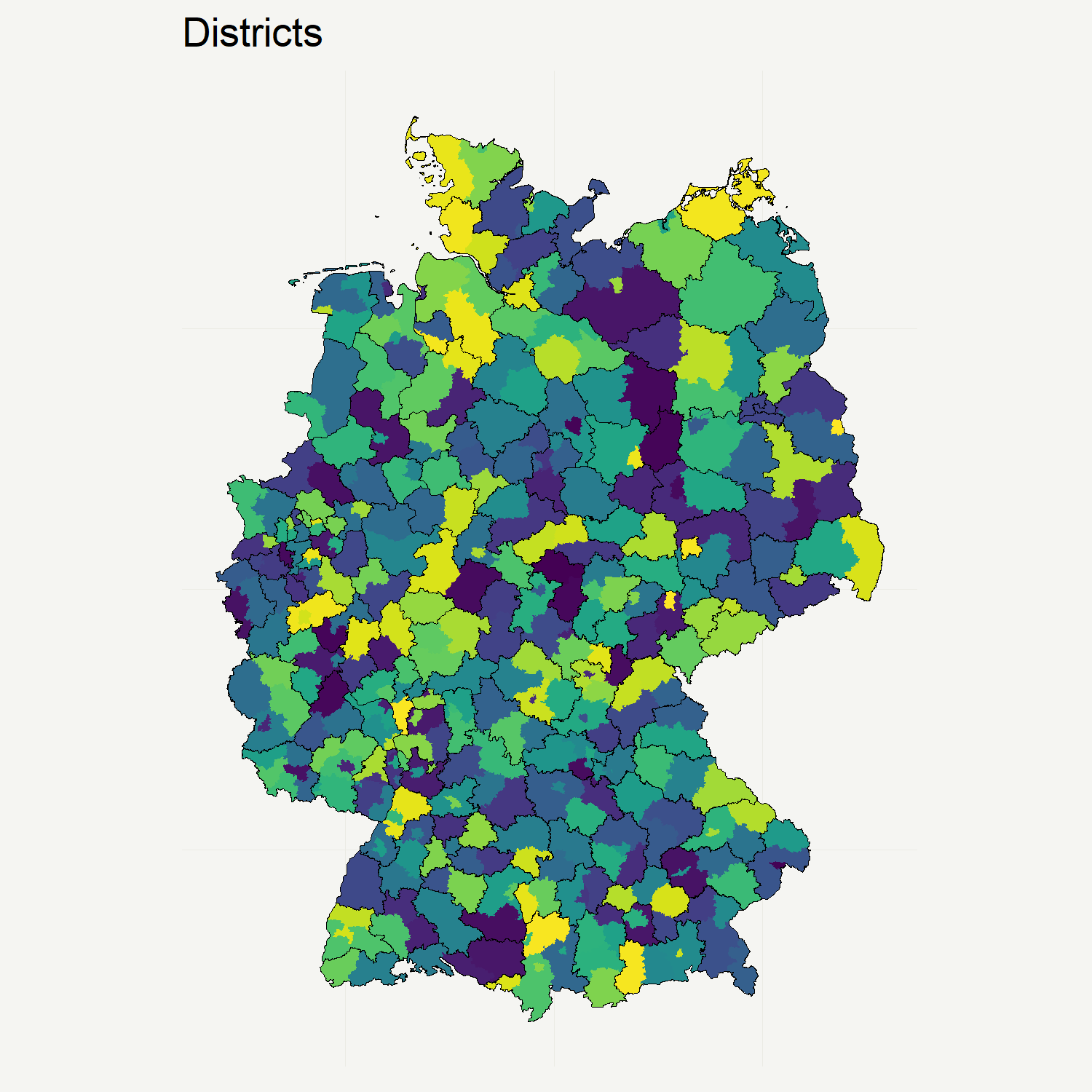}
               \end{center} \vspace{0.5cm}
               \footnotesize \textit{ Notes:} 402 Districts are depicted as coloured areas, LEA borders are drawn in black. Territorial status of 2016.
               \footnotesize 
\end{figure}

\subsubsection{Labour market definitions}

\begin{figure}[H]
\begin{center}
                 \caption{Local Labour Markets and Local Employment Agency Borders} 
                  \label{fig:map_pref}
               \vspace{-5pt}
               \centering\includegraphics[scale=0.8, clip=true, trim=1cm 0cm 1cm 1cm]{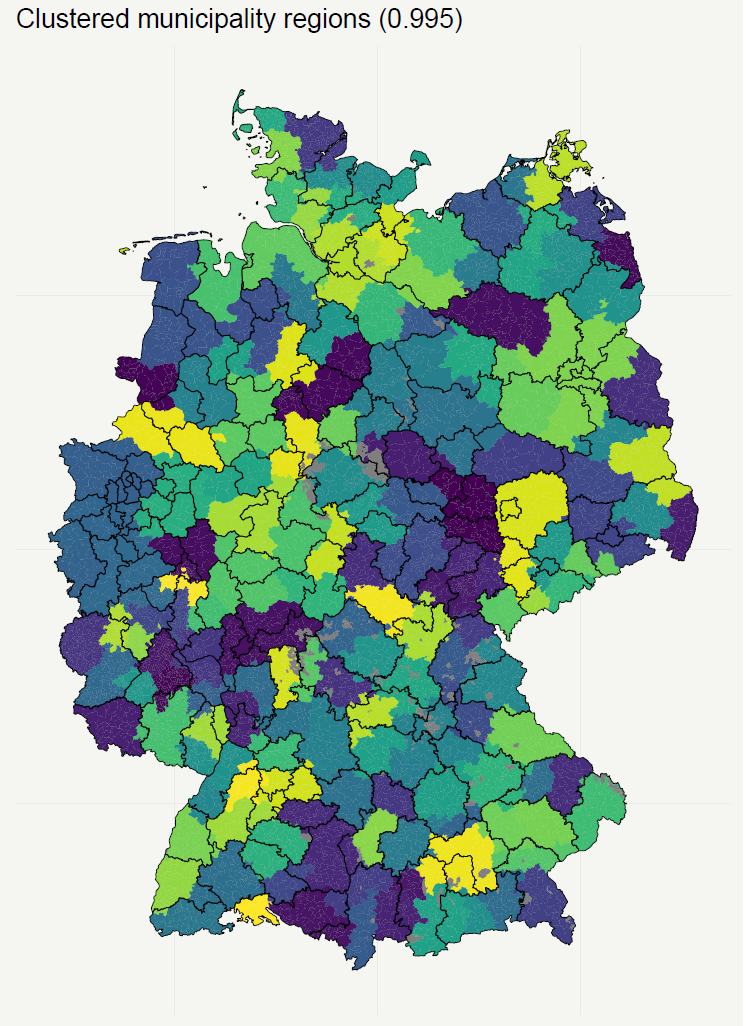}
               \end{center} \vspace{0.5cm}
               \footnotesize \textit{ Notes:} 162 Labour markets obtained by hierarchical clustering (average linkage) of 6000 municipality regions based on a dissimilarity matrix measuring the strength of commuting flows with an original cut-off value of 0.995. Labour markets are coloured areas, employment agency borders are drawn in black. Territorial status of 2016.
               \footnotesize 
\end{figure}

\subsubsection{Consistency of Labour Market Definition over Time}\label{app:consistency_time}

While it has been documented that commuting distances in Germany increased over the last two decades \citep{Dauth2018}, it is not clear whether this increase will also lead to differences in the definition of labour markets. We thus check our main definition based on commuting flows in 2016 for consistency over time and additionally define labour markets based on the commuting flows in 2000 and 2009.\footnote{We use municipality regions that have been defined based on the driving distances recorded in 2016. This might lead so some misclassification of municipalities to municipality regions in these earlier years if also driving distances between adjacent municipalities would have changed substantially over time. We argue that a potential error would be small since we only aggregate very few municipalities that are not very distant. Driving distances are more likely to be substantially affected between more distant locations, for example through larger infrastructure projects such as the construction of highways.} 


We calculate a similarity measure where for each municipality $m$, we count the number of municipalities which were assigned to the same labour market in year $t \in \{2002;2009\}$ and  year $t' \in \{2009;2016\}$ and divide it by the total number of municipalities in the market (except $m$) in year $t'$. To account for the population density, we calculate a similar measure which weights the counts by the size of the local labour force in the respective municipalities.

\begin{table}[!ht]
    \centering \caption{Consistency of Labour Market Definitions over Time}\label{tab:consitency}
    \resizebox{\linewidth}{!}{%
    \begin{tabular}{lcccccc}
    \hline\hline
    & (1)        &         (2)&        (3)&     (4)&         (5)&     (6)\\
        Year $t$-year $t'$ & All & All (weighted) & LLM centers & LLM centers (weighted) & N (year $t$) & N (year $t'$) \\ \hline
          2002-09 & 0.70 & 0.72 & 0.80 & 0.80 & 228 & 176 \\ 
        2002-16 & 0.64 & 0.67 & 0.74 & 0.73 & 228 & 161 \\ 
        2009-16 & 0.81 & 0.82 & 0.84 & 0.84 & 176 & 161 \\ 
        \hline\hline 
    \end{tabular}}
\end{table}

Table \ref{tab:consitency} depicts averages of these shares for the comparison between the 2002 and 2016 and 2009 and 2016 for our preferred labour market definition. Column (1) and (2) report raw and population weighted averages of the shares by averaging over all municipalities. Column (3) and (4) average over the largest municipalities in the labour markets, i.e. the labour market centers. We find that after weighting by population density, around 70-80\% of municipalities in a market remain the same across the three years that we consider. The shares are significantly higher when only inspecting labour market centers. We thus conclude that using our baseline definition of labour markets based on 2016 data should be sufficiently representative of the time period that we consider.


\subsubsection{Comparability to Other Definitions of Local Labour Markets for Germany}\label{app:otherdef}

\begin{table}[!ht]
    \centering \caption{Self-containment of Other Regional Delineations for Germany}\label{tab:otherdef}
    \resizebox{\linewidth}{!}{%
    \begin{tabular}{lcccccc}
    \hline  \hline
    & (1)        &         (2)&        (3)&     (4)&         (5)&     (6)\\
        Delineation & N regions & CR & ESC (mean) & ESC (sd) & ESC (min) & ESC (max) \\
        \hline
        NUTS3 (districts) & 402 & 38.1 & 58.9 & 13.6 & 20.5 & 86.7 \\
        LEA (local employment agencies) & 156 & 27.3 & 71.5 & 10.8 & 45.6 & 91.6 \\
        SPR (spatial planning regions) & 96 & 19.6 & 78.5 & 8.6 & 50.3 & 91.6 \\
        GRW (labour market regions of GRW) & 258 & 25.5 & 69.3 & 10.3 & 32.7 & 91.2 \\
        RLM (regional labour markets) & 141 & 19.9 & 75.9 & 9.4 & 42 & 91.3 \\
        LMR (labour market regions) & 50 & 10.5 & 85.7 & 5 & 71.8 & 94.6 \\
        Our definition (6000, 995) & 162 & 12.6 & 80.7 & 8.3 & 45.9 &98.3\\ 
        \hline\hline
\end{tabular}}
\begin{minipage}{\textwidth}
          \vspace{6pt}
          \footnotesize \textit{Notes:} The table displays self-containment statistics for different regional delineations in Germany. It shows the number of regions (N regions) in column 1, the commuting ratio (CR) in column 2 and the mean, standard deviation and minimum and maximum employment self-containment ratio (ESC) in columns 3 to 6. Numbers for our definition are based on data from 2016, all others are based on 2014, calculated by \cite{Wicht2020}.
     \end{minipage}
\end{table}

Table \ref{tab:otherdef} shows the number of regions, the commuting share and two measures of self-containment for German districts, LEA, and four different definitions of local labour markets: 96 Spatial planning regions, 258 regions of the GRW (labour market regions of the Joint Task of the Federal Government and the federal states dedicated to the ‘Improvement of Regional Economic Structure’), 141 regions (RLM) defined by \cite{Kosfeld2012} and 50 regions (LMR) defined by \cite{Kropp2016}. The numbers are based on calculations by  \cite{Wicht2020} based on commuting flows for the year 2014. Two points are worth mentioning: First, it stands our that administrative regions such as LEA often used as units of analysis in comparable macro analyses for Germany are characterized by a comparatively lower levels of self-containment compared to most other labour market definitions. Second, our labour market definitions have comparable or even higher levels of self-containment and lower commuting shares for comparable numbers of labour markets.

\subsection{Bias in the ARDL Model}\label{bias}

In the following we show that the error term in an ARDL model estimated in yearly changes similar to equation (\ref{eq:ardl}) is correlated with the lagged dependent variable through its influences on past values of $y_{it}$. We abstract from covariates and policy variables for this exposition.

Given a model of the following form: 

$\Delta y_{i t}=\theta \Delta y_{i,t-1}+\Delta \varepsilon_{i t} $

Backwards substitution leads to:

$\Delta y_{i, t-1}=\theta \Delta y_{i,t-2}+\Delta \varepsilon_{i, t-1}$ 

$\Delta y_{i t}=\theta\left(\theta y_{i, t-2}+\Delta \varepsilon_{i, t-1}\right)+\Delta \varepsilon_{i t}$

...

$\Delta y_{i t}=\theta\left(\theta\left(\theta\left(\theta\left(\theta y_{i, t-5}+\Delta \varepsilon_{i, t-4}\right)+\Delta \varepsilon_{i, t-3}\right)+\Delta \varepsilon_{i, t-2}\right)+\Delta \varepsilon_{i, t-1}\right)+\Delta \varepsilon_{i t}$

Since $\Delta \varepsilon_{i, t-4} =\varepsilon_{i, t-4}-\varepsilon_{i, t-8}$ and $\Delta \varepsilon_{i t} =\varepsilon_{i t}-\varepsilon_{i,t-4}$

$\Rightarrow C\left(\Delta y_{i, t-1}, \Delta \varepsilon_{i t}\right) \neq 0$.


\subsection{Descriptive Statistics}\label{descr_variation}

\begin{figure}[H]
\begin{center}
                 \caption{Quarterly Inflows into Programs} 
                  \label{fig:inflows}
               \vspace{-5pt}
               \centering\includegraphics[scale=0.8, clip=true, trim=0cm 0cm 0cm 0cm]{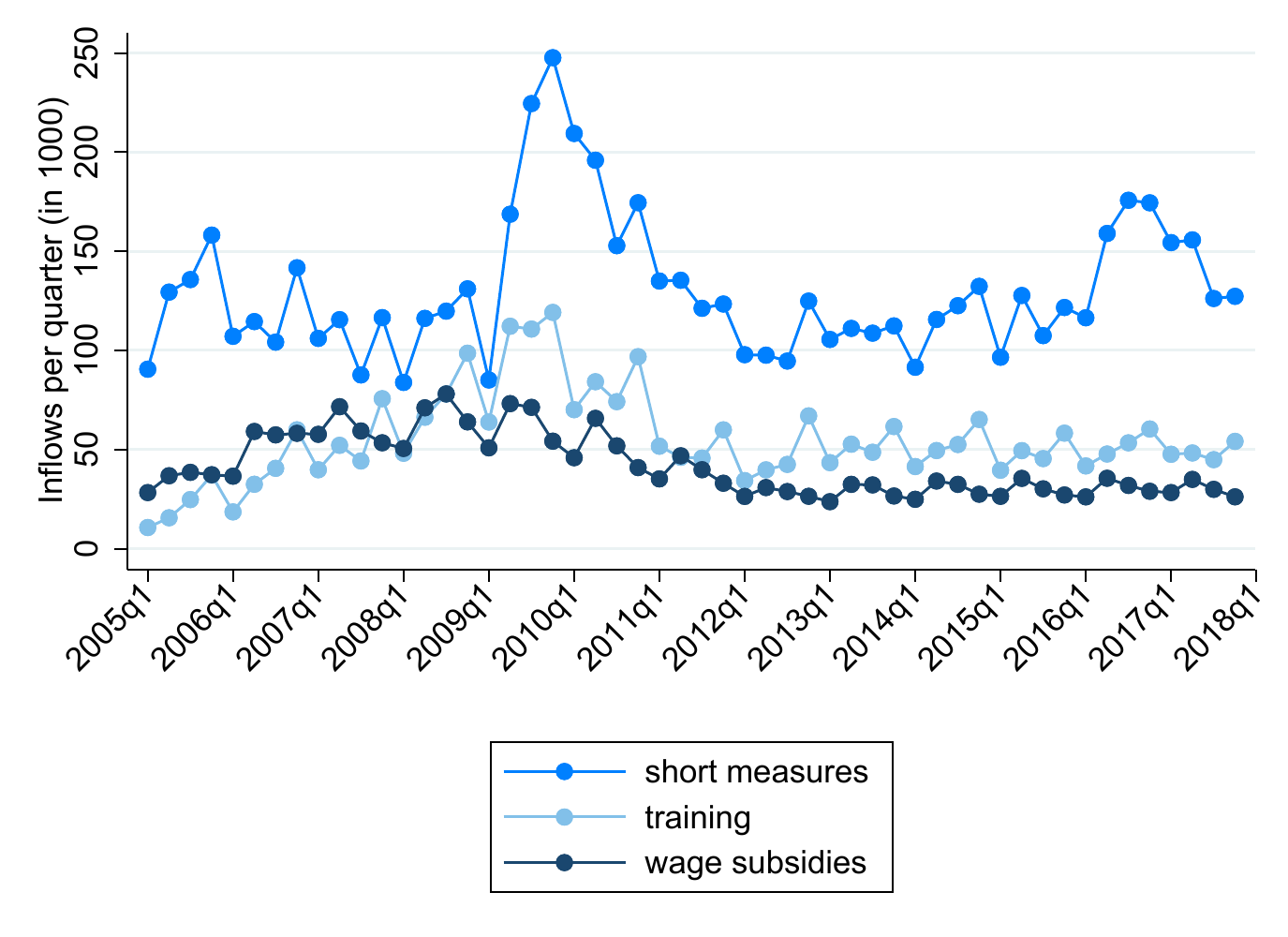}
               \end{center} \vspace{0.5cm}
               \footnotesize \textit{ Note:} This graph plots the total inflows into the three different program types at the country level. Note that due to censoring of small values at the municipality level, these numbers are slightly under-reporting the true size of inflows.
\end{figure}


\begin{figure}[H]
                \centering
                \caption{Regional Variation \label{fig:var_space}}
                \vspace{10pt}
                \begin{subfigure}[b]{0.49\textwidth}
                                \centering \caption{Unemployment rate} \label{fig:var_ue_space}
                                \includegraphics[clip=true, trim={1cm 0cm 0cm 1cm},scale=0.6]{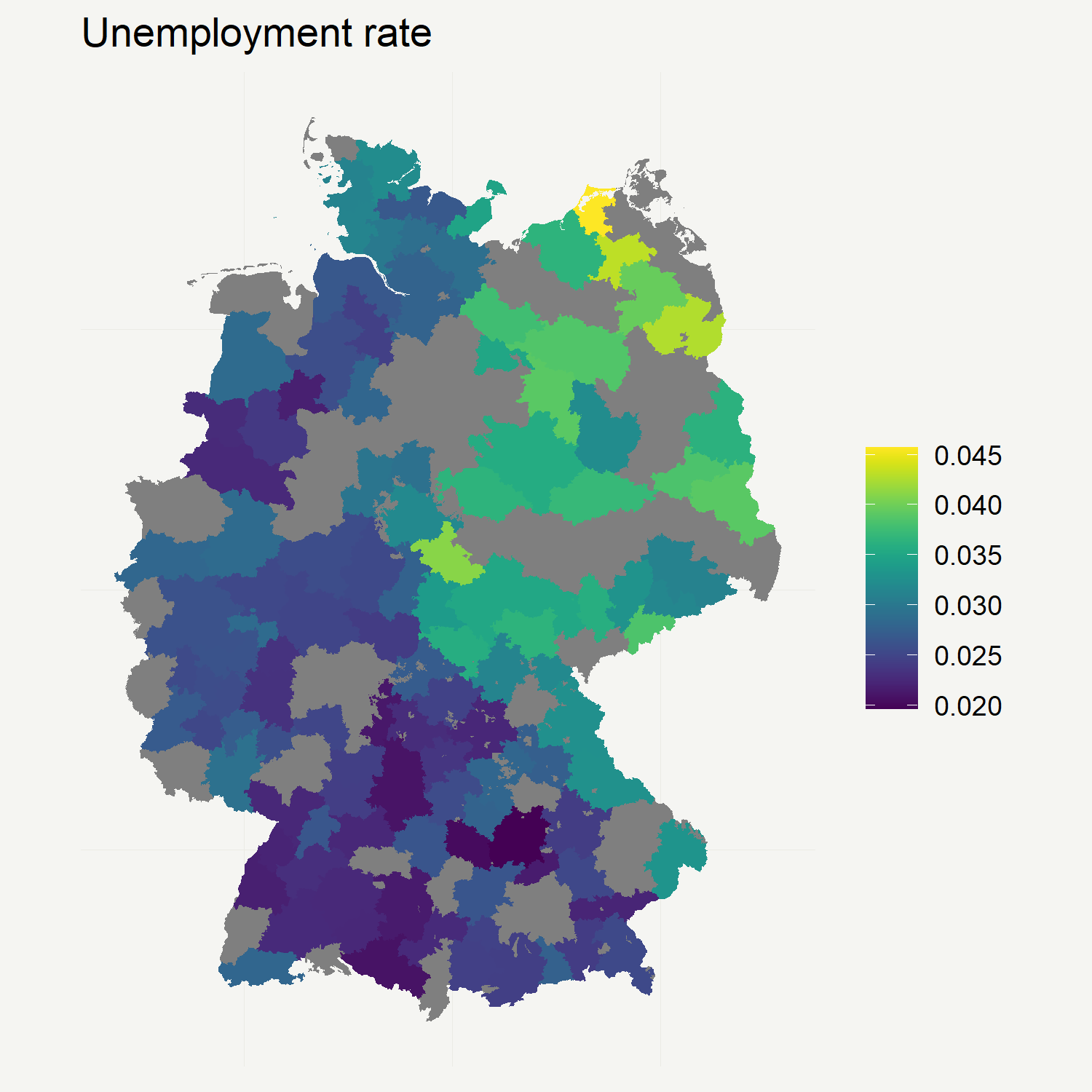}
                \end{subfigure}
                \begin{subfigure}[b]{0.49\textwidth}
                                \centering \caption{Short measures} \label{fig:var_acc_rate_tm_space}
                                \includegraphics[clip=true, trim={1cm 0cm 0cm 1cm},scale=0.6]{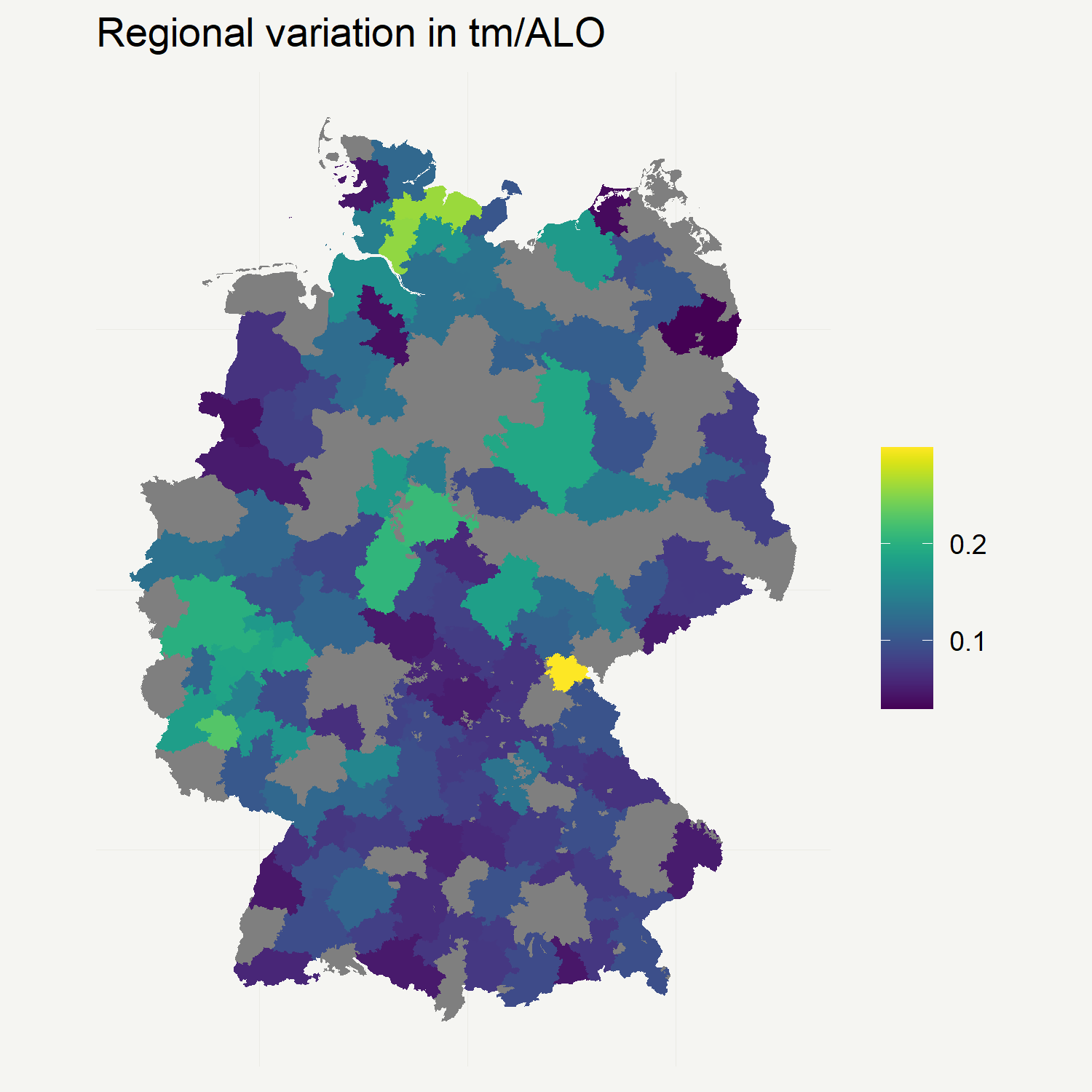}
                \end{subfigure}
               \vspace{10pt}  
                \begin{subfigure}[b]{0.49\textwidth}
                                \centering \caption{Training} \label{fig:var_acc_rate_tr_space}
                                \includegraphics[clip=true, trim={1cm 0cm 0cm 1cm},scale=0.6]{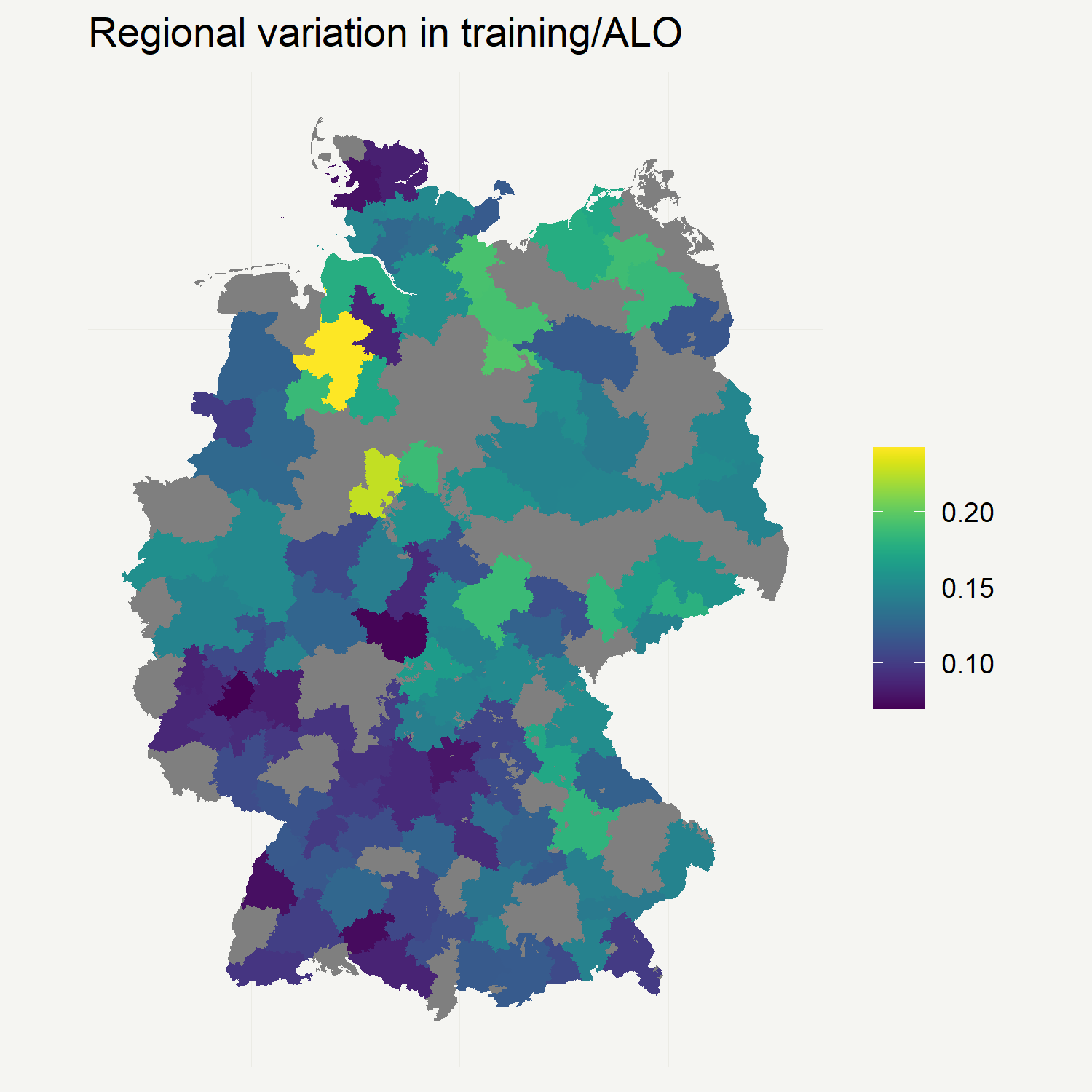}
                \end{subfigure}
                \begin{subfigure}[b]{0.49\textwidth}
                                \centering \caption{Wage subsidies} \label{fig:var_acc_rate_ws_space}
                                \includegraphics[clip=true, trim={1cm 0cm 0cm 1cm},scale=0.6]{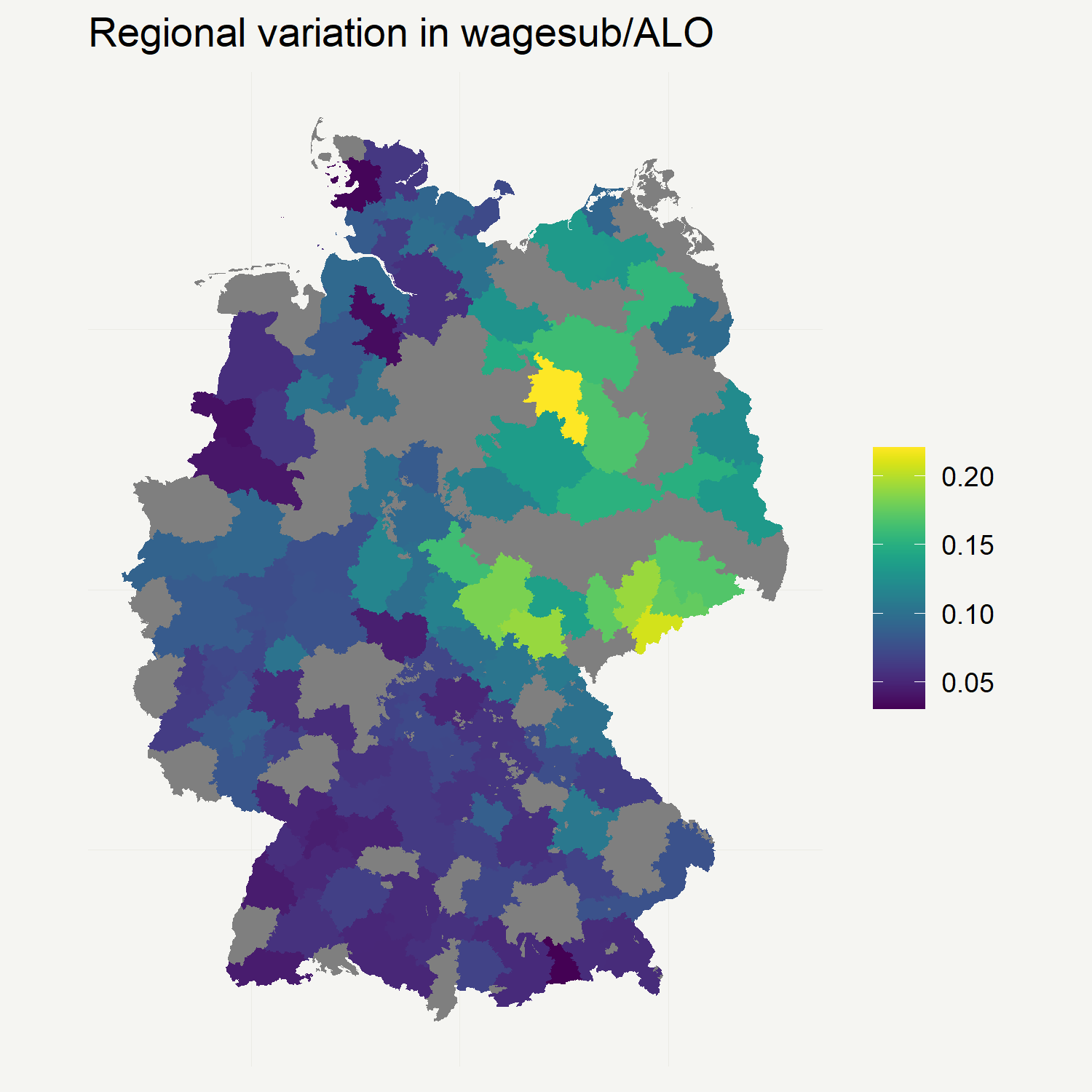}
                \end{subfigure}
                 \footnotesize \textit{ Note:} These graphs plot  the average (a) unemployment rate, (b) accommodation rate in short measures, (c) training and (d) wage subsidies across the main estimation sample of labour markets by averaging over the years 2005 to 2018. Grey areas are not included in the sample. 
\end{figure}

\begin{table}[h!]
\centering\caption{Summary Statistics - Control Variables}\label{tab:sumstats1} 
\resizebox{\linewidth}{!}{%
\begin{tabular}{lcccc}
\hline\hline
 & (1) & (2) & (3) & (4) \\
                &     mean&       sd&      min&      max\\
\hline
\multicolumn{5}{l}{\textit{Characteristics of the employed}}\\
Share German          &     0.92&     0.14&     0.00&     1.00\\
Share female          &     0.49&     0.02&     0.44&     0.58\\
Share 20-29 years*     &     0.19&     0.02&     0.10&     0.25\\
Share 30-39 years     &     0.22&     0.03&     0.15&     0.31\\
Share 40-49 years     &     0.30&     0.03&     0.20&     0.40\\
Share 50-64 years     &     0.29&     0.05&     0.19&     0.47\\
Share w/o schooling or high school degree*&     0.85&     0.05&     0.67&     0.95\\
Share w/ high school degree&     0.15&     0.05&     0.05&     0.33\\
Share  w/o vocational training*&     0.09&     0.04&     0.01&     0.23\\
Share w/ vocational training&     0.83&     0.05&     0.69&     0.94\\
Share w/ academic degree &     0.08&     0.03&     0.02&     0.20\\
Share in primary sector and manufacturing&     0.26&     0.07&     0.12&     0.45\\
Share in construction, retail trade and transportation&     0.31&     0.03&     0.22&     0.45\\
Share in services and communication&     0.12&     0.03&     0.05&     0.22\\
Share in education, government work, culture and media*&     0.31&     0.04&     0.20&     0.44\\[0.5em]
\multicolumn{5}{l}{\textit{Characteristics of the unemployed}}\\
Share  female         &     0.44&     0.05&     0.22&     0.65\\
Share  20-29 years*   &     0.23&     0.04&     0.06&     0.39\\
Share  30-39 years    &     0.19&     0.03&     0.00&     0.31\\
Share  40-49 years    &     0.21&     0.03&     0.05&     0.35\\
Share 50-64 years     &     0.38&     0.06&     0.17&     0.60\\
Share  w/o schooling or high school degree*&     0.85&     0.06&     0.00&     0.97\\
Share  w/ high school degree&     0.15&     0.06&     0.00&     0.38\\
Share  w/o vocational training*&     0.20&     0.10&     0.00&     0.42\\
Share  w/ vocational training&     0.73&     0.09&     0.52&     0.95\\
Share  w/ academic degree&     0.07&     0.04&     0.00&     0.23\\
Share vocational degree censored       &     0.00&     0.00&     0.00&     0.04\\
Share in primary sector and manufacturing&     0.23&     0.06&     0.00&     0.52\\
Share in construction, retail trade and transportation&     0.39&     0.06&     0.17&     0.70\\
Share in services and communication&     0.15&     0.04&     0.00&     0.31\\
Share in education, government work, culture and media&     0.23&     0.05&     0.00&     0.43\\[0.5em]
\multicolumn{5}{l}{\textit{Other control variables}}\\
Share in programs for long-term unemployed (\% of unemployed)&     0.24&     0.25&     0.00&     2.02\\
Share in programs for the young (\% of unemployed)&     0.05&     0.03&     0.00&     0.24\\
Share in other programs (\% of unemployed)  &     0.14&     0.11&     0.01&     0.86\\
Resident Labour Force (in 10,000)** &18.55&22.77& 1.60&137.98\\
\hline
Observations    &     5980&         &         &         \\
LLM & 115 & & & \\
\hline\hline
\end{tabular}}
\begin{minipage}{\textwidth}
          \vspace{6pt}
          \small \textit{Notes:} The table displays summary statistics for the control variables characterizing the employed and unemployed workforce for the period 2005 to 2018. Missing values were counted to the largest category. Censored values were flagged separately. * Reference categories. ** not used as control variable.
     \end{minipage}
\end{table}

\begin{table}[h!]\centering\caption{Summary Statistics - Policy and Outcome Variables}\label{tab:sumstats2} 
{\small
\begin{tabular}{lcccc}
\hline\hline
 & (1) & (2) & (3) & (4) \\
                &     mean&       sd&      min&      max\\
\hline
\multicolumn{5}{l}{\textit{Program shares (labour markets)}}\\
Further vocational training&     0.13&     0.06&     0.01&     0.51\\
Short training  &     0.11&     0.09&     0.00&     0.76\\
Wage subsidies  &     0.09&     0.06&     0.01&     0.51\\[0.5em]
\multicolumn{5}{l}{\textit{Program shares (instrument areas)}}\\
Further vocational training&     0.13&     0.05&     0.02&     0.35\\
Short training  &     0.11&     0.07&     0.01&     0.50\\
Wage subsidies  &     0.09&     0.06&     0.01&     0.40\\[0.5em]
\multicolumn{5}{l}{\textit{Outcomes}}\\
Unemployment rate (\% of LF)&     0.03&     0.01&     0.01&     0.12\\
Employment rate (\% of LF)&     0.86&     0.07&     0.57&     0.97\\
Rate of welfare recipients (\% of LF)&     0.08&     0.04&     0.00&     0.27\\
Rate of employed on benefits(\% of LF)&     0.03&     0.02&     0.01&     0.09\\
Inflow rate into employment (\% of LF)&     0.02&     0.00&     0.01&     0.05\\
Unemployment rate female&     0.03&     0.01&     0.01&     0.10\\
Unemployment rate male&     0.03&     0.01&     0.01&     0.13\\
Unemployment rate age under 30&     0.03&     0.02&     0.01&     0.14\\
Unemployment rate age 30 to 50&     0.03&     0.01&     0.01&     0.12\\
Unemployment rate age over 50&     0.04&     0.02&     0.01&     0.13\\
Unemployment rate low-educated&     0.03&     0.01&     0.00&     0.12\\
Unemployment rate highly-educated&     0.03&     0.01&     0.00&     0.11\\
Unemployment rate low-skilled&     0.05&     0.02&     0.00&     0.17\\
Unemployment rate medium skilled&     0.03&     0.01&     0.01&     0.12\\
Unemployment rate high-skilled&     0.02&     0.01&     0.00&     0.09\\
Employment rate female&     0.87&     0.06&     0.61&     0.97\\
Employment rate male&     0.85&     0.08&     0.49&     0.97\\
Employment rate age under 30 &     0.83&     0.09&     0.46&     0.97\\
Employment rate age 30 to 50 &     0.86&     0.07&     0.57&     0.97\\
Employment rate age over 50&     0.83&     0.08&     0.47&     0.96\\
Employment rate low educated&     0.86&     0.07&     0.57&     1.00\\
Employment rate high educated&     0.86&     0.07&     0.49&     1.00\\
Employment rate low-skilled&     0.74&     0.12&     0.25&     1.00\\
Employment rate medium skilled&     0.87&     0.07&     0.56&     0.98\\
Employment rate high-skilled&     0.88&     0.07&     0.52&     1.00\\
\hline
Observations    &     5980&         &         &         \\
LLM & 115 & & & \\
\hline\hline
\end{tabular}}
\begin{minipage}{\textwidth}
          \vspace{6pt}
          \small \textit{Notes:} The table displays summary statistics for the policy and outcome variables for the period 2005 to 2018. 
     \end{minipage}
\end{table}

\subsection{Further Results}\label{app_results}

\begin{table}[H]
\centering
\caption{First Stage - Unsubsidized Employment Rate}
\label{tab:fs_emp_rate}
{\small
{
\def\sym#1{\ifmmode^{#1}\else\(^{#1}\)\fi}
\begin{tabular}{l*{4}{c}}
\hline\hline
            &   SW F-stat&    SW p-val&   AP F-stat&    AP p-val\\
\hline
AR training &      115.97&       0.000&      135.14&       0.000\\
AR training (lag 1)&       86.54&       0.000&      132.59&       0.000\\
AR training (lag 2)&      110.65&       0.000&      101.31&       0.000\\
AR training (lag 3)&       84.05&       0.000&       70.68&       0.000\\
AR training (lag 4)&       81.79&       0.000&       93.56&       0.000\\
AR training (lag 5)&       72.87&       0.000&       93.35&       0.000\\
AR training (lag 6)&      115.27&       0.000&      278.97&       0.000\\
AR short measures&       61.01&       0.000&      373.57&       0.000\\
AR short measures (lag 1)&       61.58&       0.000&      125.57&       0.000\\
AR short measures (lag 2)&       85.02&       0.000&      174.44&       0.000\\
AR short measures (lag 3)&       48.39&       0.000&      142.64&       0.000\\
AR short measures (lag 4)&       61.83&       0.000&      143.11&       0.000\\
AR short measures (lag 5)&       59.79&       0.000&      149.17&       0.000\\
AR short measures (lag 6)&       65.64&       0.000&      230.23&       0.000\\
AR wage subsidies&      103.32&       0.000&      435.60&       0.000\\
AR wage subsidies (lag 1)&       93.88&       0.000&      398.92&       0.000\\
AR wage subsidies (lag 2)&       99.85&       0.000&      202.76&       0.000\\
AR wage subsidies (lag 3)&      104.09&       0.000&      243.71&       0.000\\
AR wage subsidies (lag 4)&      105.92&       0.000&      212.58&       0.000\\
AR wage subsidies (lag 5)&       82.87&       0.000&      205.46&       0.000\\
AR wage subsidies (lag 6)&      119.28&       0.000&      489.47&       0.000\\
\hline\hline
\end{tabular}

}
}
\centering
     \begin{minipage}{\textwidth}
          \vspace{6pt}
          \footnotesize \textit{Notes:} The table reports three types of diagnostic test of the first stage for the 21 excluded instruments for the model including the lagged unsubsidized employment rate. Each line refers to a separate first-stage regression for the respective accommodation rate (AR). The first two columns show the conventional F-statistic of a joint test whether all  instruments are significantly different from zero and the corresponding p-value. Columns 3 and 4 show the Sanderson-Windmeijer (2016) F statistics (SW F-stat) and p-values. Columns 5 and 6 the Angrist-Pischke (2009) F-statistics and p-values (AP F-stat).
     \end{minipage}
\end{table}

\begin{table}[H]
\centering
\caption{First Stage - Rate of Welfare Recipients}
\label{tab:fs_wf_rate}
{\small
{
\def\sym#1{\ifmmode^{#1}\else\(^{#1}\)\fi}
\begin{tabular}{l*{4}{c}}
\hline\hline
            &   SW F-stat&    SW p-val&   AP F-stat&    AP p-val\\
\hline
AR training &      114.63&       0.000&      136.00&       0.000\\
AR training (lag 1)&       86.88&       0.000&      130.84&       0.000\\
AR training (lag 2)&      104.21&       0.000&      102.30&       0.000\\
AR training (lag 3)&       83.41&       0.000&       70.29&       0.000\\
AR training (lag 4)&       81.27&       0.000&       93.45&       0.000\\
AR training (lag 5)&       73.94&       0.000&       93.18&       0.000\\
AR training (lag 6)&      114.37&       0.000&      276.01&       0.000\\
AR short measures&       61.27&       0.000&      374.94&       0.000\\
AR short measures (lag 1)&       61.63&       0.000&      124.84&       0.000\\
AR short measures (lag 2)&       85.32&       0.000&      174.57&       0.000\\
AR short measures (lag 3)&       48.28&       0.000&      142.87&       0.000\\
AR short measures (lag 4)&       61.99&       0.000&      143.35&       0.000\\
AR short measures (lag 5)&       59.24&       0.000&      148.33&       0.000\\
AR short measures (lag 6)&       65.65&       0.000&      230.49&       0.000\\
AR wage subsidies&      102.93&       0.000&      433.97&       0.000\\
AR wage subsidies (lag 1)&       90.00&       0.000&      391.05&       0.000\\
AR wage subsidies (lag 2)&      100.40&       0.000&      203.33&       0.000\\
AR wage subsidies (lag 3)&      103.86&       0.000&      245.09&       0.000\\
AR wage subsidies (lag 4)&      106.09&       0.000&      212.67&       0.000\\
AR wage subsidies (lag 5)&       83.11&       0.000&      209.75&       0.000\\
AR wage subsidies (lag 6)&      118.79&       0.000&      489.82&       0.000\\
\hline\hline
\end{tabular}

}
}
\centering
     \begin{minipage}{\textwidth}
          \vspace{6pt}
          \footnotesize \textit{Notes:} The table reports three types of diagnostic test of the first stage for the 21 excluded instruments for the model including the lagged rate of welfare recipients. Each line refers to a separate first-stage regression for the respective accommodation rate (AR). The first two columns show the conventional F-statistic of a joint test whether all  instruments are significantly different from zero and the corresponding p-value. Columns 3 and 4 show the Sanderson-Windmeijer (2016) F statistics (SW F-stat) and p-values. Columns 5 and 6 the Angrist-Pischke (2009) F-statistics and p-values (AP F-stat).
     \end{minipage}
\end{table}

\begin{table}[H]
\centering
\caption{First Stage - Rate of Employed Workers on Benefits}
\label{tab:fs_prec_work_rate}
{\small
{
\def\sym#1{\ifmmode^{#1}\else\(^{#1}\)\fi}
\begin{tabular}{l*{4}{c}}
\hline\hline
            &   SW F-stat&    SW p-val&   AP F-stat&    AP p-val\\
\hline
AR training &      112.85&       0.000&      135.23&       0.000\\
AR training (lag 1)&       85.65&       0.000&      131.05&       0.000\\
AR training (lag 2)&      104.88&       0.000&      101.51&       0.000\\
AR training (lag 3)&       80.83&       0.000&       69.02&       0.000\\
AR training (lag 4)&       83.72&       0.000&       93.24&       0.000\\
AR training (lag 5)&       74.13&       0.000&       95.76&       0.000\\
AR training (lag 6)&      113.77&       0.000&      279.21&       0.000\\
AR short measures&       62.30&       0.000&      367.49&       0.000\\
AR short measures (lag 1)&       62.63&       0.000&      126.95&       0.000\\
AR short measures (lag 2)&       86.57&       0.000&      176.02&       0.000\\
AR short measures (lag 3)&       48.54&       0.000&      143.43&       0.000\\
AR short measures (lag 4)&       62.22&       0.000&      143.37&       0.000\\
AR short measures (lag 5)&       60.21&       0.000&      150.02&       0.000\\
AR short measures (lag 6)&       65.98&       0.000&      229.79&       0.000\\
AR wage subsidies&      103.35&       0.000&      437.69&       0.000\\
AR wage subsidies (lag 1)&       89.69&       0.000&      387.55&       0.000\\
AR wage subsidies (lag 2)&       99.87&       0.000&      214.19&       0.000\\
AR wage subsidies (lag 3)&      101.56&       0.000&      236.65&       0.000\\
AR wage subsidies (lag 4)&      110.38&       0.000&      215.68&       0.000\\
AR wage subsidies (lag 5)&       79.02&       0.000&      198.36&       0.000\\
AR wage subsidies (lag 6)&      119.59&       0.000&      515.27&       0.000\\
\hline\hline
\end{tabular}

}
}
\centering
     \begin{minipage}{\textwidth}
          \vspace{6pt}
          \footnotesize \textit{Notes:} The table reports three types of diagnostic test of the first stage for the 21 excluded instruments for the model including the lagged rate of employed workers on benefits. Each line refers to a separate first-stage regression for the respective accommodation rate (AR). The first two columns show the conventional F-statistic of a joint test whether all  instruments are significantly different from zero and the corresponding p-value. Columns 3 and 4 show the Sanderson-Windmeijer (2016) F statistics (SW F-stat) and p-values. Columns 5 and 6 the Angrist-Pischke (2009) F-statistics and p-values (AP F-stat).
     \end{minipage}
\end{table}

\begin{figure}
                \centering
                \caption{Cumulative Marginal Effects for Females  \label{fig:cum_effects_iv_alo_rate_female}}
                \vspace{10pt}
                \begin{subfigure}[b]{0.49\textwidth}
                                \centering \caption*{Training} \subcaption*{Unemployment rate} 
                                \includegraphics[clip=true, trim={0cm 0cm 0cm 0cm},scale=0.50]{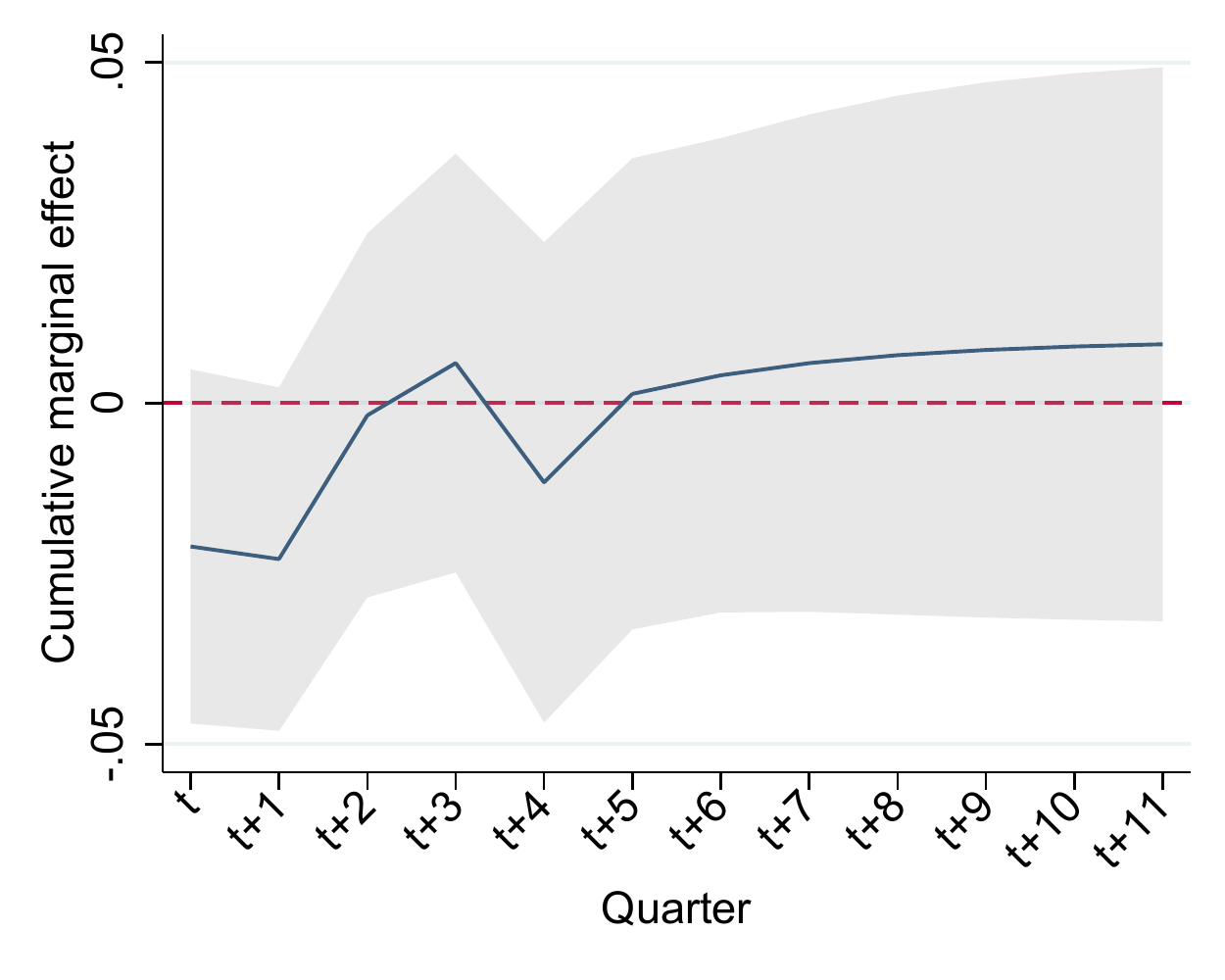}
                \end{subfigure}
                \begin{subfigure}[b]{0.49\textwidth}
                                \centering \caption*{Training} \subcaption*{Employment rate} 
                                \includegraphics[clip=true, trim={0cm 0cm 0cm 0cm},scale=0.50]{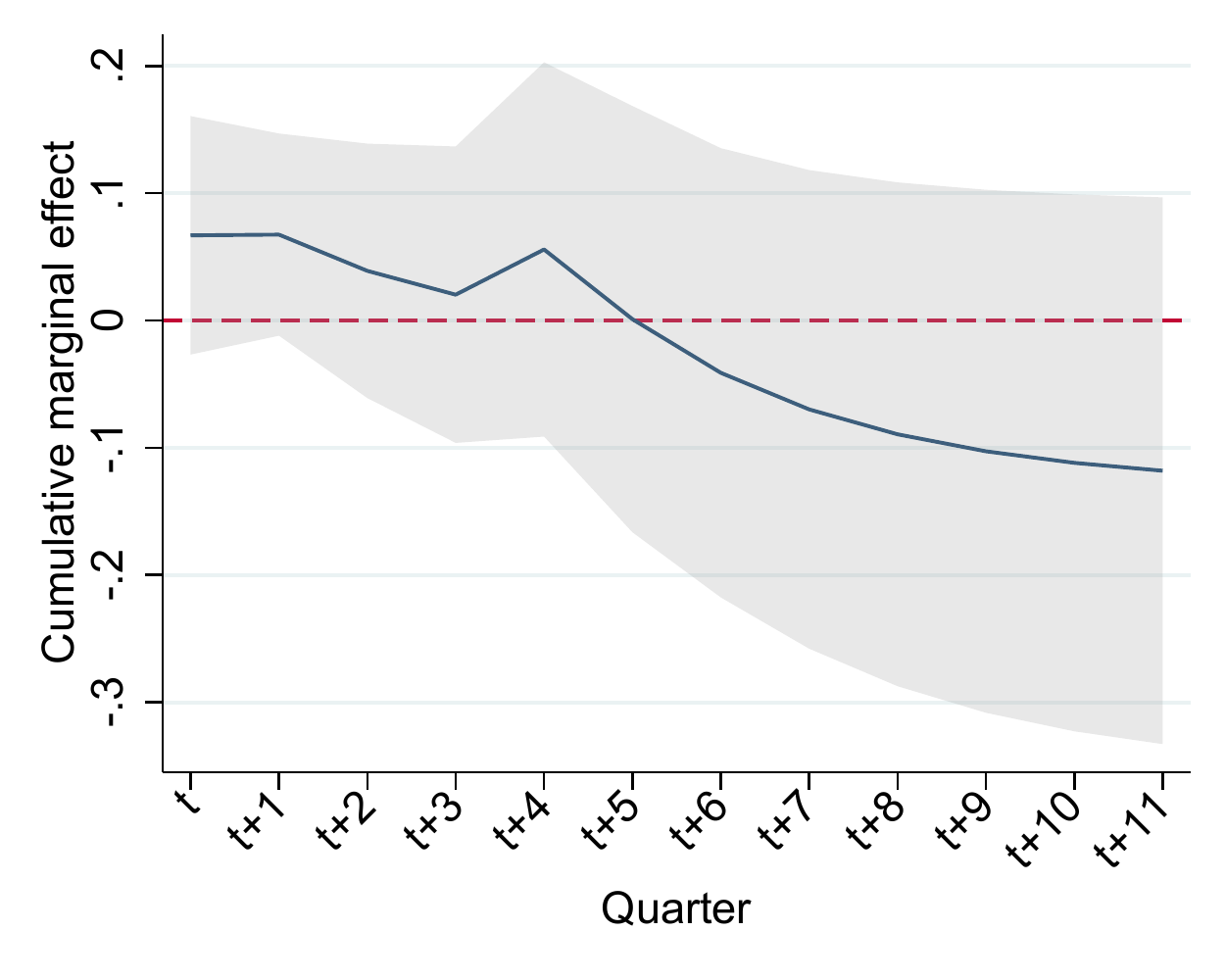}
                \end{subfigure}
                \vspace{10pt}    
                \newline
                \begin{subfigure}[b]{0.49\textwidth}
                                \centering \caption*{Short measures} \subcaption*{Unemployment rate} 
                                \includegraphics[clip=true, trim={0cm 0cm 0cm 0cm},scale=0.50]{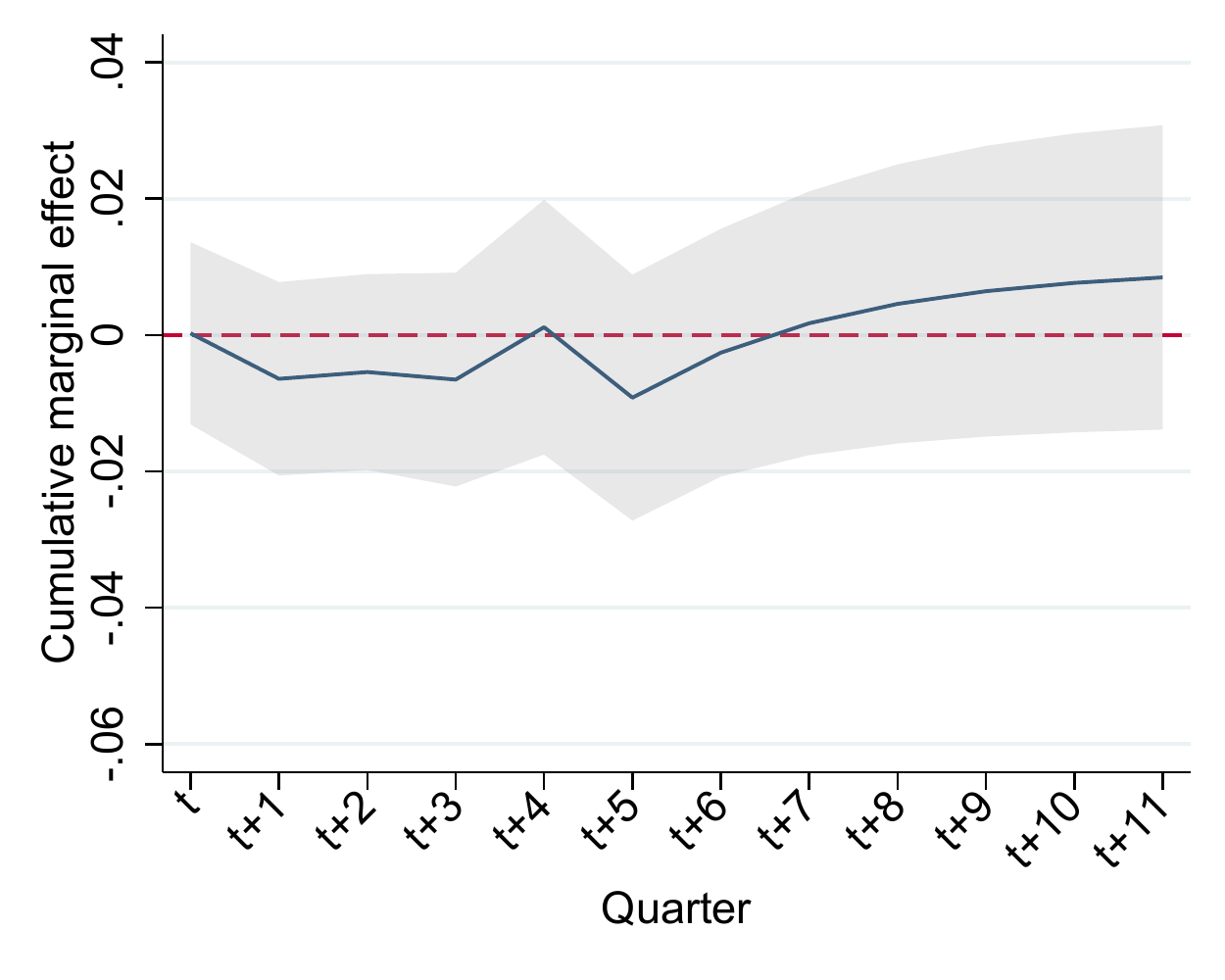}
                \end{subfigure}
                \vspace{10pt}    
                \begin{subfigure}[b]{0.49\textwidth}
                                \centering \caption*{Short measures}  \subcaption*{Employment rate} 
                                \includegraphics[clip=true, trim={0cm 0cm 0cm 0cm},scale=0.50]{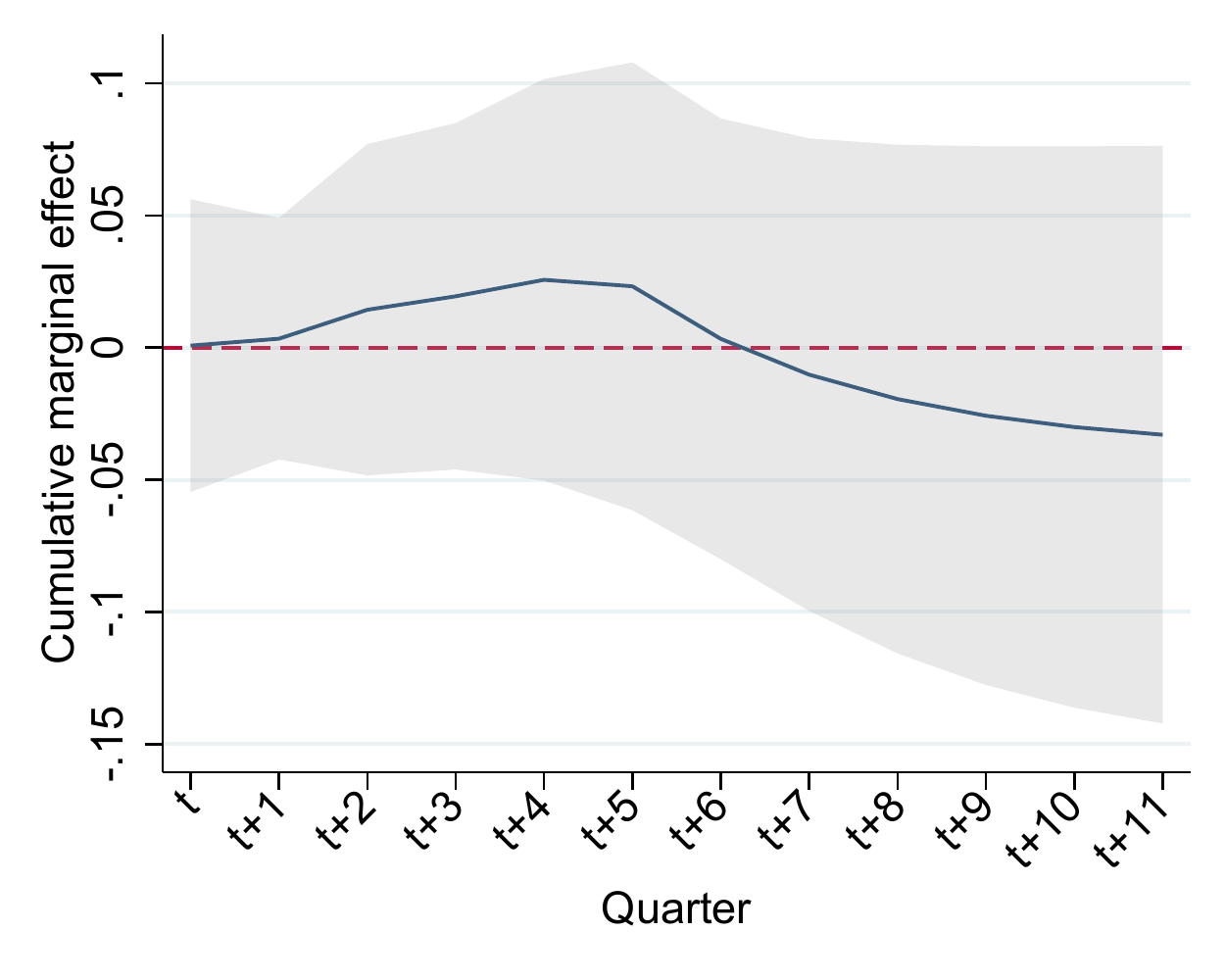}
                \end{subfigure}
                \vspace{10pt}
                \newline
                \begin{subfigure}[b]{0.49\textwidth}
                                \centering \caption*{Wage subsidies} \subcaption*{Unemployment rate} 
                                \includegraphics[clip=true, trim={0cm 0cm 0cm 0cm},scale=0.50]{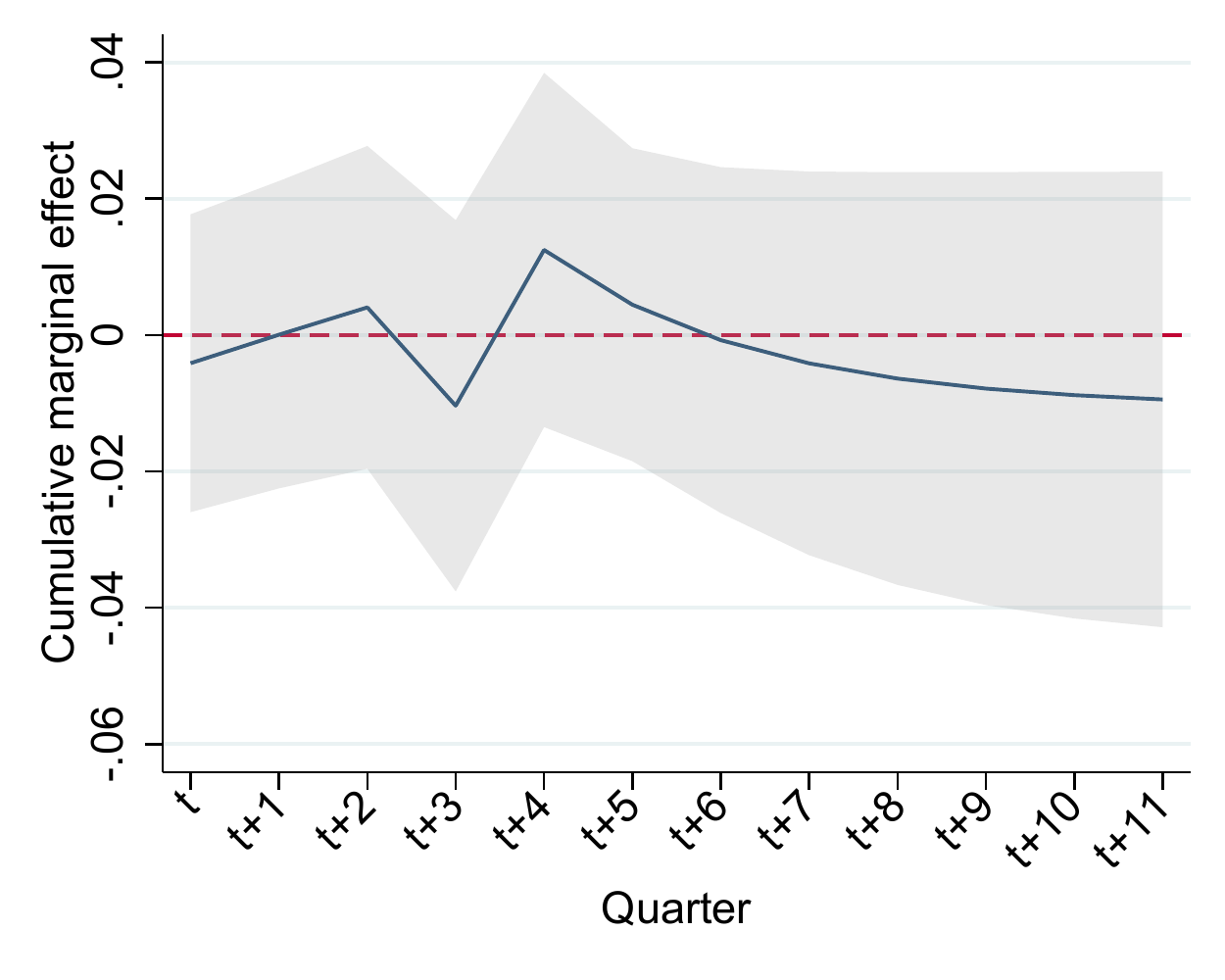}
                \end{subfigure}
                \vspace{10pt}    
                \begin{subfigure}[b]{0.49\textwidth}
                                \centering \caption*{Wage subsidies}  \subcaption*{Employment rate} 
                                \includegraphics[clip=true, trim={0cm 0cm 0cm 0cm},scale=0.50]{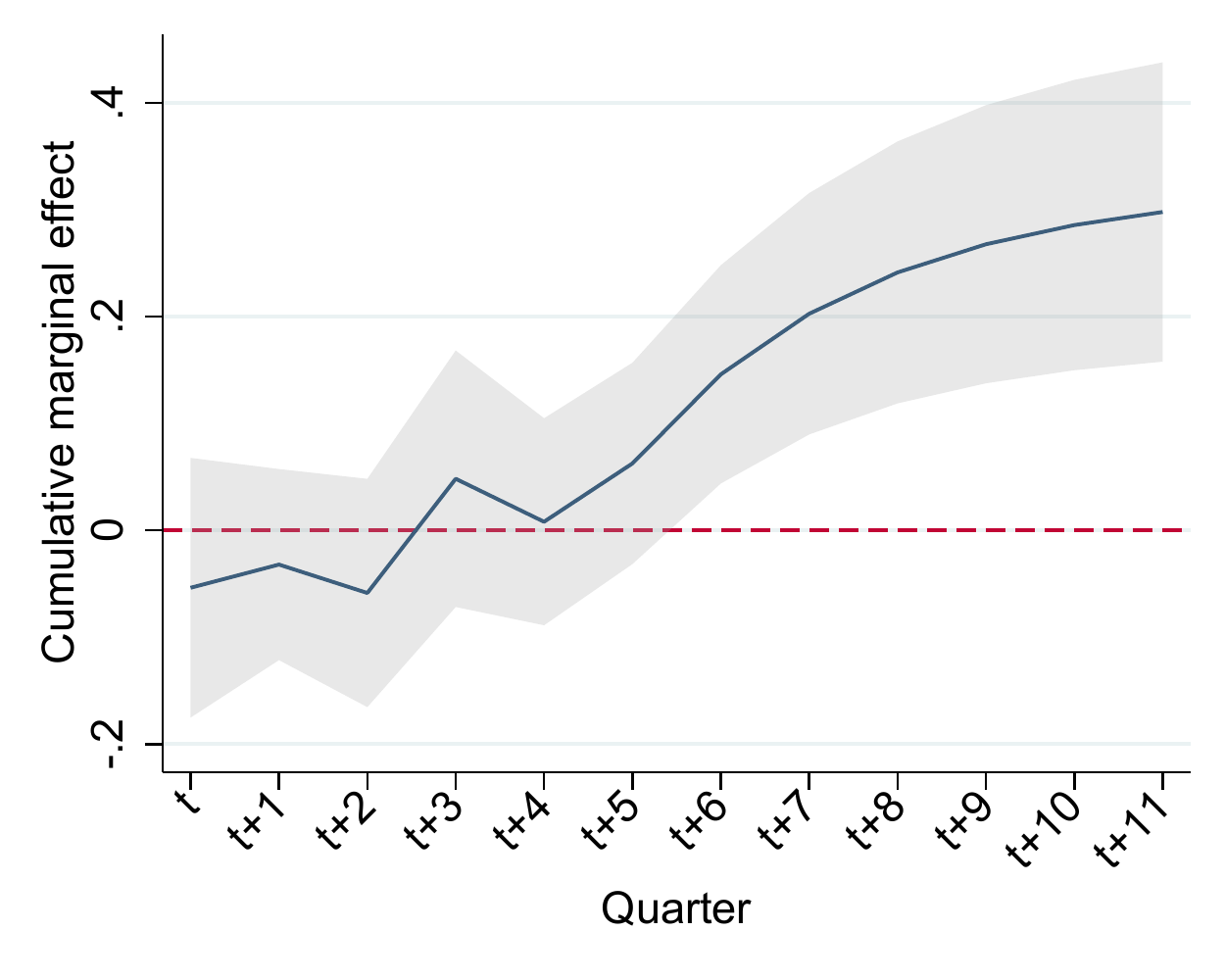}
                \end{subfigure}
                \vspace{10pt}
                \begin{minipage}{\textwidth}
                                \footnotesize \textit{Notes:} These graphs show the cumulative marginal effects of the three types of ALMP, i.e. training, short measures and wage subsidies on the unemployment rate and unsubsidized employment rate. 95\% confidence intervals are shown as grey areas. The effects are based on the ARDL model estimated by 2SLS. Program variables are included with 6 lags, main sample restrictions apply.  Standard errors obtained by a cross-sectional bootstrap (499 replications).
                \end{minipage}
\end{figure}

\begin{figure}
                \centering
                \caption{Cumulative Marginal Effects for Males \label{fig:cum_effects_iv_alo_rate_male}}
                \vspace{10pt}
                \begin{subfigure}[b]{0.49\textwidth}
                                \centering \caption*{Training} \subcaption*{Unemployment rate} 
                                \includegraphics[clip=true, trim={0cm 0cm 0cm 0cm},scale=0.50]{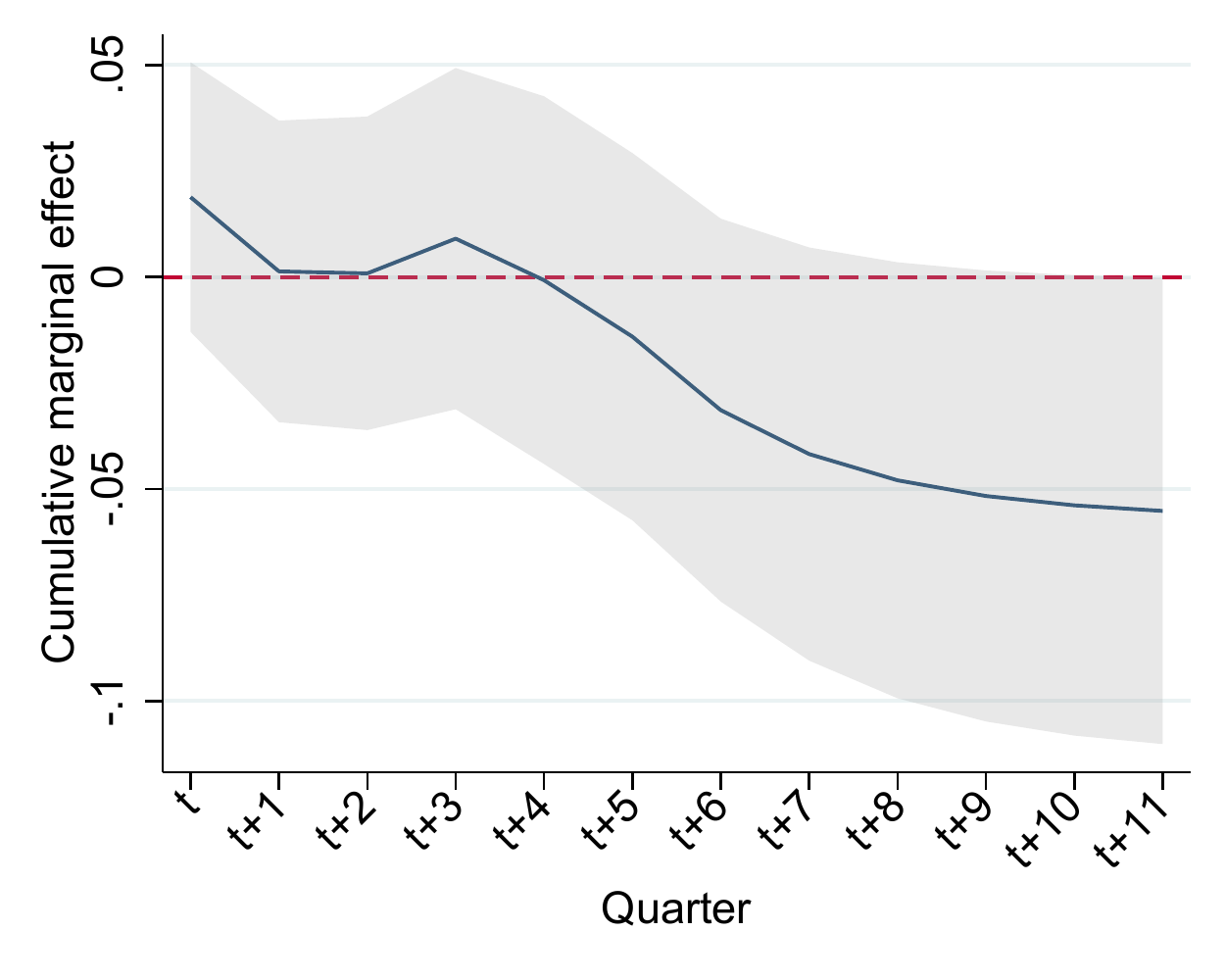}
                \end{subfigure}
                \begin{subfigure}[b]{0.49\textwidth}
                                \centering \caption*{Training} \subcaption*{Employment rate} 
                                \includegraphics[clip=true, trim={0cm 0cm 0cm 0cm},scale=0.50]{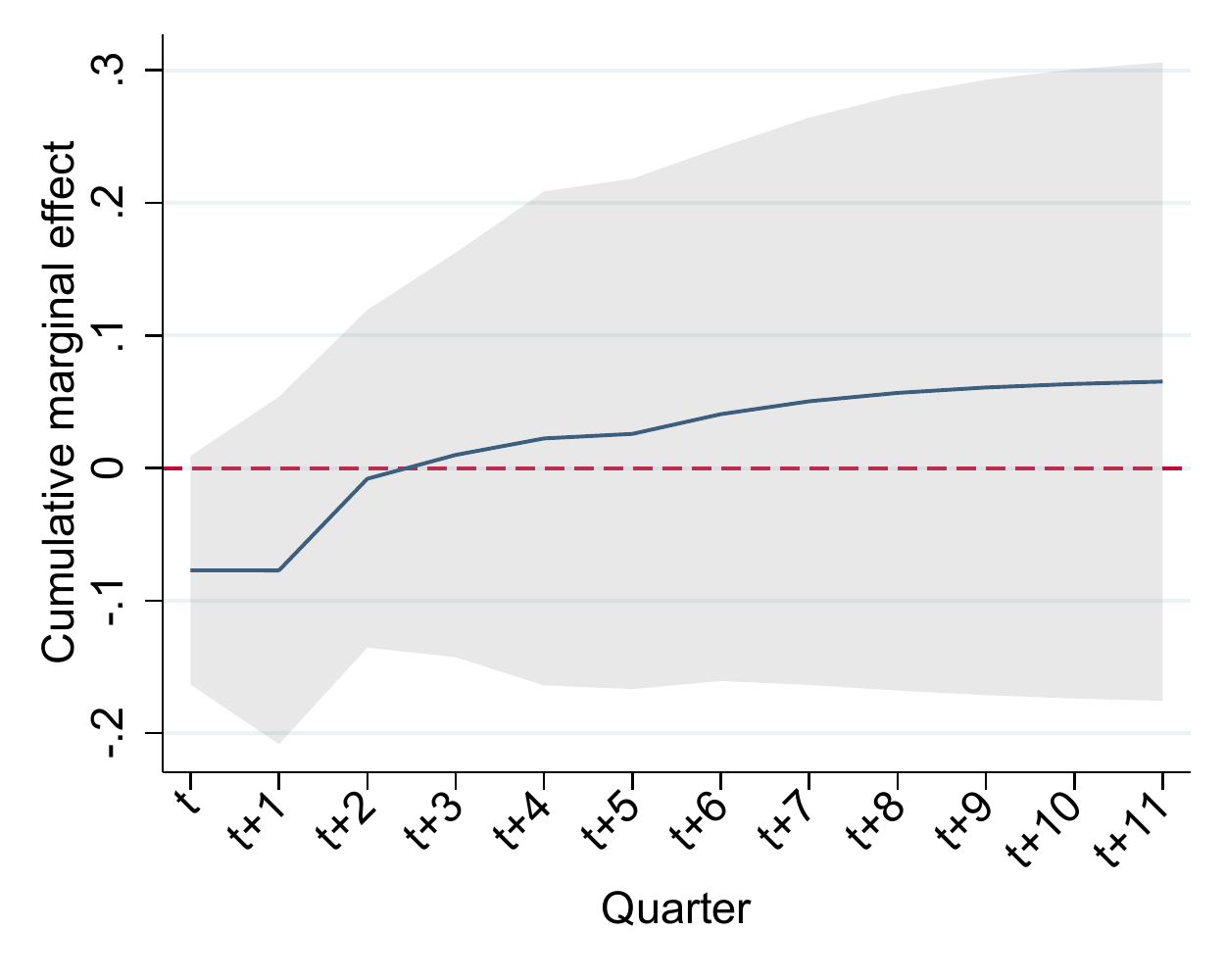}
                \end{subfigure}
                \vspace{10pt}    
                \newline
                \begin{subfigure}[b]{0.49\textwidth}
                                \centering \caption*{Short measures} \subcaption*{Unemployment rate} 
                                \includegraphics[clip=true, trim={0cm 0cm 0cm 0cm},scale=0.50]{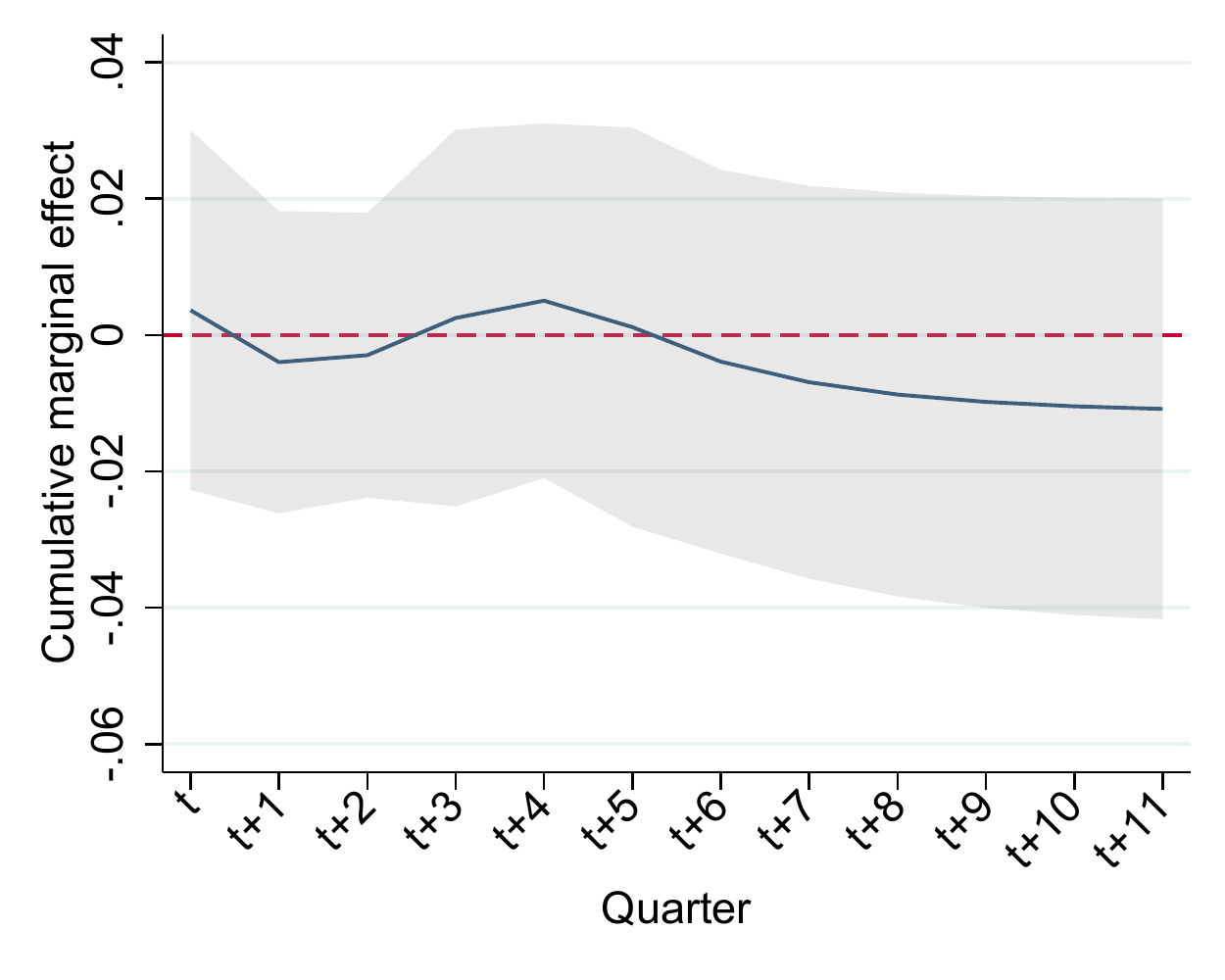}
                \end{subfigure}
                \vspace{10pt}    
                \begin{subfigure}[b]{0.49\textwidth}
                                \centering \caption*{Short measures}  \subcaption*{Employment rate} 
                                \includegraphics[clip=true, trim={0cm 0cm 0cm 0cm},scale=0.50]{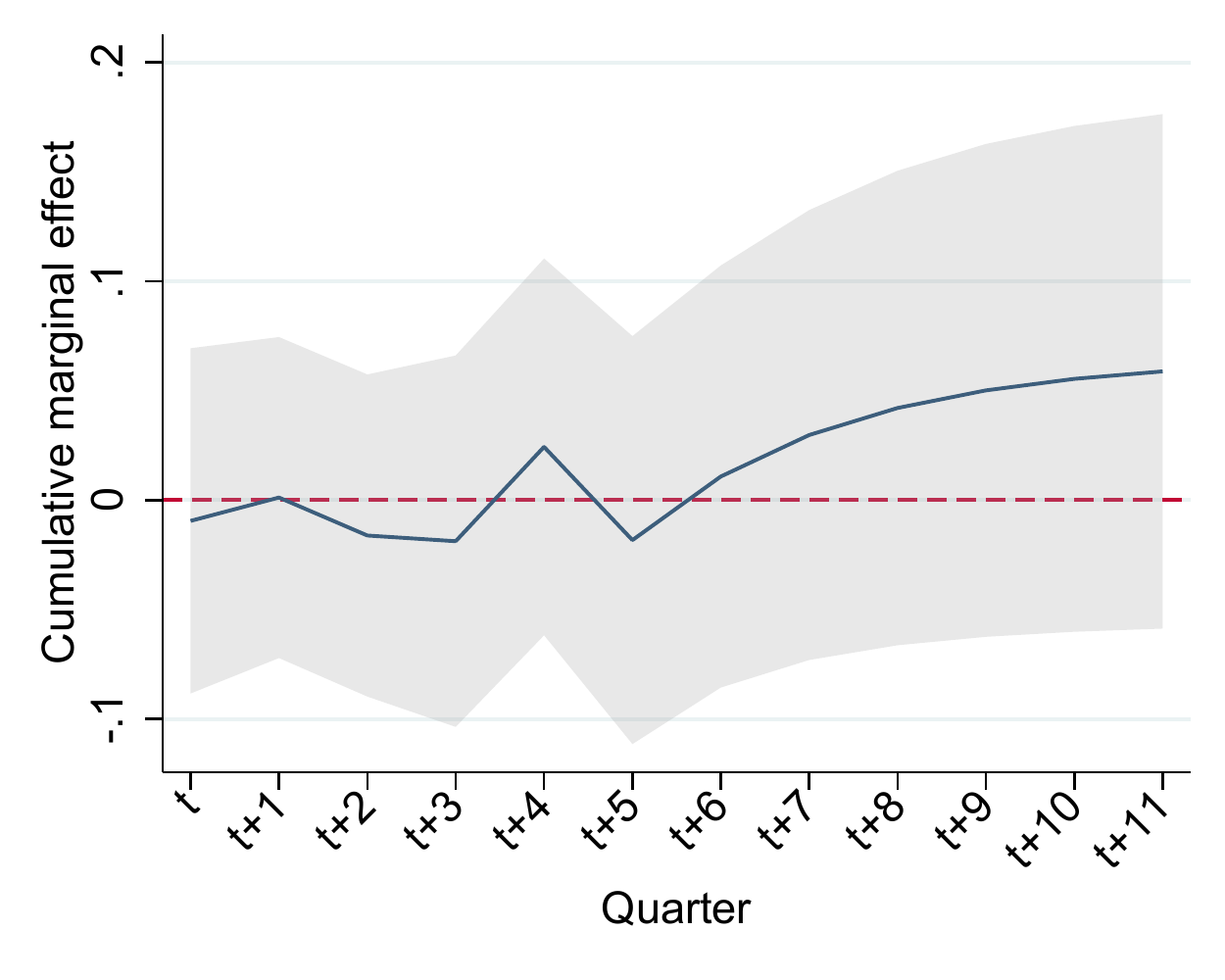}
                \end{subfigure}
                \vspace{10pt}
                \newline
                \begin{subfigure}[b]{0.49\textwidth}
                                \centering \caption*{Wage subsidies} \subcaption*{Unemployment rate} 
                                \includegraphics[clip=true, trim={0cm 0cm 0cm 0cm},scale=0.50]{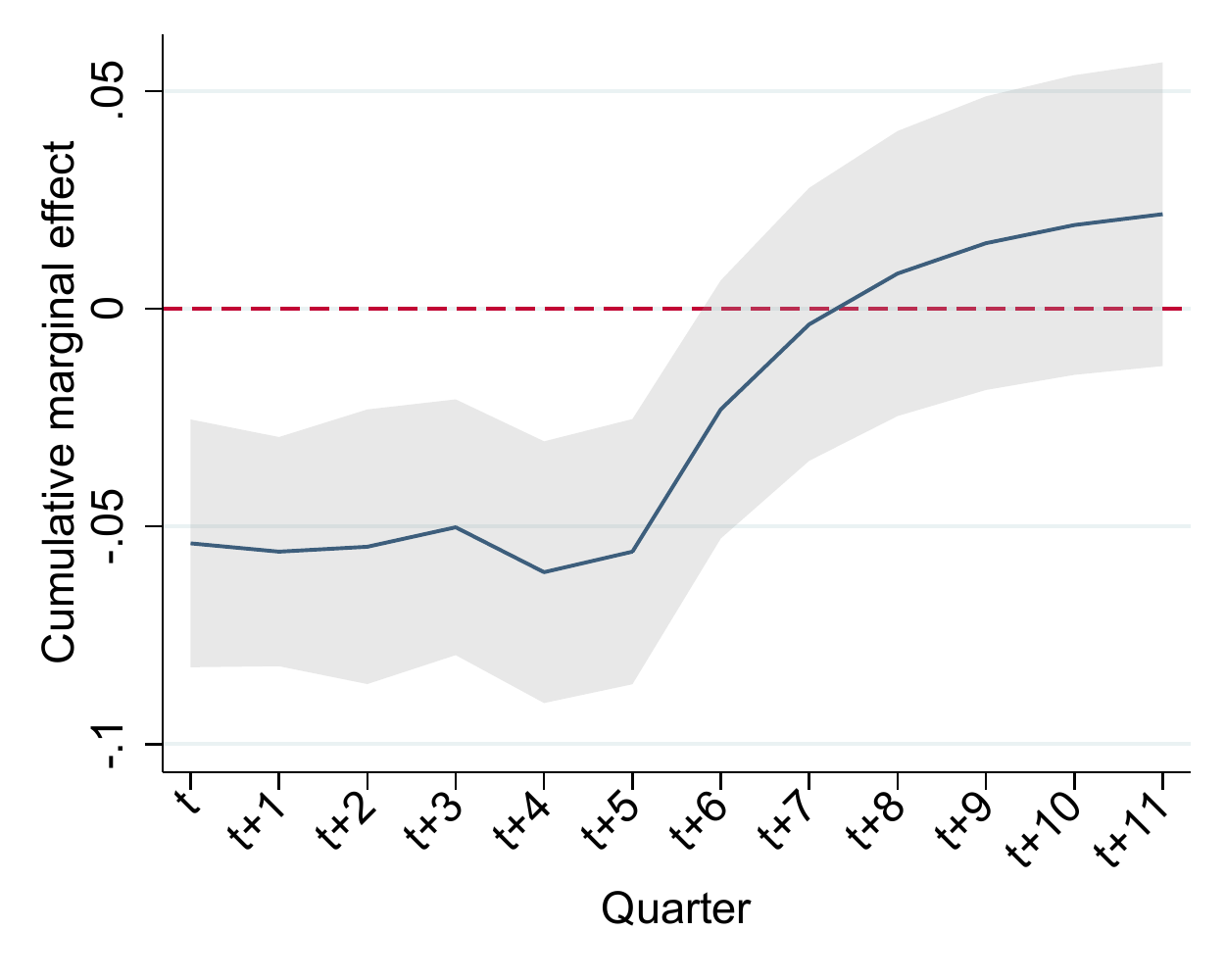}
                \end{subfigure}
                \vspace{10pt}    
                \begin{subfigure}[b]{0.49\textwidth}
                                \centering \caption*{Wage subsidies}  \subcaption*{Employment rate} 
                                \includegraphics[clip=true, trim={0cm 0cm 0cm 0cm},scale=0.50]{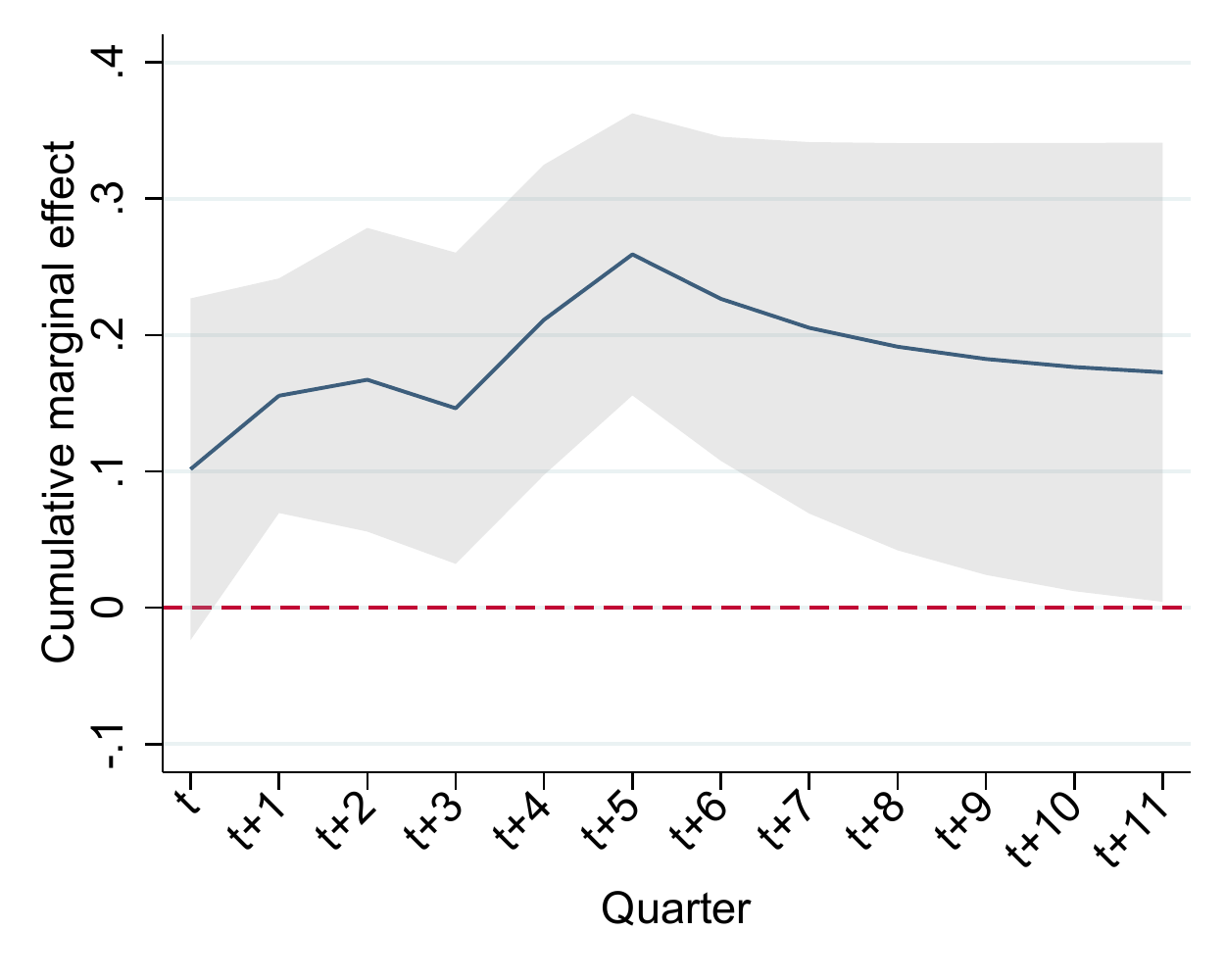}
                \end{subfigure}
                \vspace{10pt}
                \begin{minipage}{\textwidth}
                                \footnotesize \textit{Notes:} These graphs show the cumulative marginal effects of the three types of ALMP, i.e. training, short measures and wage subsidies on the unemployment rate and unsubsidized employment rate. 95\% confidence intervals are shown as grey areas. The effects are based on the ARDL model estimated by 2SLS. Program variables are included with 6 lags, main sample restrictions apply.  Standard errors obtained by a cross-sectional bootstrap (499 replications).
                \end{minipage}
\end{figure}

\begin{figure}
                \centering
                \caption{Cumulative Marginal Effects for Low-skilled \label{fig:cum_effects_iv_alo_rate_lowskilled}}
                \vspace{10pt}
                \begin{subfigure}[b]{0.49\textwidth}
                                \centering \caption*{Training} \subcaption*{Unemployment rate} 
                                \includegraphics[clip=true, trim={0cm 0cm 0cm 0cm},scale=0.50]{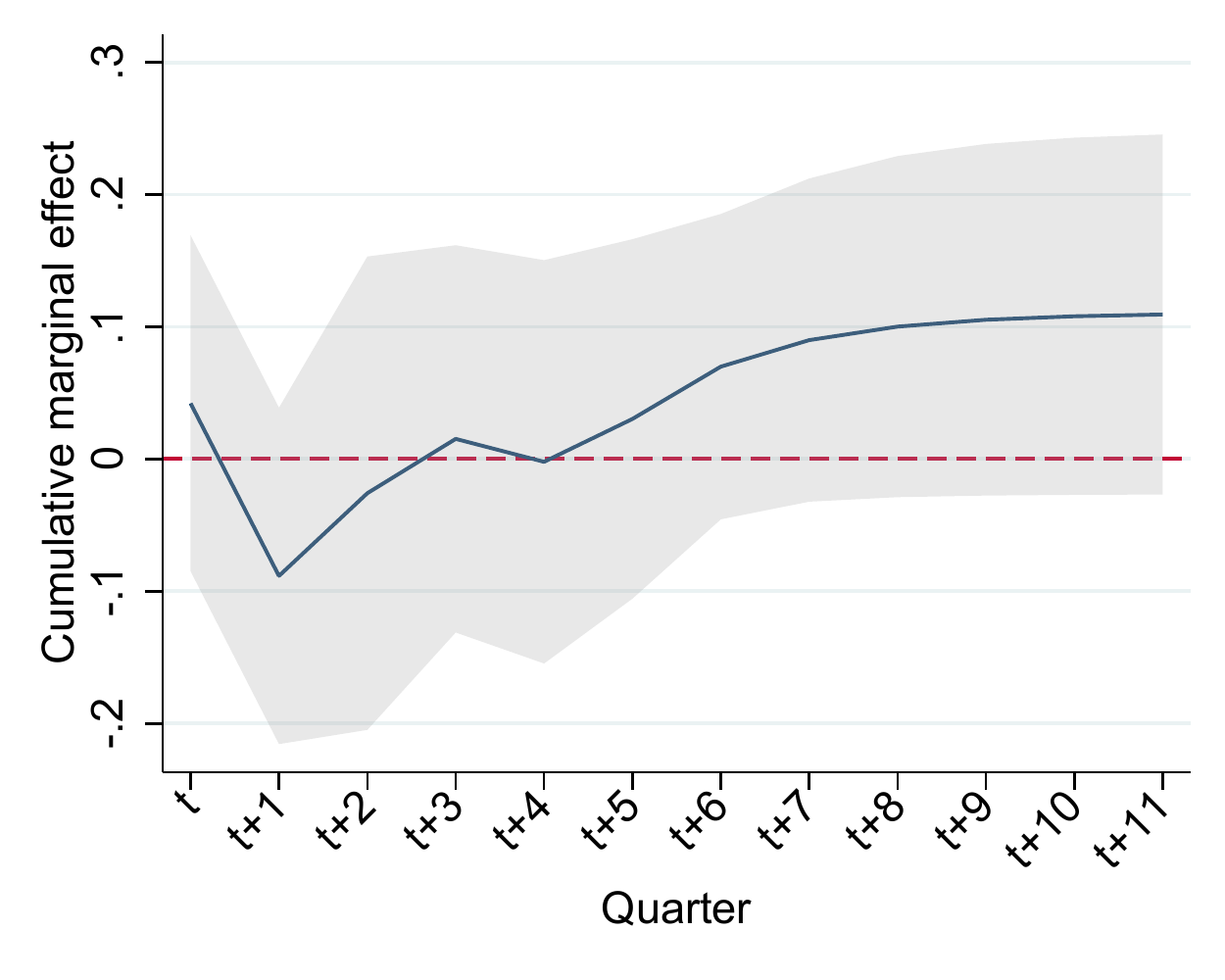}
                \end{subfigure}
                \begin{subfigure}[b]{0.49\textwidth}
                                \centering \caption*{Training} \subcaption*{Employment rate} 
                                \includegraphics[clip=true, trim={0cm 0cm 0cm 0cm},scale=0.50]{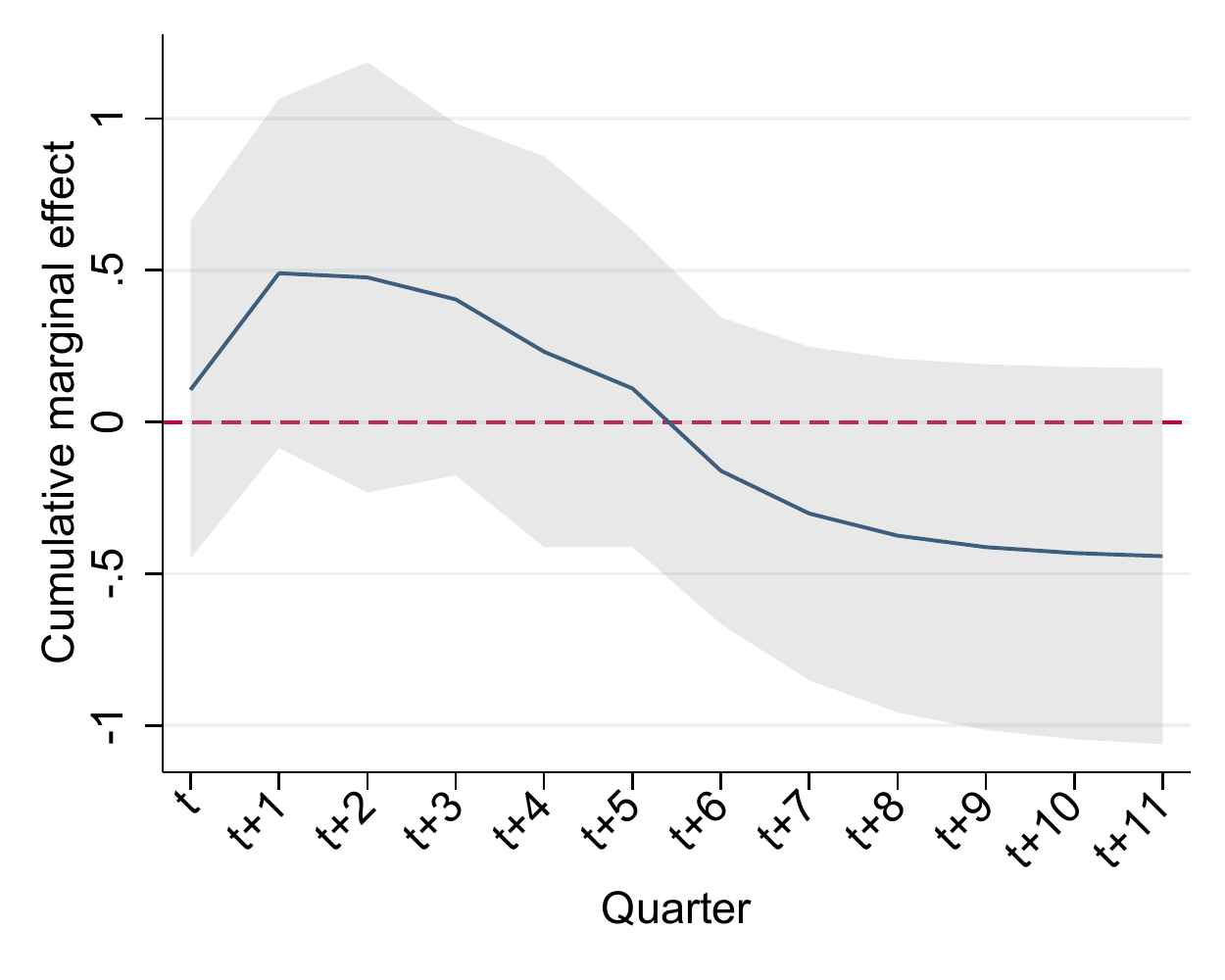}
                \end{subfigure}
                \vspace{10pt}    
                \newline
                \begin{subfigure}[b]{0.49\textwidth}
                                \centering \caption*{Short measures} \subcaption*{Unemployment rate} 
                                \includegraphics[clip=true, trim={0cm 0cm 0cm 0cm},scale=0.50]{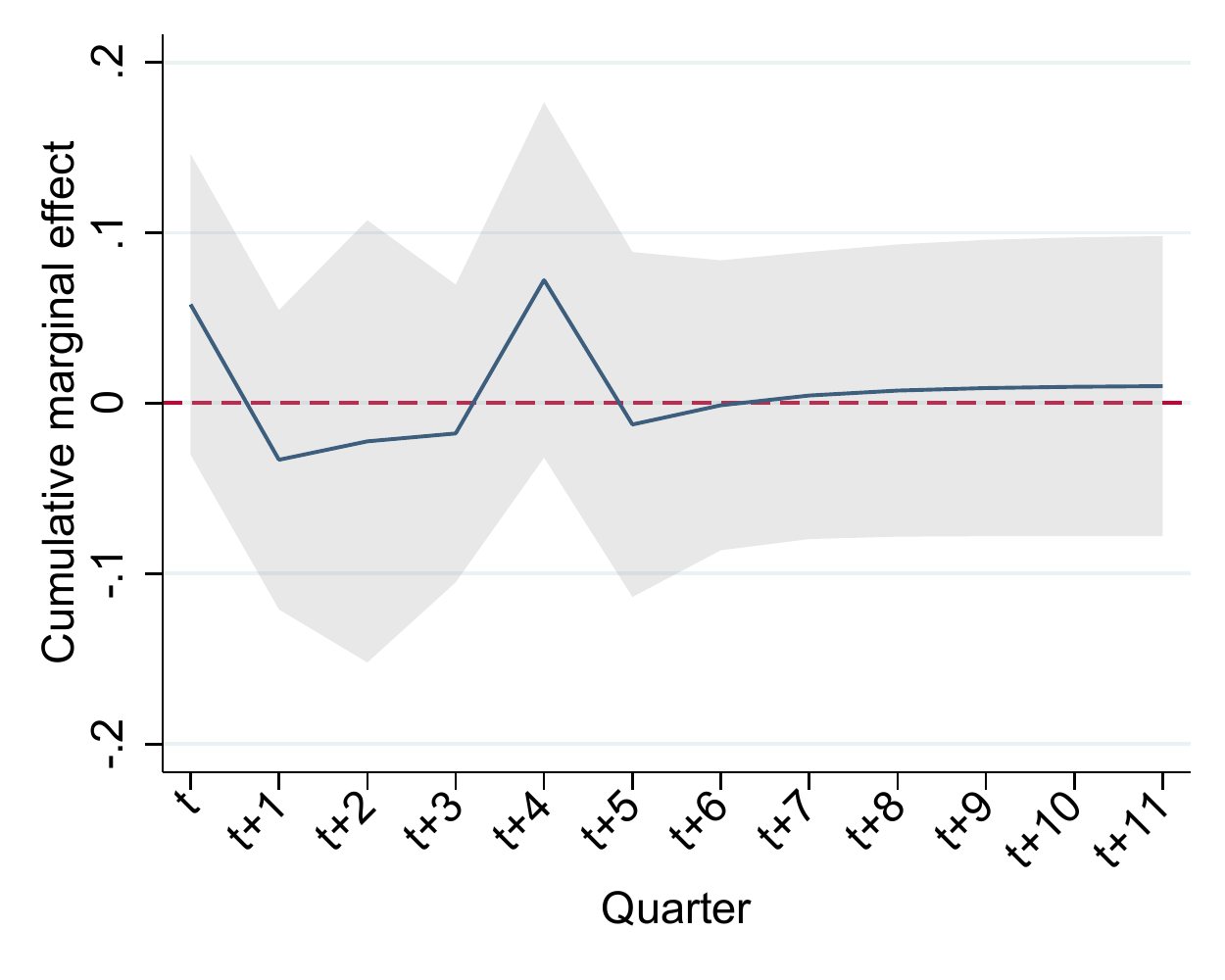}
                \end{subfigure}
                \vspace{10pt}    
                \begin{subfigure}[b]{0.49\textwidth}
                                \centering \caption*{Short measures}  \subcaption*{Employment rate} 
                                \includegraphics[clip=true, trim={0cm 0cm 0cm 0cm},scale=0.50]{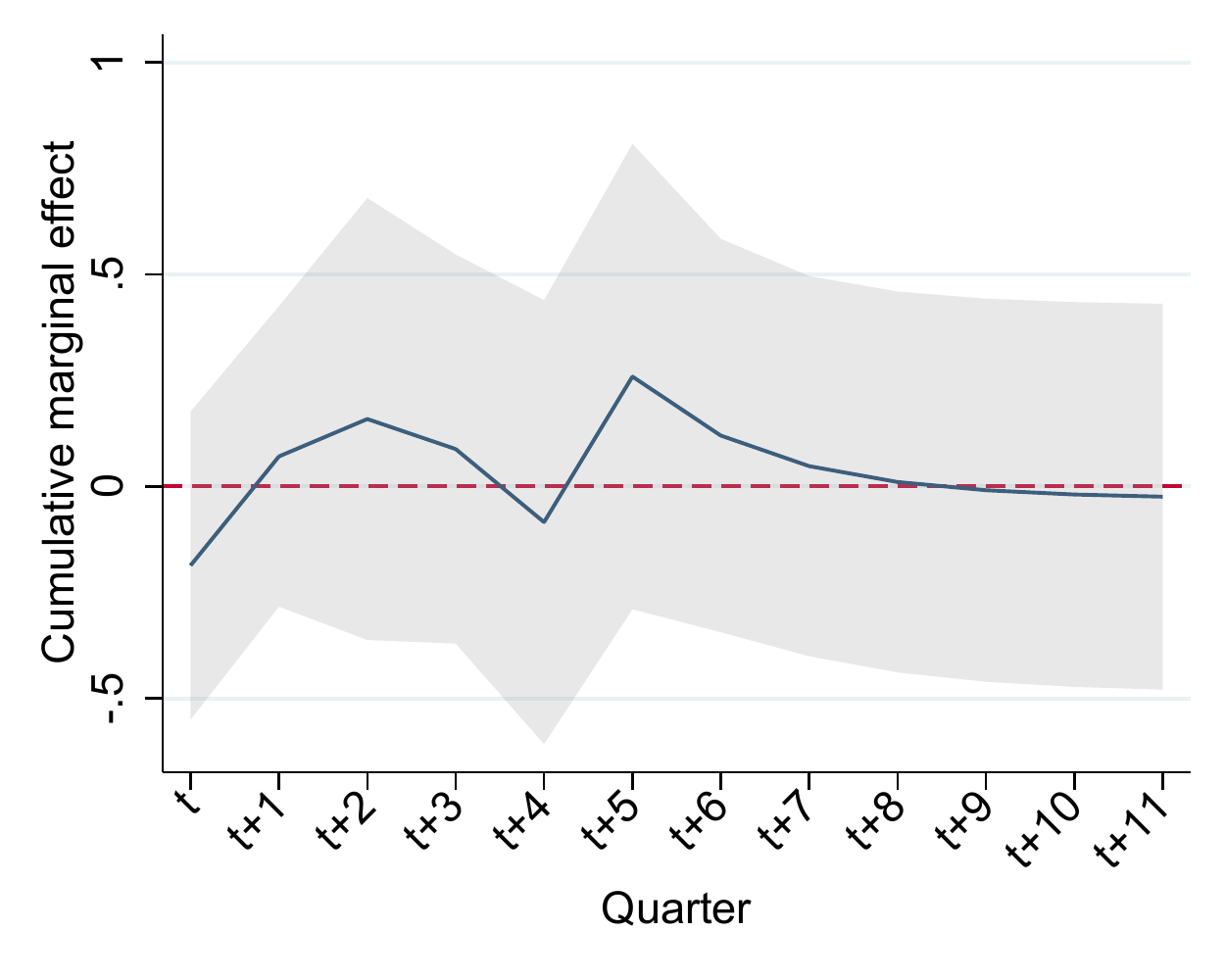}
                \end{subfigure}
                \vspace{10pt}
                \newline
                \begin{subfigure}[b]{0.49\textwidth}
                                \centering \caption*{Wage subsidies} \subcaption*{Unemployment rate} 
                                \includegraphics[clip=true, trim={0cm 0cm 0cm 0cm},scale=0.50]{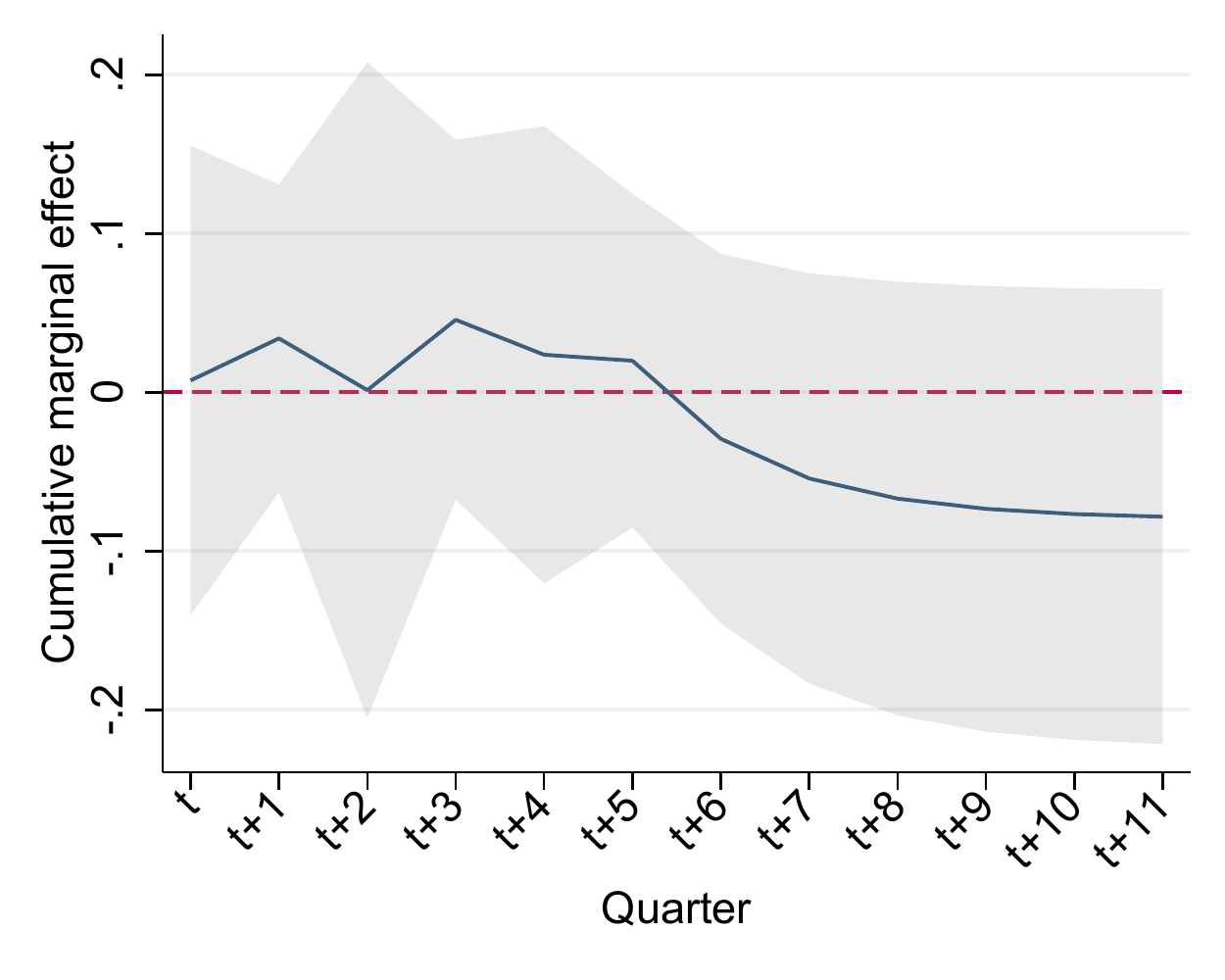}
                \end{subfigure}
                \vspace{10pt}    
                \begin{subfigure}[b]{0.49\textwidth}
                                \centering \caption*{Wage subsidies}  \subcaption*{Employment rate} 
                                \includegraphics[clip=true, trim={0cm 0cm 0cm 0cm},scale=0.50]{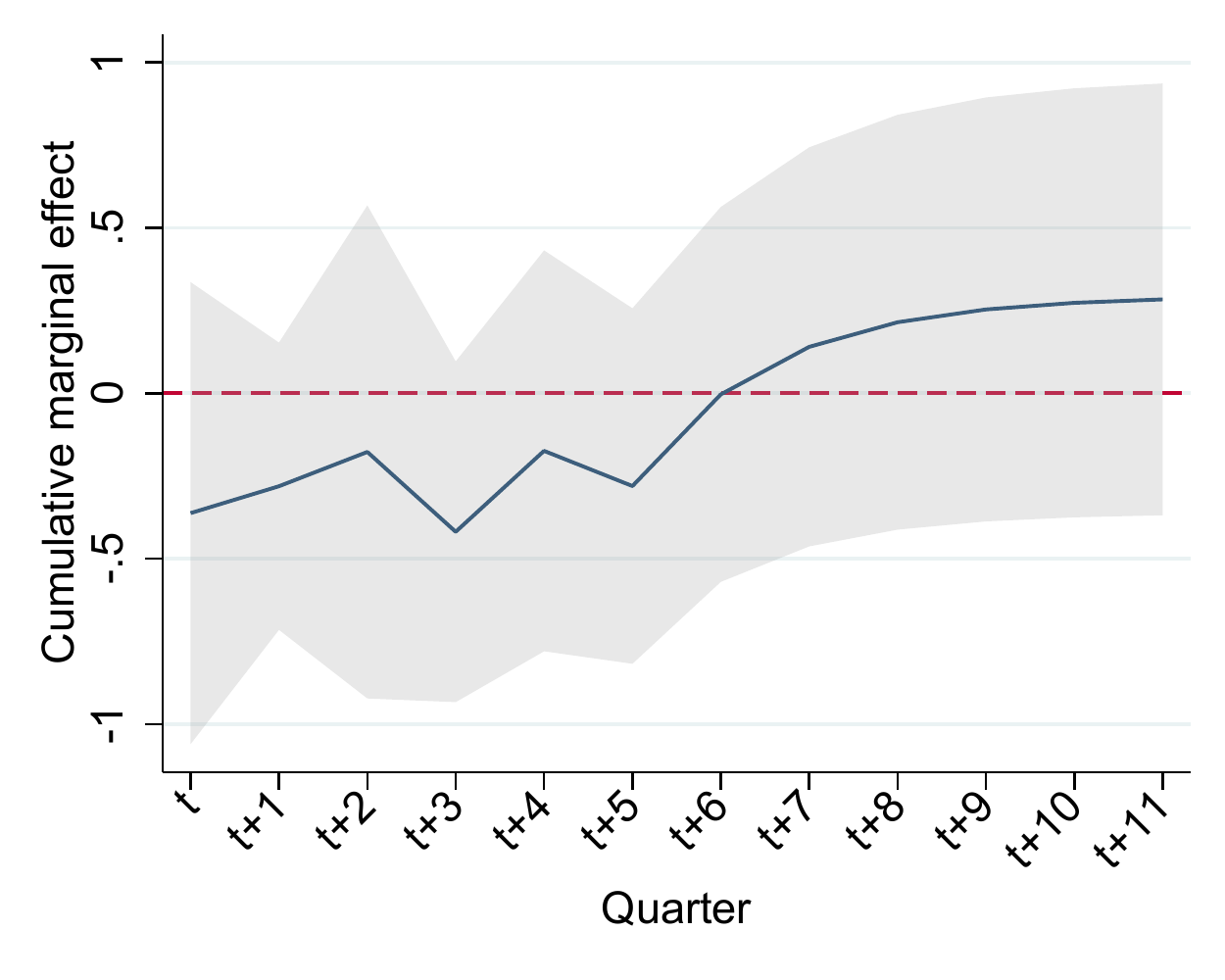}
                \end{subfigure}
                \vspace{10pt}
                \begin{minipage}{\textwidth}
                                \footnotesize \textit{Notes:} These graphs show the cumulative marginal effects of the three types of ALMP, i.e. training, short measures and wage subsidies on the unemployment rate and unsubsidized employment rate. 95\% confidence intervals are shown as grey areas. The effects are based on the ARDL model estimated by 2SLS. Program variables are included with 6 lags, main sample restrictions apply.  Standard errors obtained by a cross-sectional bootstrap (499 replications).
                \end{minipage}
\end{figure}

\begin{figure}
                \centering
                \caption{Cumulative Marginal Effects for Medium-skilled \label{fig:cum_effects_iv_alo_rate_midskilled}}
                \vspace{10pt}
                \begin{subfigure}[b]{0.49\textwidth}
                                \centering \caption*{Training} \subcaption*{Unemployment rate} 
                                \includegraphics[clip=true, trim={0cm 0cm 0cm 0cm},scale=0.50]{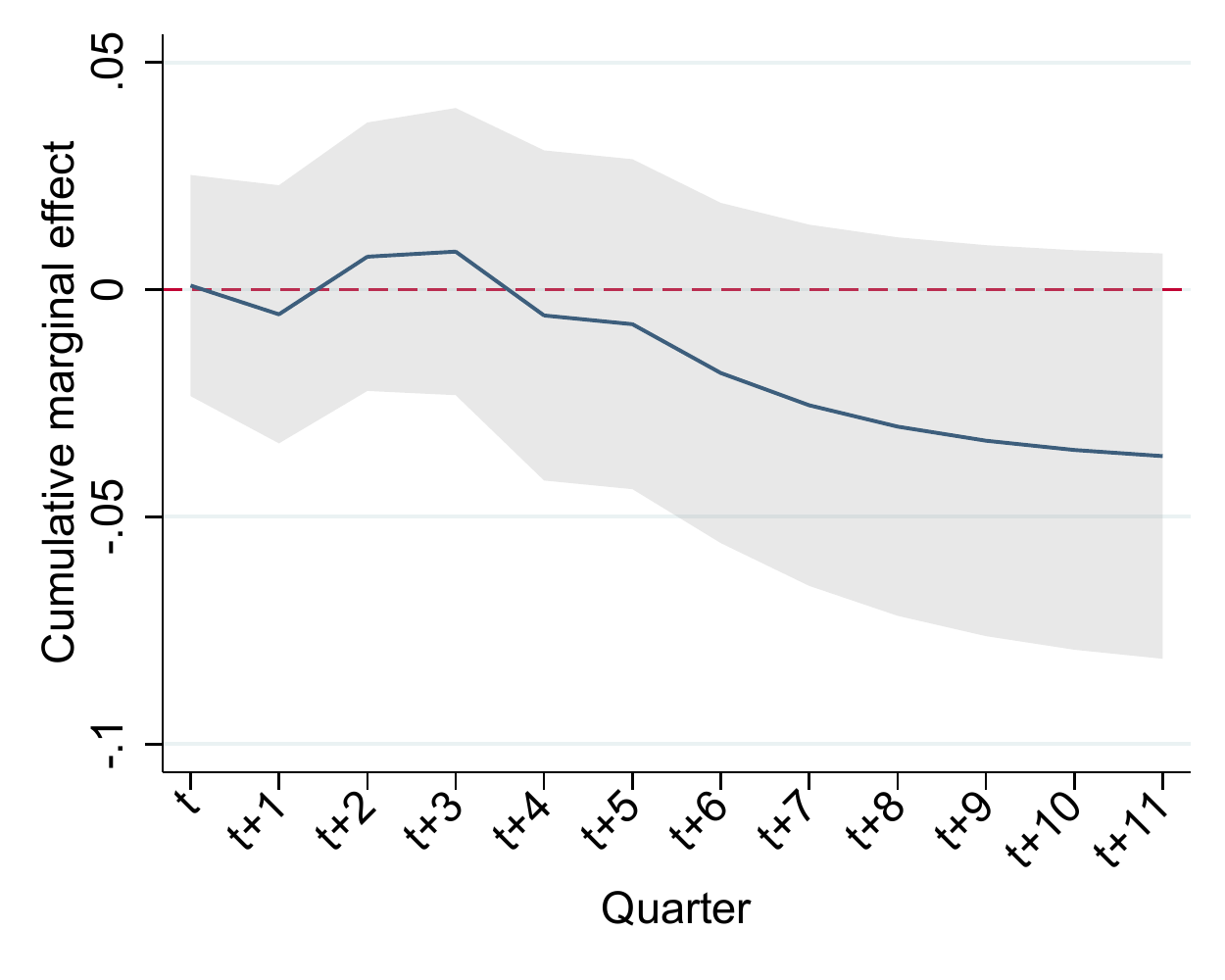}
                \end{subfigure}
                \begin{subfigure}[b]{0.49\textwidth}
                                \centering \caption*{Training} \subcaption*{Unsubsidized employment rate} 
                                \includegraphics[clip=true, trim={0cm 0cm 0cm 0cm},scale=0.50]{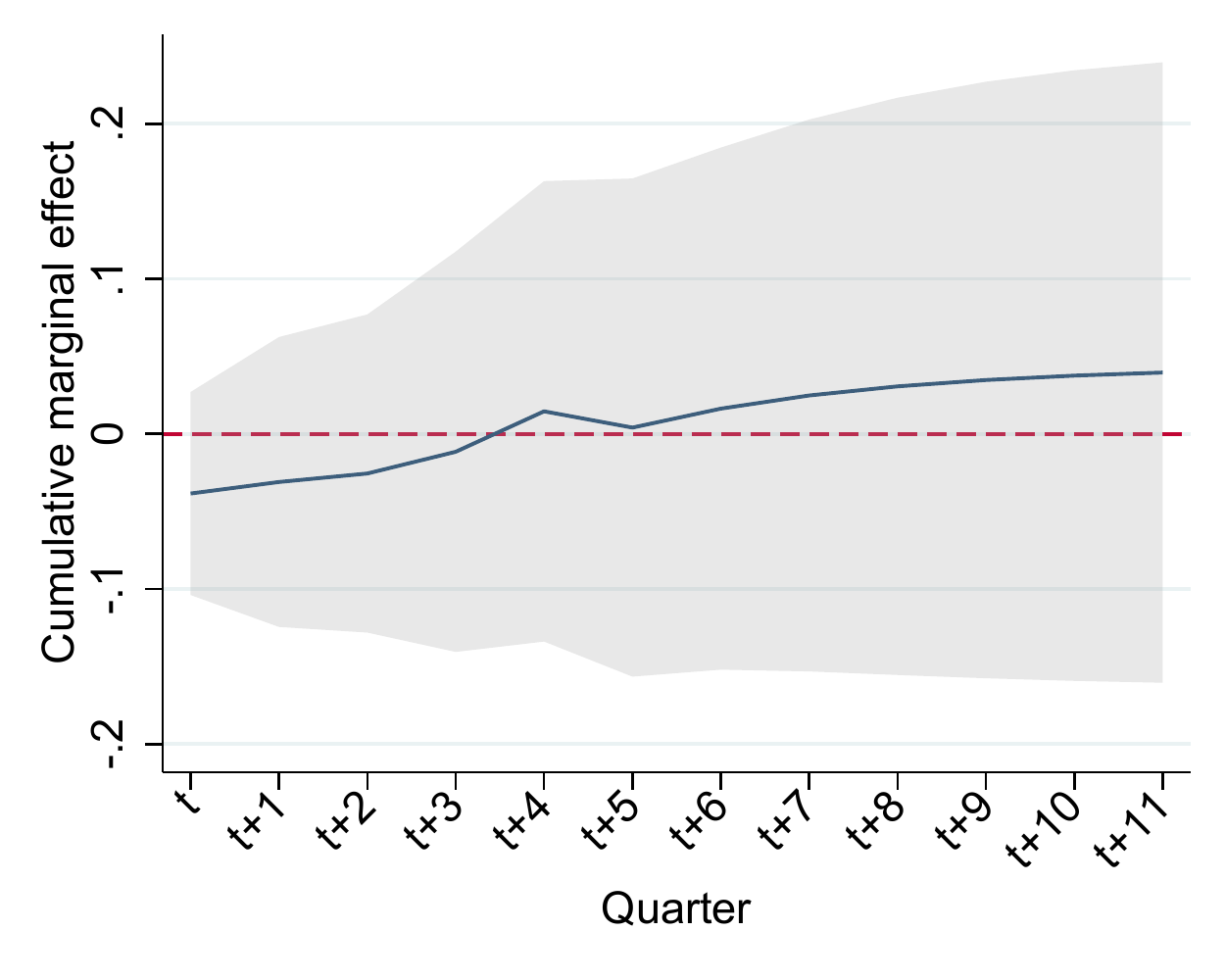}
                \end{subfigure}
                \vspace{10pt}    
                \newline
                \begin{subfigure}[b]{0.49\textwidth}
                                \centering \caption*{Short measures} \subcaption*{Unemployment rate} 
                                \includegraphics[clip=true, trim={0cm 0cm 0cm 0cm},scale=0.50]{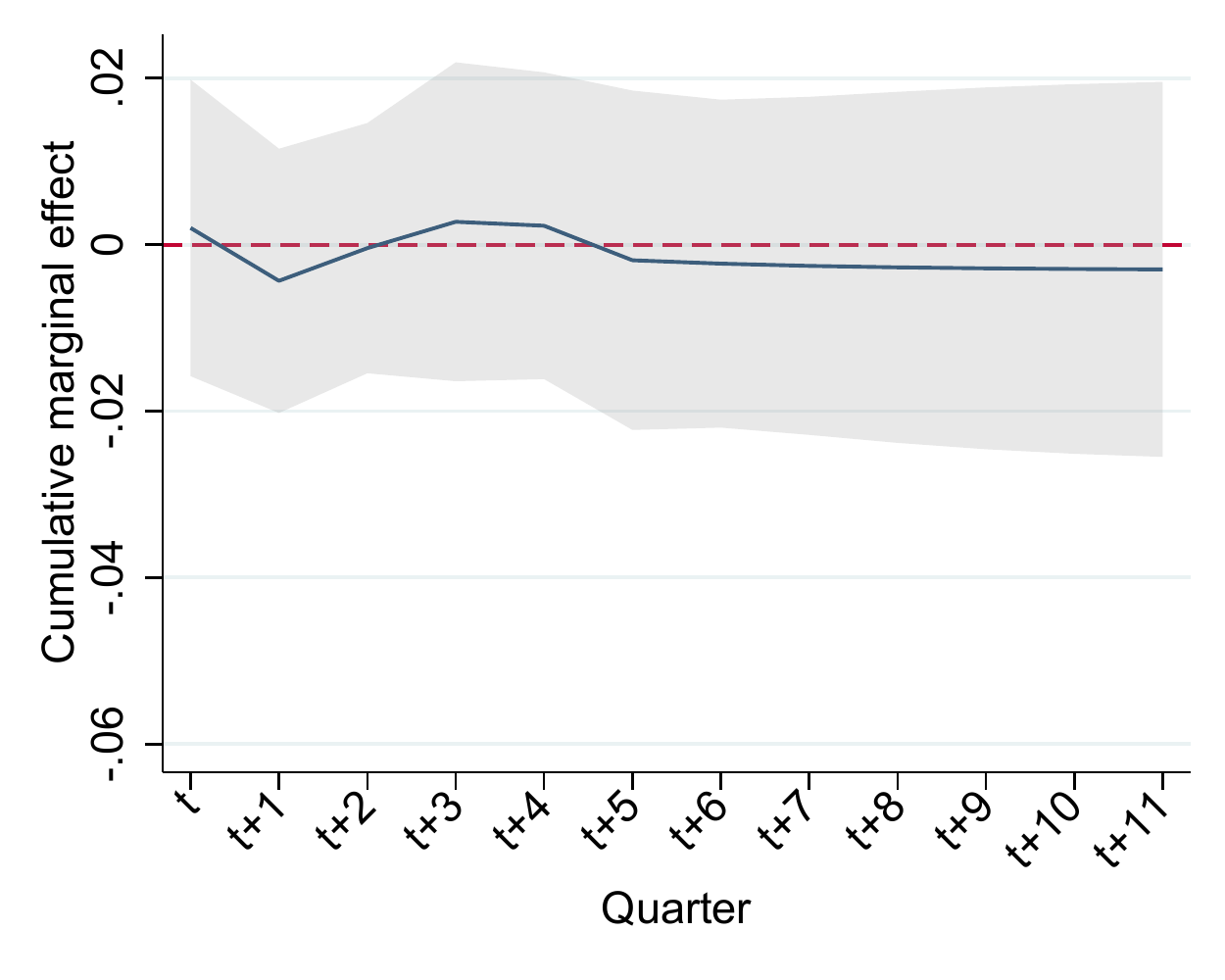}
                \end{subfigure}
                \vspace{10pt}    
                \begin{subfigure}[b]{0.49\textwidth}
                                \centering \caption*{Short measures}  \subcaption*{Unsubsidized employment rate} 
                                \includegraphics[clip=true, trim={0cm 0cm 0cm 0cm},scale=0.50]{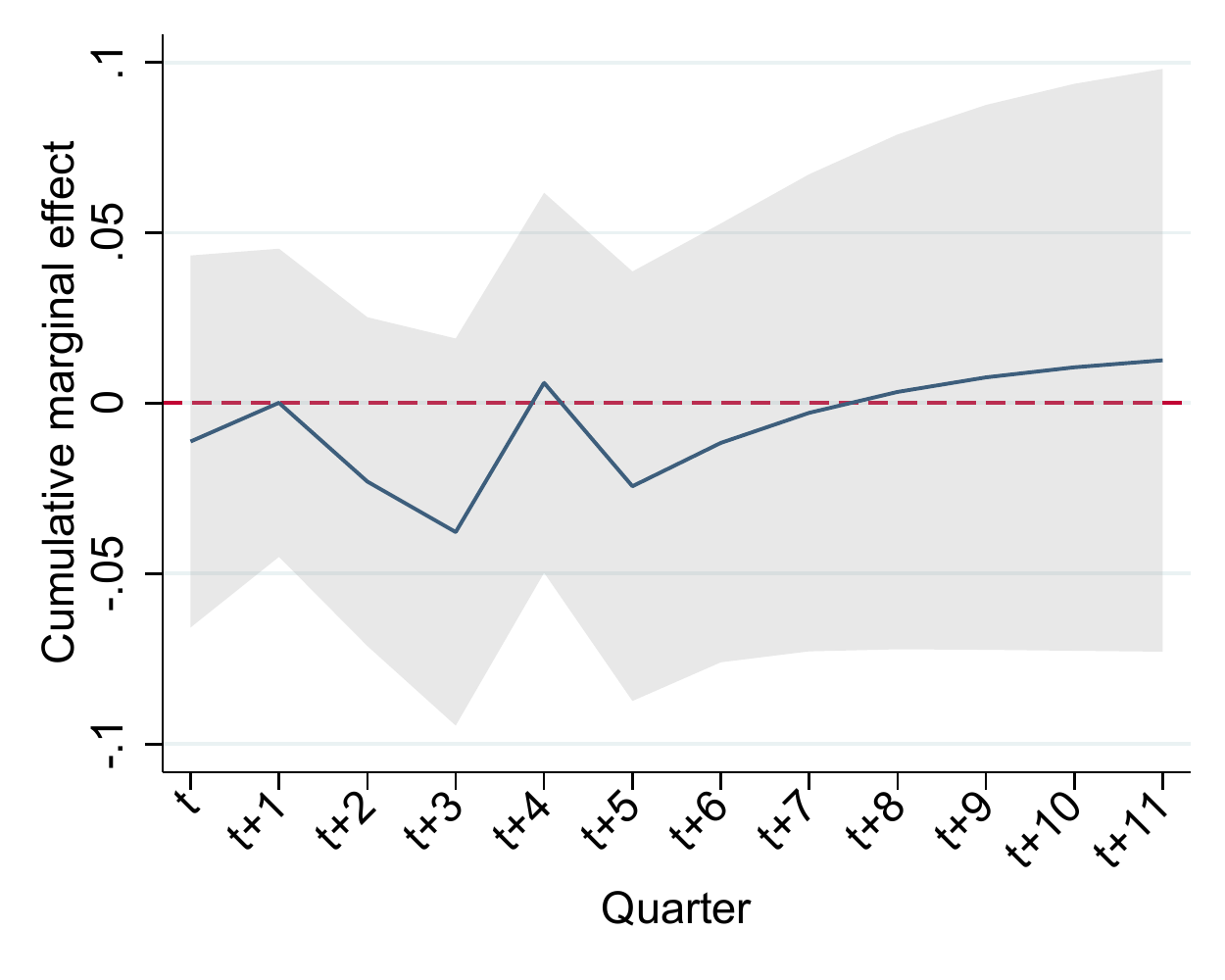}
                \end{subfigure}
                \vspace{10pt}
                \newline
                \begin{subfigure}[b]{0.49\textwidth}
                                \centering \caption*{Wage subsidies} \subcaption*{Unemployment rate} 
                                \includegraphics[clip=true, trim={0cm 0cm 0cm 0cm},scale=0.50]{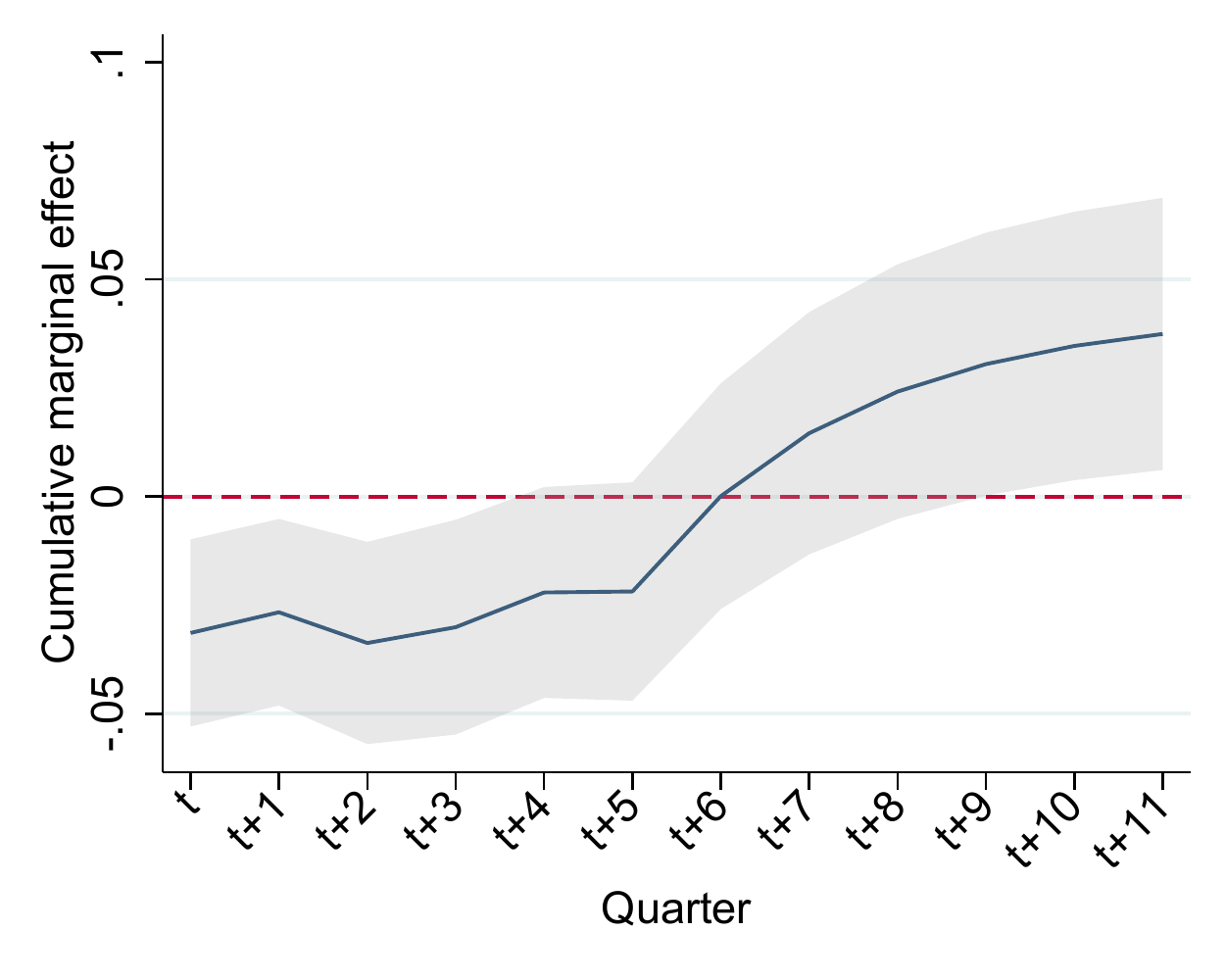}
                \end{subfigure}
                \vspace{10pt}    
                \begin{subfigure}[b]{0.49\textwidth}
                                \centering \caption*{Wage subsidies}  \subcaption*{Unsubsidized employment rate} 
                                \includegraphics[clip=true, trim={0cm 0cm 0cm 0cm},scale=0.50]{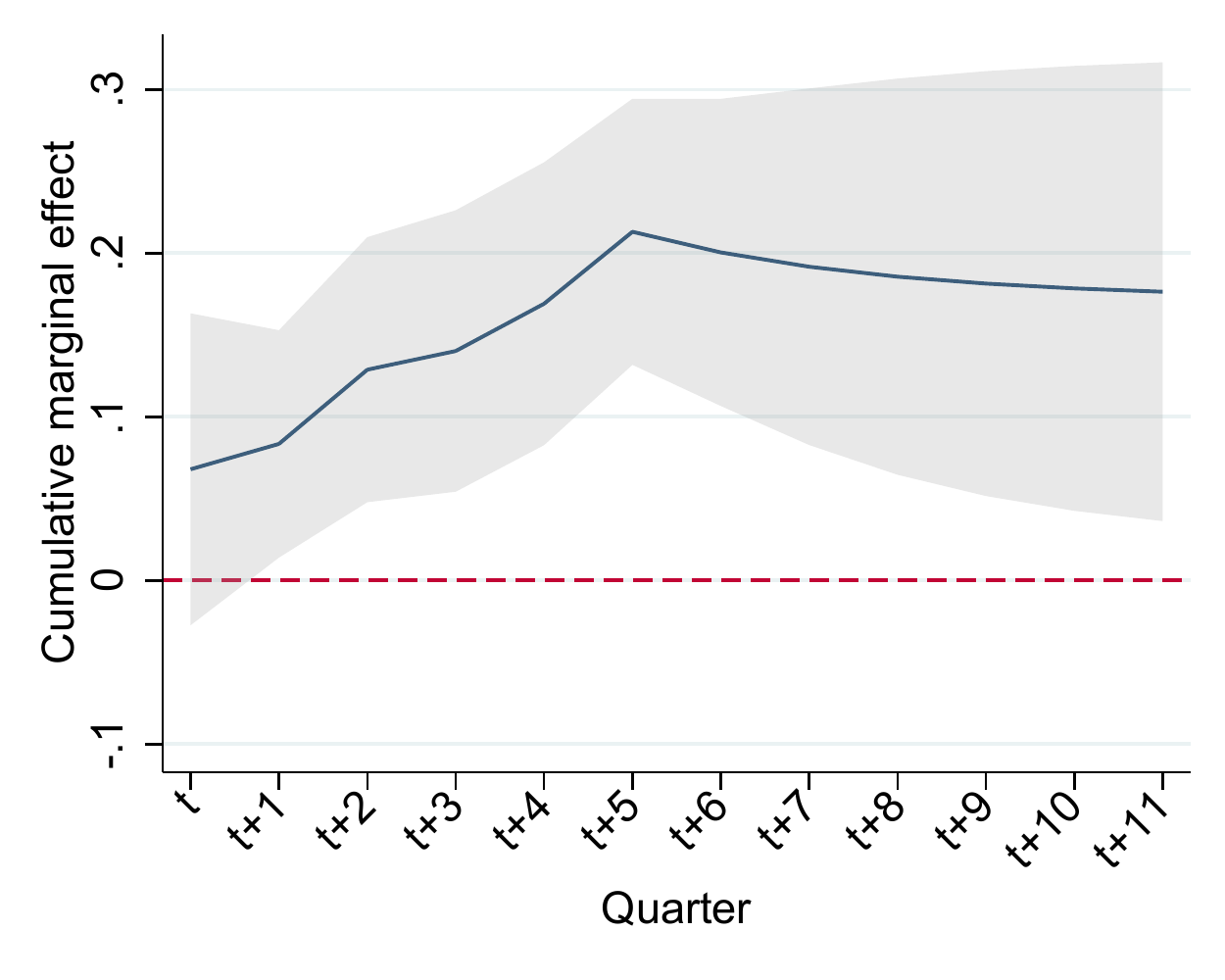}
                \end{subfigure}
                \vspace{10pt}
                \begin{minipage}{\textwidth}
                                \footnotesize \textit{Notes:} These graphs show the cumulative marginal effects of the three types of ALMP, i.e. training, short measures and wage subsidies on the unemployment rate and unsubsidized employment rate. 95\% confidence intervals are shown as grey areas. The effects are based on the ARDL model estimated by 2SLS. Program variables are included with 6 lags, main sample restrictions apply.  Standard errors obtained by a cross-sectional bootstrap (499 replications).
                \end{minipage}
\end{figure}

\begin{figure}
                \centering
                \caption{Cumulative Marginal Effects for High-skilled \label{fig:cum_effects_iv_alo_rate_highskilled}}
                \vspace{10pt}
                \begin{subfigure}[b]{0.49\textwidth}
                                \centering \caption*{Training} \subcaption*{Unemployment rate} 
                                \includegraphics[clip=true, trim={0cm 0cm 0cm 0cm},scale=0.50]{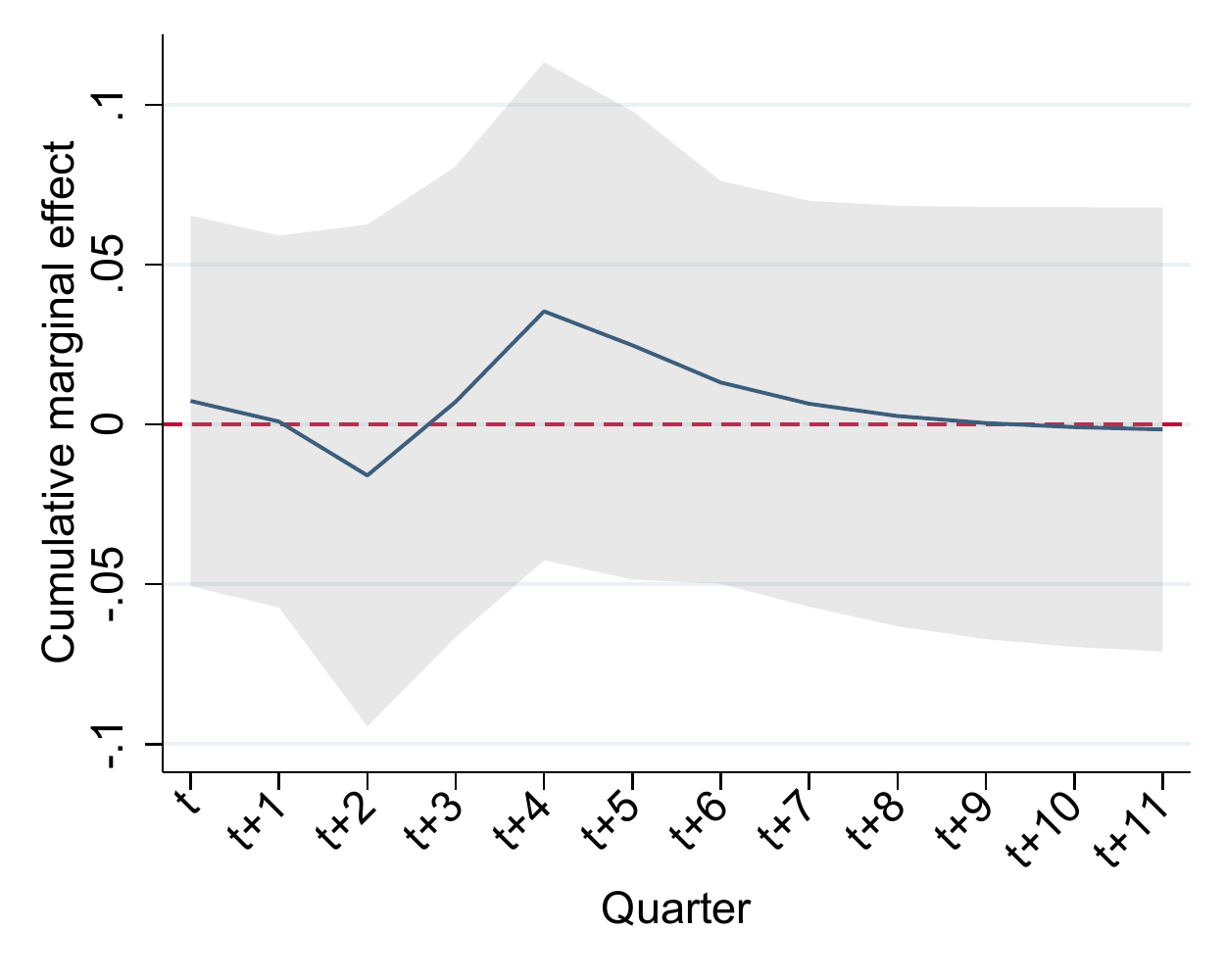}
                \end{subfigure}
                \begin{subfigure}[b]{0.49\textwidth}
                                \centering \caption*{Training} \subcaption*{Unsubsidized employment rate} 
                                \includegraphics[clip=true, trim={0cm 0cm 0cm 0cm},scale=0.50]{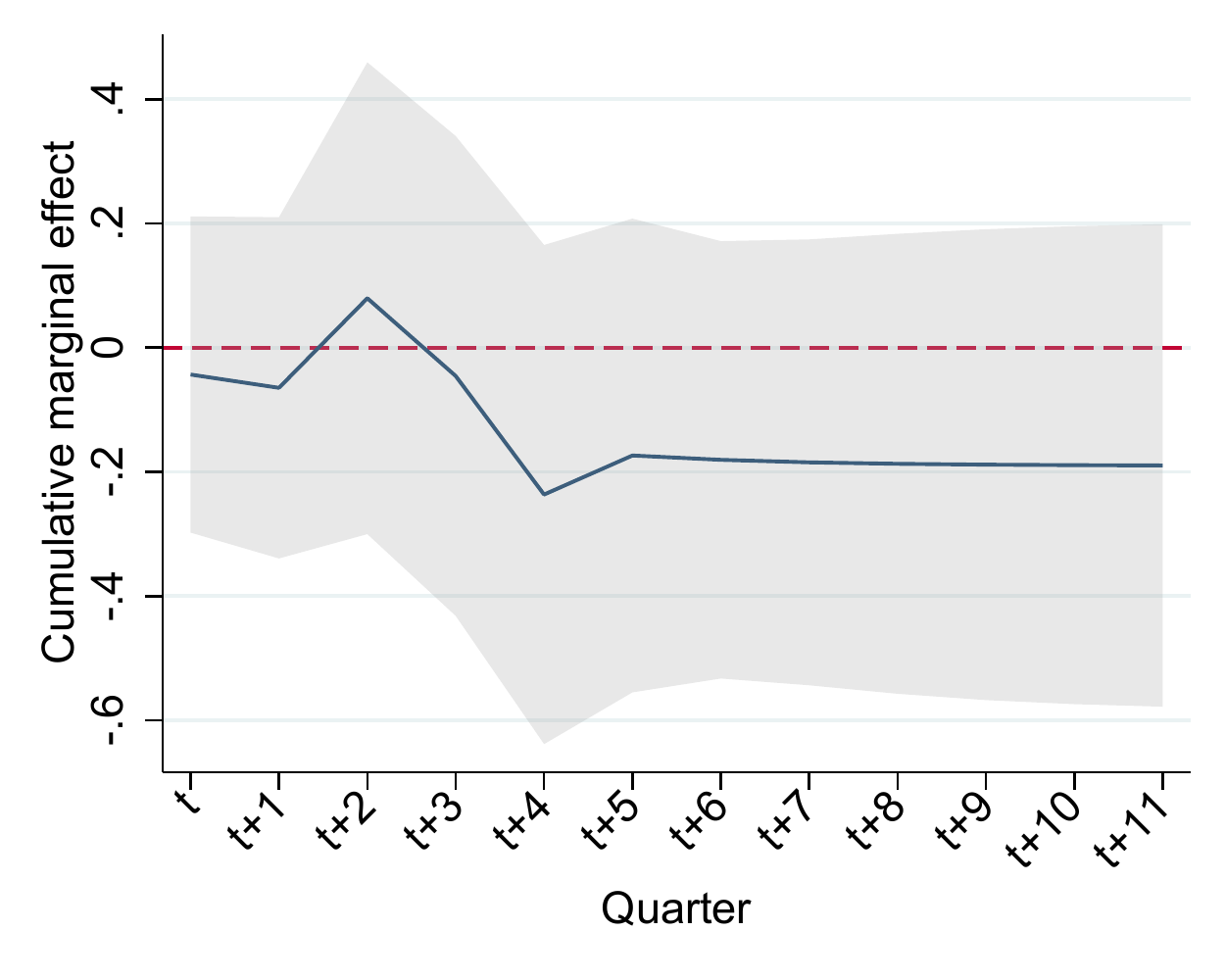}
                \end{subfigure}
                \vspace{10pt}    
                \newline
                \begin{subfigure}[b]{0.49\textwidth}
                                \centering \caption*{Short measures} \subcaption*{Unemployment rate} 
                                \includegraphics[clip=true, trim={0cm 0cm 0cm 0cm},scale=0.50]{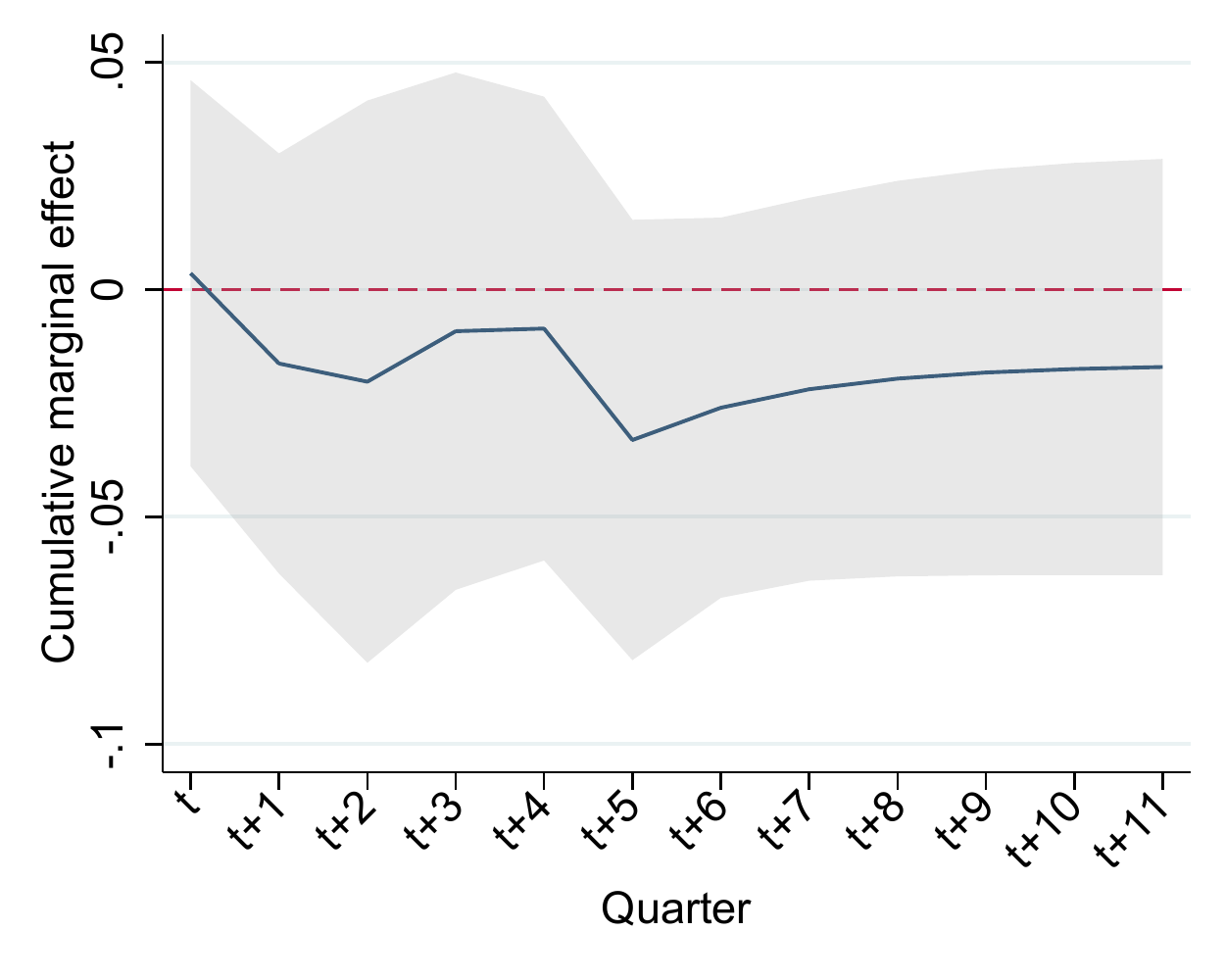}
                \end{subfigure}
                \vspace{10pt}    
                \begin{subfigure}[b]{0.49\textwidth}
                                \centering \caption*{Short measures}  \subcaption*{Unsubsidized employment rate} 
                                \includegraphics[clip=true, trim={0cm 0cm 0cm 0cm},scale=0.50]{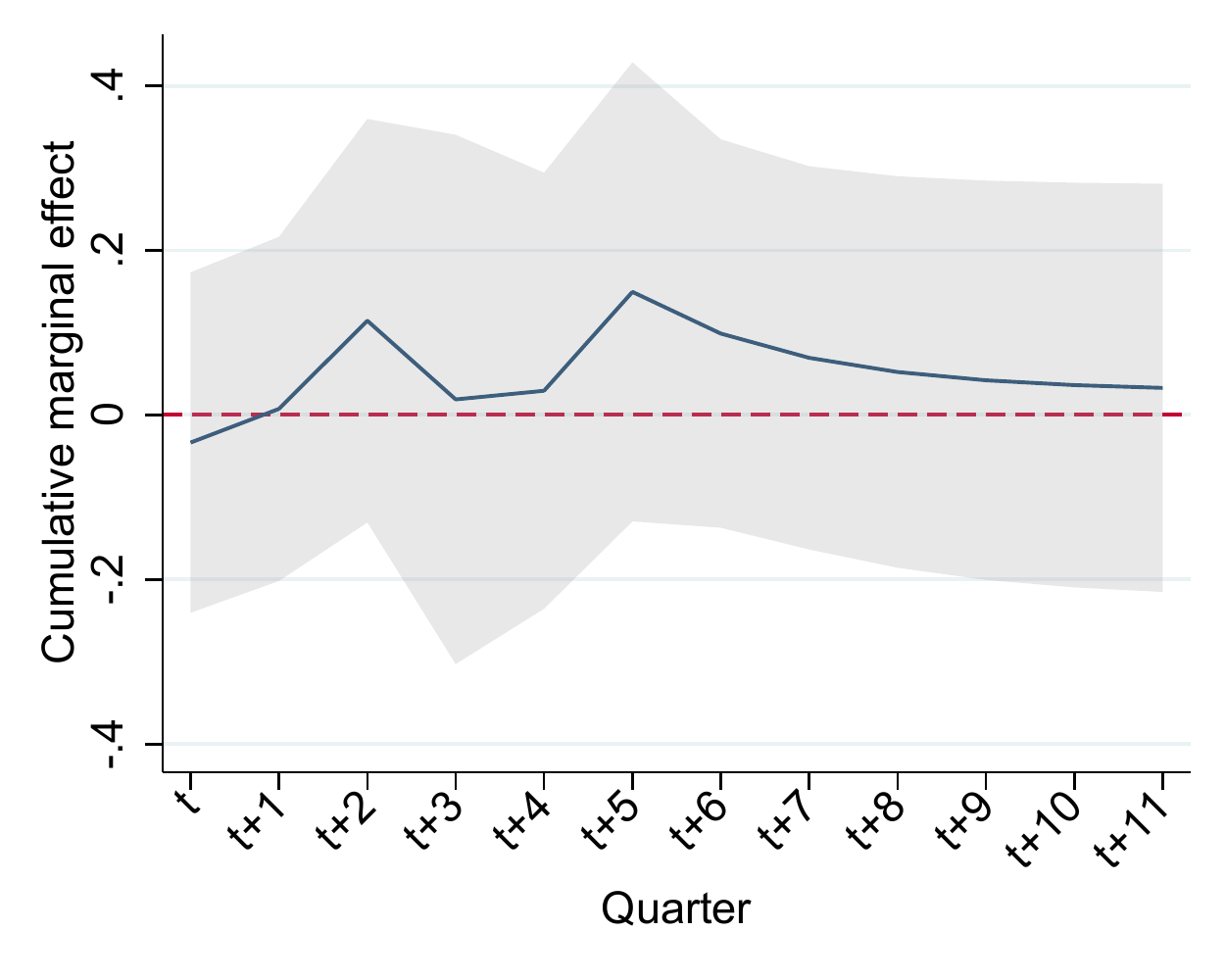}
                \end{subfigure}
                \vspace{10pt}
                \newline
                \begin{subfigure}[b]{0.49\textwidth}
                                \centering \caption*{Wage subsidies} \subcaption*{Unemployment rate} 
                                \includegraphics[clip=true, trim={0cm 0cm 0cm 0cm},scale=0.50]{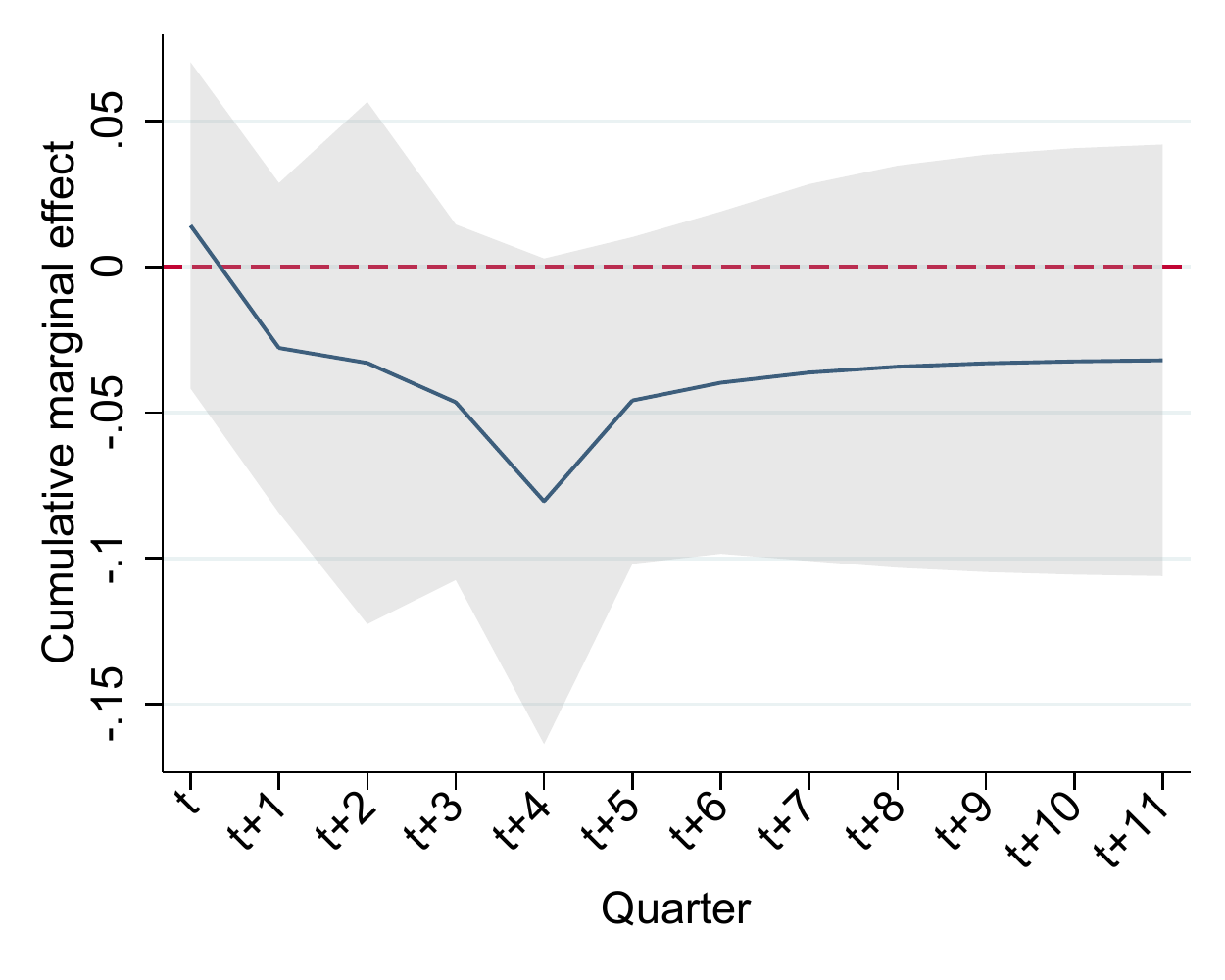}
                \end{subfigure}
                \vspace{10pt}    
                \begin{subfigure}[b]{0.49\textwidth}
                                \centering \caption*{Wage subsidies}  \subcaption*{Unsubsidized employment rate} 
                                \includegraphics[clip=true, trim={0cm 0cm 0cm 0cm},scale=0.50]{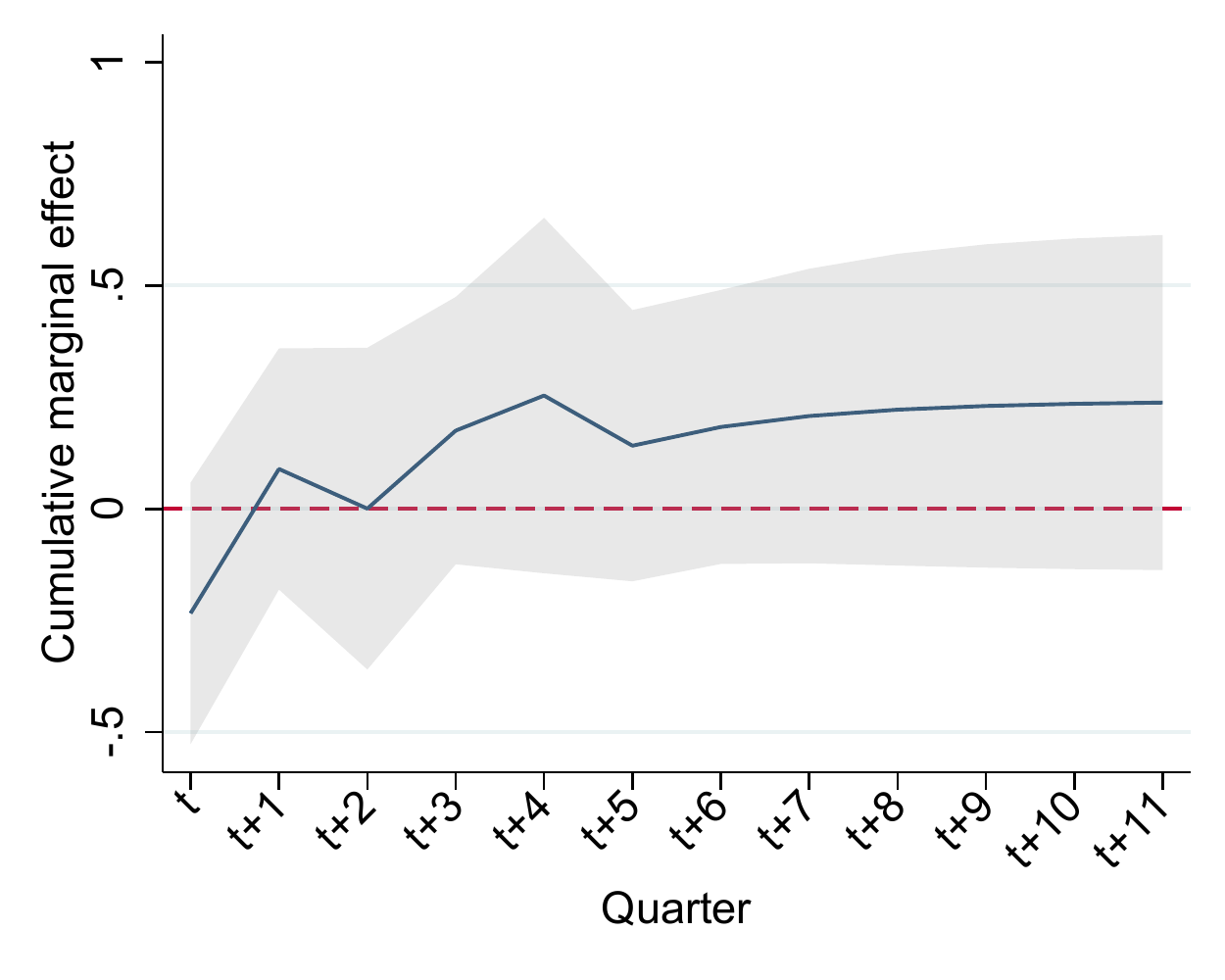}
                \end{subfigure}
                \vspace{10pt}
                \begin{minipage}{\textwidth}
                                \footnotesize \textit{Notes:} These graphs show the cumulative marginal effects of the three types of ALMP, i.e. training, short measures and wage subsidies on the unemployment rate and unsubsidized employment rate. 95\% confidence intervals are shown as grey areas. The effects are based on the ARDL model estimated by 2SLS. Program variables are included with 6 lags, main sample restrictions apply.  Standard errors obtained by a cross-sectional bootstrap (499 replications).
                \end{minipage}
\end{figure}

\begin{figure}
                \centering
                \caption{Cumulative Marginal Effects for Ages under 30 \label{fig:cum_effects_iv_alo_rate_u_30}}
                \vspace{10pt}
                \begin{subfigure}[b]{0.49\textwidth}
                                \centering \caption*{Training} \subcaption*{Unemployment rate} 
                                \includegraphics[clip=true, trim={0cm 0cm 0cm 0cm},scale=0.50]{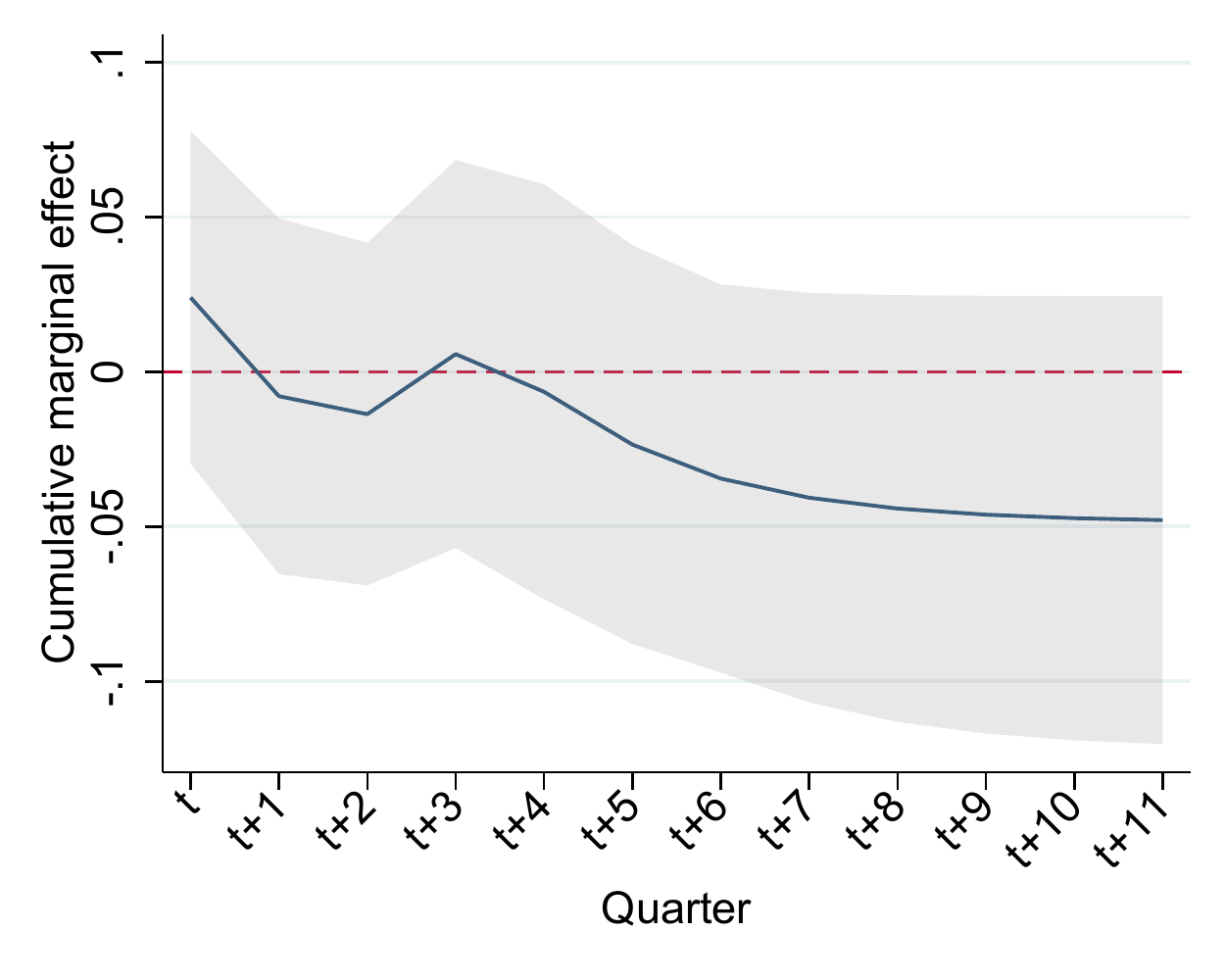}
                \end{subfigure}
                \begin{subfigure}[b]{0.49\textwidth}
                                \centering \caption*{Training} \subcaption*{Unsubsidized employment rate} 
                                \includegraphics[clip=true, trim={0cm 0cm 0cm 0cm},scale=0.50]{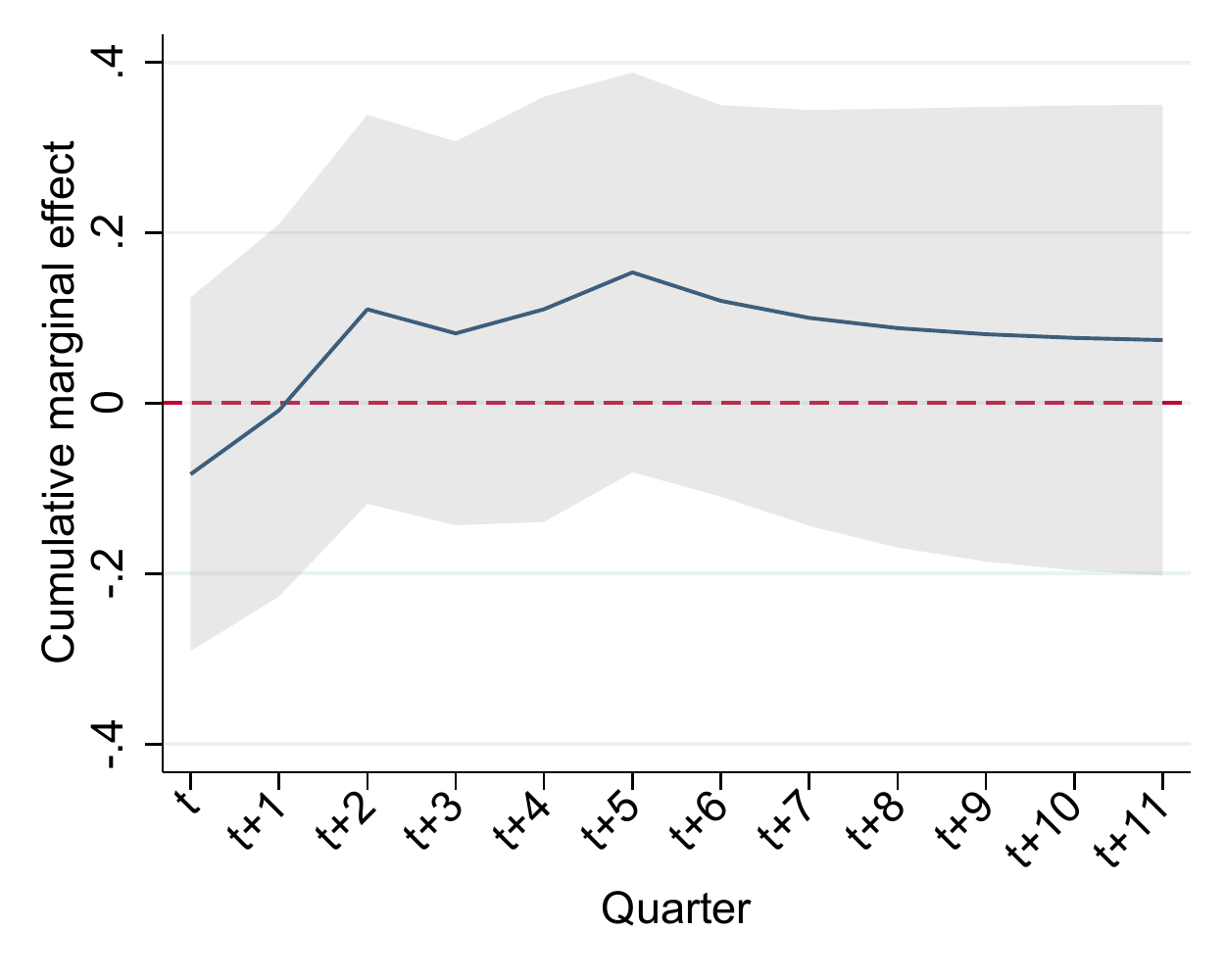}
                \end{subfigure}
                \vspace{10pt}    
                \newline
                \begin{subfigure}[b]{0.49\textwidth}
                                \centering \caption*{Short measures} \subcaption*{Unemployment rate} 
                                \includegraphics[clip=true, trim={0cm 0cm 0cm 0cm},scale=0.50]{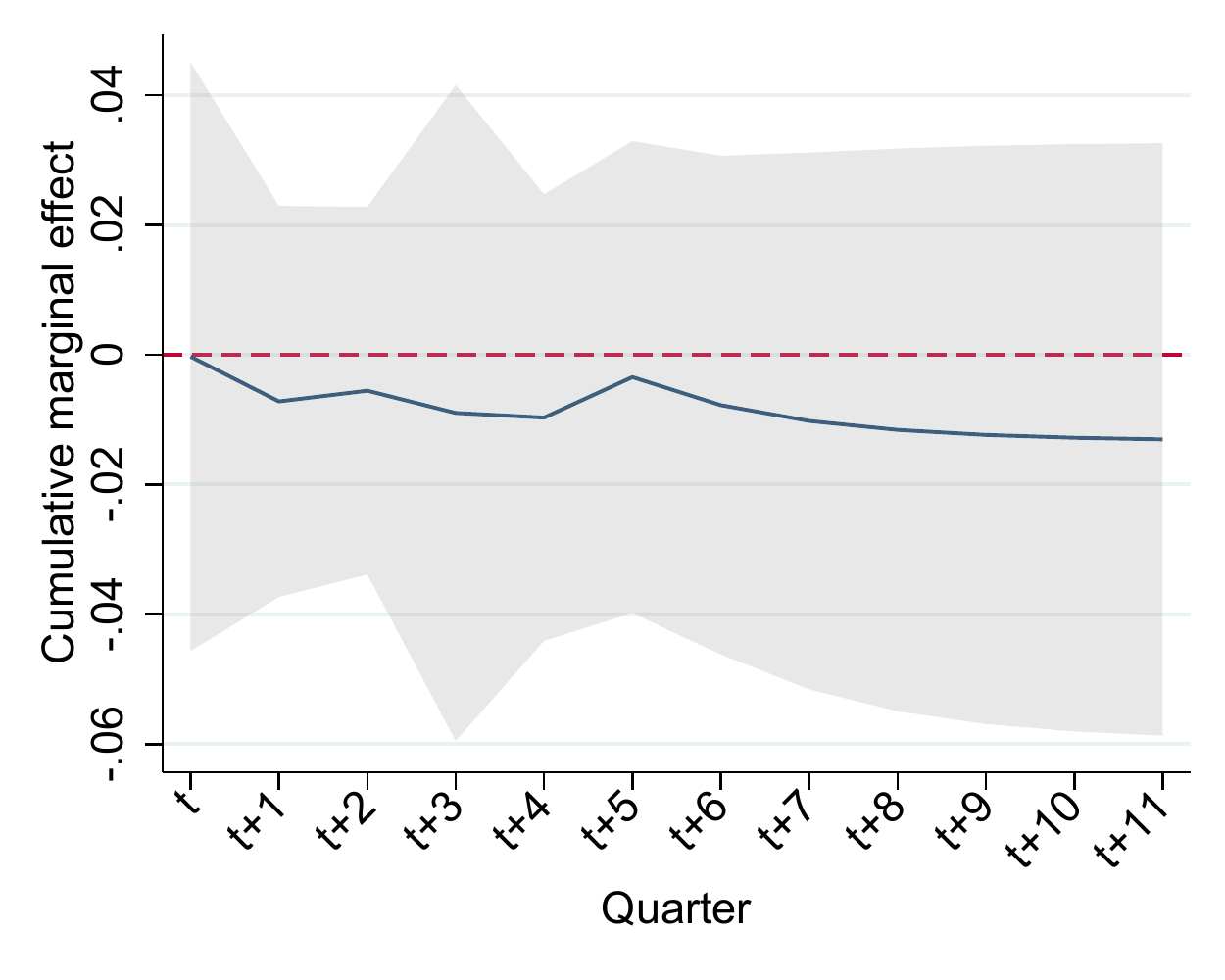}
                \end{subfigure}
                \vspace{10pt}    
                \begin{subfigure}[b]{0.49\textwidth}
                                \centering \caption*{Short measures}  \subcaption*{Unsubsidized employment rate} 
                                \includegraphics[clip=true, trim={0cm 0cm 0cm 0cm},scale=0.50]{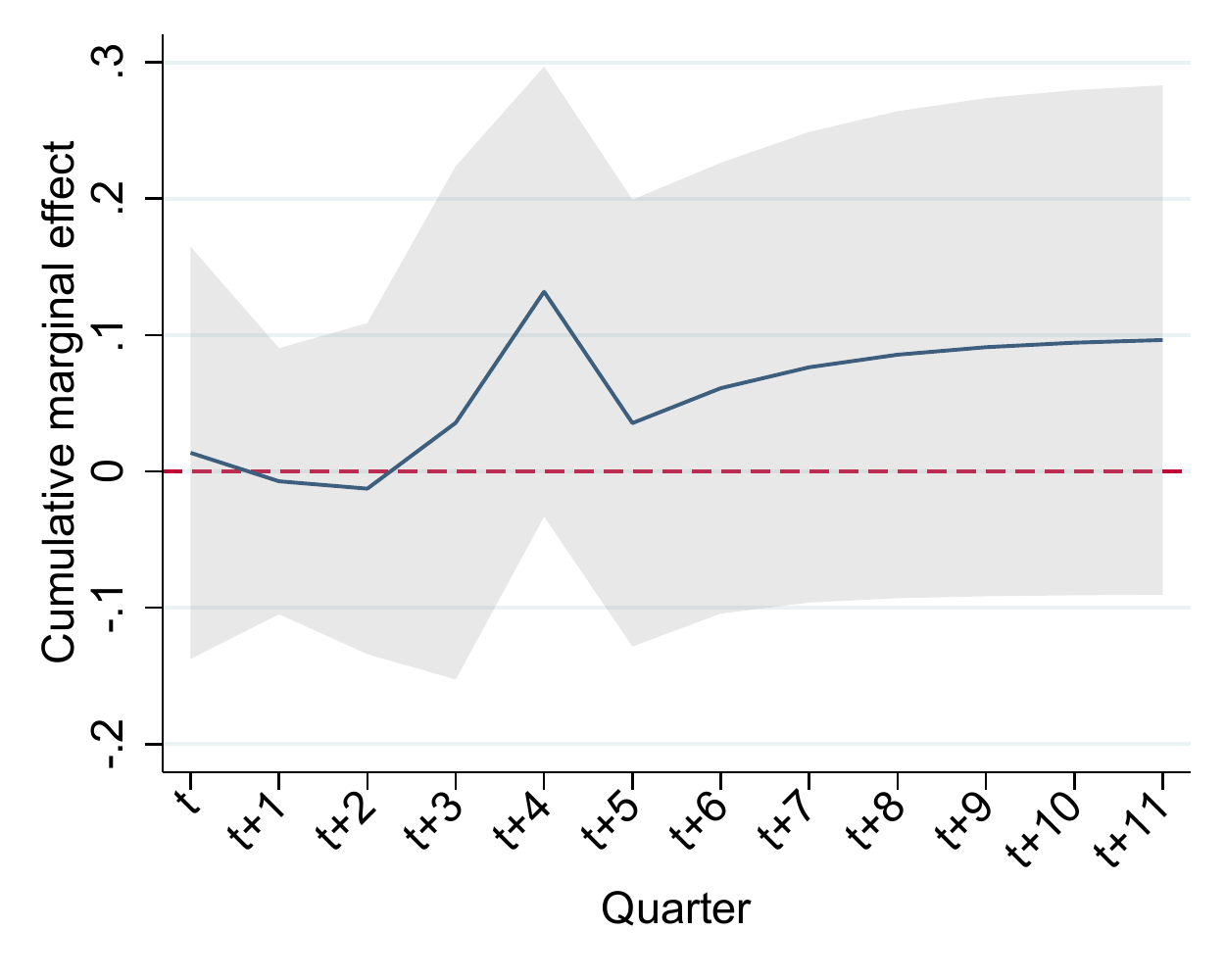}
                \end{subfigure}
                \vspace{10pt}
                \newline
                \begin{subfigure}[b]{0.49\textwidth}
                                \centering \caption*{Wage subsidies} \subcaption*{Unemployment rate} 
                                \includegraphics[clip=true, trim={0cm 0cm 0cm 0cm},scale=0.50]{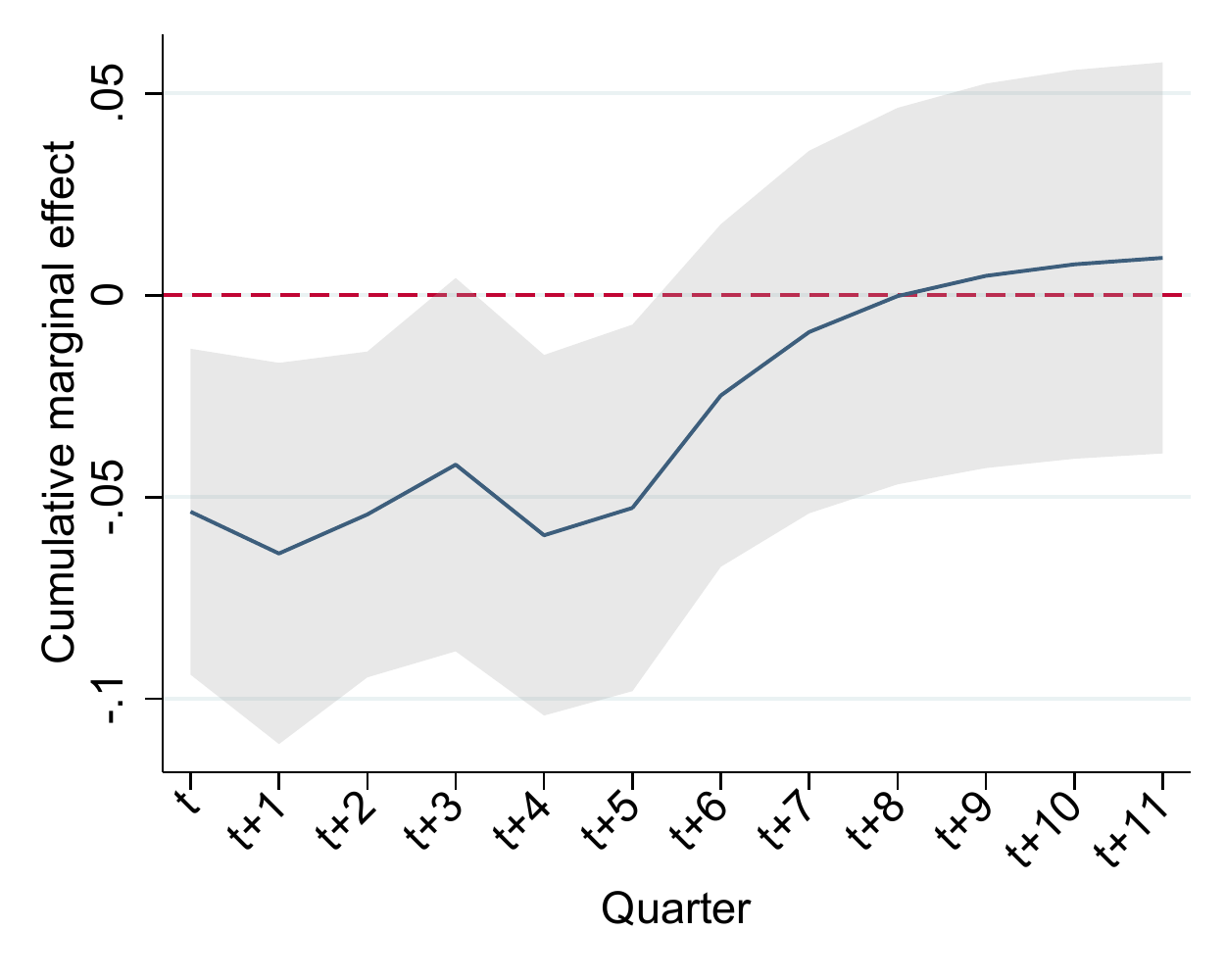}
                \end{subfigure}
                \vspace{10pt}    
                \begin{subfigure}[b]{0.49\textwidth}
                                \centering \caption*{Wage subsidies}  \subcaption*{Unsubsidized employment rate} 
                                \includegraphics[clip=true, trim={0cm 0cm 0cm 0cm},scale=0.50]{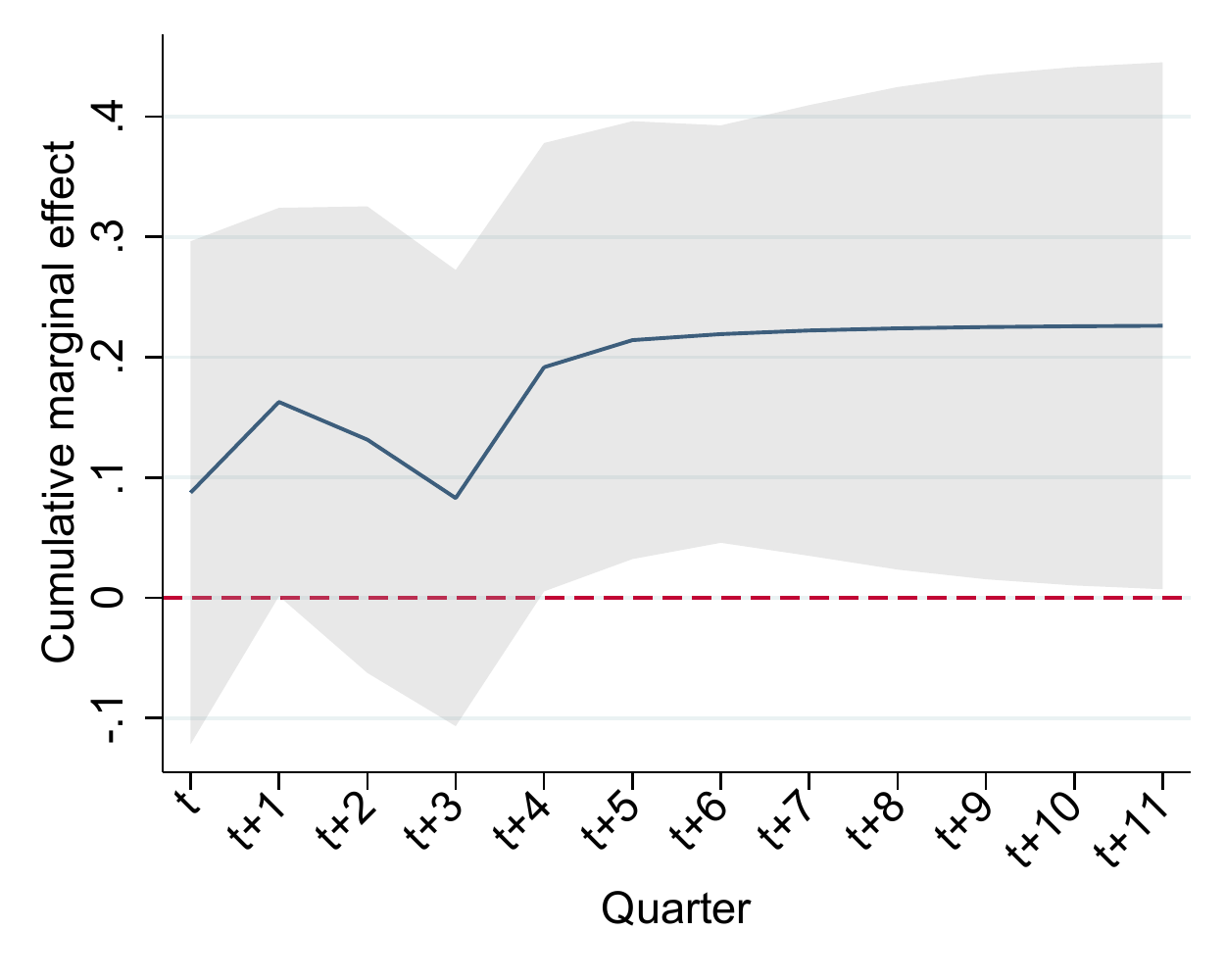}
                \end{subfigure}
                \vspace{10pt}
                \begin{minipage}{\textwidth}
                                \footnotesize \textit{Notes:} These graphs show the cumulative marginal effects of the three types of ALMP, i.e. training, short measures and wage subsidies on the unemployment rate and unsubsidized employment rate. 95\% confidence intervals are shown as grey areas. The effects are based on the ARDL model estimated by 2SLS. Program variables are included with 6 lags, main sample restrictions apply.  Standard errors obtained by a cross-sectional bootstrap (499 replications).
                \end{minipage}
\end{figure}

\begin{figure}
                \centering
                \caption{Cumulative Marginal Effects for Ages 30-50 \label{fig:cum_effects_iv_alo_rate_30_50}}
                \vspace{10pt}
                \begin{subfigure}[b]{0.49\textwidth}
                                \centering \caption*{Training} \subcaption*{Unemployment rate} 
                                \includegraphics[clip=true, trim={0cm 0cm 0cm 0cm},scale=0.50]{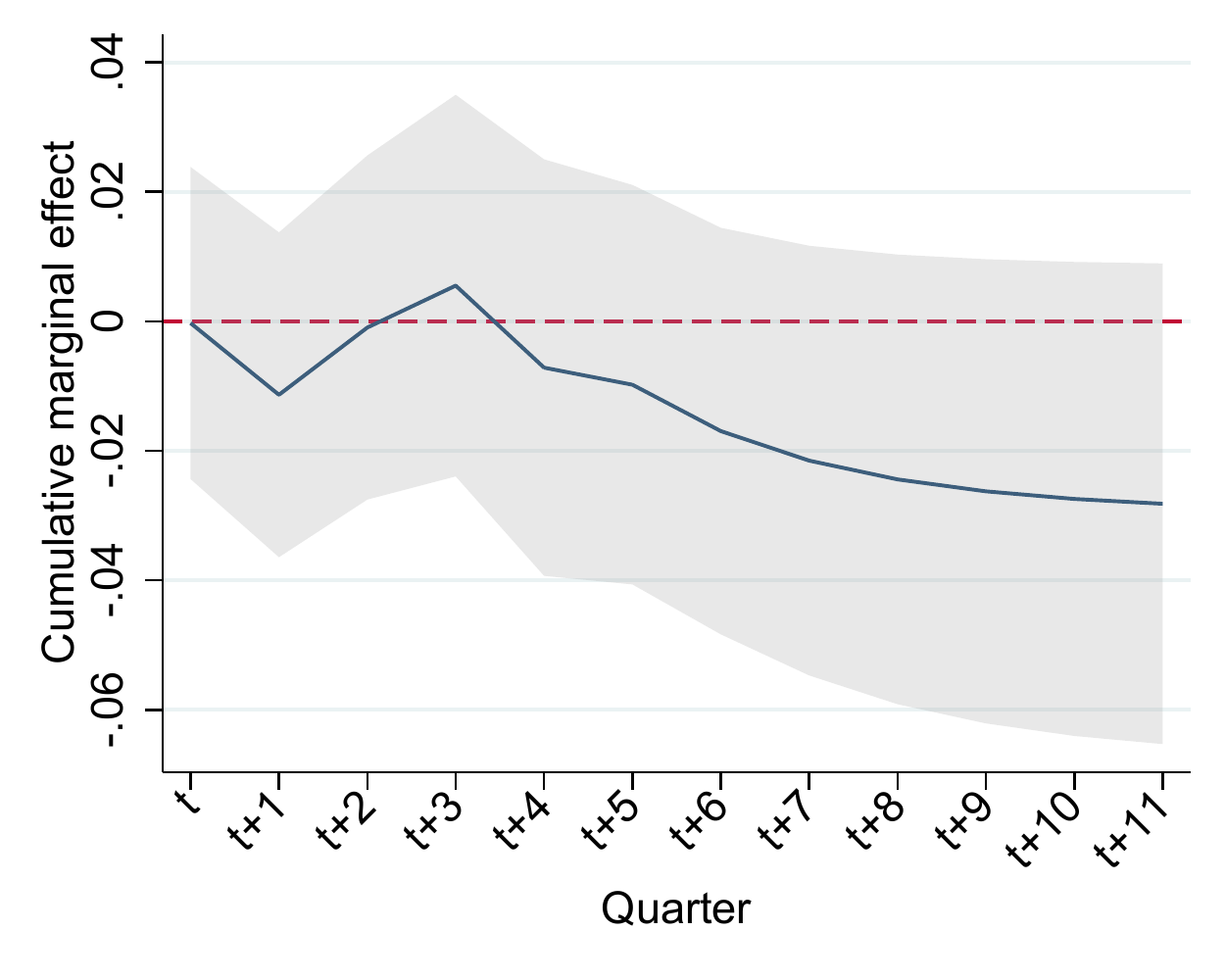}
                \end{subfigure}
                \begin{subfigure}[b]{0.49\textwidth}
                                \centering \caption*{Training} \subcaption*{Unsubsidized employment rate} 
                                \includegraphics[clip=true, trim={0cm 0cm 0cm 0cm},scale=0.50]{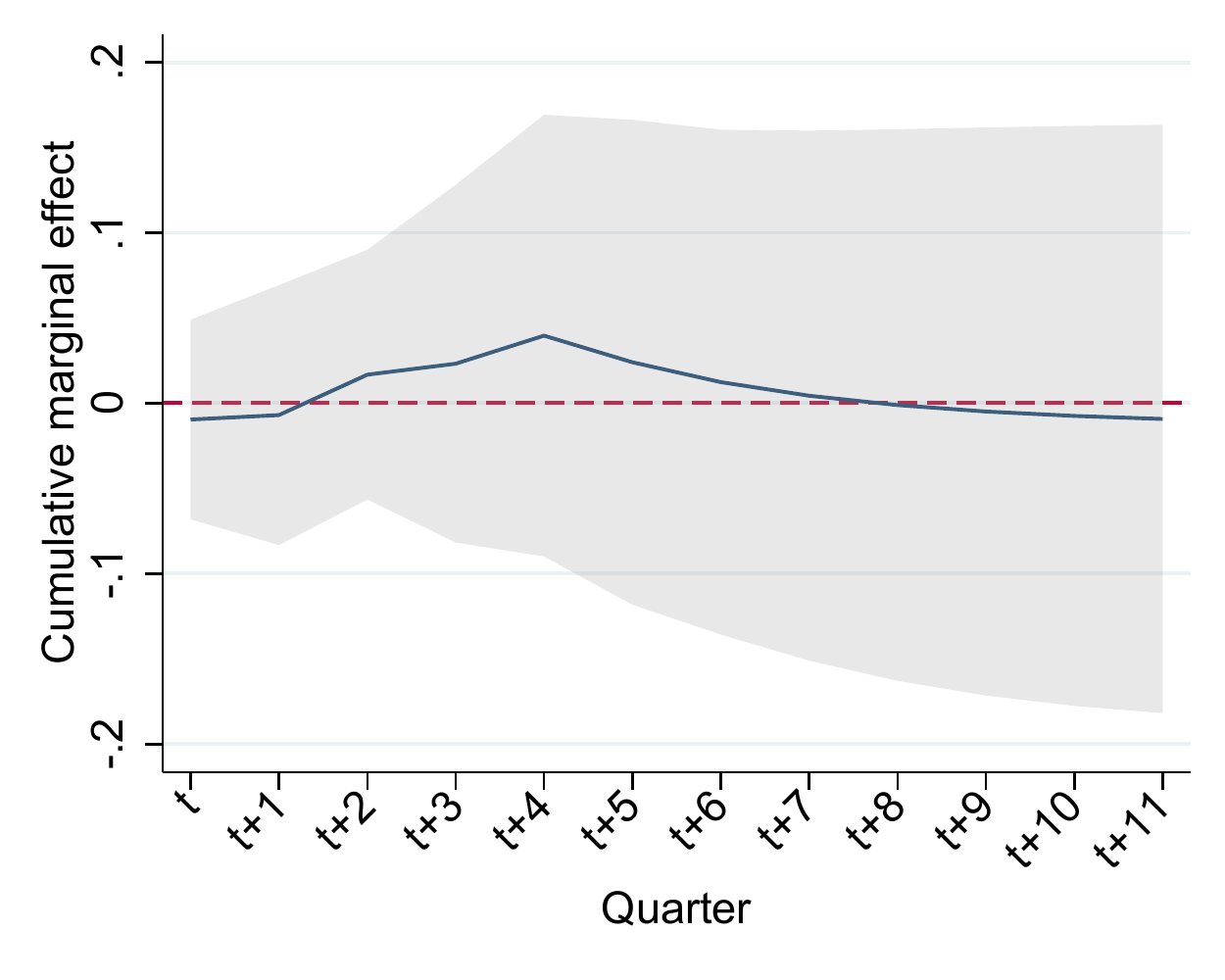}
                \end{subfigure}
                \vspace{10pt}    
                \newline
                \begin{subfigure}[b]{0.49\textwidth}
                                \centering \caption*{Short measures} \subcaption*{Unemployment rate} 
                                \includegraphics[clip=true, trim={0cm 0cm 0cm 0cm},scale=0.50]{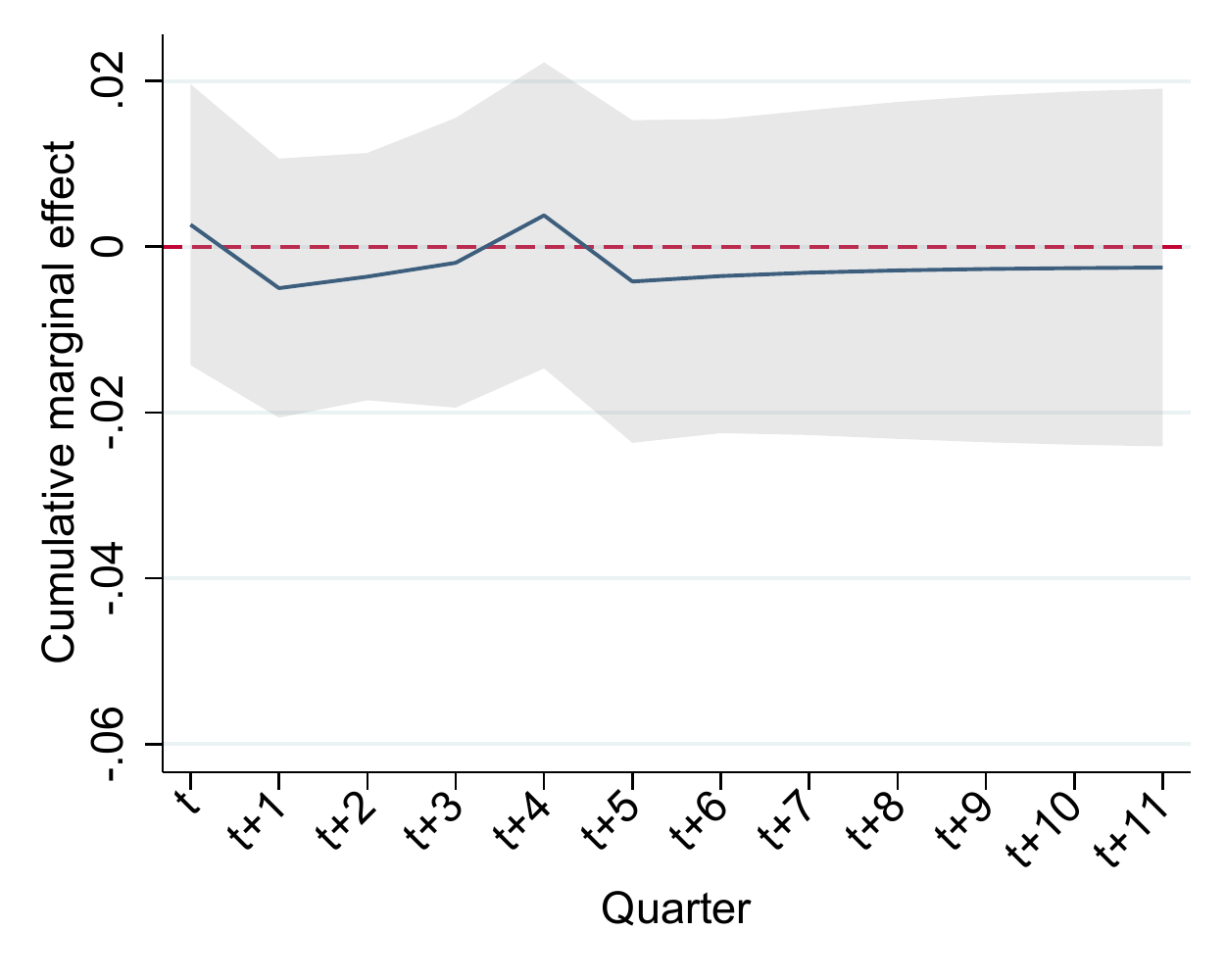}
                \end{subfigure}
                \vspace{10pt}    
                \begin{subfigure}[b]{0.49\textwidth}
                                \centering \caption*{Short measures}  \subcaption*{Unsubsidized employment rate} 
                                \includegraphics[clip=true, trim={0cm 0cm 0cm 0cm},scale=0.50]{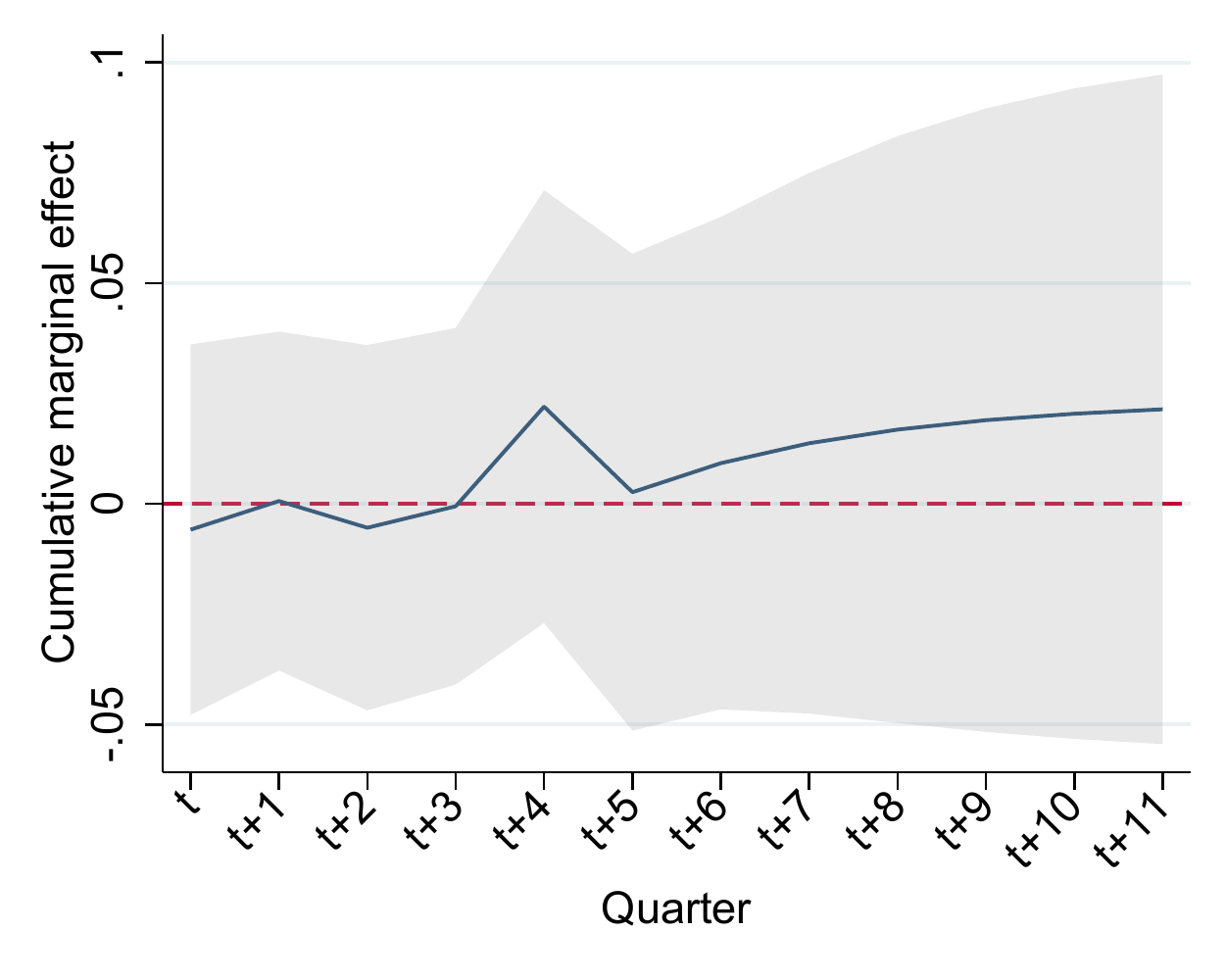}
                \end{subfigure}
                \vspace{10pt}
                \newline
                \begin{subfigure}[b]{0.49\textwidth}
                                \centering \caption*{Wage subsidies} \subcaption*{Unemployment rate} 
                                \includegraphics[clip=true, trim={0cm 0cm 0cm 0cm},scale=0.50]{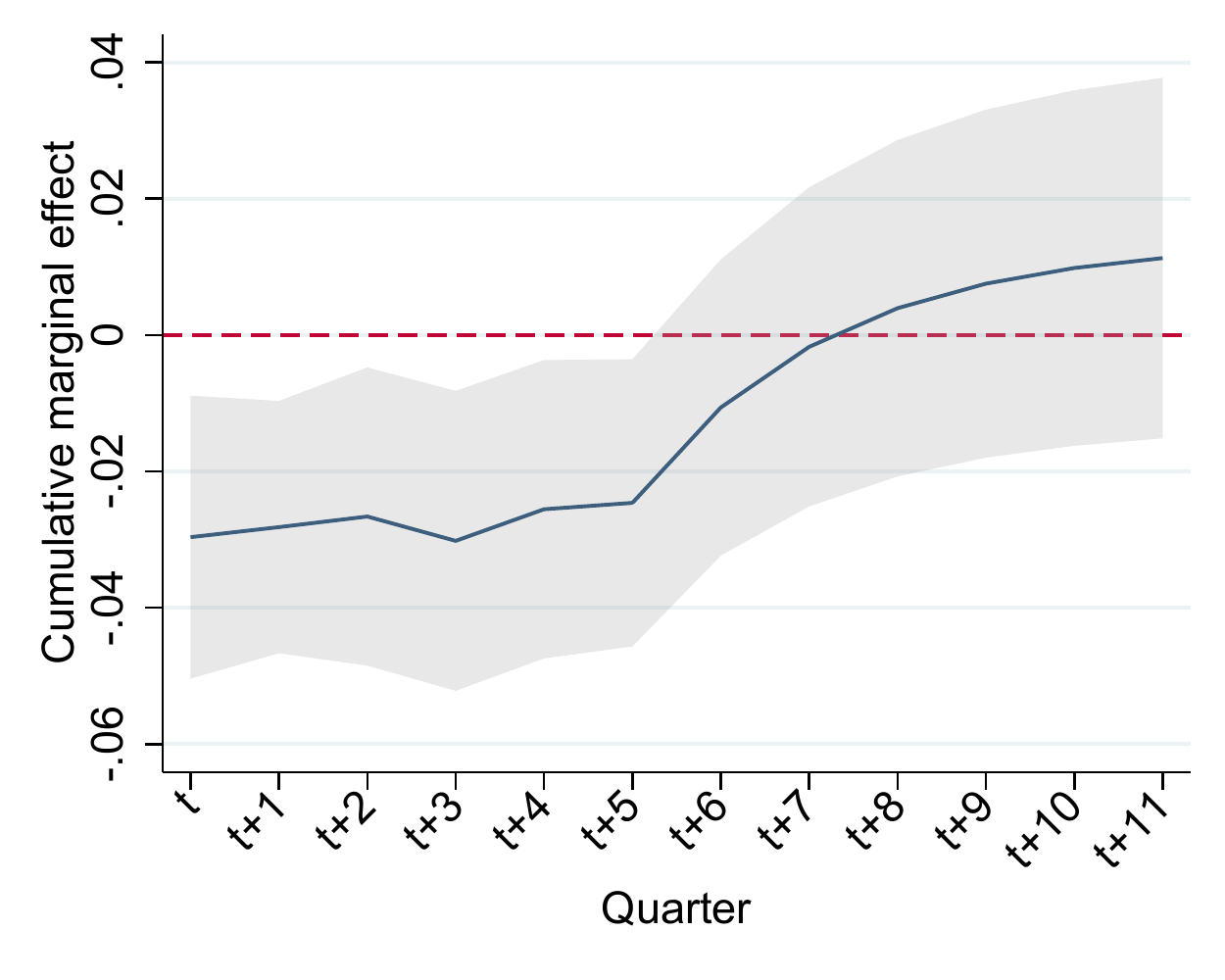}
                \end{subfigure}
                \vspace{10pt}    
                \begin{subfigure}[b]{0.49\textwidth}
                                \centering \caption*{Wage subsidies}  \subcaption*{Unsubsidized employment rate} 
                                \includegraphics[clip=true, trim={0cm 0cm 0cm 0cm},scale=0.50]{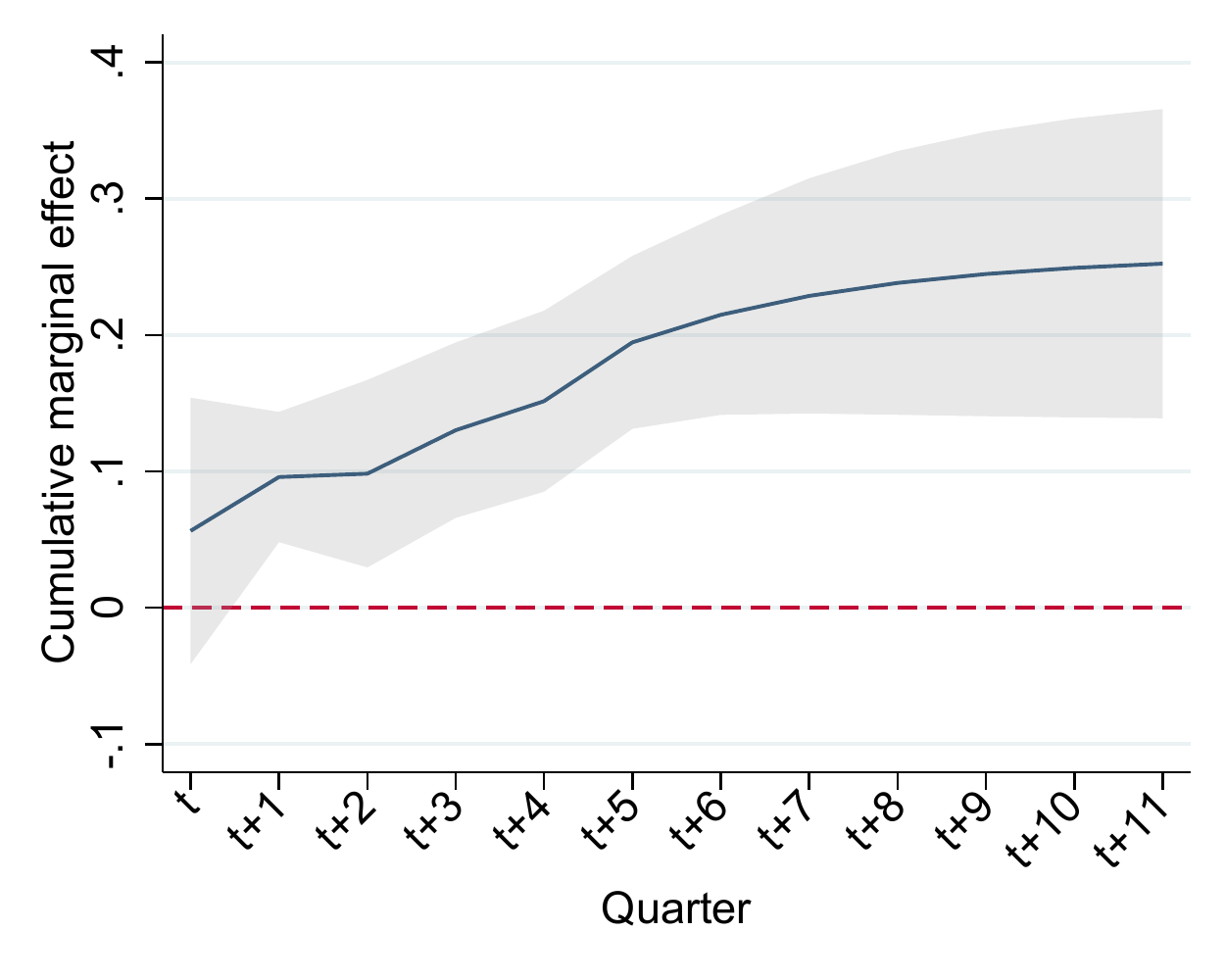}
                \end{subfigure}
                \vspace{10pt}
                \begin{minipage}{\textwidth}
                                \footnotesize \textit{Notes:} These graphs show the cumulative marginal effects of the three types of ALMP, i.e. training, short measures and wage subsidies on the unemployment rate and unsubsidized employment rate. 95\% confidence intervals are shown as grey areas. The effects are based on the ARDL model estimated by 2SLS. Program variables are included with 6 lags, main sample restrictions apply.  Standard errors obtained by a cross-sectional bootstrap (499 replications).
                \end{minipage}
\end{figure}

\begin{figure}
                \centering
                \caption{Cumulative Marginal Effects for Ages over 50 \label{fig:cum_effects_iv_alo_rate_o_50}}
                \vspace{10pt}
                \begin{subfigure}[b]{0.49\textwidth}
                                \centering \caption*{Training} \subcaption*{Unemployment rate} 
                                \includegraphics[clip=true, trim={0cm 0cm 0cm 0cm},scale=0.50]{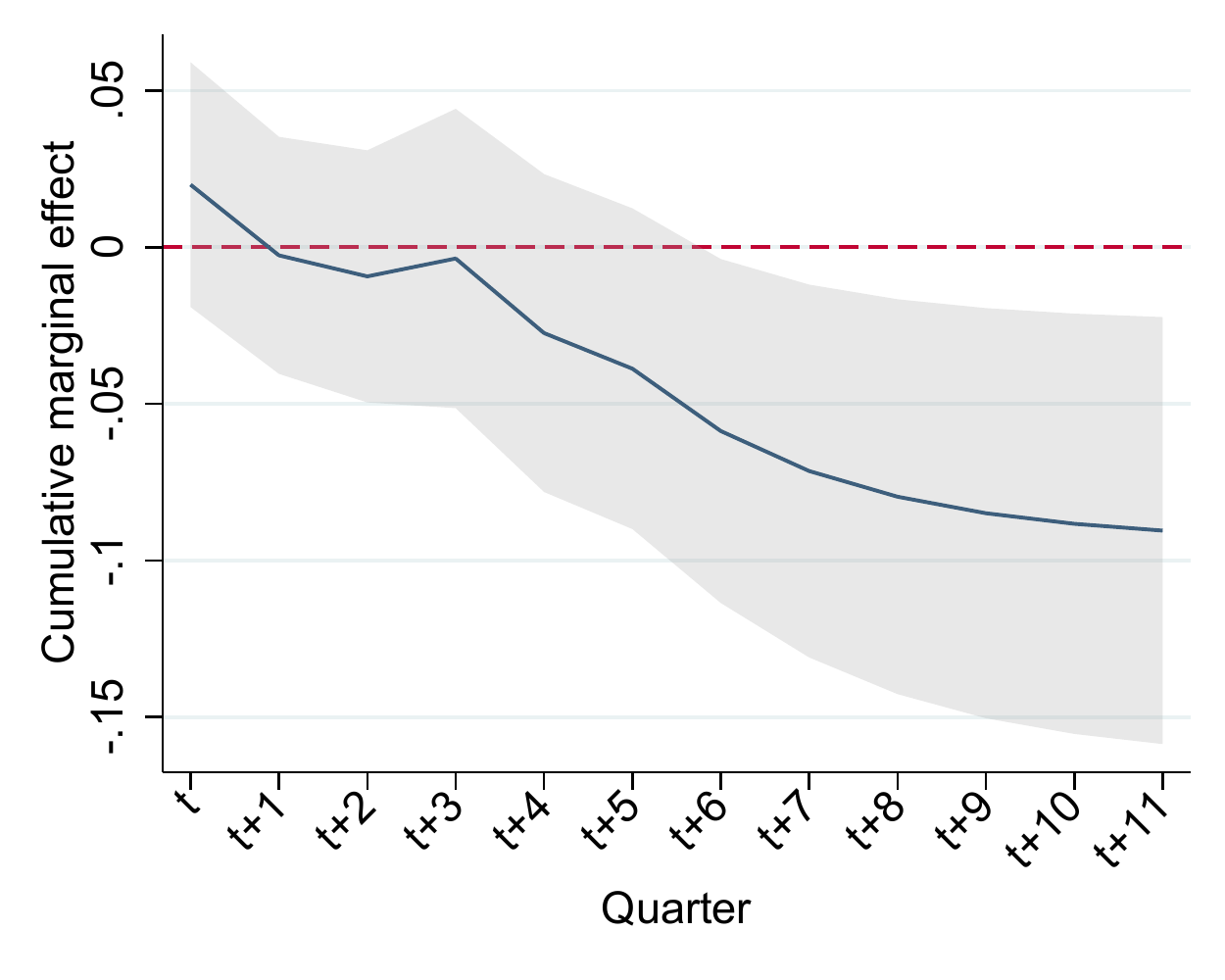}
                \end{subfigure}
                \begin{subfigure}[b]{0.49\textwidth}
                                \centering \caption*{Training} \subcaption*{Unsubsidized employment rate} 
                                \includegraphics[clip=true, trim={0cm 0cm 0cm 0cm},scale=0.50]{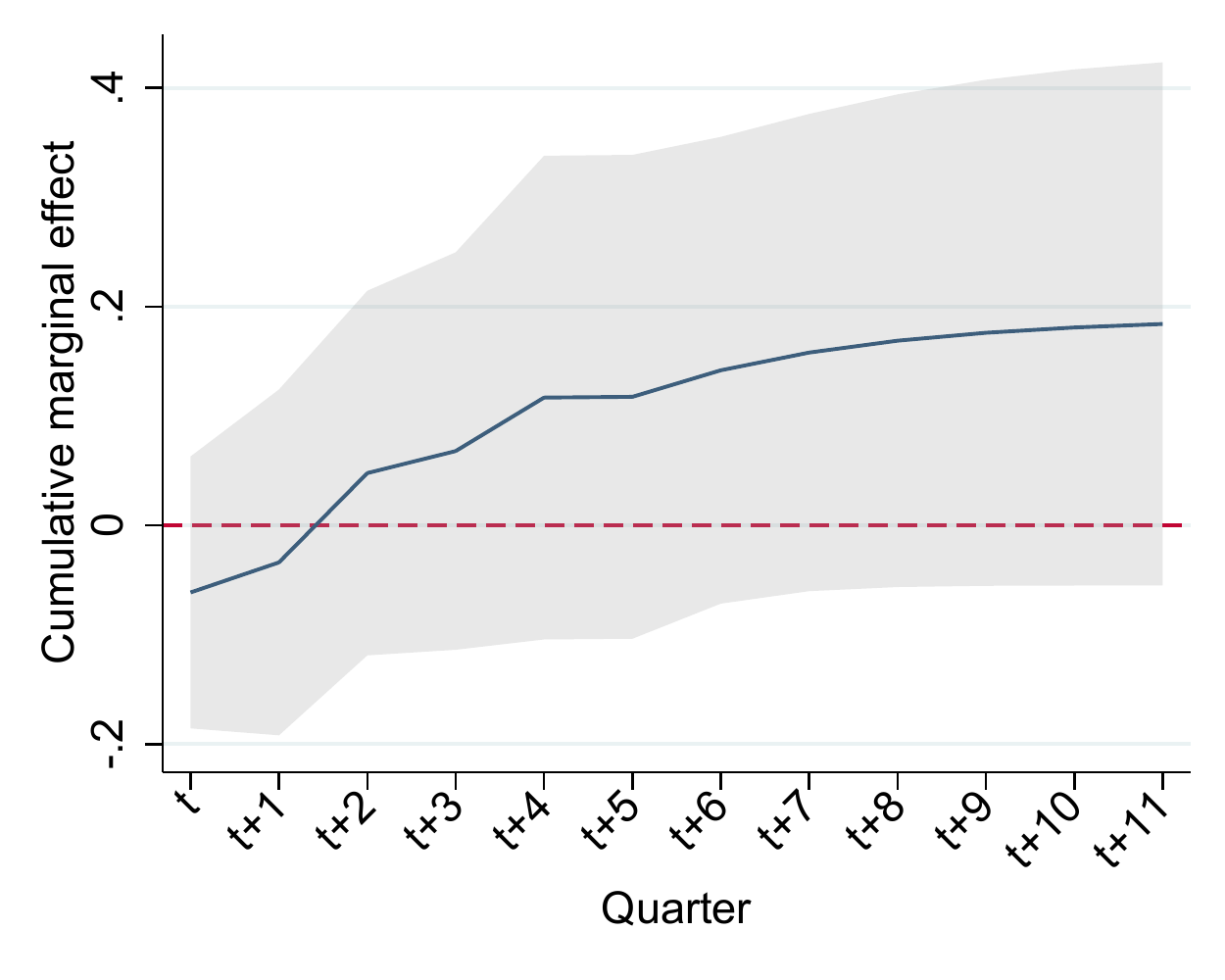}
                \end{subfigure}
                \vspace{10pt}    
                \newline
                \begin{subfigure}[b]{0.49\textwidth}
                                \centering \caption*{Short measures} \subcaption*{Unemployment rate} 
                                \includegraphics[clip=true, trim={0cm 0cm 0cm 0cm},scale=0.50]{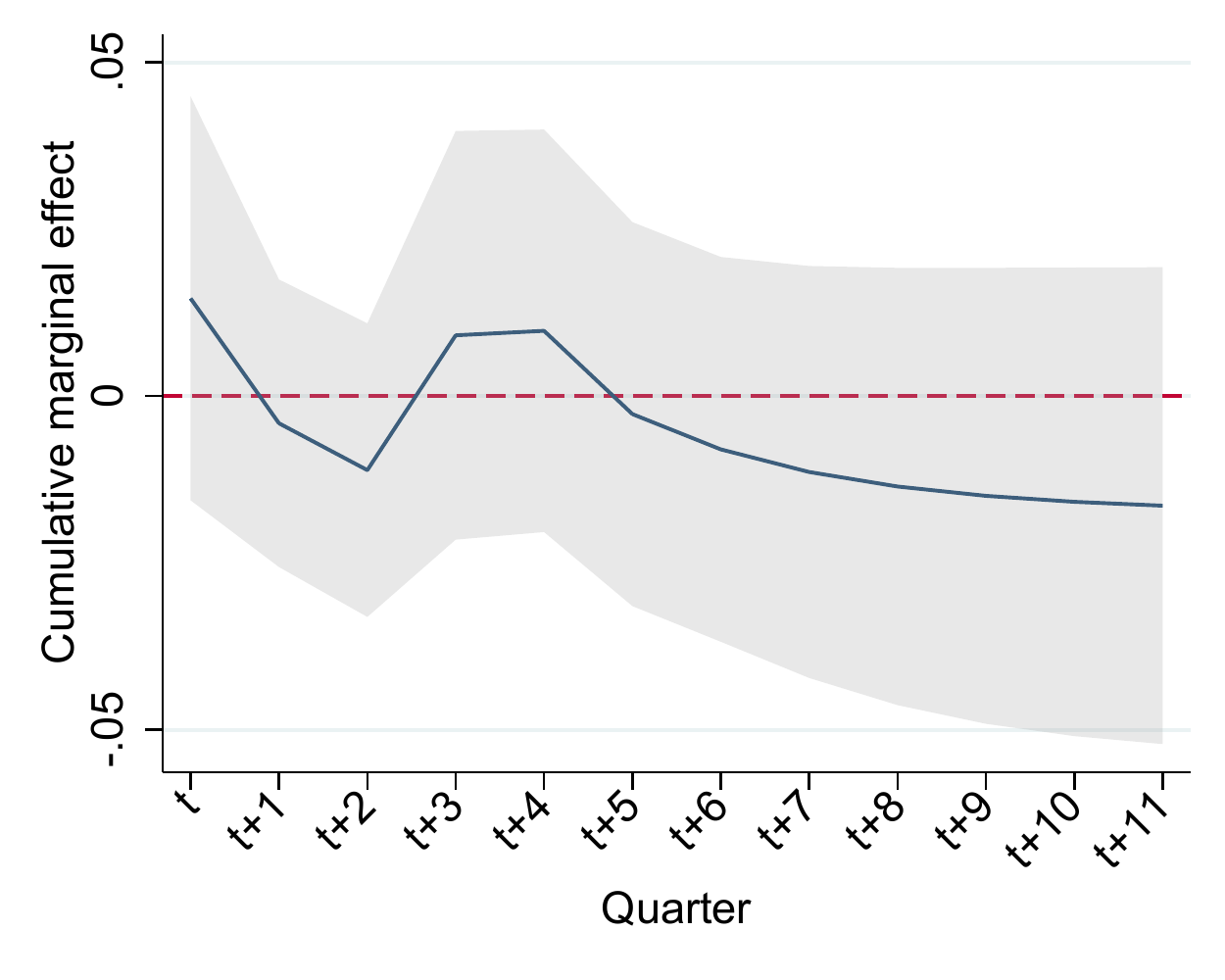}
                \end{subfigure}
                \vspace{10pt}    
                \begin{subfigure}[b]{0.49\textwidth}
                                \centering \caption*{Short measures}  \subcaption*{Unsubsidized employment rate} 
                                \includegraphics[clip=true, trim={0cm 0cm 0cm 0cm},scale=0.50]{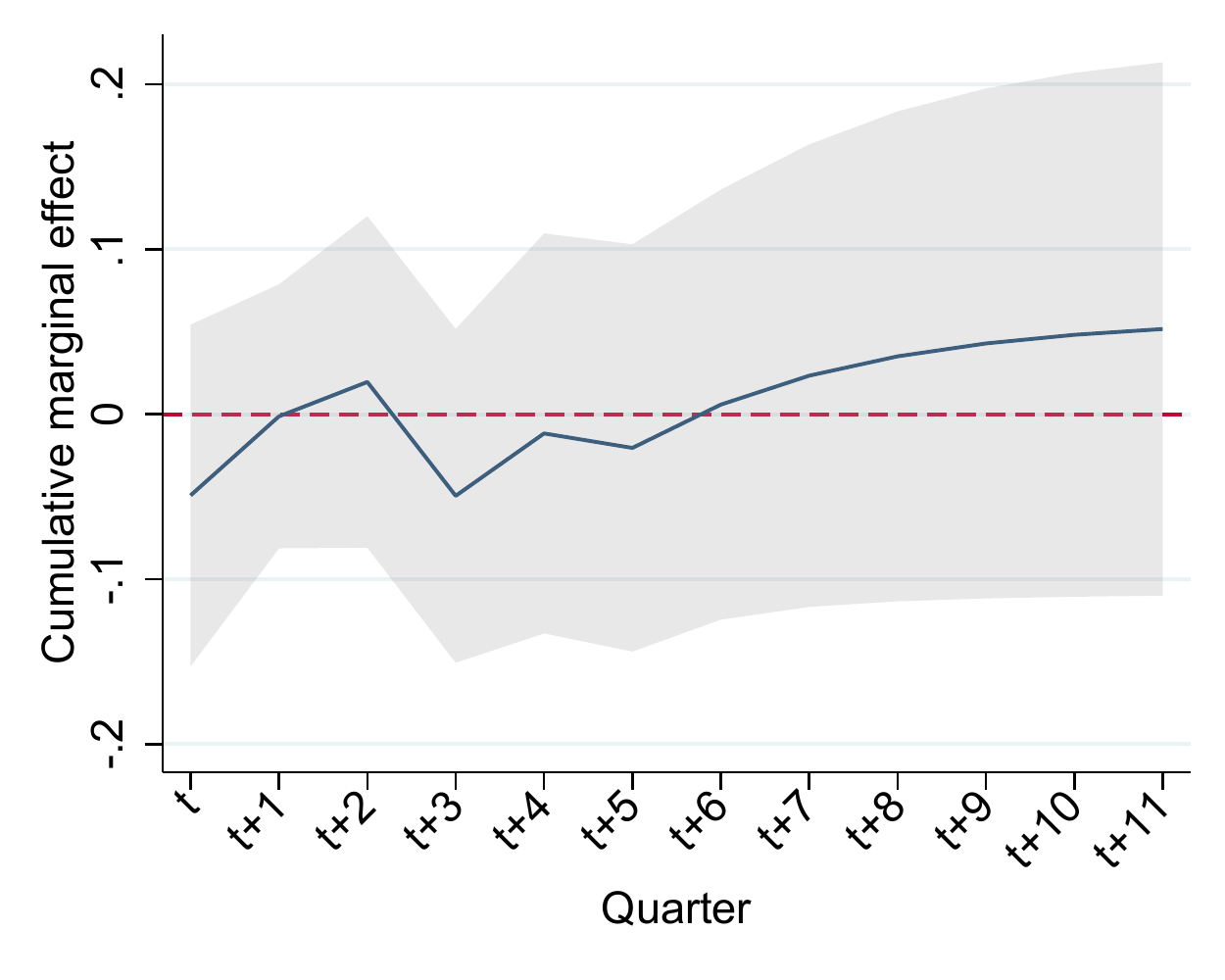}
                \end{subfigure}
                \vspace{10pt}
                \newline
                \begin{subfigure}[b]{0.49\textwidth}
                                \centering \caption*{Wage subsidies} \subcaption*{Unemployment rate} 
                                \includegraphics[clip=true, trim={0cm 0cm 0cm 0cm},scale=0.50]{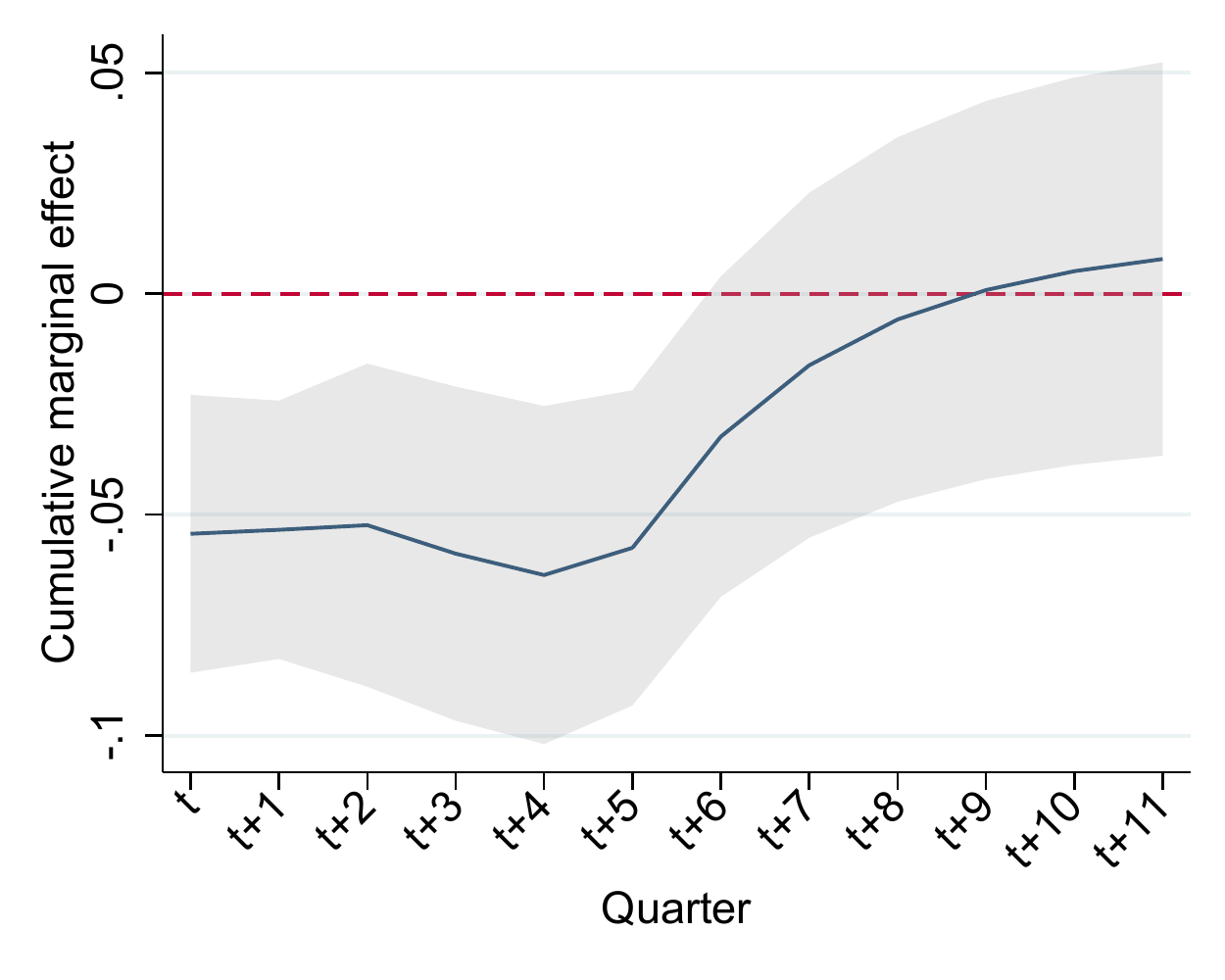}
                \end{subfigure}
                \vspace{10pt}    
                \begin{subfigure}[b]{0.49\textwidth}
                                \centering \caption*{Wage subsidies}  \subcaption*{Unsubsidized employment rate} 
                                \includegraphics[clip=true, trim={0cm 0cm 0cm 0cm},scale=0.50]{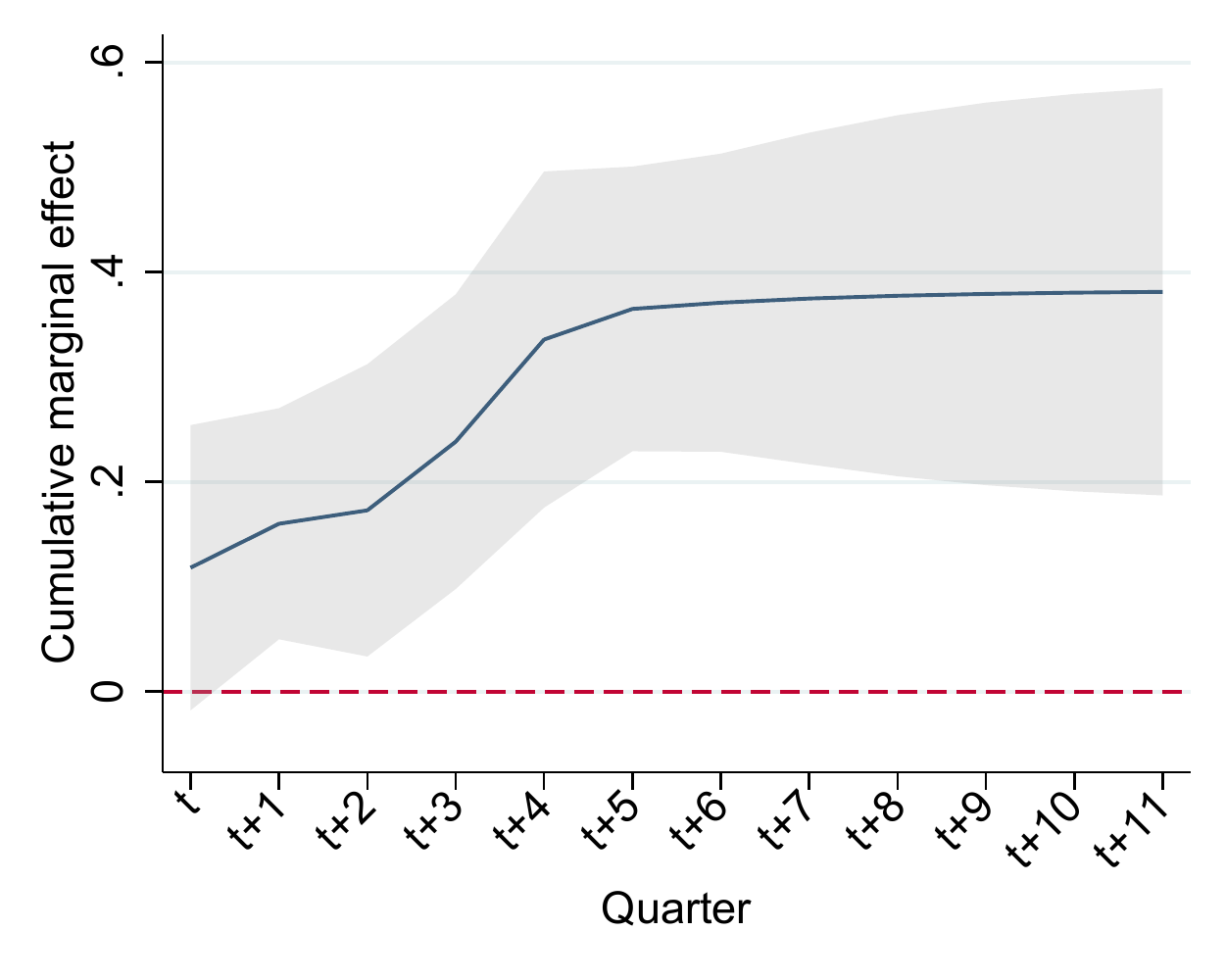}
                \end{subfigure}
                \vspace{10pt}
                \begin{minipage}{\textwidth}
                                \footnotesize \textit{Notes:} These graphs show the cumulative marginal effects of the three types of ALMP, i.e. training, short measures and wage subsidies on the unemployment rate and unsubsidized employment rate. 95\% confidence intervals are shown as grey areas. The effects are based on the ARDL model estimated by 2SLS. Program variables are included with 6 lags, main sample restrictions apply.  Standard errors obtained by a cross-sectional bootstrap (499 replications).
                \end{minipage}
\end{figure}



\begin{landscape}

\begin{table}[H]
\centering
\caption{Effects Based on the Distributed Lag Model}
\label{tab:dl_alo_emp}
{\small
{
\def\sym#1{\ifmmode^{#1}\else\(^{#1}\)\fi}
\begin{tabular}{l| c c c | c c c | c c c | c c c}
\hline\hline
&\multicolumn{3}{c}{Unemployment rate} &\multicolumn{3}{c}{Unsub. employment rate} &\multicolumn{3}{c}{Rate on welfare} &\multicolumn{3}{c}{Rate of emp. on benefits} \\ 
&\multicolumn{1}{c}{(1)} &\multicolumn{1}{c}{(2)} &\multicolumn{1}{c}{(3)}&\multicolumn{1}{c}{(4)} &\multicolumn{1}{c}{(5)} &\multicolumn{1}{c}{(6)} 
&\multicolumn{1}{c}{(7)} &\multicolumn{1}{c}{(8)} &\multicolumn{1}{c}{(9)}&\multicolumn{1}{c}{(10)} &\multicolumn{1}{c}{(11)} &\multicolumn{1}{c}{(12)} \\ 
\\ 
\hline
            &      effect&          se&       p-val&      effect&          se&       p-val&      effect&          se&       p-val&      effect&          se&       p-val\\
Training(st)&       0.004&       0.012&       0.753&       0.028&       0.069&       0.691&      -0.034&       0.058&       0.555&       0.000&       0.017&       0.978\\
Training(lt)&      -0.016&       0.014&       0.270&       0.020&       0.045&       0.659&      -0.006&       0.036&       0.870&      -0.008&       0.016&       0.628\\
Short measures (st)&      -0.000&       0.008&       0.989&      -0.016&       0.026&       0.540&       0.016&       0.021&       0.441&       0.002&       0.008&       0.754\\
Short measures (lt)&      -0.008&       0.008&       0.351&       0.001&       0.022&       0.964&       0.015&       0.016&       0.369&      -0.008&       0.007&       0.264\\
Wage subsidies (st)&      -0.038&       0.011&       0.001&       0.031&       0.071&       0.661&      -0.018&       0.062&       0.774&      -0.009&       0.016&       0.575\\
Wage subsidies (lt)&      -0.014&       0.010&       0.145&       0.219&       0.035&       0.000&      -0.160&       0.028&       0.000&      -0.074&       0.013&       0.000\\
\\
\hline
Quarter-Year FEs&      $\checkmark$ & & &  $\checkmark$ & &  &  $\checkmark$ & & &  $\checkmark$ & &     \\
Additional Controls&      $\checkmark$ & & &  $\checkmark$ & &   & $\checkmark$ & & &  $\checkmark$ & &     \\
Observations& 5980 & & &  5980 & & & 5980 & & &  5980 & &       \\
LLM & 115 & & &  115 & & & 115 & & &  115 & &       \\
\hline\hline
\end{tabular}
}
}
\centering
     \begin{minipage}{20cm}
          \vspace{6pt}
          \footnotesize \textit{Notes:} This table shows the short-term (st) and long-term (lt) effects of the three types of ALMP, i.e. training, short measures (SM) and wage subsdies (wagesub) on the unemployment rate (columns 1-3), unsubsidized employment rate (columns 4-6), rate of welfare recipients (clumns 7-9) and rate of employed workers on benefits (columns 10-12). Program variables are included with 6 lags. The model is estimated for the time period 2010 to 2018.  Standard errors (se)  are obtained by a cross-sectional bootstrap (499 replications).
     \end{minipage}
\end{table}

\end{landscape}


\begin{figure}[H]
                \centering
                \caption{Effect Patterns for Different Lag Order I \label{fig:lt_effects_rob_lags}}
                \vspace{10pt}
                \begin{subfigure}[b]{0.49\textwidth}
                                \centering \caption*{Training} \subcaption*{Unemployment rate} 
                                \includegraphics[clip=true, trim={0cm 0cm 0cm 0cm},scale=0.50]{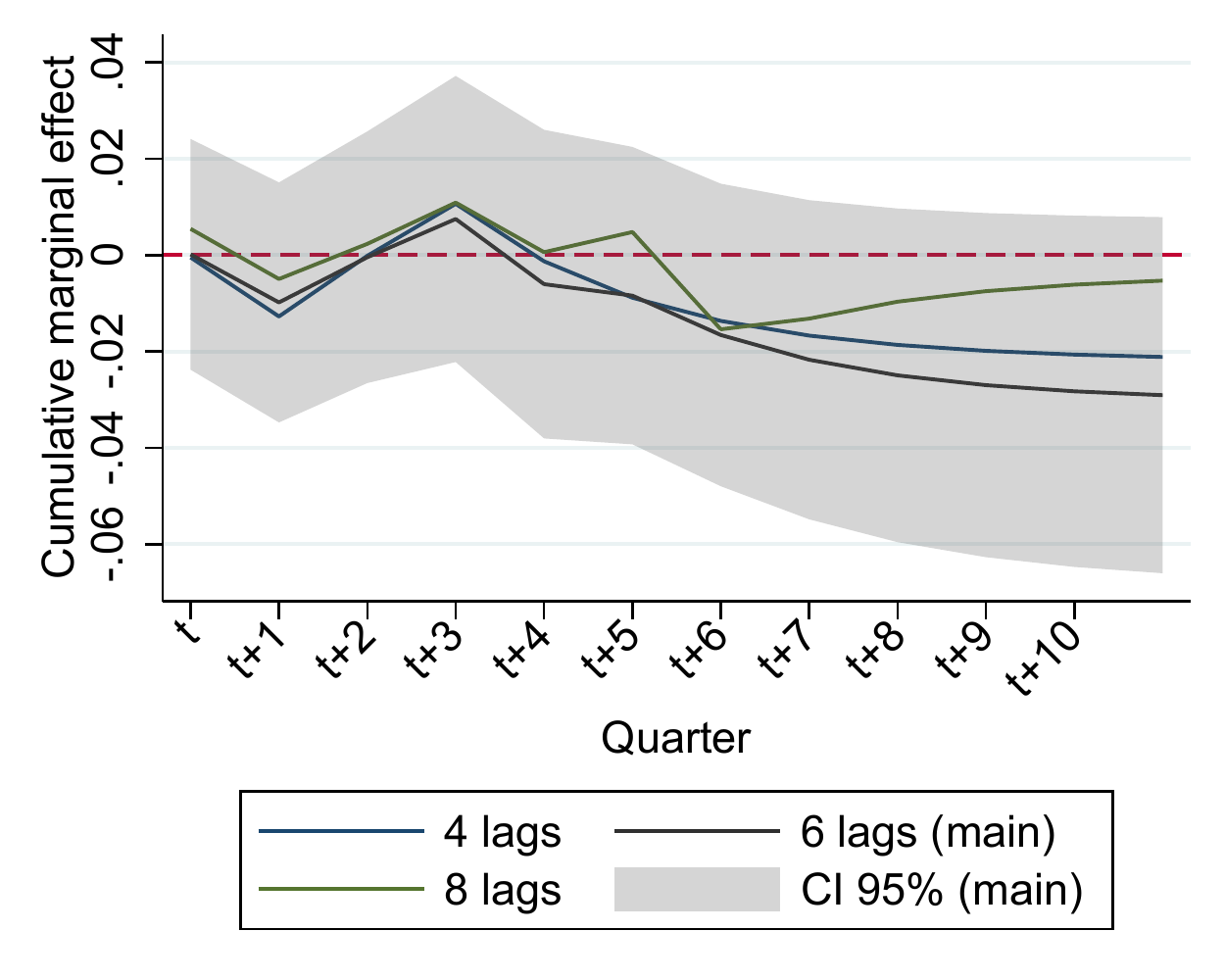}
                \end{subfigure}
                \begin{subfigure}[b]{0.49\textwidth}
                                \centering \caption*{Training} \subcaption*{Unsubsidised employment rate} 
                                \includegraphics[clip=true, trim={0cm 0cm 0cm 0cm},scale=0.50]{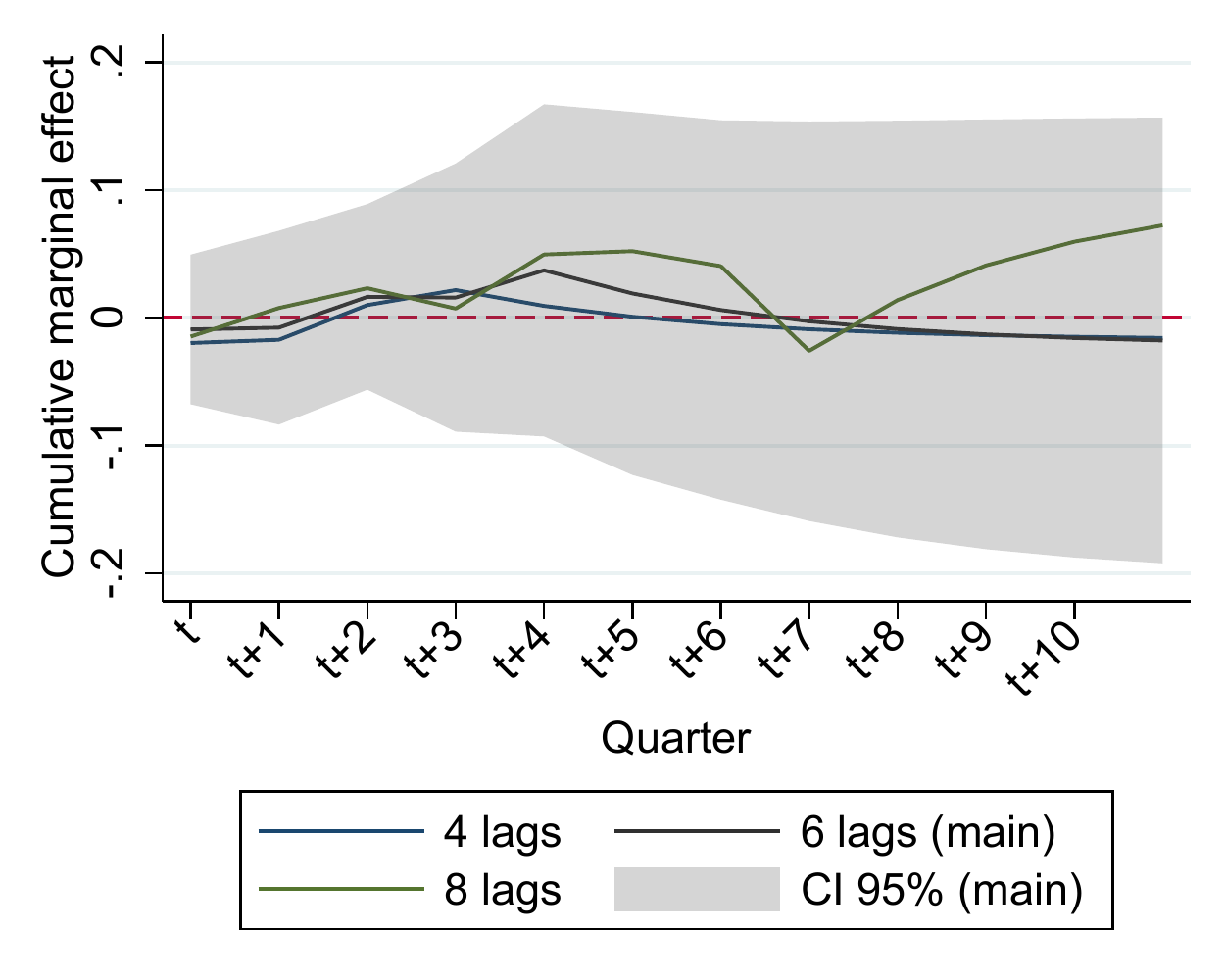}
                \end{subfigure}
                \vspace{10pt}    
                \newline
                \begin{subfigure}[b]{0.49\textwidth}
                                \centering \caption*{Short measures} \subcaption*{Unemployment rate} 
                                \includegraphics[clip=true, trim={0cm 0cm 0cm 0cm},scale=0.50]{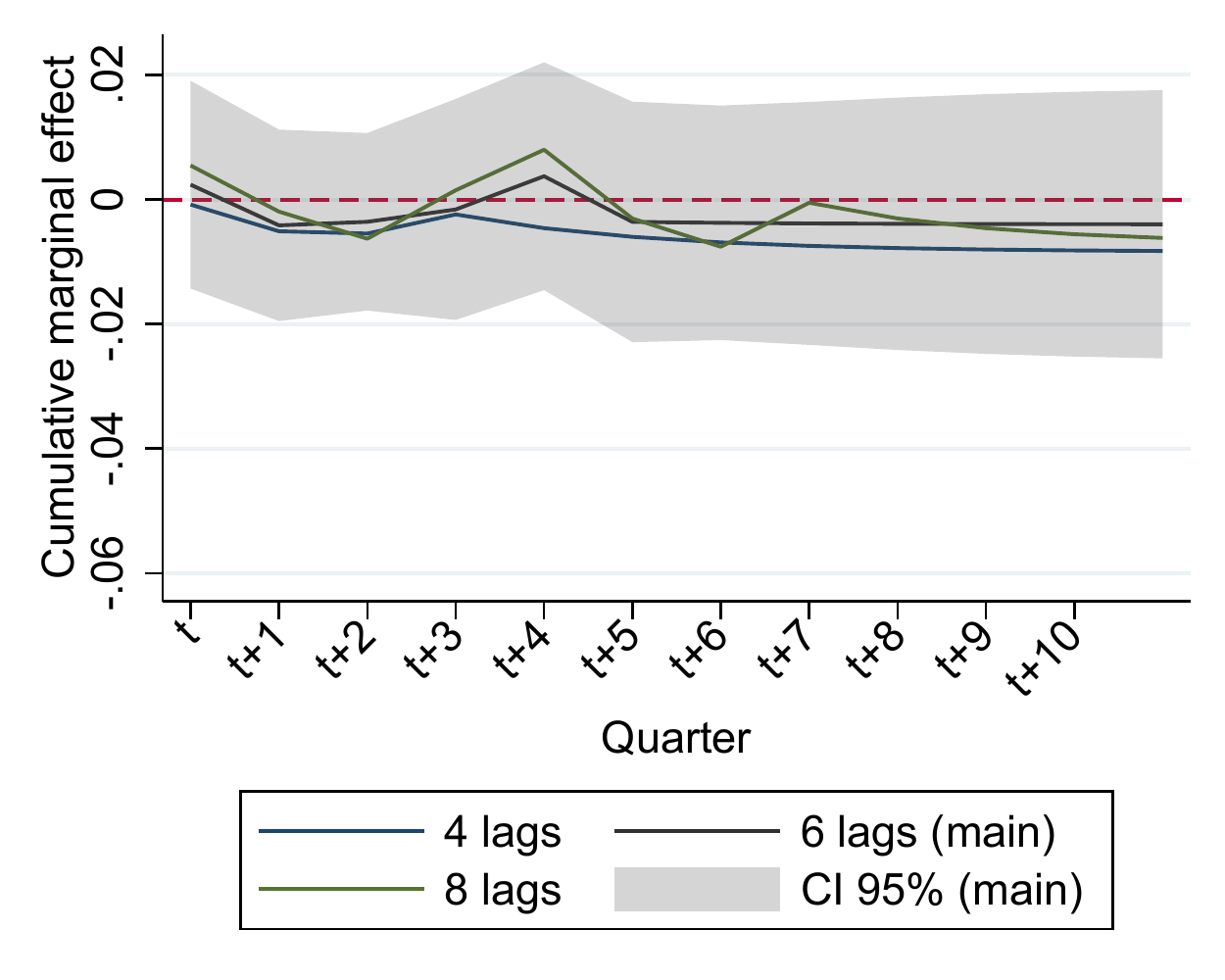}
                \end{subfigure}
                \vspace{10pt}    
                \begin{subfigure}[b]{0.49\textwidth}
                                \centering \caption*{Short measures}  \subcaption*{Unsubsidised employment rate} 
                                \includegraphics[clip=true, trim={0cm 0cm 0cm 0cm},scale=0.50]{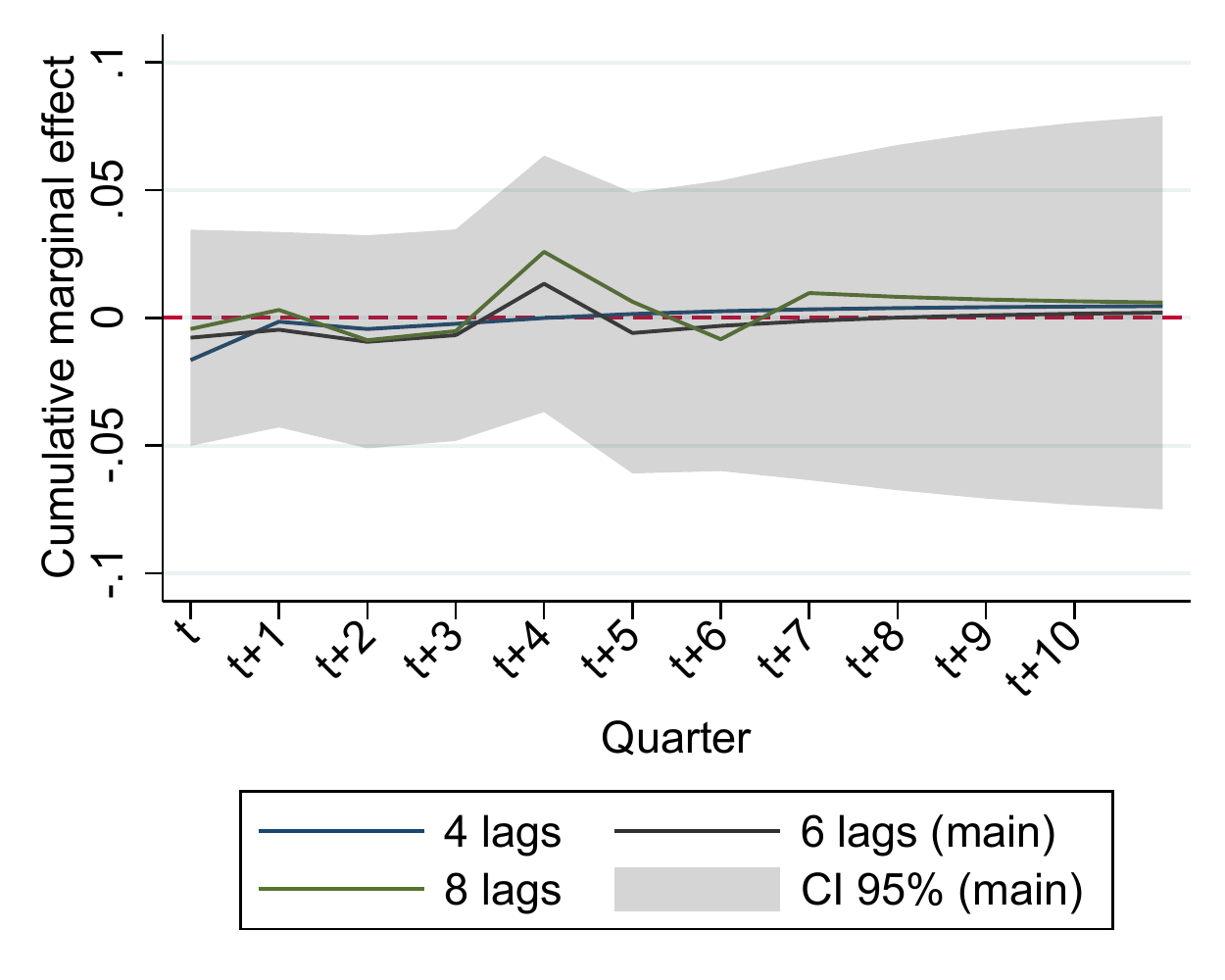}
                \end{subfigure}
                \vspace{10pt}
                \newline
                \begin{subfigure}[b]{0.49\textwidth}
                                \centering \caption*{Wage subsidies} \subcaption*{Unemployment rate} 
                                \includegraphics[clip=true, trim={0cm 0cm 0cm 0cm},scale=0.50]{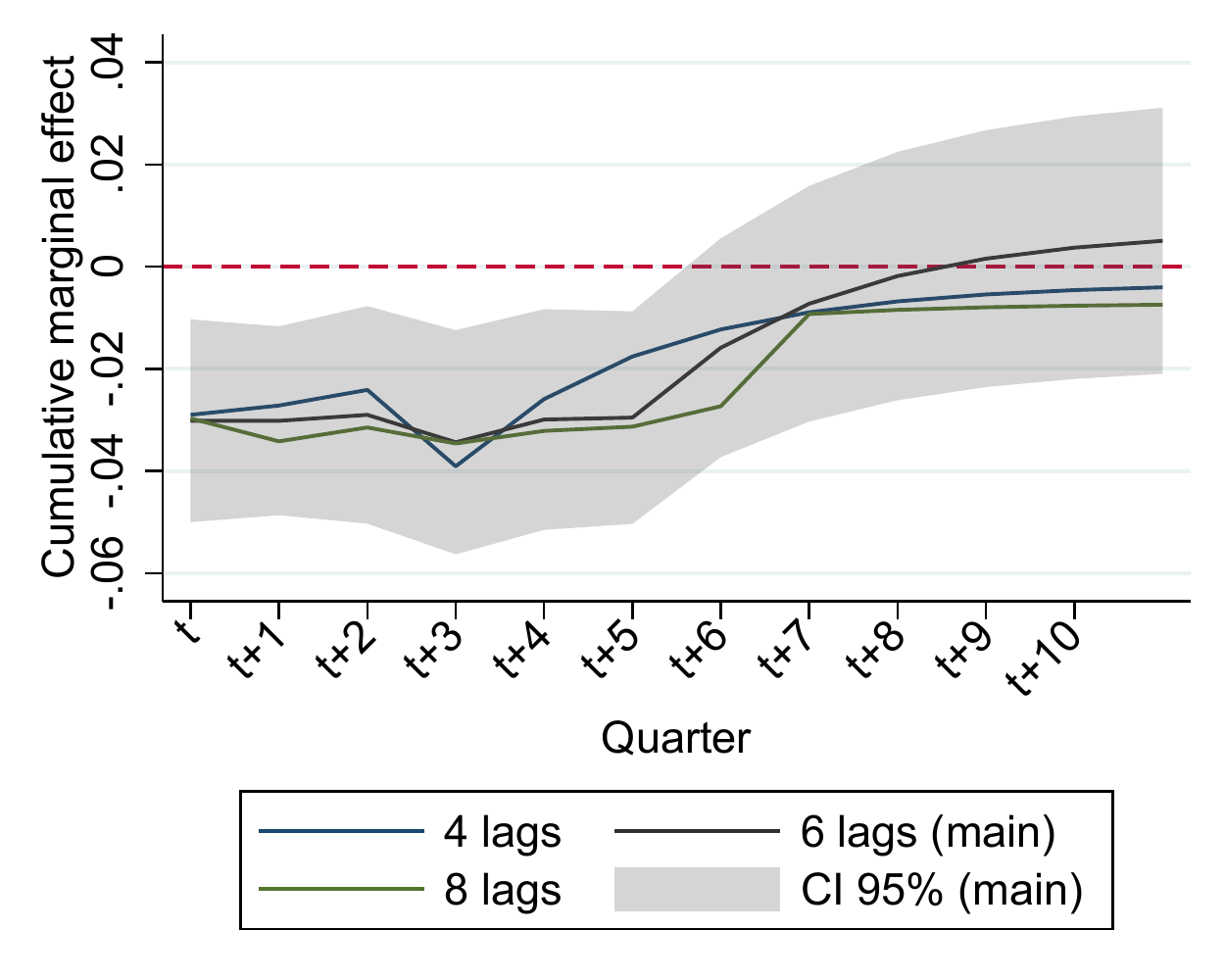}
                \end{subfigure}
                \vspace{10pt}    
                \begin{subfigure}[b]{0.49\textwidth}
                                \centering \caption*{Wage subsidies}  \subcaption*{Unsubsidised employment rate} 
                                \includegraphics[clip=true, trim={0cm 0cm 0cm 0cm},scale=0.50]{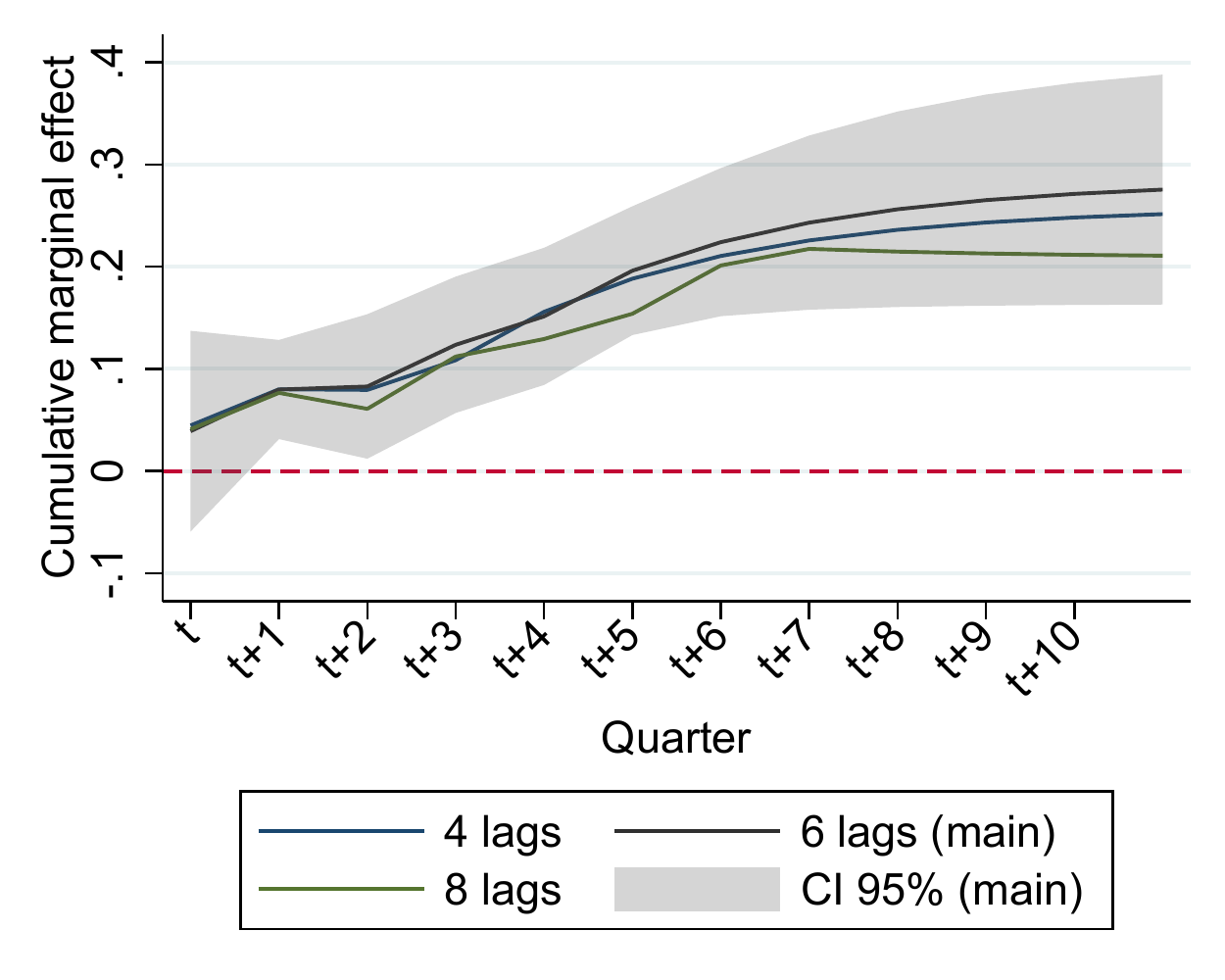}
                \end{subfigure}
                \vspace{10pt}
                \begin{minipage}{\textwidth}
                                \footnotesize \textit{Notes:} These graphs show the cumulative marginal effects of the three types of ALMP, i.e. training, short measures and wage subsidies on the unemployment and unsubsidised employment rates. We show the effects for three models based on different lag orders. 95\% confidence intervals for the model including policy variables with 6 lags are shown as grey areas. The effects are based on the ARDL model estimated by 2SLS. Main sample restrictions apply. Standard errors obtained by a cross-sectional bootstrap (499 replications).
                \end{minipage}
\end{figure}

\begin{figure}[H]
                \centering
                \caption{Effect Patterns for Different Lag order II \label{fig:lt_effects_rob_lags2}}
                \vspace{10pt}
                \begin{subfigure}[b]{0.49\textwidth}
                                \centering \caption*{Training} \subcaption*{Rate of welfare recipients} 
                                \includegraphics[clip=true, trim={0cm 0cm 0cm 0cm},scale=0.50]{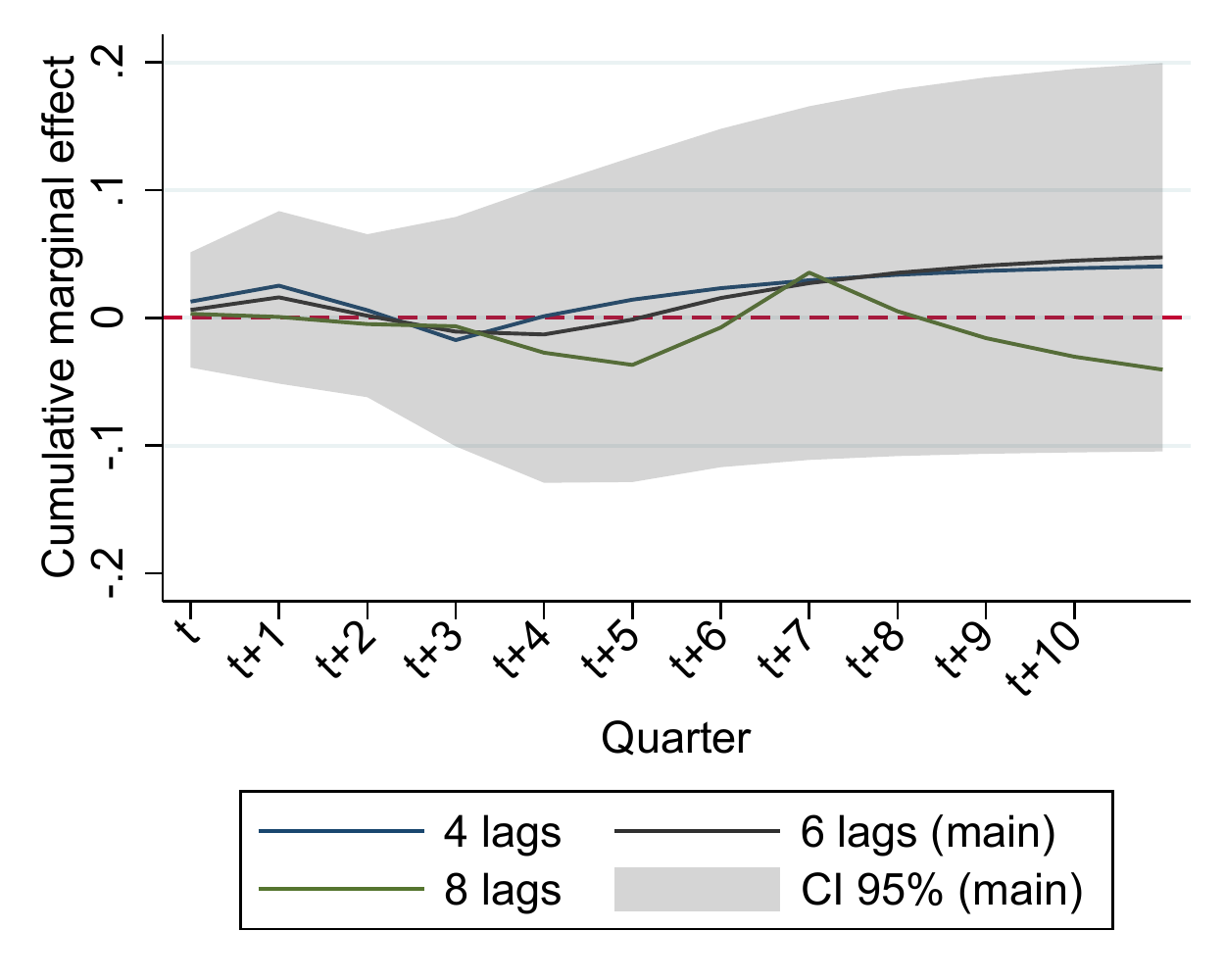}
                \end{subfigure}
                \begin{subfigure}[b]{0.49\textwidth}
                                \centering \caption*{Training} \subcaption*{Rate of employed workers on benefits} 
                                \includegraphics[clip=true, trim={0cm 0cm 0cm 0cm},scale=0.50]{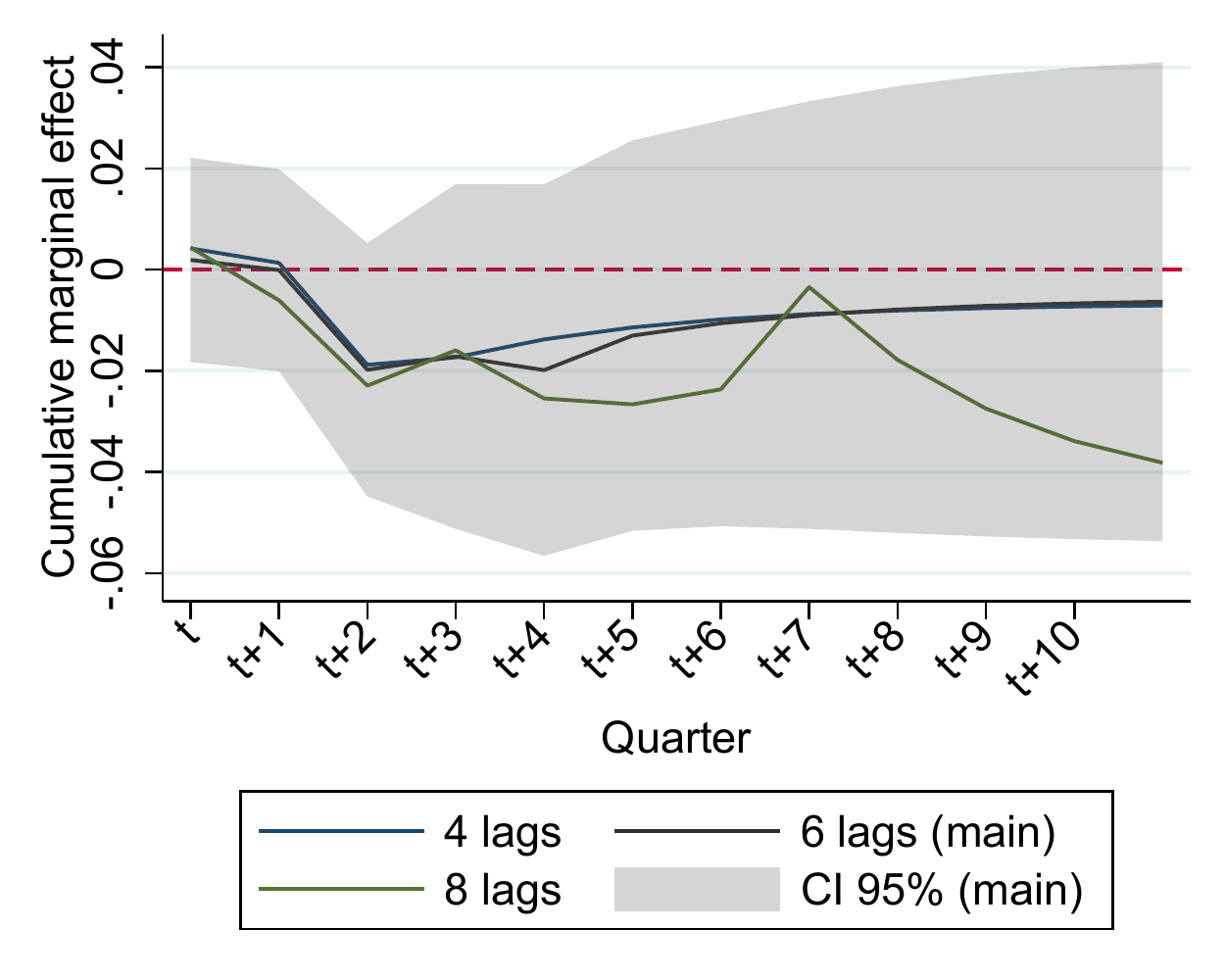}
                \end{subfigure}
                \vspace{10pt}    
                \newline
                \begin{subfigure}[b]{0.49\textwidth}
                                \centering \caption*{Short measures} \subcaption*{Rate of welfare recipients} 
                                \includegraphics[clip=true, trim={0cm 0cm 0cm 0cm},scale=0.50]{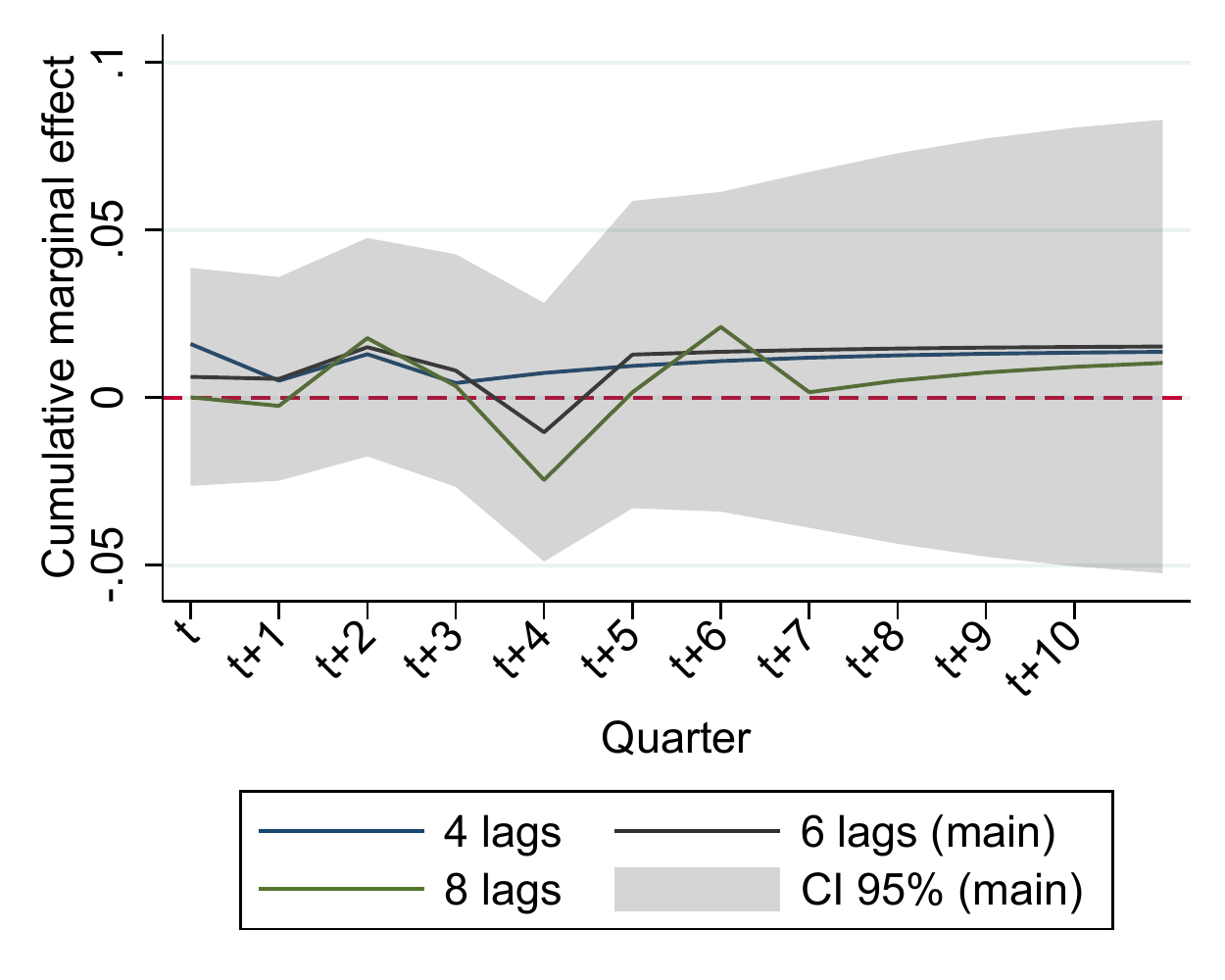}
                \end{subfigure}
                \vspace{10pt}    
                \begin{subfigure}[b]{0.49\textwidth}
                                \centering \caption*{Short measures}  \subcaption*{Rate of employed workers on benefits} 
                                \includegraphics[clip=true, trim={0cm 0cm 0cm 0cm},scale=0.50]{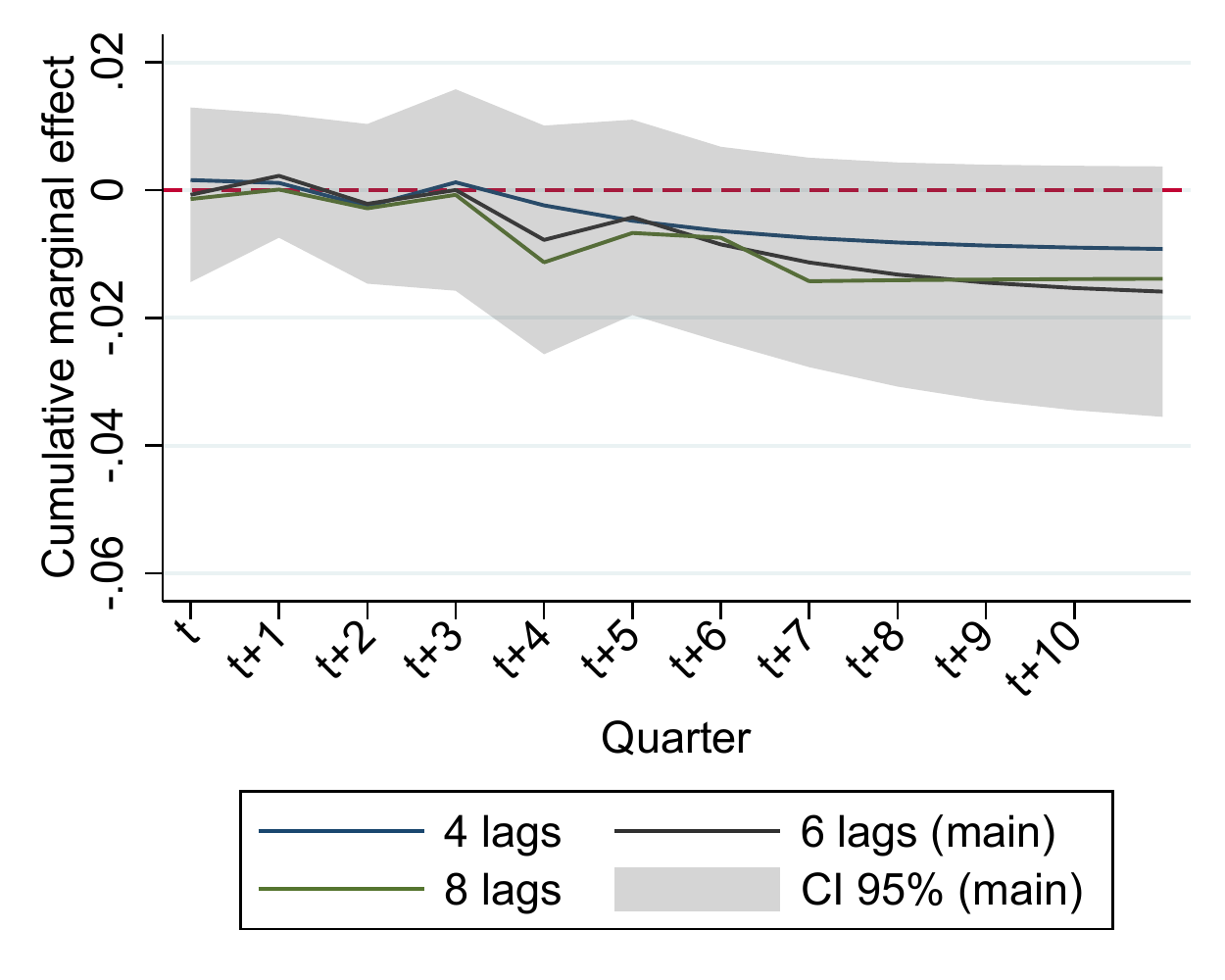}
                \end{subfigure}
                \vspace{10pt}
                \newline
                \begin{subfigure}[b]{0.49\textwidth}
                                \centering \caption*{Wage subsidies} \subcaption*{Rate of welfare recipients} 
                                \includegraphics[clip=true, trim={0cm 0cm 0cm 0cm},scale=0.50]{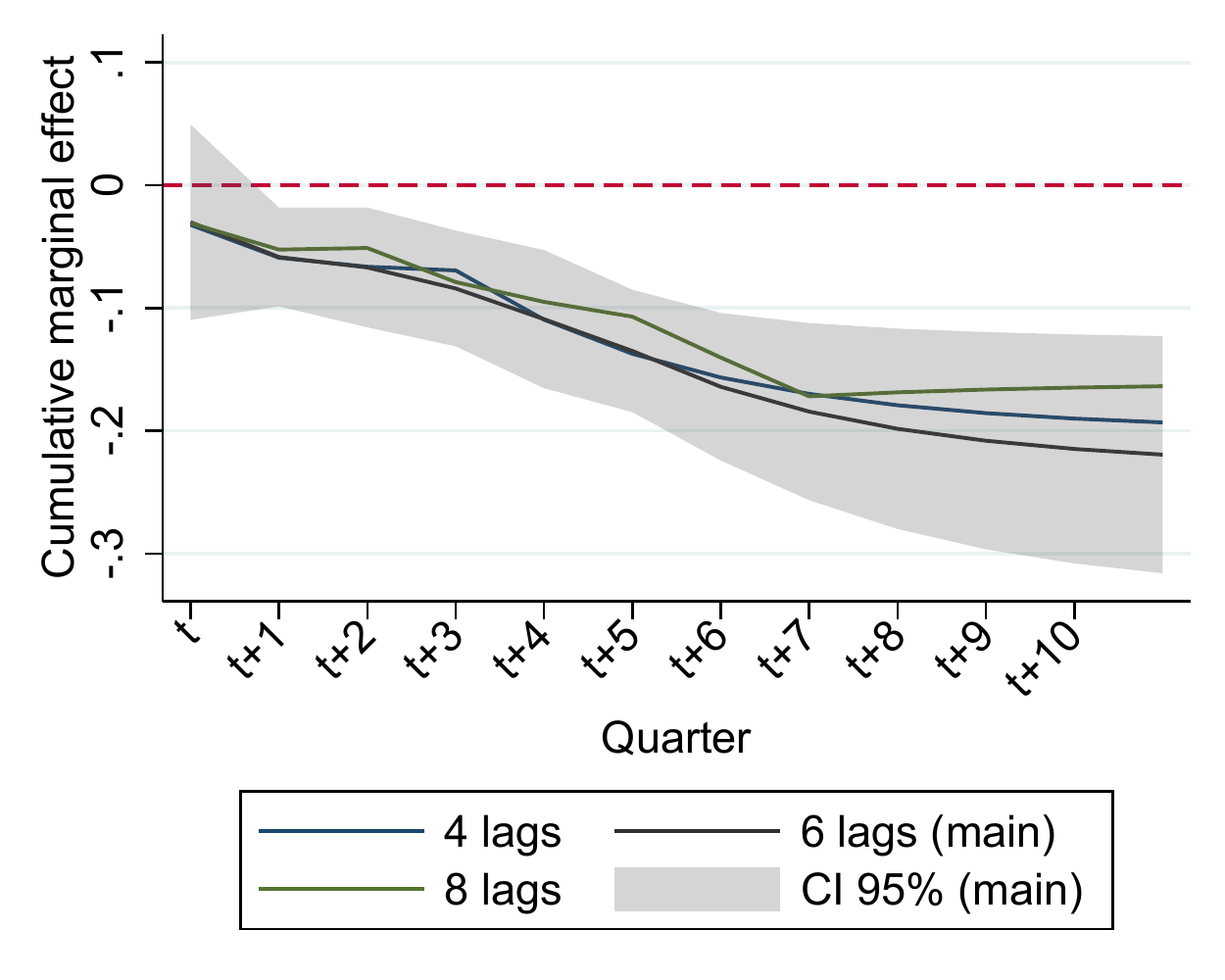}
                \end{subfigure}
                \vspace{10pt}    
                \begin{subfigure}[b]{0.49\textwidth}
                                \centering \caption*{Wage subsidies}  \subcaption*{Rate of employed workers on benefits} 
                                \includegraphics[clip=true, trim={0cm 0cm 0cm 0cm},scale=0.50]{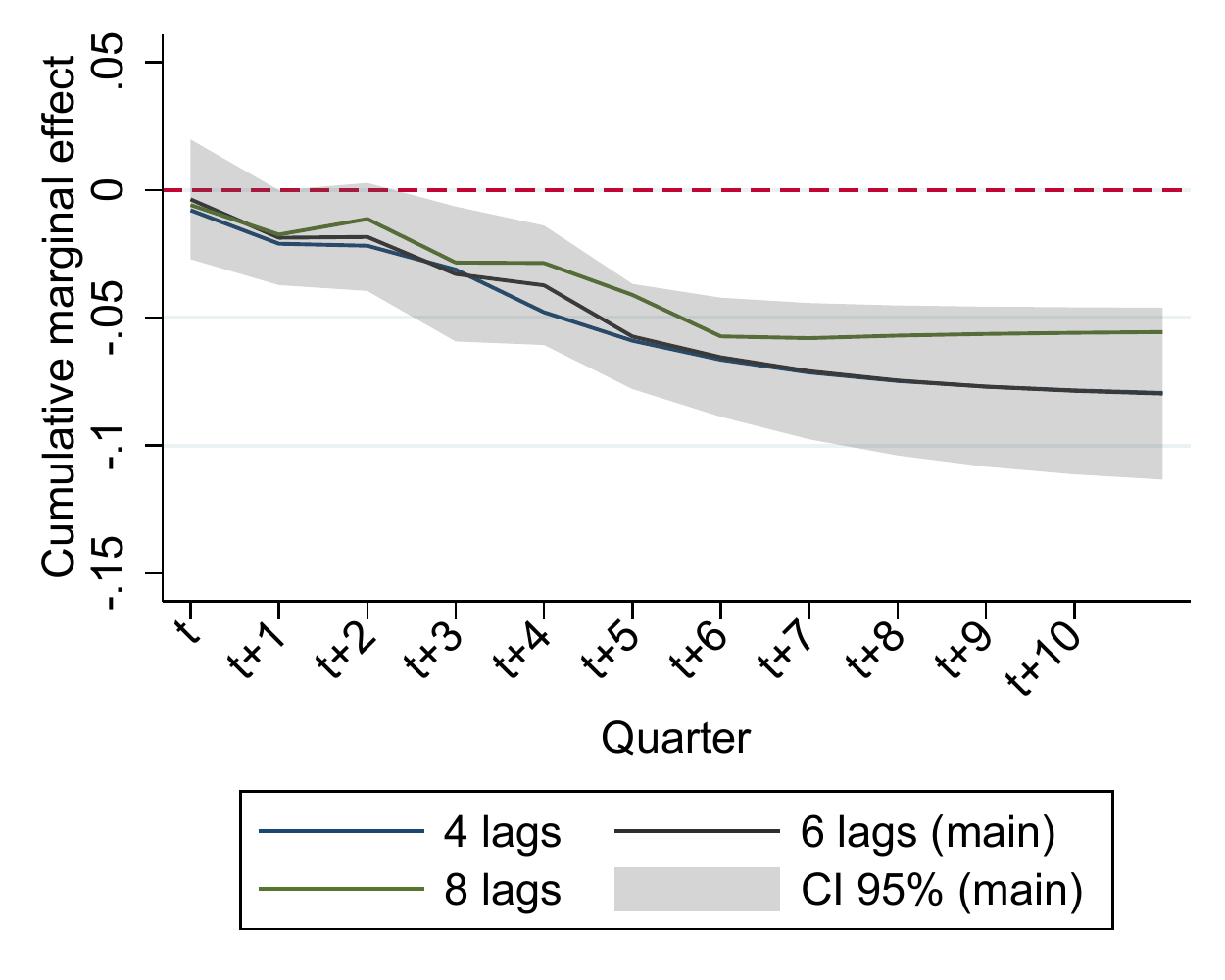}
                \end{subfigure}
                \vspace{10pt}
                \begin{minipage}{\textwidth}
                                \footnotesize \textit{Notes:} These graphs show the cumulative marginal effects of the three types of ALMP, i.e. training, short measures and wage subsidies on the rate of welfare recipients and employed workers on benefits. We show the effects for three models based on different lag orders. 95\% confidence intervals for the model including policy variables with 6 lags are shown as grey areas. The effects are based on the ARDL model estimated by 2SLS. Main sample restrictions apply. Standard errors obtained by a cross-sectional bootstrap (499 replications).
                \end{minipage}
\end{figure}

\begin{figure}[H]
                \centering
                \caption{Effect Patterns for Different LLM Definitions I \label{fig:lt_effects_rob_llm}}
                \vspace{10pt}
                \begin{subfigure}[b]{0.49\textwidth}
                                \centering \caption*{Training} \subcaption*{Unemployment rate} 
                                \includegraphics[clip=true, trim={0cm 0cm 0cm 0cm},scale=0.50]{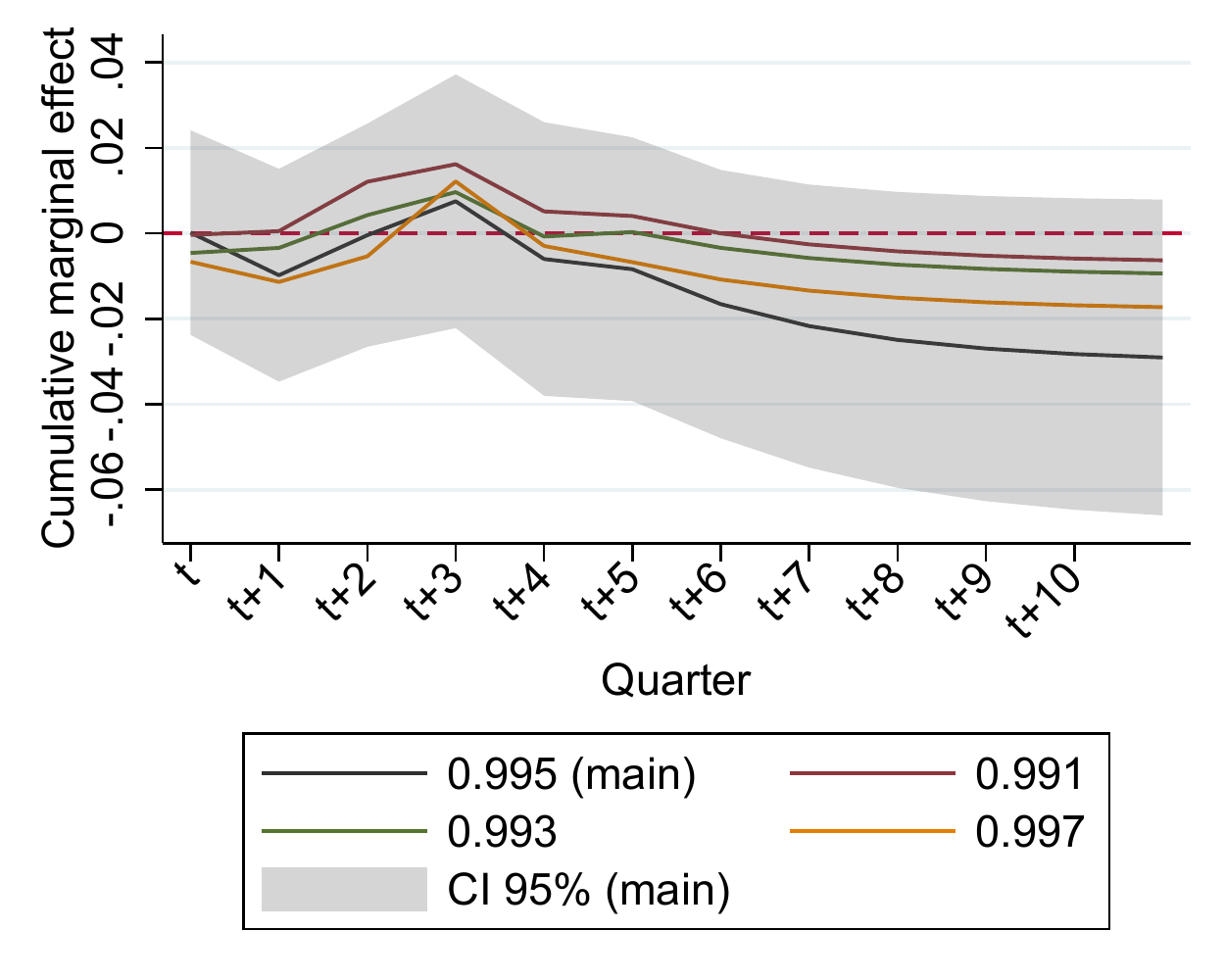}
                \end{subfigure}
                \begin{subfigure}[b]{0.49\textwidth}
                                \centering \caption*{Training} \subcaption*{Employment rate} 
                                \includegraphics[clip=true, trim={0cm 0cm 0cm 0cm},scale=0.50]{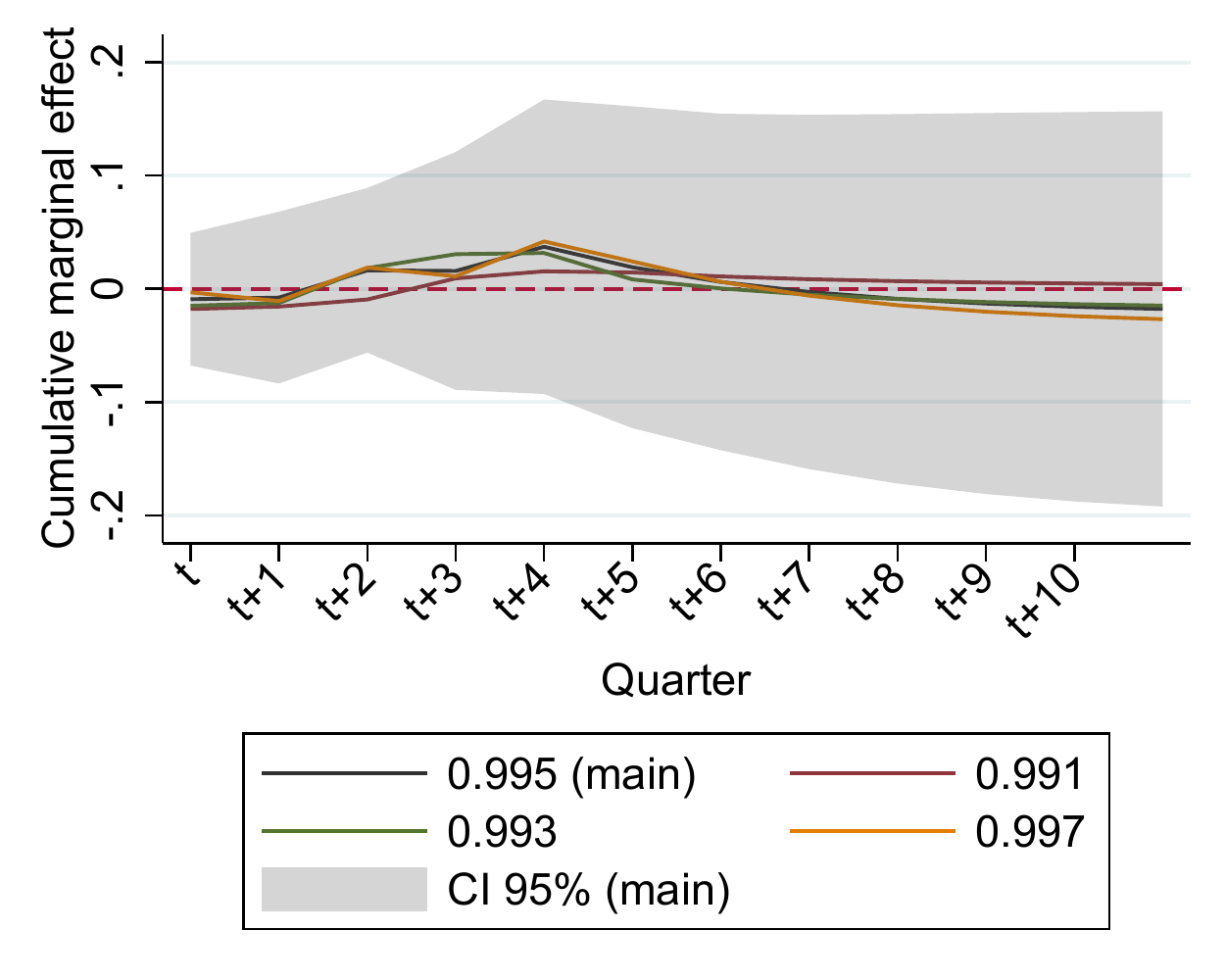}
                \end{subfigure}
                \vspace{10pt}    
                \newline
                \begin{subfigure}[b]{0.49\textwidth}
                                \centering \caption*{Short measures} \subcaption*{Unemployment rate} 
                                \includegraphics[clip=true, trim={0cm 0cm 0cm 0cm},scale=0.50]{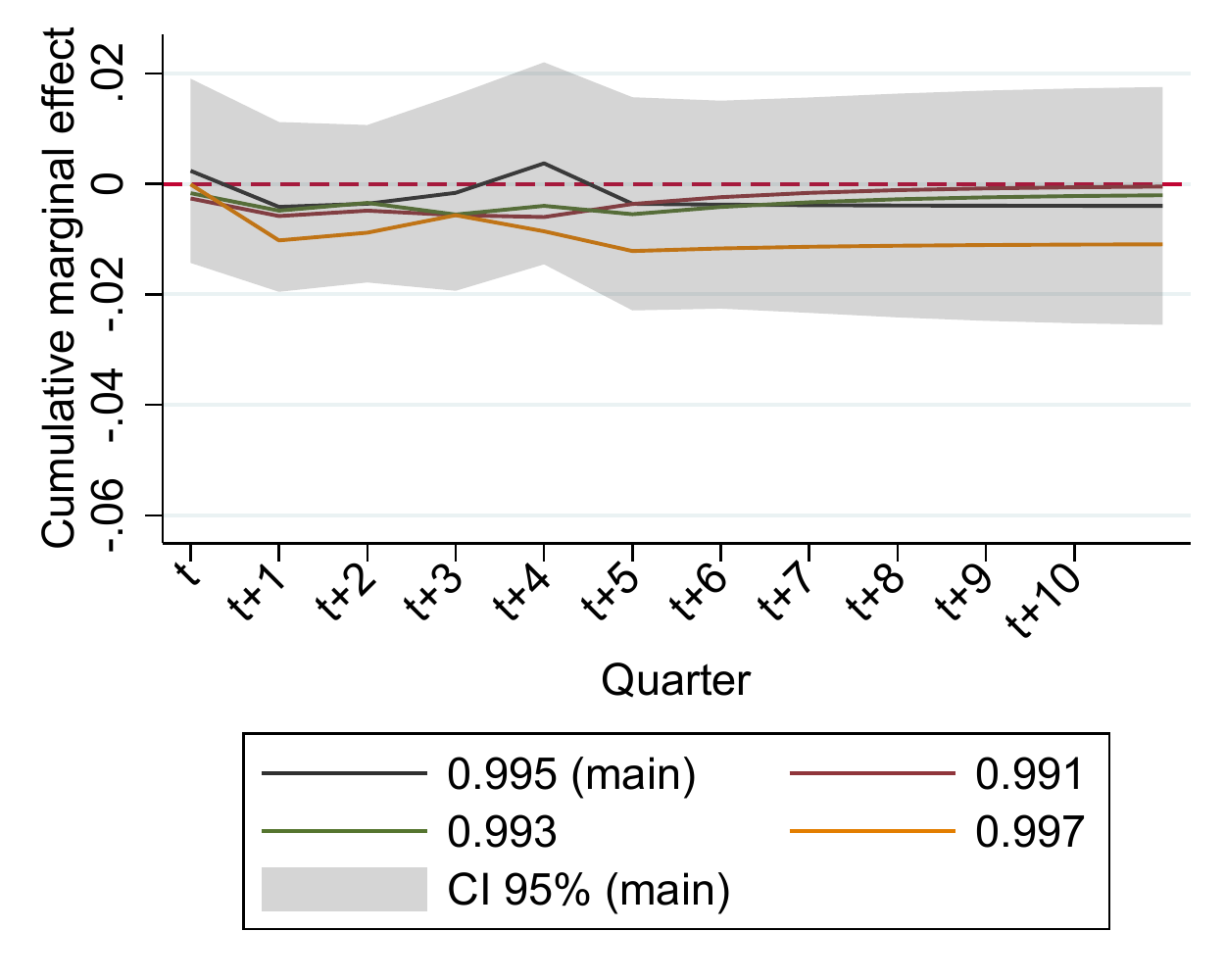}
                \end{subfigure}
                \vspace{10pt}    
                \begin{subfigure}[b]{0.49\textwidth}
                                \centering \caption*{Short measures}  \subcaption*{Employment rate} 
                                \includegraphics[clip=true, trim={0cm 0cm 0cm 0cm},scale=0.50]{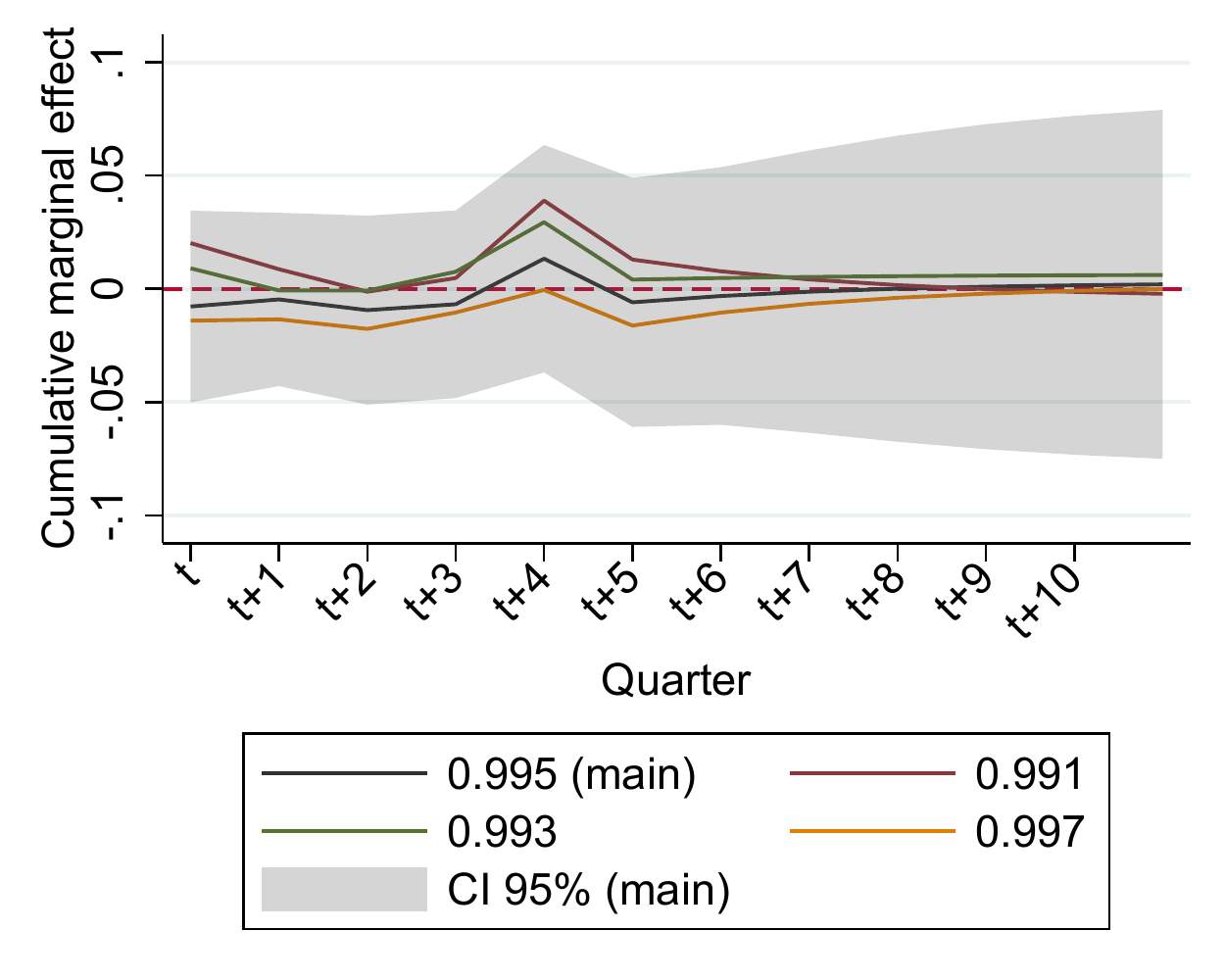}
                \end{subfigure}
                \vspace{10pt}
                \newline
                \begin{subfigure}[b]{0.49\textwidth}
                                \centering \caption*{Wage subsidies} \subcaption*{Unemployment rate} 
                                \includegraphics[clip=true, trim={0cm 0cm 0cm 0cm},scale=0.50]{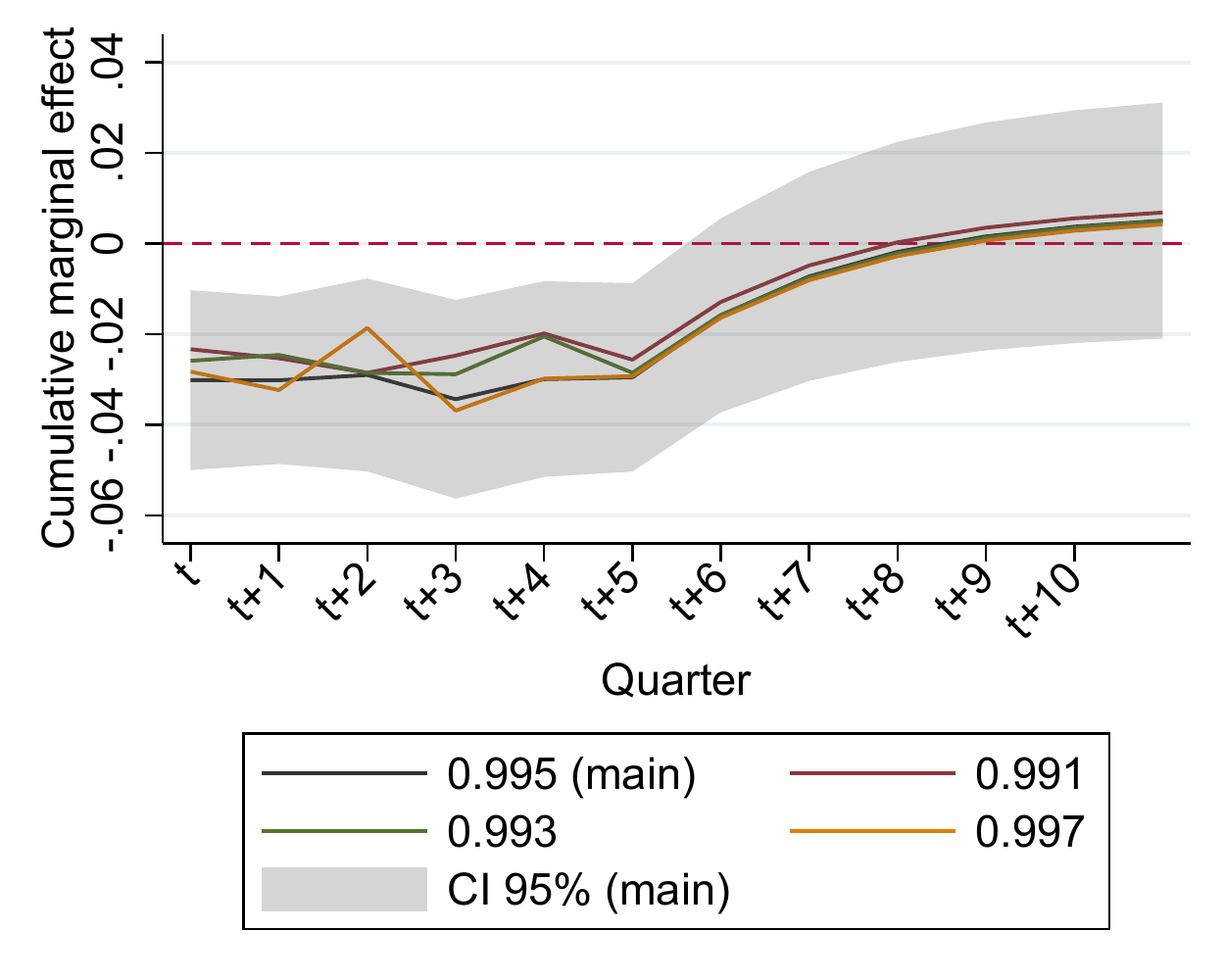}
                \end{subfigure}
                \vspace{10pt}    
                \begin{subfigure}[b]{0.49\textwidth}
                                \centering \caption*{Wage subsidies}  \subcaption*{Employment rate} 
                                \includegraphics[clip=true, trim={0cm 0cm 0cm 0cm},scale=0.50]{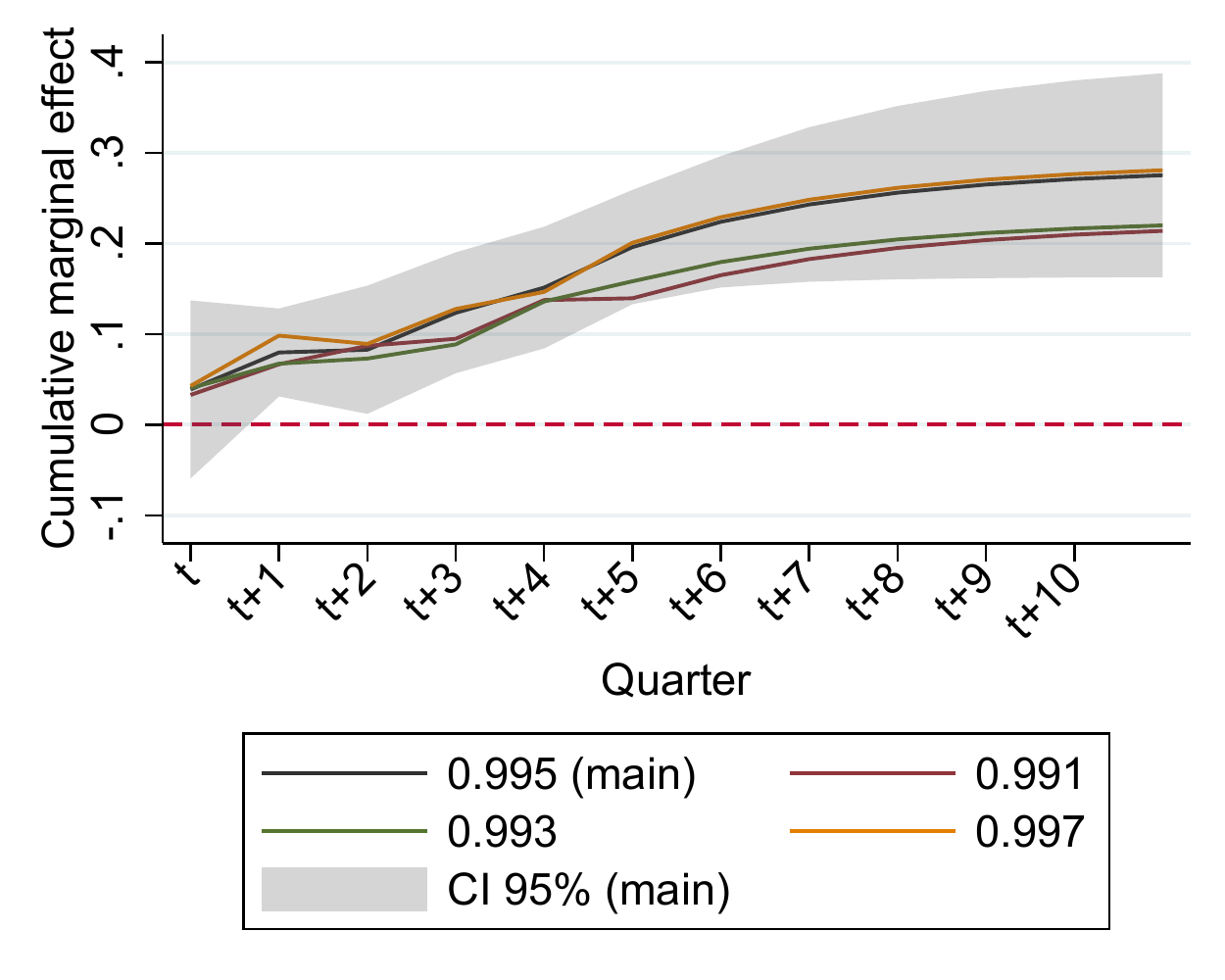}
                \end{subfigure}
                \vspace{10pt}
                \begin{minipage}{\textwidth}
                                \footnotesize \textit{Notes:} These graphs show the cumulative marginal effects of the three types of ALMP, i.e. training, short measures and wage subsidies on the unemployment rate and unsubsidized employment rate. We show the effects for three models based on different labour market definitions. 95\% confidence intervals for the main definition are shown as grey areas. The effects are based on the ARDL model estimated by 2SLS. Main sample restrictions apply. Standard errors obtained by a cross-sectional bootstrap (499 replications).
                \end{minipage}
\end{figure}

\begin{figure}[H]
                \centering
                \caption{Effect Patterns for Different LLM Definitions II \label{fig:lt_effects_rob_llm2}}
                \vspace{10pt}
                \begin{subfigure}[b]{0.49\textwidth}
                                \centering \caption*{Training} \subcaption*{Rate of welfare recipients} 
                                \includegraphics[clip=true, trim={0cm 0cm 0cm 0cm},scale=0.50]{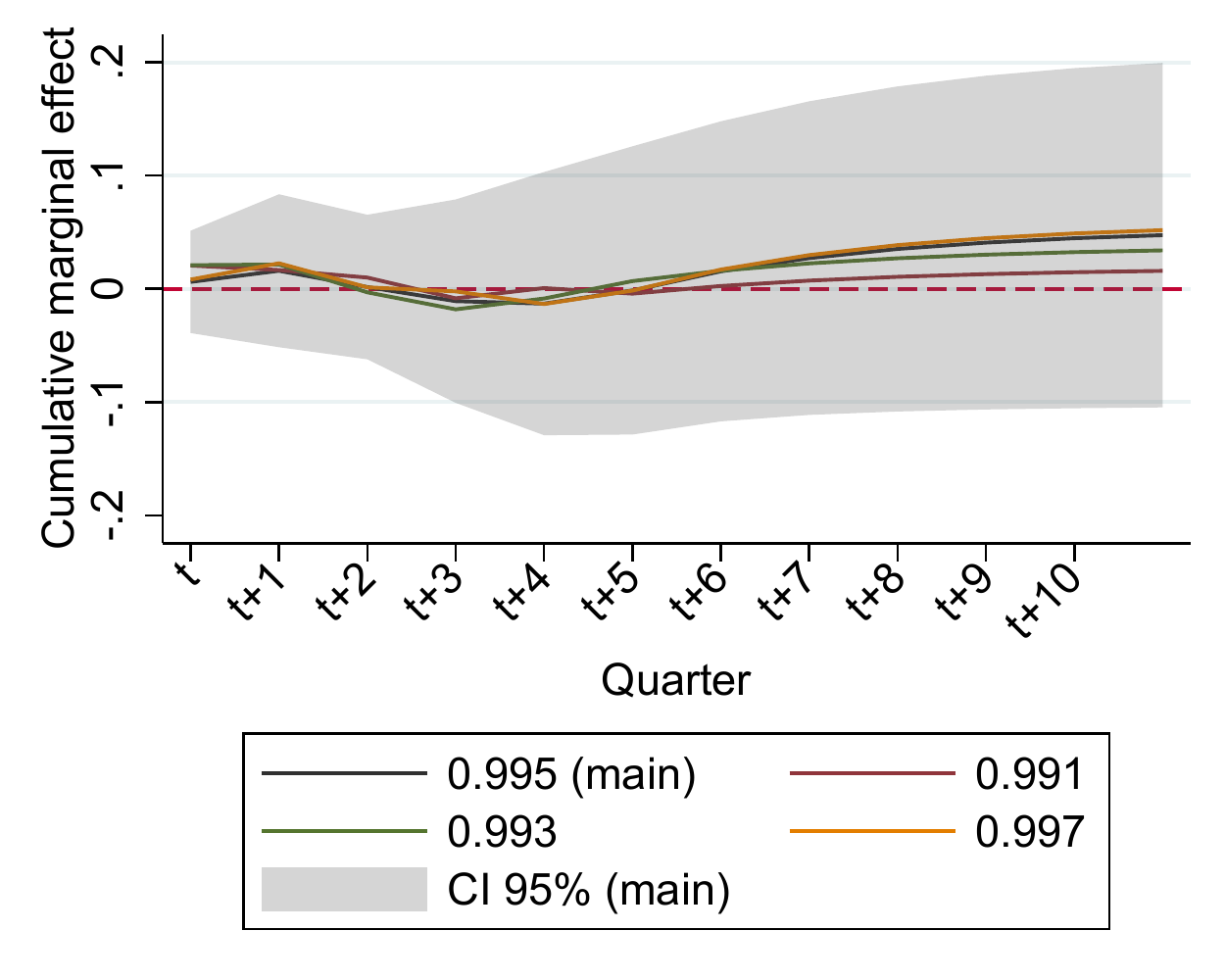}
                \end{subfigure}
                \begin{subfigure}[b]{0.49\textwidth}
                                \centering \caption*{Training} \subcaption*{Rate of employed workers on benefits} 
                                \includegraphics[clip=true, trim={0cm 0cm 0cm 0cm},scale=0.50]{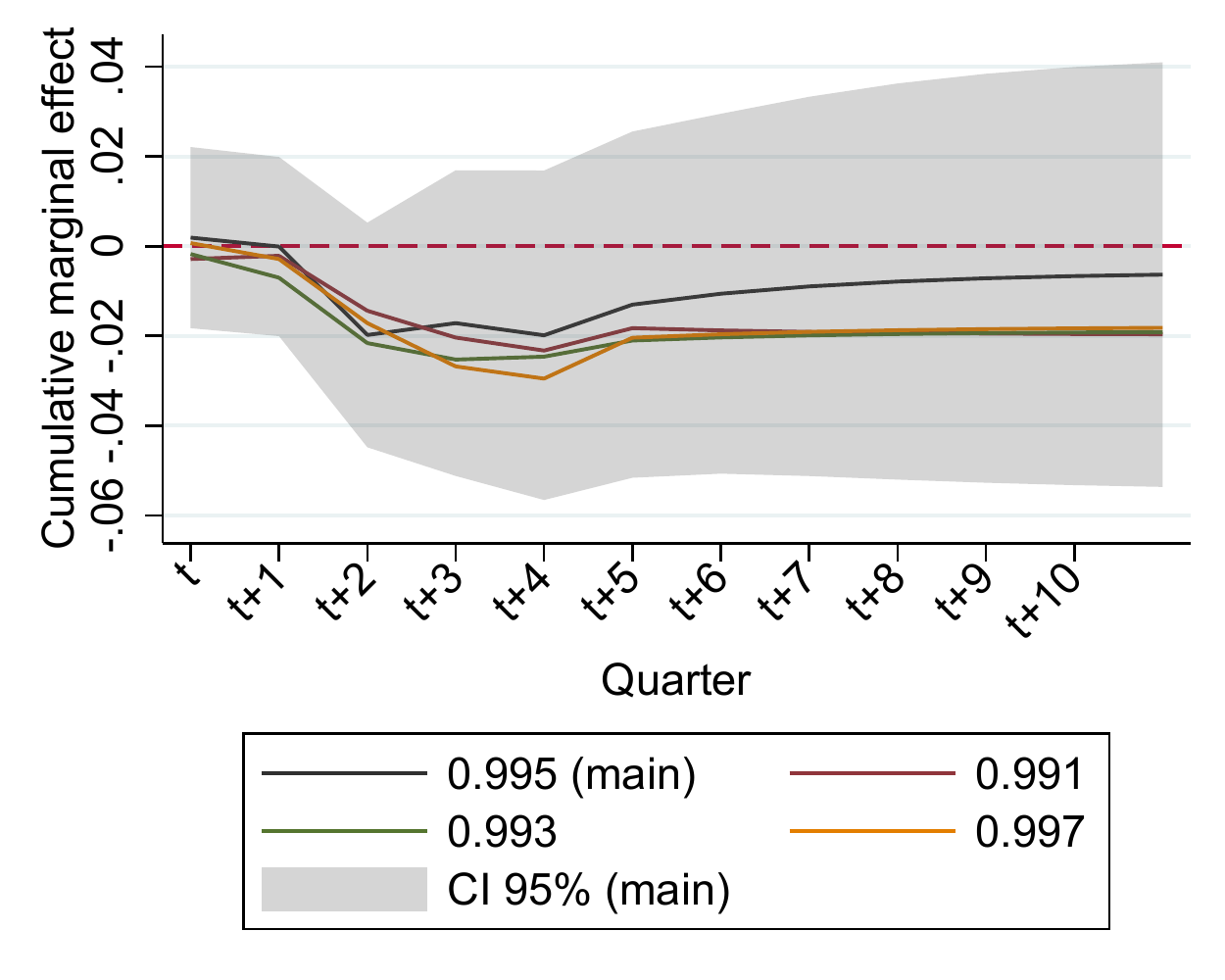}
                \end{subfigure}
                \vspace{10pt}    
                \newline
                \begin{subfigure}[b]{0.49\textwidth}
                                \centering \caption*{Short measures} \subcaption*{Rate of welfare recipients} 
                                \includegraphics[clip=true, trim={0cm 0cm 0cm 0cm},scale=0.50]{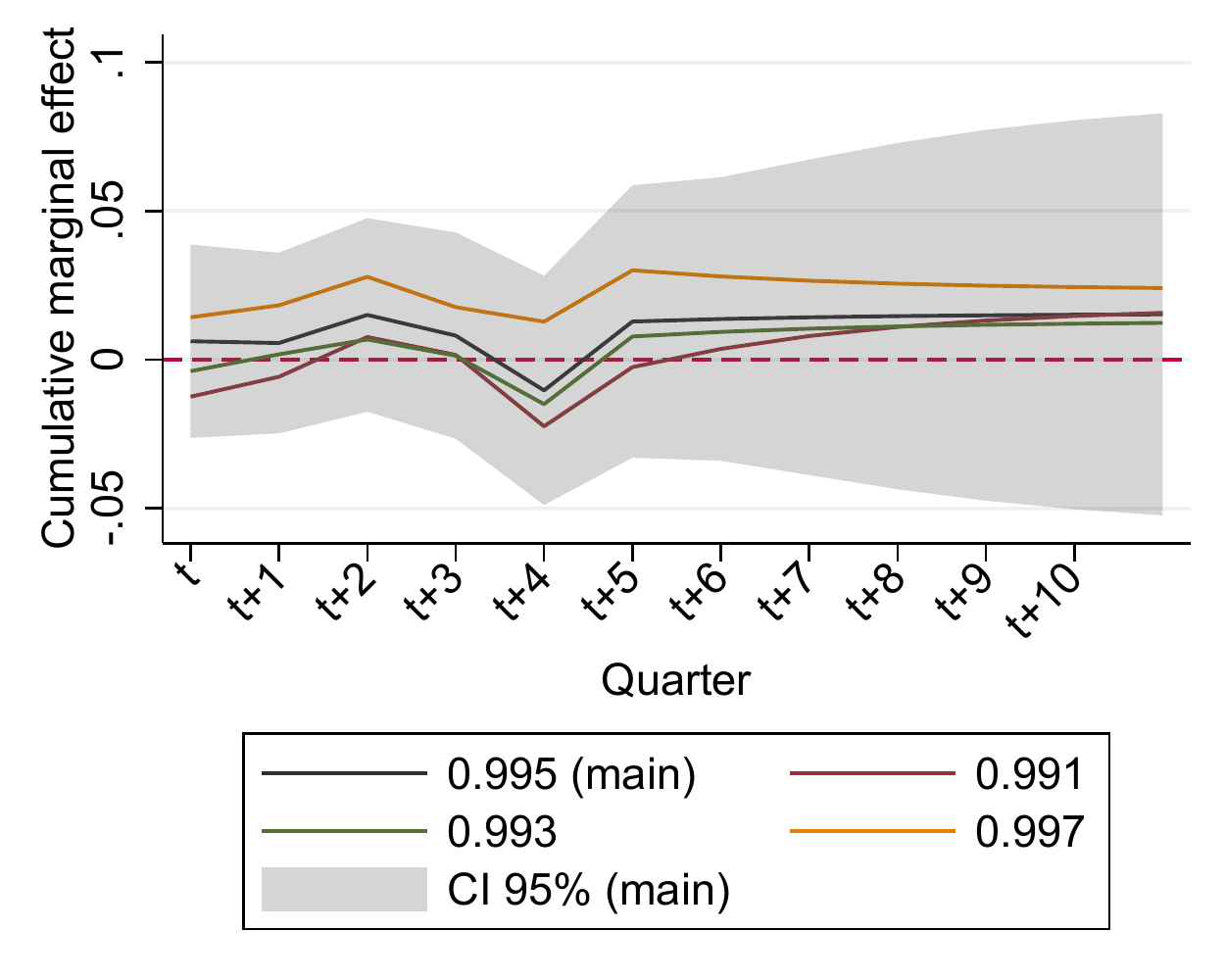}
                \end{subfigure}
                \vspace{10pt}    
                \begin{subfigure}[b]{0.49\textwidth}
                                \centering \caption*{Short measures}  \subcaption*{Rate of employed workers on benefits} 
                                \includegraphics[clip=true, trim={0cm 0cm 0cm 0cm},scale=0.50]{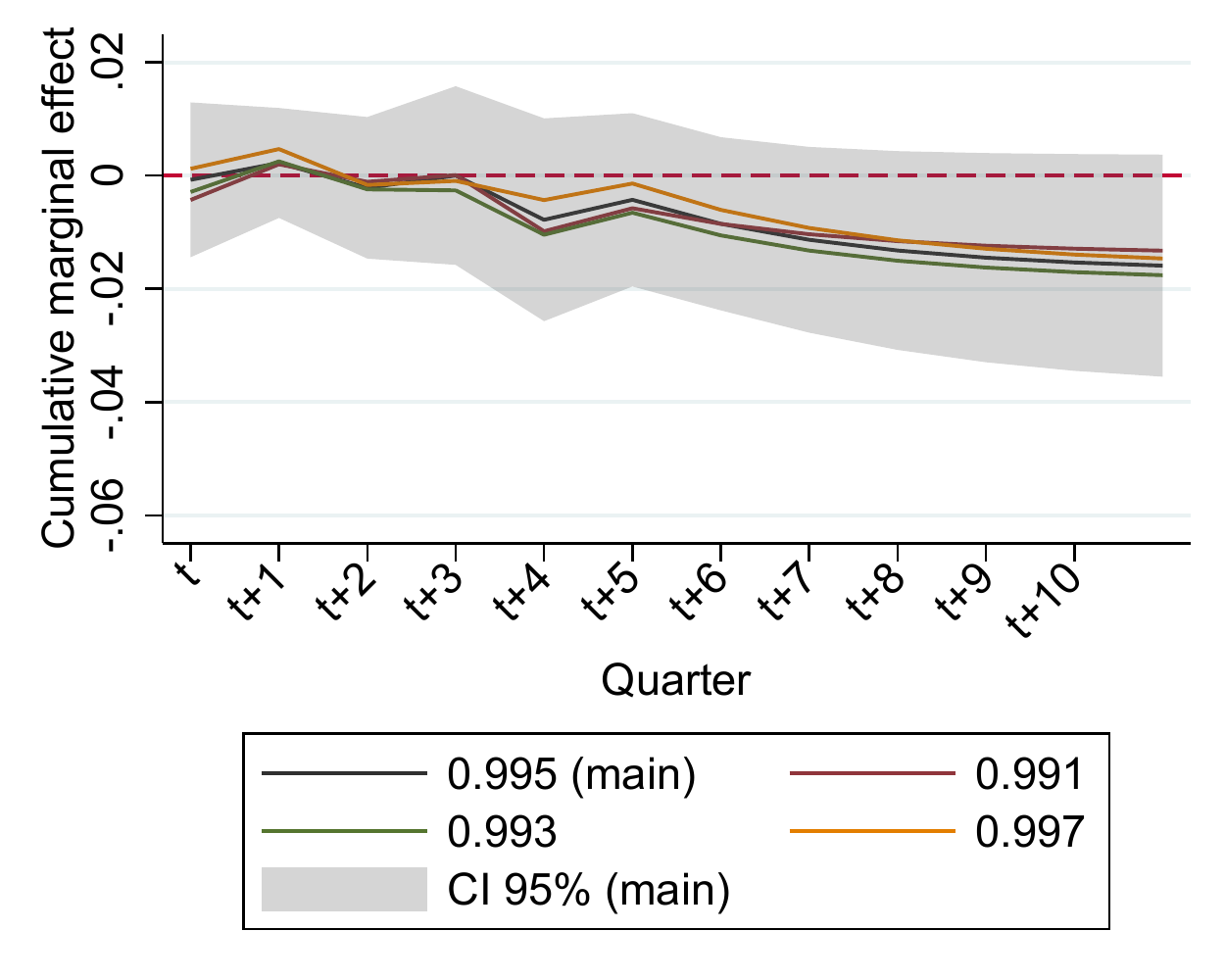}
                \end{subfigure}
                \vspace{10pt}
                \newline
                \begin{subfigure}[b]{0.49\textwidth}
                                \centering \caption*{Wage subsidies} \subcaption*{Rate of welfare recipients} 
                                \includegraphics[clip=true, trim={0cm 0cm 0cm 0cm},scale=0.50]{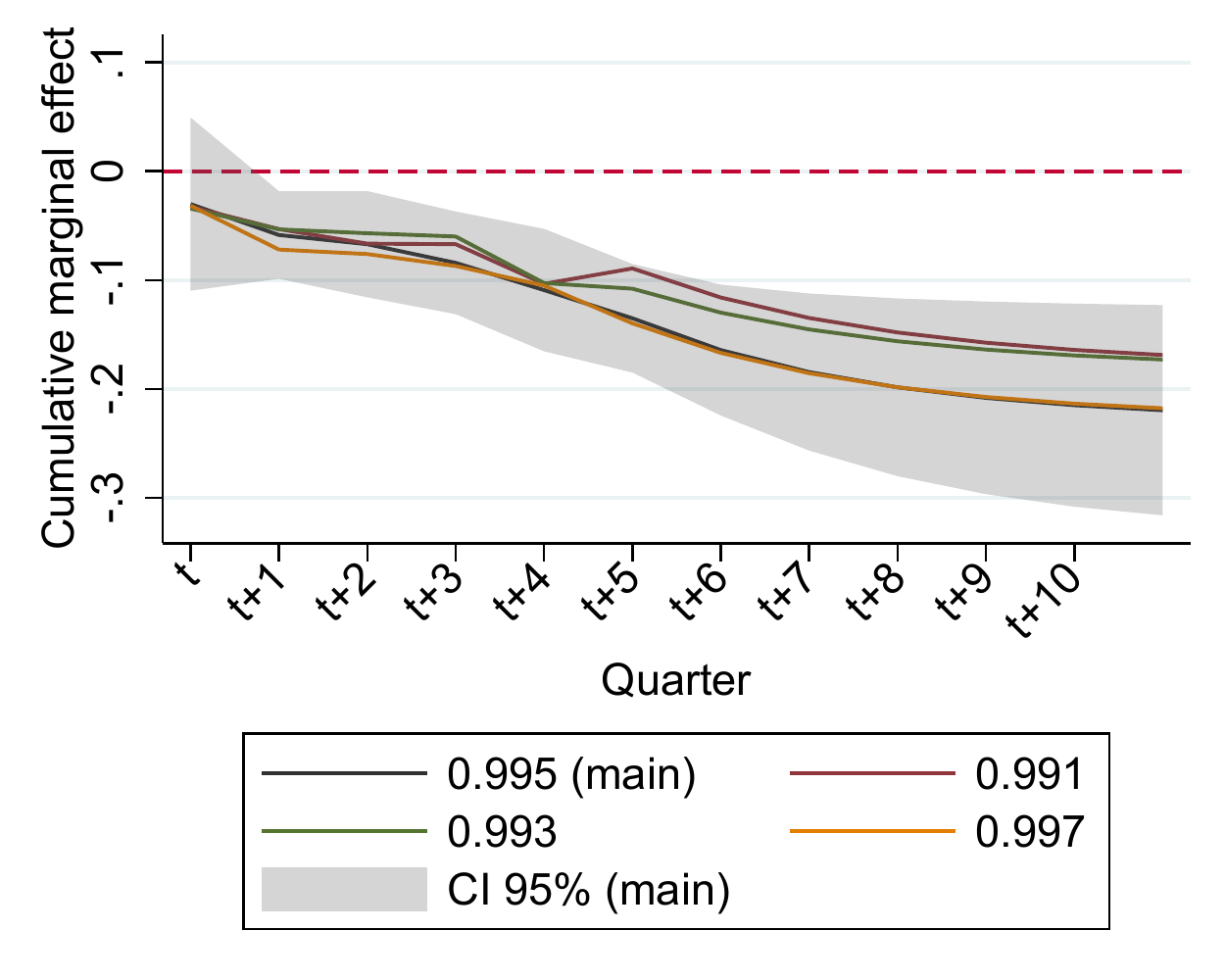}
                \end{subfigure}
                \vspace{10pt}    
                \begin{subfigure}[b]{0.49\textwidth}
                                \centering \caption*{Wage subsidies}  \subcaption*{Rate of employed workers on benefits} 
                                \includegraphics[clip=true, trim={0cm 0cm 0cm 0cm},scale=0.50]{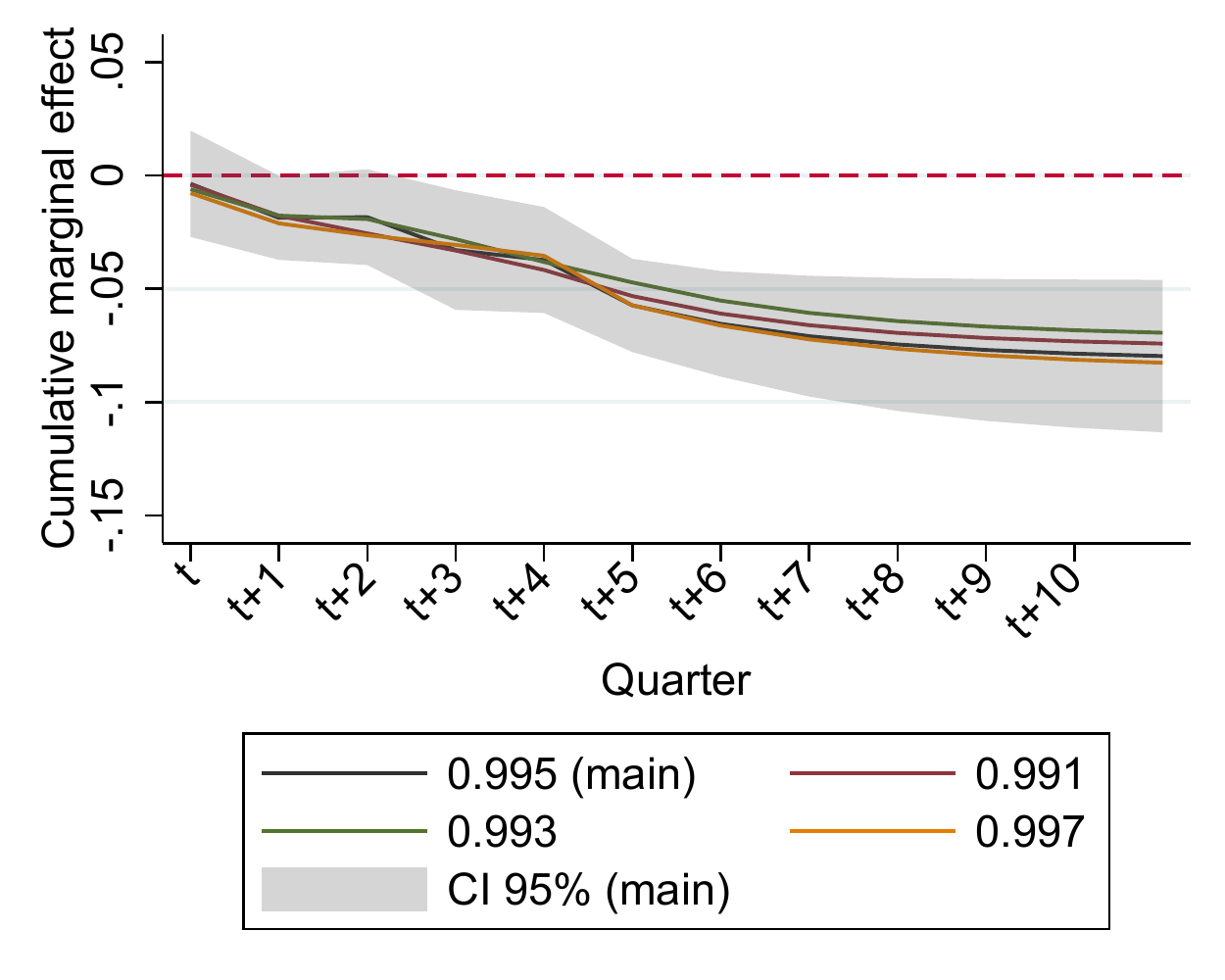}
                \end{subfigure}
                \vspace{10pt}
                \begin{minipage}{\textwidth}
                                \footnotesize \textit{Notes:} These graphs show the cumulative marginal effects of the three types of ALMP, i.e. training, short measures and wage subsidies on the rate of welfare recipients and employed workers on benefits. We show the effects for three models based on different labour market definitions. 95\% confidence intervals for the main definition are shown as grey areas. The effects are based on the ARDL model estimated by 2SLS. Main sample restrictions apply. Standard errors obtained by a cross-sectional bootstrap (499 replications).
                \end{minipage}
\end{figure}


\begin{figure}[H]
                 \centering
                 \caption{Overlap Criteria \label{fig:overlap_crit}}
                 \vspace{10pt}
                  \begin{subfigure}[b]{0.8\textwidth}
                                 \centering \caption*{(a)}
                                 \includegraphics[clip=true, trim={0cm 0cm 0cm 0cm},scale=0.87]{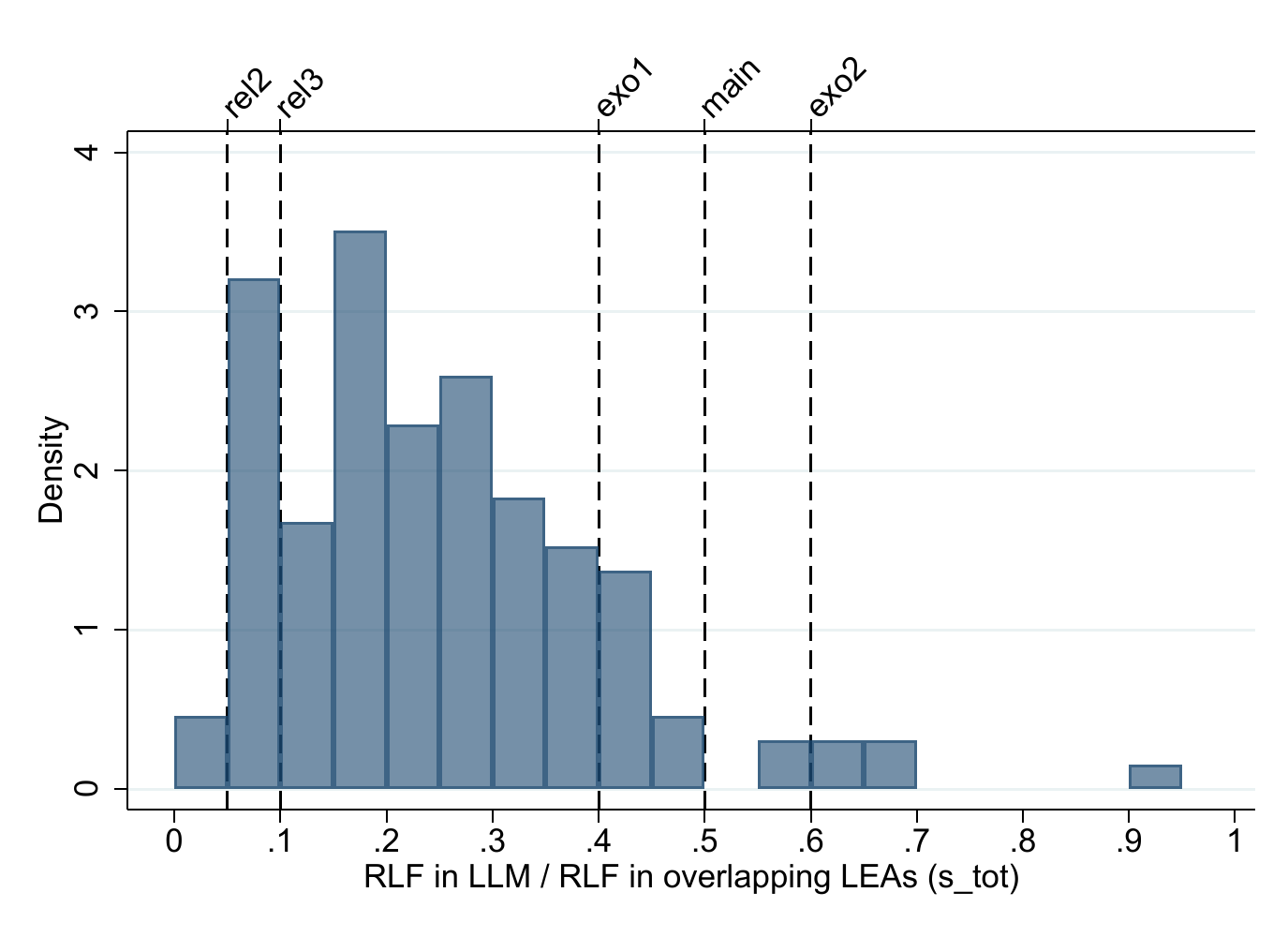}
                 \end{subfigure}
                 \vspace{10pt}  
                 \begin{subfigure}[b]{0.8\textwidth}
                                 \centering \caption*{(b)}
                                 \includegraphics[clip=true, trim={0cm 0cm 0cm 0cm},scale=0.87]{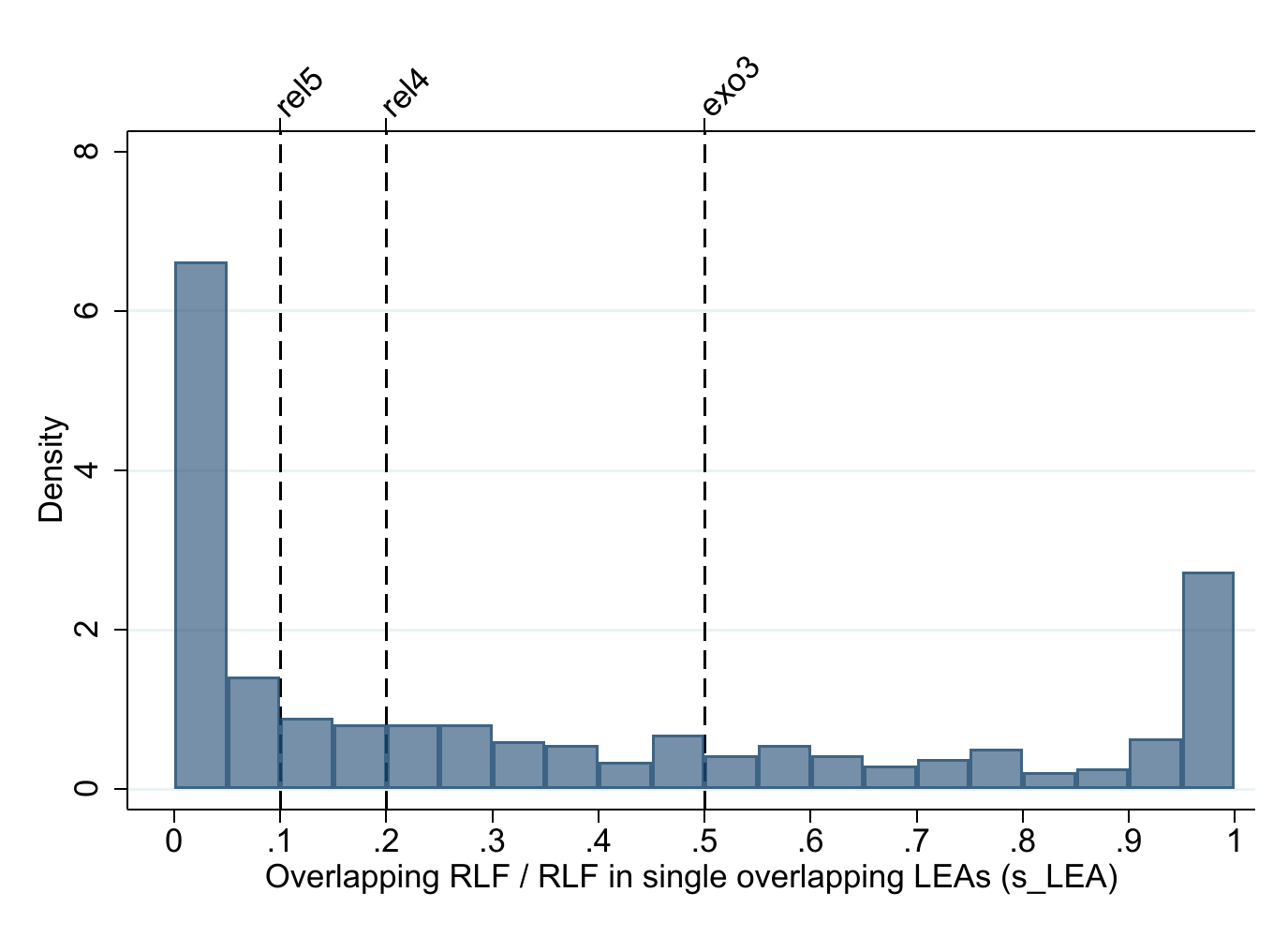}
                 \end{subfigure}
                 \begin{minipage}{\textwidth}
                  \footnotesize \textit{Notes:} Panel (a) plots the distribution of the RLF in the labour market as share of the total RLF in all partially overlapping agencies across all labour markets that overlap with at least two agencies (131). Panel (b) plots the distribution of the overlapping RLF between each of these markets and agencies as share of the RLF in the agency for all market-agency combinations (468). Notice that the overlap measure in panel (b) is always strictly above zero. The minimum overlap is 0.0002.
                 \end{minipage}
\end{figure}

\begin{figure}[H]
                \centering
                \caption{Size of Subsamples (Relevance and Exogeneity Criteria)\label{fig:N_rob_IV}}
                \includegraphics[clip=true, trim={0cm 0cm 0cm 0cm},scale=0.70]{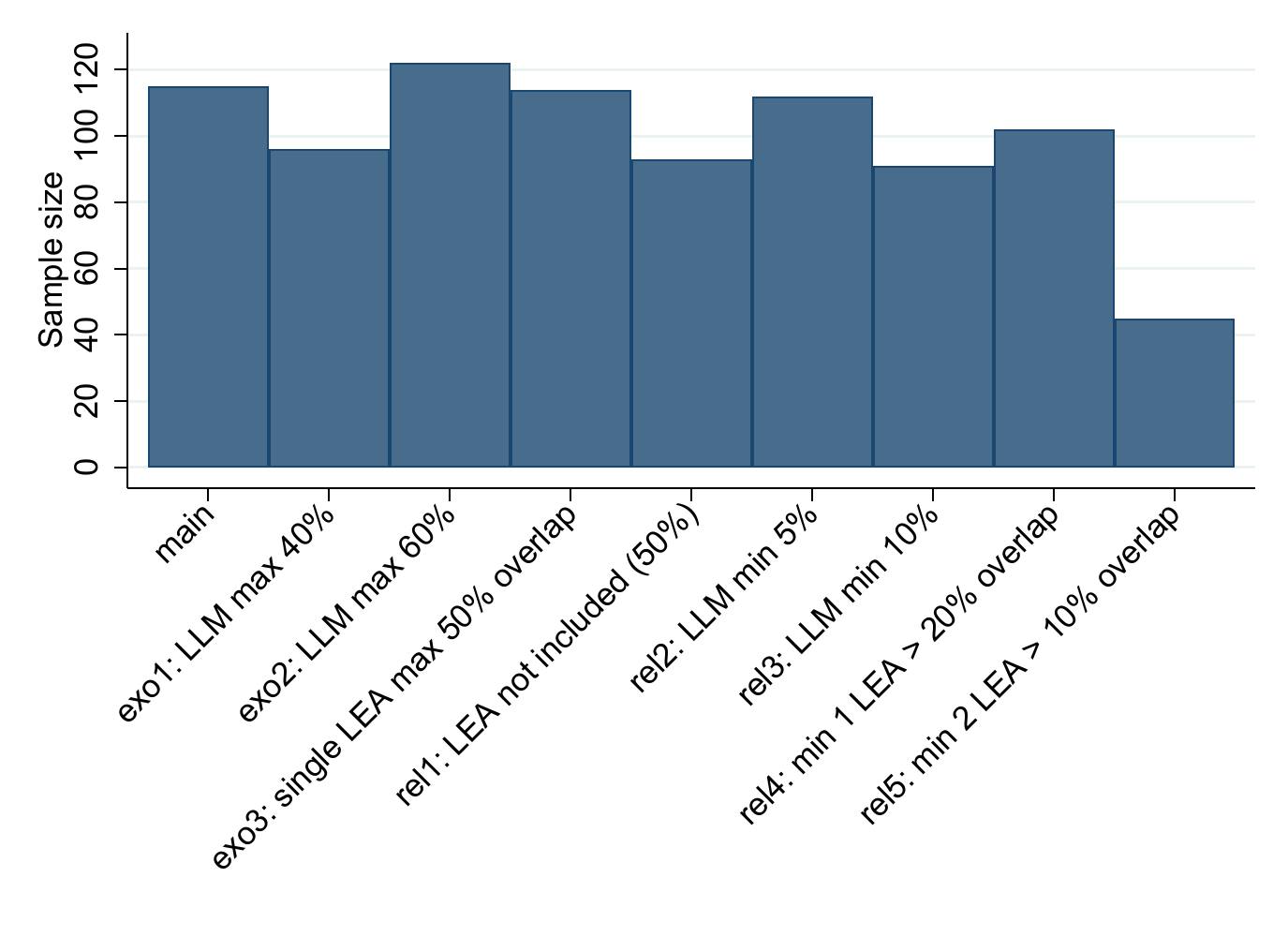}
 \begin{minipage}{\textwidth}
  \footnotesize \textit{Notes:} The figure plots the sample size (number of labour markets) after imposing different overlap criteria targeting relevance and exogeneity.
 \end{minipage}
\end{figure}

\begin{figure}[H]
                \centering
                \caption{Long-term Effects for Different Subsamples (Relevance and Exogeneity Criteria) \label{fig:lt_effects_rob_IV}}
                \vspace{10pt}
                \begin{subfigure}[b]{0.49\textwidth}
                                \centering \caption*{Training} \subcaption*{Unemployment rate} 
                                \includegraphics[clip=true, trim={0cm 0cm 0cm 0cm},scale=0.50]{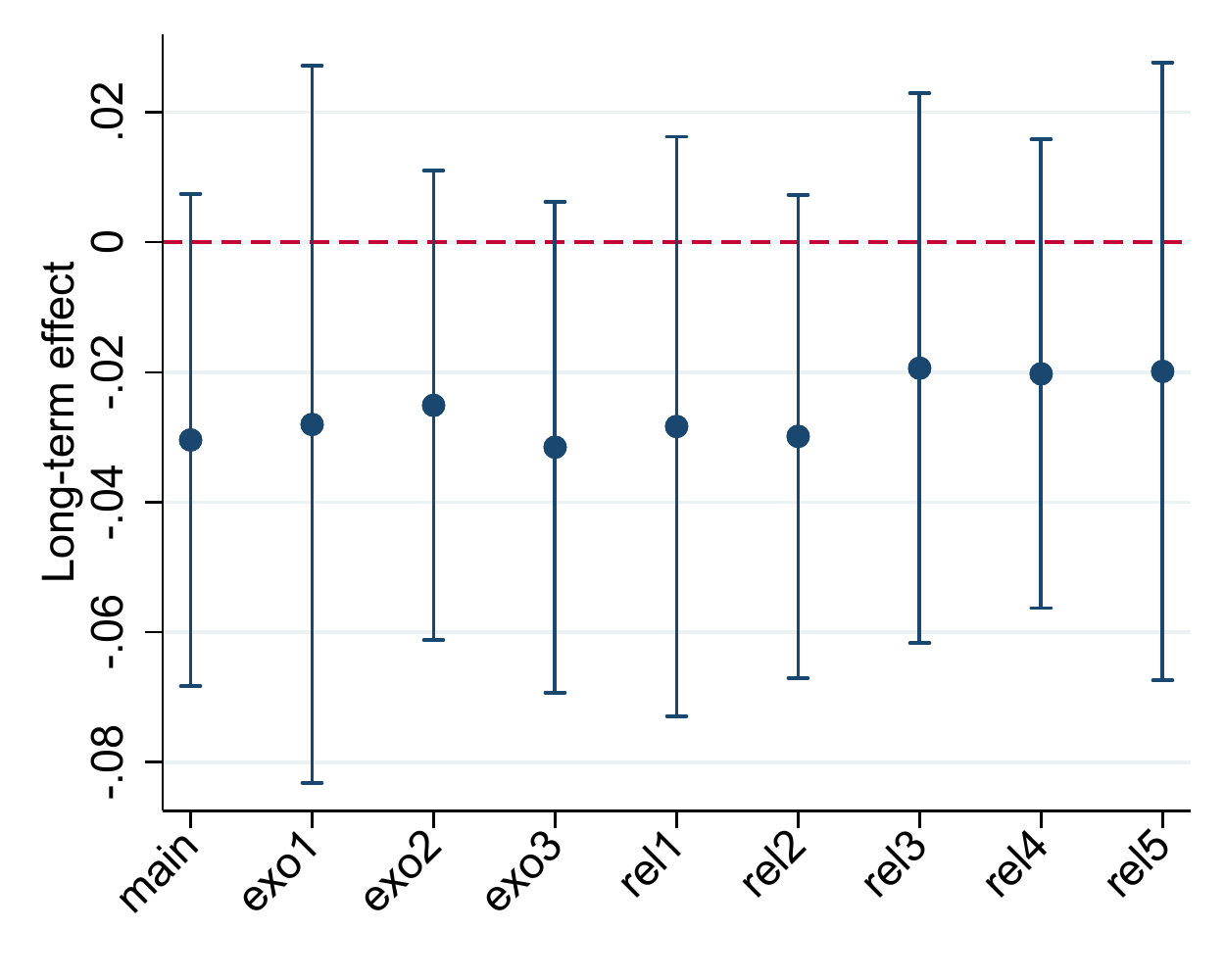}
                \end{subfigure}
                \begin{subfigure}[b]{0.49\textwidth}
                                \centering \caption*{Training} \subcaption*{Unsubsidzied employment rate} 
                                \includegraphics[clip=true, trim={0cm 0cm 0cm 0cm},scale=0.50]{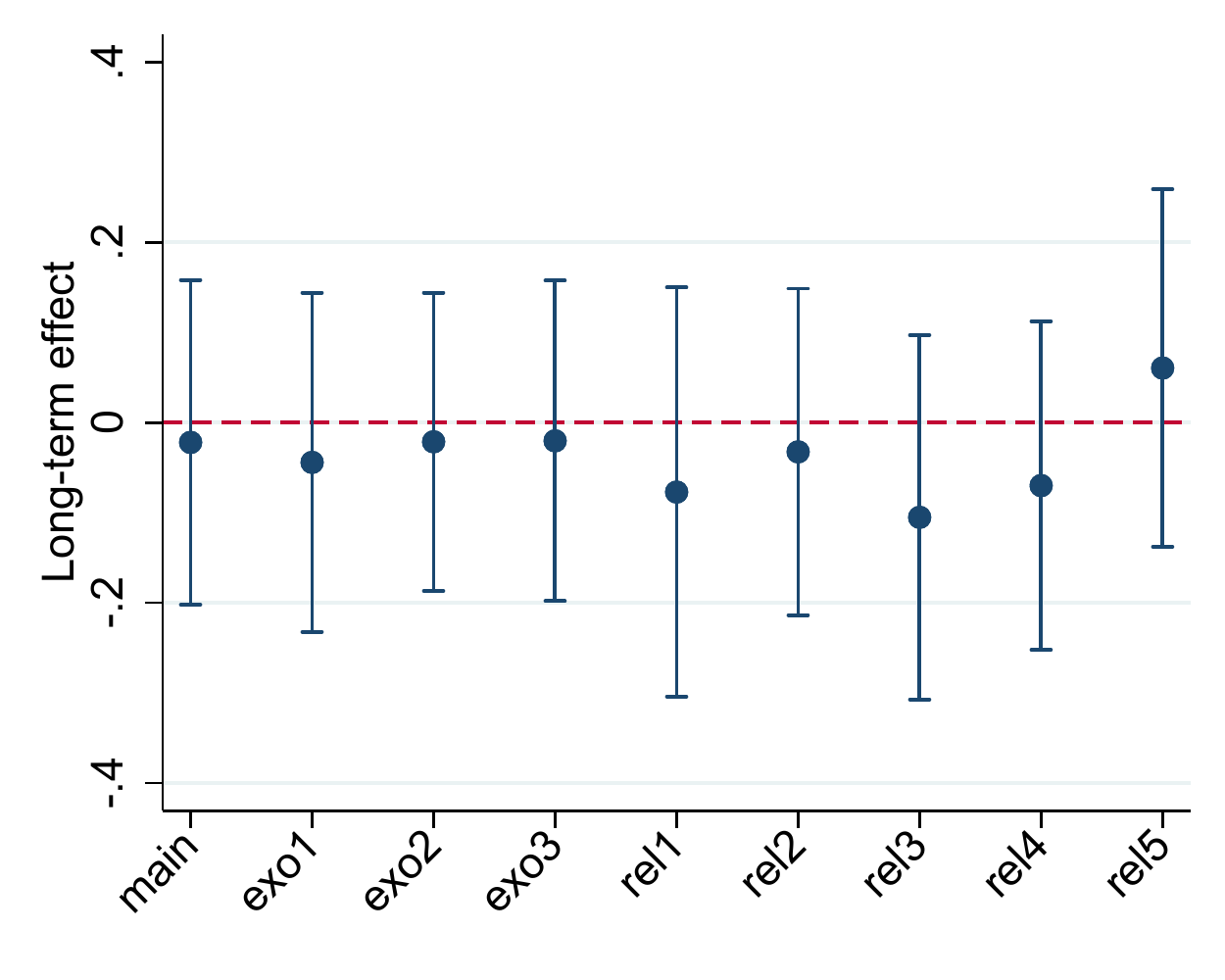}
                \end{subfigure}
                \vspace{10pt}    
                \newline
                \begin{subfigure}[b]{0.49\textwidth}
                                \centering \caption*{Short measures} \subcaption*{Unemployment rate} 
                                \includegraphics[clip=true, trim={0cm 0cm 0cm 0cm},scale=0.50]{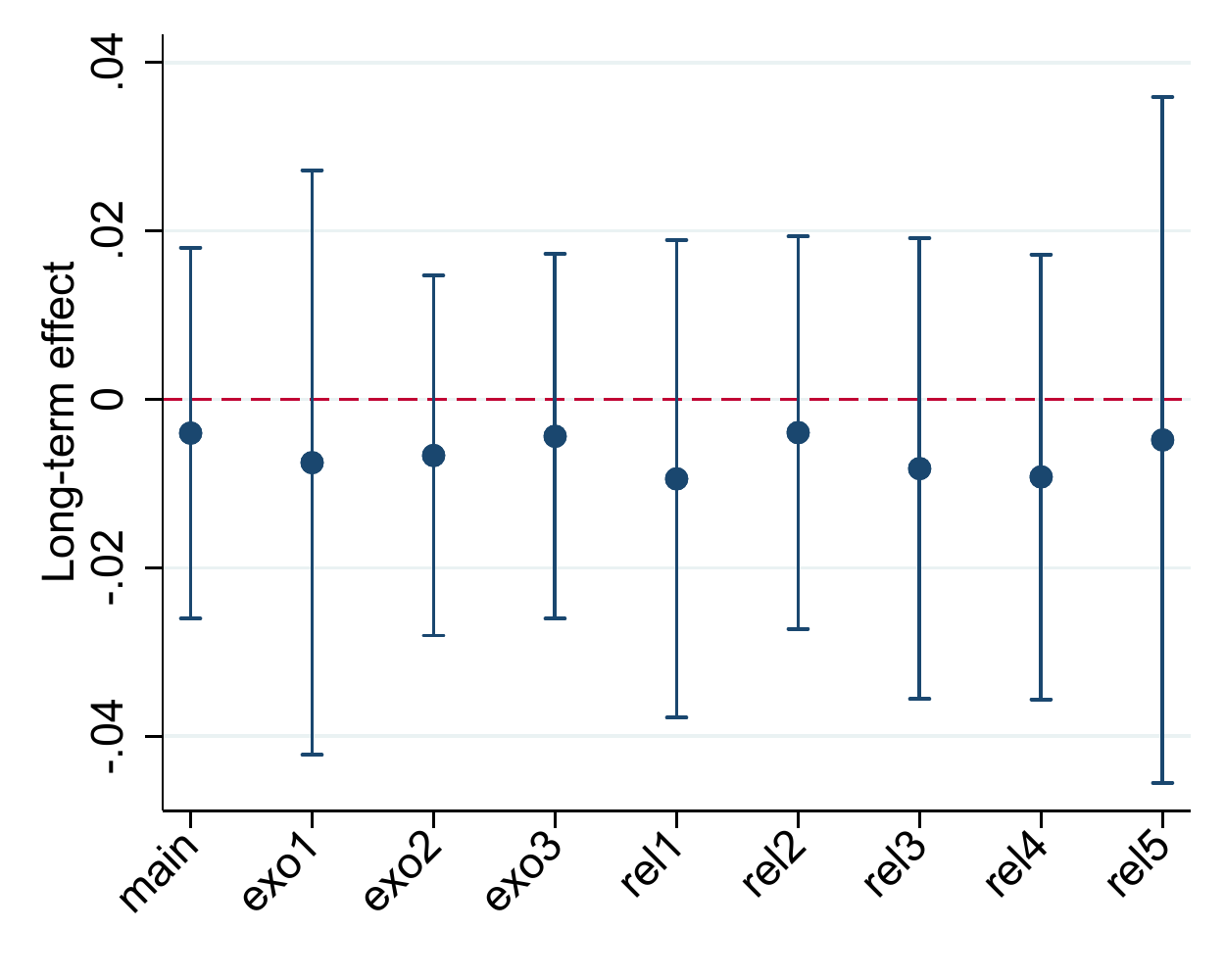}
                \end{subfigure}
                \vspace{10pt}    
                \begin{subfigure}[b]{0.49\textwidth}
                                \centering \caption*{Short measures}  \subcaption*{Unsubsidzied employment rate} 
                                \includegraphics[clip=true, trim={0cm 0cm 0cm 0cm},scale=0.50]{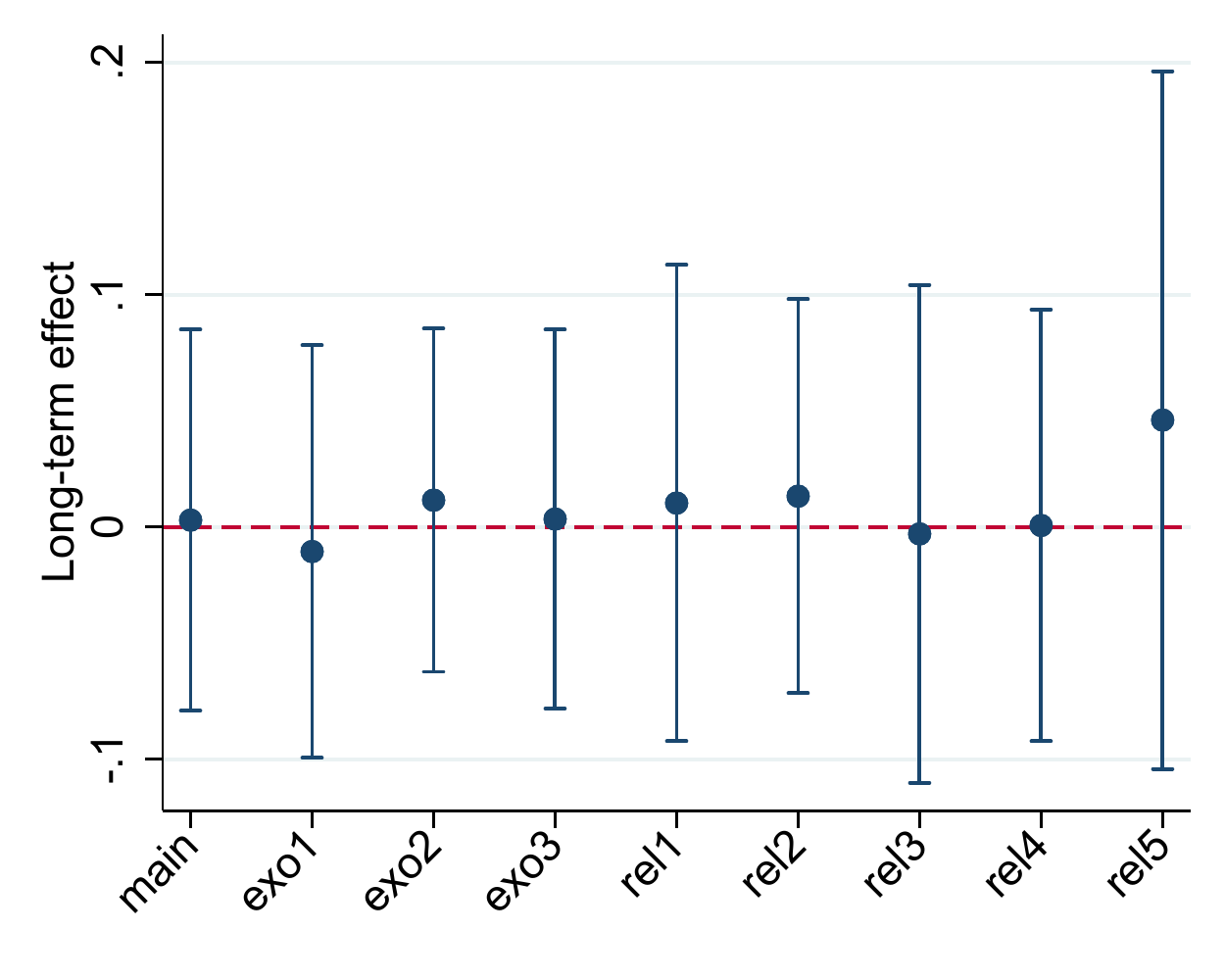}
                \end{subfigure}
                \vspace{10pt}
                \newline
                \begin{subfigure}[b]{0.49\textwidth}
                                \centering \caption*{Wage subsidies} \subcaption*{Unemployment rate} 
                                \includegraphics[clip=true, trim={0cm 0cm 0cm 0cm},scale=0.50]{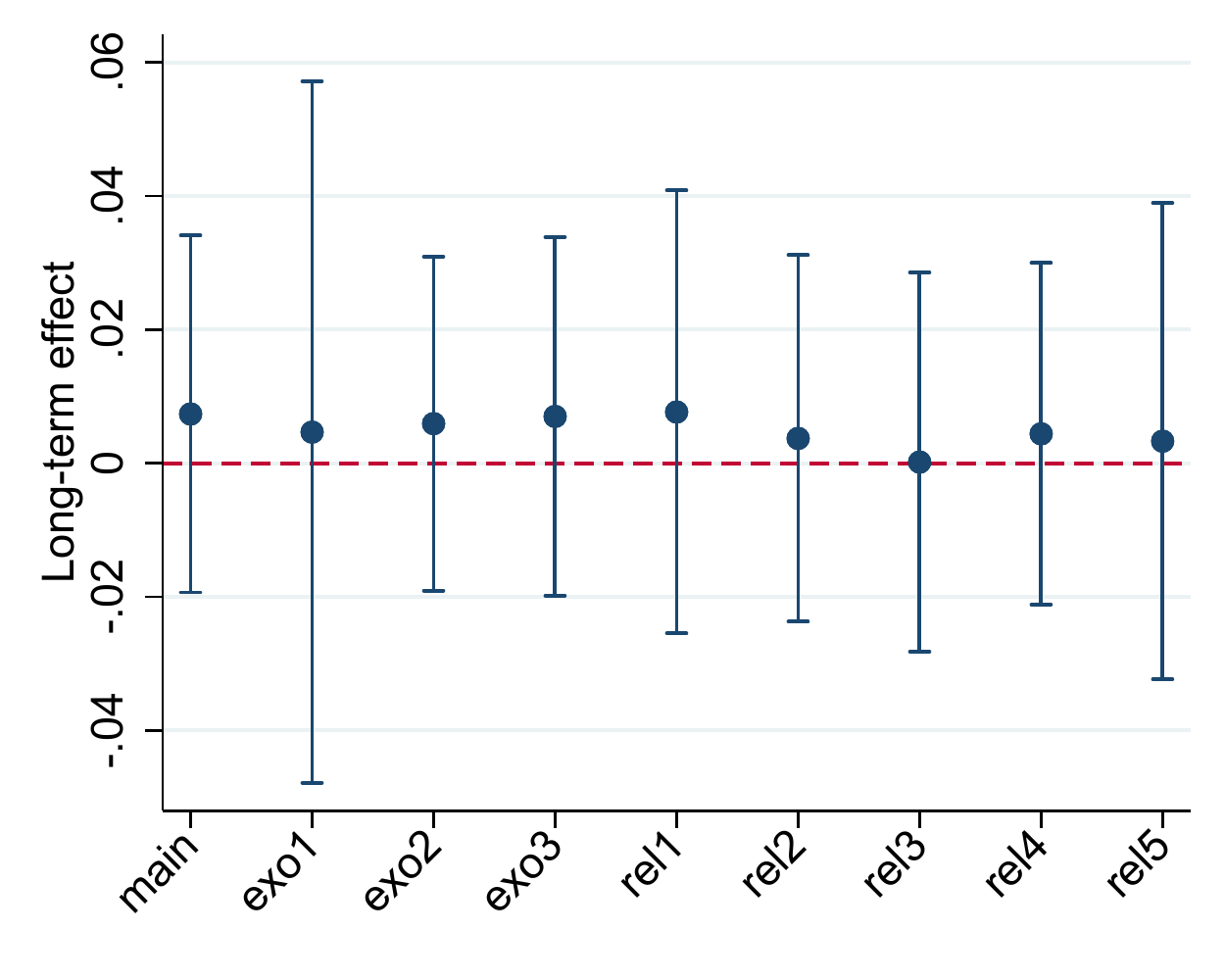}
                \end{subfigure}
                \vspace{10pt}    
                \begin{subfigure}[b]{0.49\textwidth}
                                \centering \caption*{Wage subsidies}  \subcaption*{Unsubsidzied employment rate} 
                                \includegraphics[clip=true, trim={0cm 0cm 0cm 0cm},scale=0.50]{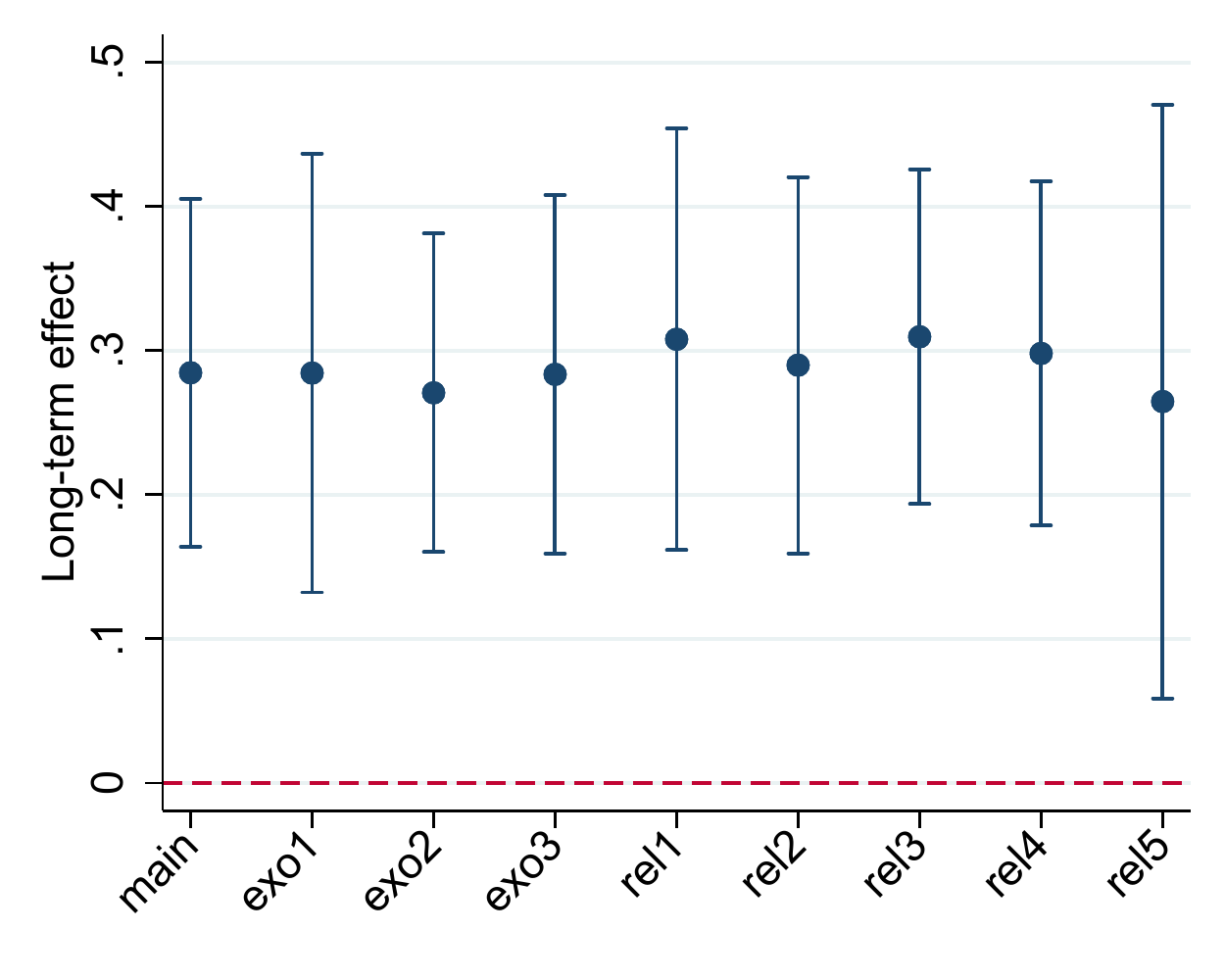}
                \end{subfigure}
                \vspace{10pt}
                \begin{minipage}{\textwidth}
                                \footnotesize \textit{Notes:} This graphs show the long-term effects and the corresponding 96\% CI of the three types of ALMP, i.e. training, short measures and wage subsidies on the unemployment rate and unsubsidized employment rate. The effects on the x-axis are obtained by separate regressions based on different sample restrictions. The effects are based on the ARDL model estimated by 2SLS. Program variables are included with 6 lags.  Standard errors obtained by a cross-sectional bootstrap (499 replications). 
                \end{minipage}
\end{figure}

\begin{figure}[H]
                \centering
                \caption{Long-term Effects for Different Subsamples (Relevance and Exogeneity Criteria) II \label{fig:lt_effects_rob_IV2}}
                \vspace{10pt}
                \begin{subfigure}[b]{0.49\textwidth}
                                \centering \caption*{Training} \subcaption*{Rate of welfare recipients} 
                                \includegraphics[clip=true, trim={0cm 0cm 0cm 0cm},scale=0.50]{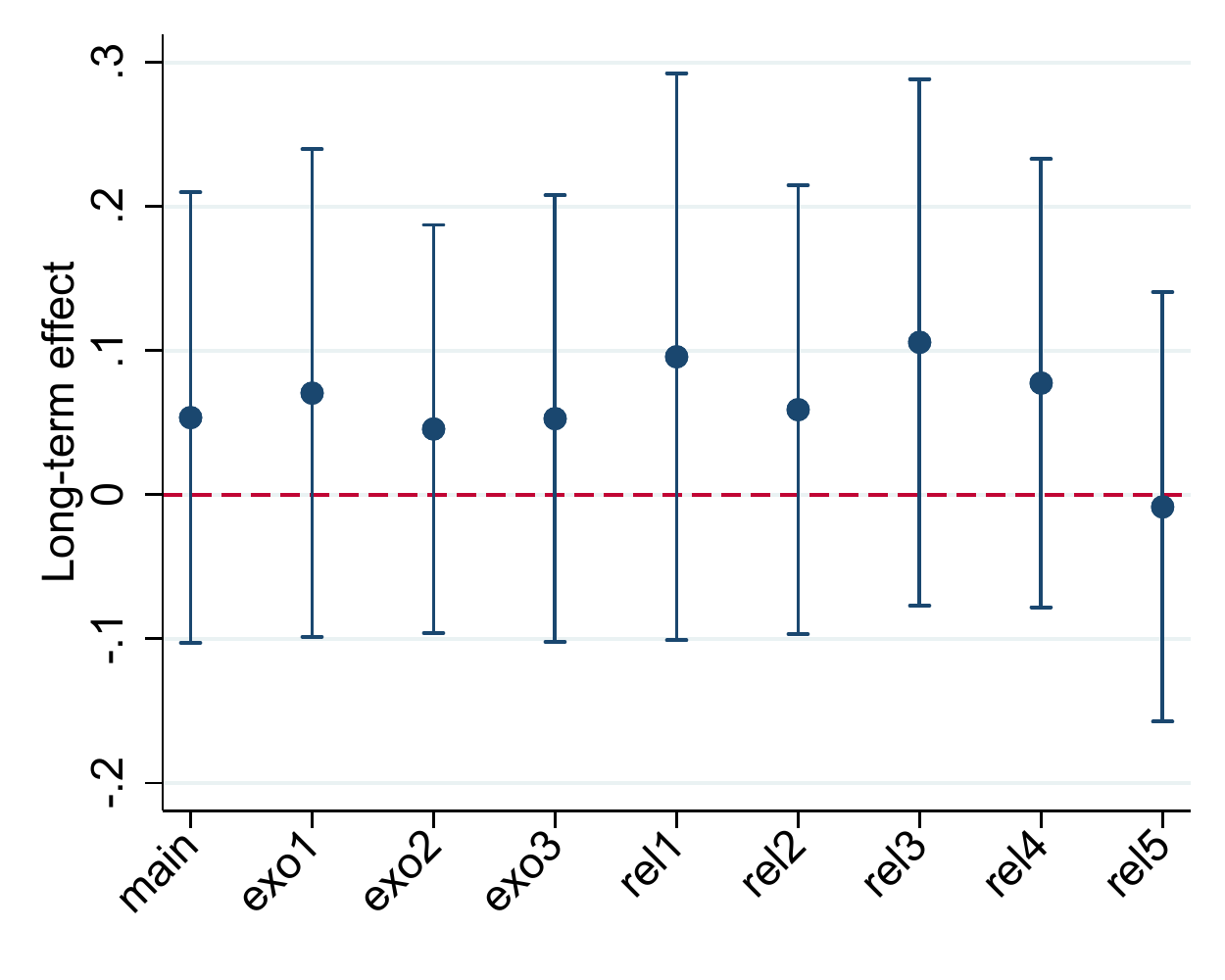}
                \end{subfigure}
                \begin{subfigure}[b]{0.49\textwidth}
                                \centering \caption*{Training} \subcaption*{Rate of employed workers on benefits} 
                                \includegraphics[clip=true, trim={0cm 0cm 0cm 0cm},scale=0.50]{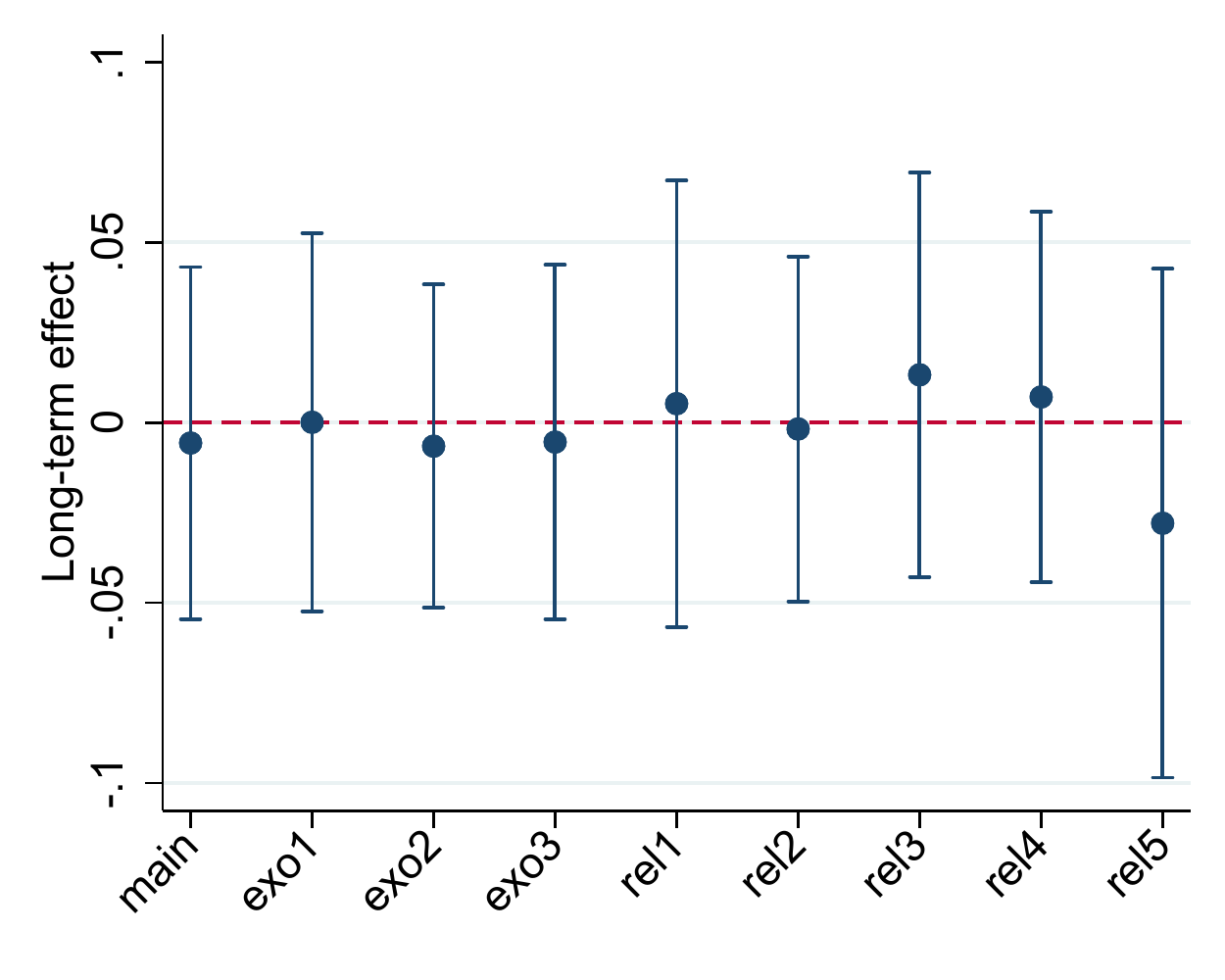}
                \end{subfigure}
                \vspace{10pt}    
                \newline
                \begin{subfigure}[b]{0.49\textwidth}
                                \centering \caption*{Short measures} \subcaption*{Rate of welfare recipients} 
                                \includegraphics[clip=true, trim={0cm 0cm 0cm 0cm},scale=0.50]{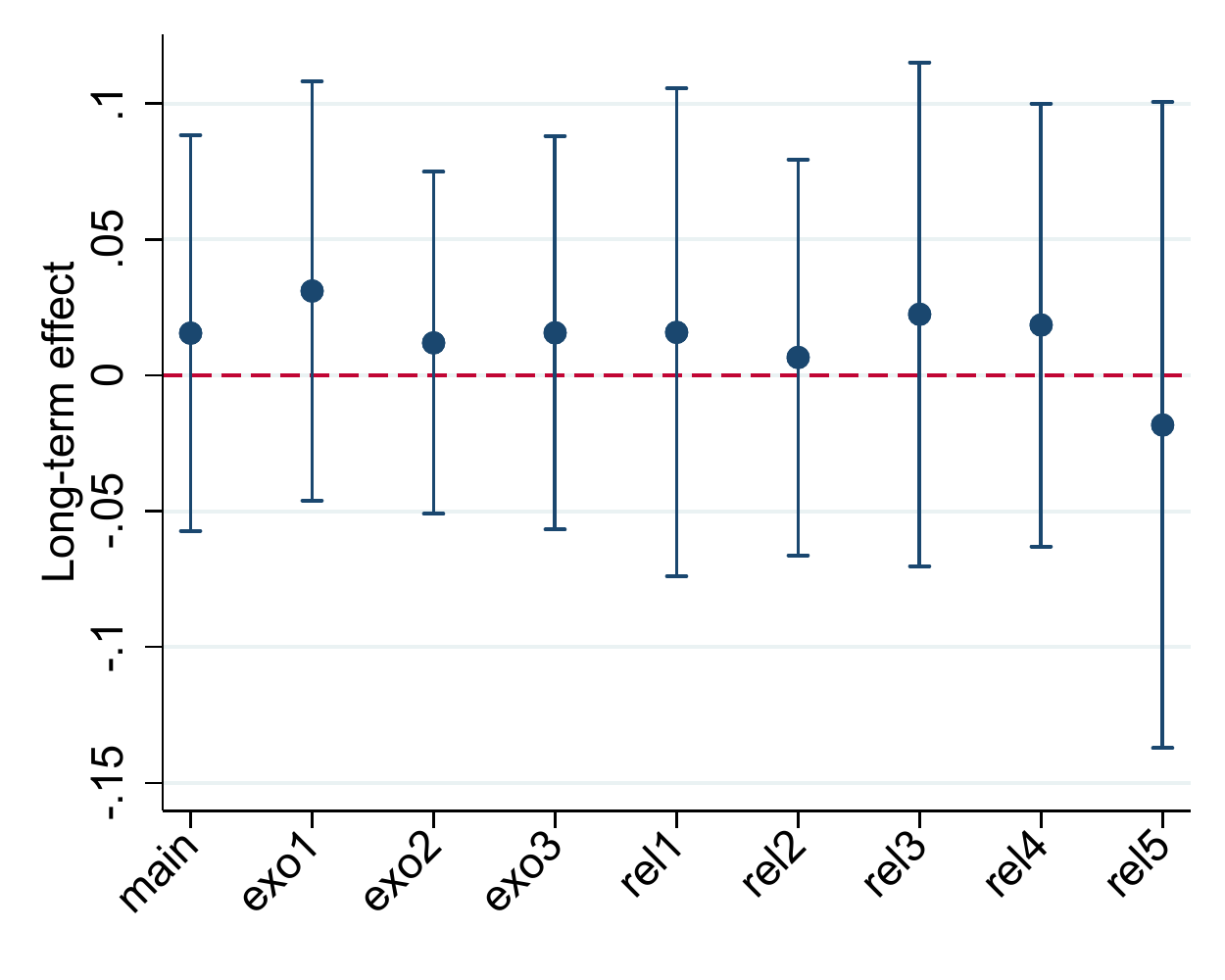}
                \end{subfigure}
                \vspace{10pt}    
                \begin{subfigure}[b]{0.49\textwidth}
                                \centering \caption*{Short measures}  \subcaption*{Rate of employed workers on benefits} 
                                \includegraphics[clip=true, trim={0cm 0cm 0cm 0cm},scale=0.50]{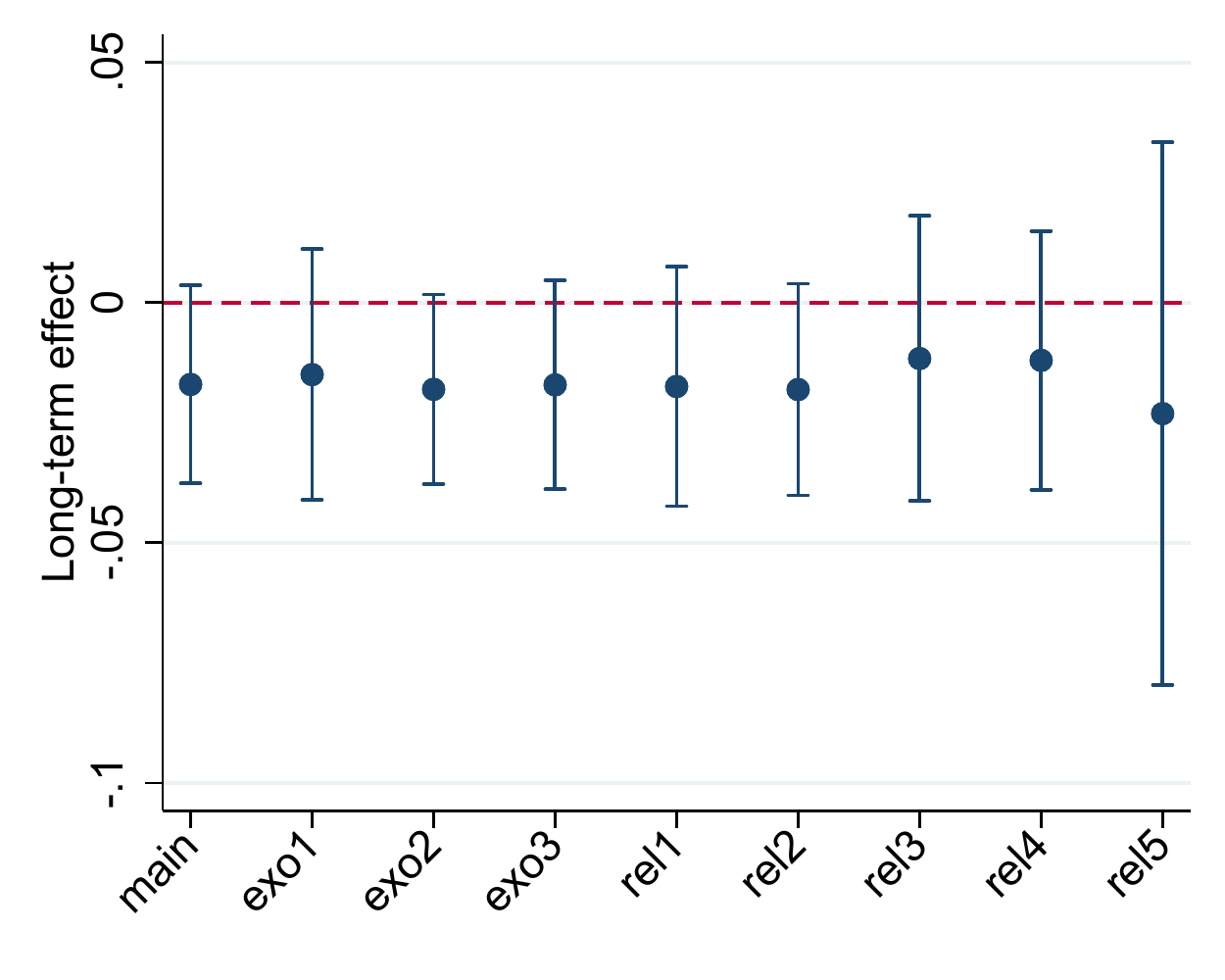}
                \end{subfigure}
                \vspace{10pt}
                \newline
                \begin{subfigure}[b]{0.49\textwidth}
                                \centering \caption*{Wage subsidies} \subcaption*{Rate of welfare recipients} 
                                \includegraphics[clip=true, trim={0cm 0cm 0cm 0cm},scale=0.50]{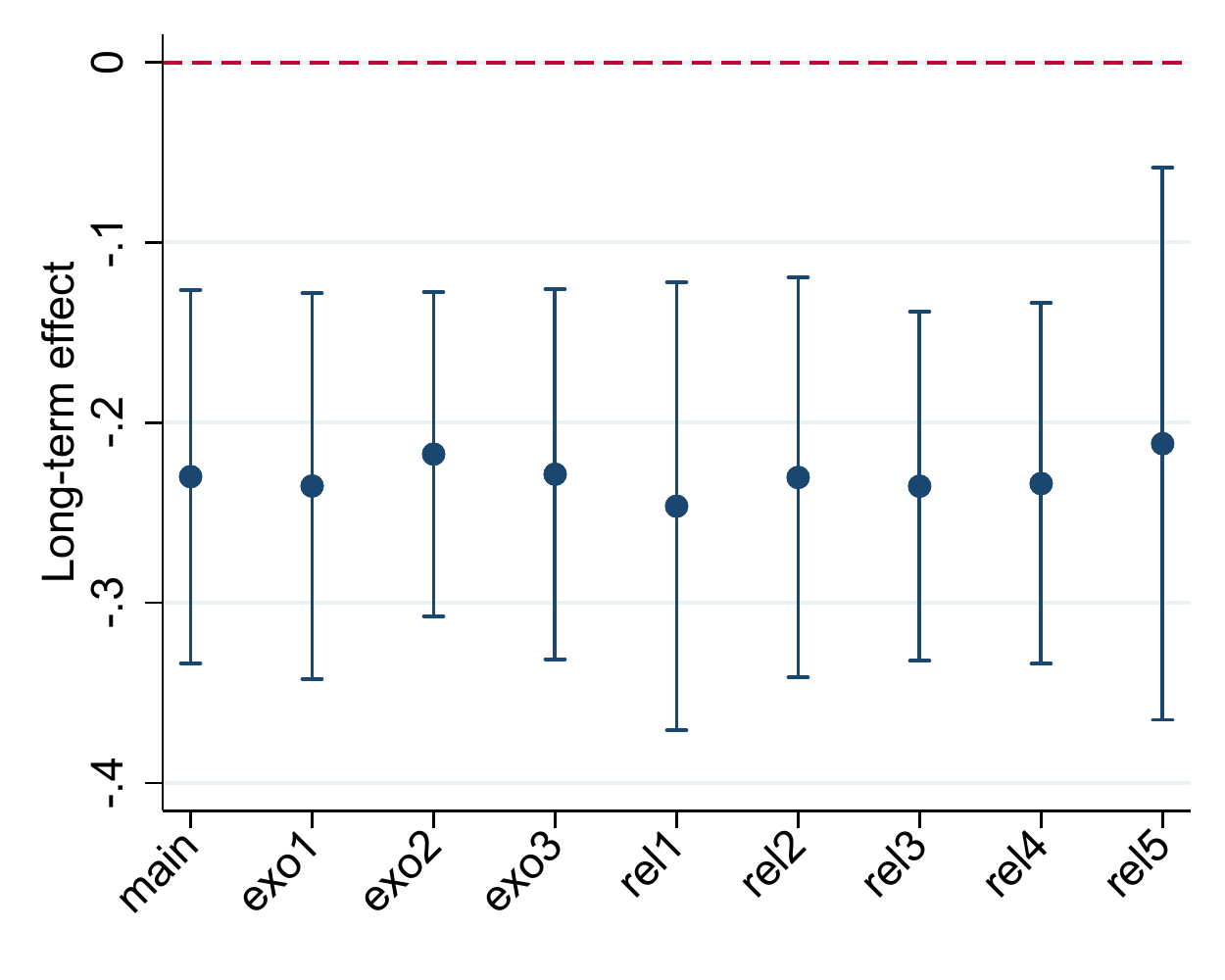}
                \end{subfigure}
                \vspace{10pt}    
                \begin{subfigure}[b]{0.49\textwidth}
                                \centering \caption*{Wage subsidies}  \subcaption*{Rate of employed workers on benefits} 
                                \includegraphics[clip=true, trim={0cm 0cm 0cm 0cm},scale=0.50]{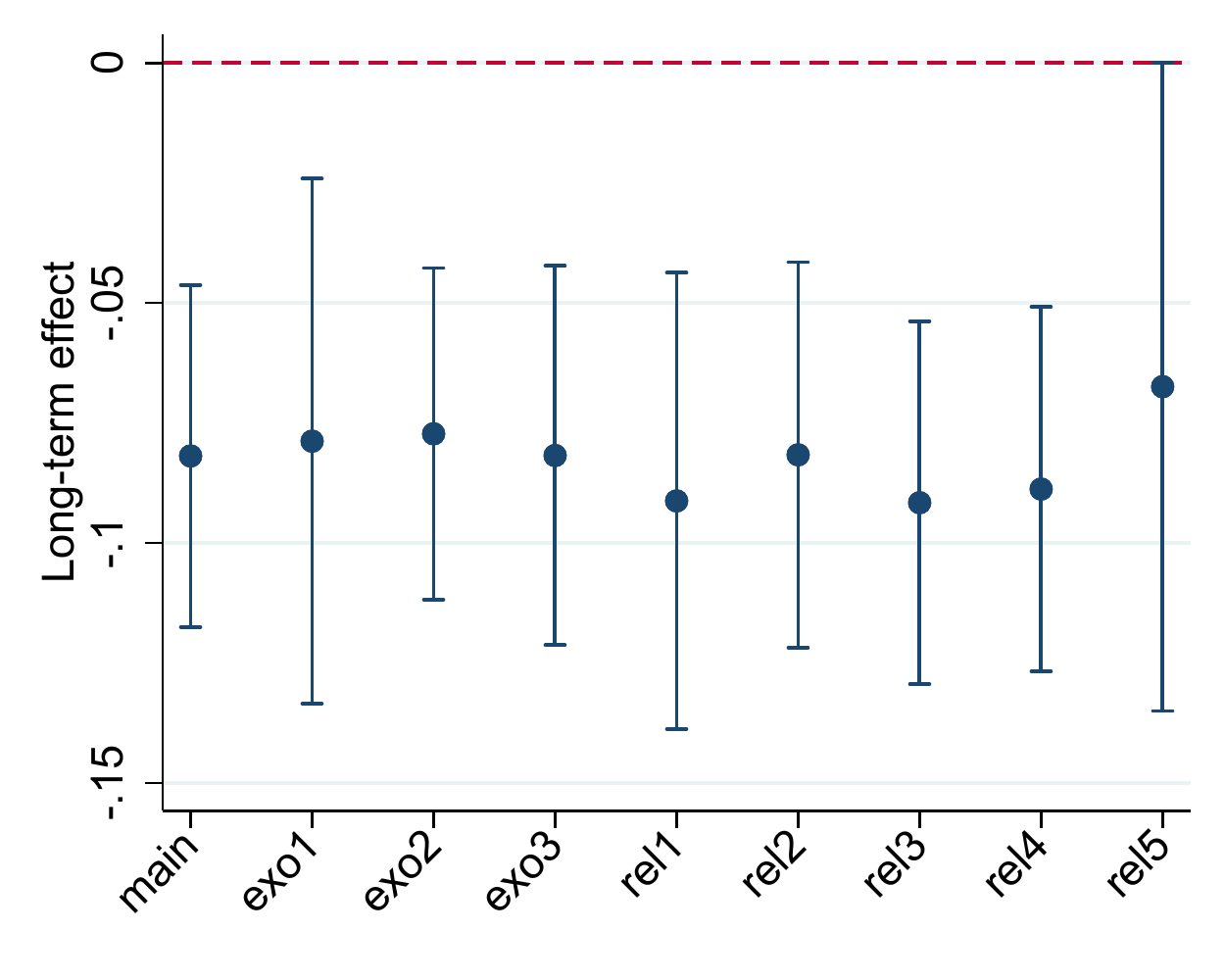}
                \end{subfigure}
                \vspace{10pt}
                \begin{minipage}{\textwidth}
                                \footnotesize \textit{Notes:} This graphs show the long-term effects and the corresponding 96\% CI of the three types of ALMP, i.e. training, short measures and wage subsidies on the unemployment rate and unsubsidized employment rate. The effects on the x-axis are obtained by separate regressions based on different sample restrictions. The effects are based on the ARDL model estimated by 2SLS. Program variables are included with 6 lags.  Standard errors obtained by a cross-sectional bootstrap (499 replications). 
                \end{minipage}
\end{figure}

\begin{landscape}
\begin{table}[H]
\centering
\caption{Effects Based on the Period 2010-2018}
\label{tab:post2010_alo_emp}
{\small
{
\def\sym#1{\ifmmode^{#1}\else\(^{#1}\)\fi}
\begin{tabular}{l| c c c | c c c | c c c | c c c}
\hline\hline
&\multicolumn{3}{c}{Unemployment rate} &\multicolumn{3}{c}{Unsub. employment rate} &\multicolumn{3}{c}{Rate on welfare} &\multicolumn{3}{c}{Rate of emp. on benefits} \\ 
&\multicolumn{1}{c}{(1)} &\multicolumn{1}{c}{(2)} &\multicolumn{1}{c}{(3)}&\multicolumn{1}{c}{(4)} &\multicolumn{1}{c}{(5)} &\multicolumn{1}{c}{(6)} 
&\multicolumn{1}{c}{(7)} &\multicolumn{1}{c}{(8)} &\multicolumn{1}{c}{(9)}&\multicolumn{1}{c}{(10)} &\multicolumn{1}{c}{(11)} &\multicolumn{1}{c}{(12)} \\ 
\\ 
\hline
            &      effect&          se&       p-val&      effect&          se&       p-val&      effect&          se&       p-val&      effect&          se&       p-val\\
Lagged dependent variable&       0.568&       0.239&       0.018&       0.662&       0.204&       0.001&       0.697&       0.060&       0.000&       0.633&       0.086&       0.000\\
Training(st)&       0.003&       0.070&       0.965&      -0.040&       0.202&       0.843&       0.019&       0.098&       0.850&       0.017&       0.072&       0.814\\
Training(lt)&      -0.019&       0.042&       0.657&      -0.006&       0.153&       0.967&       0.051&       0.249&       0.838&      -0.020&       0.052&       0.700\\
Short measures (st)&       0.010&       0.030&       0.741&      -0.024&       0.077&       0.760&       0.008&       0.031&       0.794&       0.005&       0.022&       0.825\\
Short measures (lt)&       0.002&       0.021&       0.919&      -0.020&       0.067&       0.760&       0.034&       0.113&       0.761&      -0.017&       0.022&       0.445\\
Wage subsidies (st)&      -0.039&       0.056&       0.490&       0.056&       0.176&       0.751&      -0.005&       0.126&       0.968&      -0.034&       0.091&       0.705\\
Wage subsidies (lt)&      -0.048&       0.026&       0.062&       0.152&       0.074&       0.039&      -0.098&       0.058&       0.092&      -0.022&       0.040&       0.571\\
\\
\hline
Quarter-Year FEs&      $\checkmark$ & & &  $\checkmark$ & &  &  $\checkmark$ & & &  $\checkmark$ & &     \\
Additional Controls&      $\checkmark$ & & &  $\checkmark$ & &   & $\checkmark$ & & &  $\checkmark$ & &     \\
Observations& 3680 & & &  3680 & & & 3680 & & &  3680 & &       \\
LLM & 115 & & &  115 & & & 115 & & &  115 & &       \\
\hline\hline
\end{tabular}
}
}
\centering
     \begin{minipage}{20cm}
          \vspace{6pt}
          \footnotesize \textit{Notes:}  This table shows the effect on the lagged dependent variable, the short-term (st) and long-term (lt) effects of the three types of ALMP, i.e. training, short measures (SM) and wage subsdies (wagesub) on the unemployment rate (columns 1-3), unsubsidized employment rate (columns 4-6), rate of welfare recipients (clumns 7-9) and rate of employed workers on benefits (columns 10-12). The model is estimated for the time period 2010 to 2018. Standard errors (se)  are obtained by a cross-sectional bootstrap (499 replications).
     \end{minipage}
\end{table}
\end{landscape}


\end{document}